%% file: thesis.tex
\documentclass[12pt]{report}
\usepackage{manthesis,axodraw,epsfig,amssymb}
\usepackage[]{graphicx}
\usepackage{bm}
\usepackage{amsmath}
\usepackage{txfonts}
\usepackage{cite}

\newcommand{\boldr}{{\bf x}}
\newcommand{\boldrho}{{\bf y}}

\begin{document}
\title{The Renormalisation Group and Applications in Few-Body Systems}
\author{Thomas Barford}
\supervisor{Dr. Michael Birse}
\department{Department of Physics and Astronomy}
\submissiondate{2004}
\degree{Doctor of Philosophy}
\faculty{Faculty of Science and Engineering}
\institution{University of Manchester}
\maketitle{yes}{yes}{
This thesis is about effective field theories. We study
the distorted wave renormalisation group (DWRG),
a tool for constructing power-counting schemes in systems
where some diagrams must be summed to all orders.
We solve the DWRG equation for a variety of long-range
potentials including the Coulomb, Yukawa and inverse square.
We derive established results in the case of the Yukawa and
Coulomb potentials and new results
in the case of the inverse square potential. We generalise
the DWRG to systems of three bodies and use it to find
the power-counting for the three-body force in the KSW EFT.
This power-counting corresponds to perturbations in the
renormalisation group about a limit-cycle. We also derive
equations for the LO and NLO KSW EFT distorted waves for three bosons
and for three nucleons in the ${}^3S_1$ channel. The physical
results in the nuclear system agree well, once the LO three-body
force is determined, with predictions of more sophisticated potential
models.}
\preface{Declaration}
\input{misc/declaration.tex}
\cnotice

\preface{Acknowledgements}
I would like to thank my supervisor Mike Birse for
for all the help and inspiration he has offered me over my
three years. Mike has taught me the difference between
thinking like a physicist, as I should, rather than
like a mathematician, as I tend to!

I would also like to thank everyone else in the University of
Manchester theoretical physics department for making an environment
in which the mind of physicist can be moulded. In particular, I would
like to thank my office-mates Gav and Rob for many hours of table
tennis and other procrastinations and also for their friendship.

I must also thank my mum who started me on the particle physics warpath
at an early age with her continuing love of popular science and the
rest of my family for putting up with me.

Finally, I must thank my wife, Clare; for all her patience as I
fiddled with the content of my thesis and failed to be finished when
I said I would; and for all her love.

\preface{Autobiographical Note}
The author, Thomas Barford was born in Northampton in October
1977. He stayed in that grand town of rugby and Carlsberg until
starting a Masters degree in Mathematics and Physics at the University
of Warwick in 1996. Fours years and 273 hangovers later, having gained a first,
he started a PhD in the theoretical physics
at the University of Manchester. He has recently married a marvellous
girl called Clare and they are very happy.

\preface{Publications}

T.~Barford and M.~C.~Birse,\\
``A renormalisation-group approach to two-body scattering with long-range
forces,''\\
AIP Conf.\ Proc.\  {\bf 603} (2001) 229
[arXiv:nucl-th/0108024].\\
\\
T.~Barford and M.~C.~Birse,\\
``A renormalisation group approach to two-body scattering in the presence  of
long-range forces,''\\
Phys.\ Rev.\ C {\bf 67} (2003) 064006
[arXiv:hep-ph/0206146].\\
\\
T.~Barford and M.~C.~Birse,\\
``An effective field theory for the three-body system,''\\
Few Body Syst.\ Suppl.\  {\bf 14} (2003) 123
[arXiv:nucl-th/0210084].\\
\\

\preface{Dedication}
To Bella.

\chapter{Introduction}
From an historical perspective,
the hierarchy of scales to be found in the universe
has controlled the development of physics
since the middle ages. Time and again the
theories of the past have been found to be only effective
theories that approximate the physical world.

The first great post-Renaissance physics theory,
Newtonian mechanics, is still used extensively in
many areas, yet it has been `known' for a long time that
it is only an approximation. The advent of relativistic
mechanics brought fresh insight into the nature
of the universe as well as explaining some mysterious
behaviour in the orbit of the planet Mercury, yet its
use in explaining the path of a cricket ball struck through
mid-wicket has never been advocated. Relativistic mechanics
is a far more complicated beast than its `lesser' cousin,
in many cases the pay-off for the extra work involved in
a calculation is simply not worth it. The difference in the
path of that cricket ball as explained by Newtonian and
relativistic mechanics would be unmeasurable.

Even though Newtonian mechanics is not fundamental to the nature
of the universe as we know it now, it is still an
{\it effective theory}; it explains what we see
at speeds far below the speed of light, $c$. In quantifying this
statement we might say that the difference between any
classical or relativistic calculation, in which the
typical speed is of order $v$, is of order $v^2/c^2$. We may then
consider any problem, and knowing how accurately we
want the result, chose between a full relativistic calculation,
a simple classical calculation or maybe some middle ground in
which the classical result is corrected to some order
in $v^2/c^2$.

This kind of idea pervades physics today but nowhere is its
importance as pronounced as in particle physics. Although the
domain of particle physics may off-handedly be described as the
study of the very small, the range of energies that are scrutinised,
from the neutrino mass to the unification scale, far outstrips
any other area of physics.

The first great triumph of quantum mechanics,
the theory of the very small, was the calculation of the
spectra of the hydrogen atom. However, the need for corrections to this
result was soon apparent and there followed calculations for the
fine structure, Lamb shift and hyperfine structure, each of which
relied upon some previously unconsidered higher scale physics:
spin-orbit coupling, relativistic corrections, loop diagrams in
QED and dipole interactions. Despite all the corrections made
in this calculation there are yet more that can be made,
such as corrections due to the finite charge radius of the proton. 
An estimate of the magnitude of such a correction can be made in
a similar way to the estimate of the relativistic corrections
above. The corrections are expected to be of the order of
$r_p^2/a_0^2$ where $r_p\sim 10^{-16}$m is the charge radius of the proton
and $a_0=5.29\times 10^{-11}$m is the Bohr radius and an estimate of the
typical distance of a ground state electron from the proton. Hence we expect
a correction of the order $\Delta E\sim 10^{-9}E$. The key point here is
the seemingly innocuous term `of the order of'. What do we mean by `of the
order of'?

Arguments, like the one above, that involve using a separation of scales
and expressions like `of the order of' are implicitly assuming
`naturalness'. The assumption of naturalness in these arguments is
absolutely essential, without it nothing can be said about corrections
such as relativistic corrections. In a nutshell one might say that
the assumption of naturalness means that the coefficient, $b$,
of for example the relativistic correction $\delta=bv^2/c^2$, is
of order unity. That is, the number $b$, which is governed by the dynamics
and geometry of the problem, is not so large or small to
significantly change the size of the correction so that expressions
like `of the order of' apply.

Naturalness is not something that can necessarily be taken for granted,
and indeed some fluke of the parameters may
change the face of the problem, resulting in unexpected effects
such as resonances.

It is only comparatively recently that physicists have sought to
use the separations of scales in particle physics quantitatively
to produce what have become known as effective field theories
(EFTs) \cite{eck,bkm95,uvkrev,eft1,eft2,border}.
In an EFT we seek to write down a low-energy equivalent
of some `true' higher energy theory. The concept was first
introduced by Weinberg \cite{weinchpt} in an attempt to use
the near chiral symmetry of quantum chromodynamics (QCD)
and the resulting separation of
scales between the pion mass, $m_\pi$, and all other QCD scales to
construct a low energy theory equivalent to QCD. Since
this initial spark this idea has grown to become known as 
chiral perturbation theory (ChPT).

ChPT is defined by an effective Lagrangian that contains only those
degrees of freedom explicit at the energies of interest,
$Q\sim m_\pi$, i.e. pion fields and depending upon the particular
problem, nucleon fields, hyper-nucleon fields etc. The fields are
coupled by effective vertexes that contain the effects of all
higher energy physics, of order $\Lambda_0$,
not included in the Lagrangian, which is said to
be `integrated out'. To include all of the integrated-out physics, the
Lagrangian must include all terms not forbidden by the symmetries of
the higher energy physics, invariably resulting in an infinite number
of terms. To handle such a formidable Lagrangian the terms must be
organised according to some `power-counting' that assigns each
term an order in an expansion in powers of $Q/\Lambda_0$. Any
observable can be calculated to arbitrary accuracy
by using the power-counting to include all the necessary terms.

In principle the effective couplings in the ChPT Lagrangian
can be found by solving QCD, however, in practice they are determined
by matching theory to a few experimental values, making the theory predictive
in other areas.

In particle physics,
the expression EFT was once synonymous with ChPT but it has since
become an umbrella term for any number of theories in particle and
atomic physics that subscribe to the same philosophy. Although EFTs are
predominantly found in nuclear physics, where QCD is
non-perturbative and hence almost impossible, they have been applied
to areas as diverse as quantum gravity and string theory. 

In studying nuclear systems at even lower energies than the pion mass,
$Q\ll m_\pi$,
we may integrate out the pionic degrees of freedom and
deal with the pionless EFT in which the only degrees of freedom are
the nucleon fields themselves. In this theory the underlying
physics is that of the pion and the higher energy scale for the
theory is the pion mass.
The non-relativistic pionless EFT Lagrangian for nucleons of mass $M$
is given by,
\begin{eqnarray}
&&\displaystyle{
{\cal L}=N^{\dagger}\left(i\partial_0+\frac{\nabla^2}{2M}\right)N
+\left[\frac{C_0}{4}|N|^4
+\frac{C_2}{4}\left(N^\dagger(\overrightarrow\nabla
-\overleftarrow\nabla)^2N^\dagger\right)N^2+{\rm H.c.})
+\ldots\right]}\nonumber\\
&&\displaystyle{\qquad\qquad+\left[\frac{D_0}{36}|N|^6+\frac{D_2}{36}
\left(N^\dagger(\overrightarrow\nabla-
\overleftarrow\nabla)^2N^\dagger\right)N^\dagger N^3+{\rm H.c.})
+\ldots\right].}
\end{eqnarray}
It contains four point couplings denoted by $C_{2n}$ where $2n$ is
the order of the field derivative at the vertex. Similarly,
since they are not forbidden it contains six point couplings, denoted by
$D_{2n}$, and eight point vertexes, ten point vertexes, etc.
Relativistic corrections may be included by introducing higher order
derivatives in the kinetic term \cite{uvk}.

Ignoring the six and higher
point vertexes, which are not relevant to a two nucleon problem, 
we may immediately construct a power-counting to organise the terms
in the effective Lagrangian using naive dimensional analysis,
which inspired the power-counting in ChPT. 
The four point vertex couplings, $C_{2n}$, have dimension $-2-2n$. Since
they describe physics that occurs at the order of the pion mass we
may expect,
\begin{equation}
C_{2n}\sim\frac{1}{Mm_\pi^{2n+1}}.
\end{equation}
In addition each loop between the vertexes contributes
a term $\sim MQ/4\pi$. The order of
any diagram involving any number of interactions can be evaluated
by simply combining $n$ vertex terms with the $n-1$ loops in between.
A diagram in which the $i^{\rm th}$ vertex contains $v_i$ derivatives
will occur at order $(Q/m_\pi)^d$ where,
\begin{equation}
d=n-1+\sum_{i=1}^n v_i.
\end{equation}
This power-counting is known as the Weinberg scheme \cite{wein1}.
Unfortunately, when this organisation of the terms is used to
describe the two-nucleon system it fails. The problem is the
implicit assumption of naturalness.

Empirically, we observe in the two nucleon system a
low-energy bound state, the deuteron, which has binding momentum
$\gamma_d=45.7$Mev$\ll m_\pi$. The existence of this and the
low energy resonances in the other isospin channels alerts
us at once to some fine tuning of the parameters in the EFT.
This fine tuning means that the naive dimensional analysis that
results in the Weinberg scheme is no longer applicable. Some
other power-counting must be found if we are to
handle the infinite number of terms in the effective Lagrangian
and produce the low-energy resonances \cite{wein1,ksw,uvk,geg,egm,bcp}.

A new power-counting that resolved the problem was first introduced
by Kaplan, Savage and Wise (KSW) \cite{ksw} using the power divergence
subtraction scheme (PDS) but found independently by van Kolck \cite{uvk}.
The PDS scheme introduces a new scale, $\mu$,
which may be chosen to produce the fine tuning effects seen in the
two nucleon system. Using the PDS scheme the scaling of the coefficients
in the effective Lagrangian may be given as \cite{ksw}
\begin{equation}
C_{2n}\sim\frac{1}{M\mu^{n+1}m_\pi^n}.
\end{equation}
By choosing $\mu\sim Q$ we introduce the new low-energy scale
required. Subsequently the term $C_0$ occurs at order
$Q^{-1}$ and so all diagrams containing only the $C_0$ vertex must be
summed to all orders. All other interactions scale with
the negative powers of $m_\pi$ and so may be dealt with perturbatively
provided each diagram is dressed with $C_0$ interaction bubbles.
(Since an arbitrary number of $C_0$ interactions before or after the
diagram does not change its order.) The subsequent power-counting is given by,
\begin{equation}
d=-2+\sum_{i=1}^n v_i.
\end{equation}
This KSW power-counting is able to produce the low-energy resonances 
of order $\mu$ seen in nuclear systems.

The origin of these two different power-counting schemes
can, in one way, be understood
by using the renormalisation group (RG) \cite{bmr}. The effective
Lagrangian contains, by its very definition, an infinite
number of non-renormalisable couplings. These result in divergent
loop integrals which must be regularised, with a sharp
momentum cut-off, $\Lambda$, for example, and then renormalised
by absorbing all cut-off dependence into the effective couplings.
In the context of EFTs the cut-off, $\Lambda$, is far more than a UV
regulator, it `floats' between the low-energy physics and the high-energy
physics and allows us to control the introduction of high-energy
physics and construct power-counting schemes.
After rescaling all low-energy scales in terms of $\Lambda$,
we are led to the concept of Wilson's continuous RG \cite{wilson}
initially conceived for use in condensed matter theory.
The RG describes how the couplings in the theory change
as $\Lambda$ is varied.

The use of a sharp momentum cut-off leads us to a very intuitive
idea of the RG flow. As $\Lambda$ gets smaller, more and more of the
high energy physics is `integrated out' and absorbed into the
couplings of the EFT. Finally as $\Lambda$ reaches zero, all high
energy physics is removed and all that remains is the rescaled low-energy
physics, embodied in an infra-red fixed point of the RG. By
perturbing about the IR fixed point with powers of $\Lambda$ we
introduce dimensioned constants of integration that must scale
with the high-energy scale $m_\pi$. In this way we
can re-introduce high-energy physics into the couplings and create
a power-counting scheme associated with that fixed point.

In chapter \ref{Ch1} we will study the RG for the two-body
pionless EFT and reproduce the results of Birse {\it et al} \cite{bmr}
that reveal two fixed point solutions, a so-called trivial fixed point,
which is associated with the Weinberg power-counting scheme and a
non-trivial fixed point, which is associated with the KSW power-counting
scheme. The constants of integration in the energy dependent
perturbations around the
trivial fixed point are in one-to-one correspondence with the terms
in the Born expansion. Those around the non-trivial fixed point are
in one-to-one correspondence with the effective range expansion \cite{ere}.

The two-nucleon pionless EFT is a simple example that demonstrates
the use of the RG in determining power-counting schemes.
However, it is clearly not applicable for the interaction of two
protons because of their electromagnetic interaction.
Photon exchange between two protons has the characteristic energy scale,
\begin{equation}
\kappa=\frac{\alpha M}{2}=3.42\text{MeV},
\end{equation}
where $\alpha$ is the fine structure constant and $M$ is the
mass of the proton. In an EFT for protons below the pion mass
$\kappa$ must be considered a low energy scale. All single photon exchange
diagrams must, therefore, be included to all orders.
However, as with the KSW power-counting, summing some
diagrams to all orders and dressing others with `Coulomb bubbles'
may completely change the power-counting.

In chapter 3 we will introduce the
distorted wave renormalisation group (DWRG) \cite{tbmb} and use it
to study, as a first example, proton-proton scattering.
The DWRG allows us to sum some low energy physics and
absorb it into the fixed points of the RG. In particular we study a
system of two particles interacting via some known long-range potential,
equivalent to the sum of all known non-perturbative diagrams in the EFT,
and a short-range potential, equivalent to all the shorter range
interactions. By working in the basis of the distorted waves (DWs) of
the long-range potential we show how to separate the effects of the
long and short range potentials and then to construct a power-counting
for the shorter range interactions.

The long range potential for the first example in chapter 3
will be the Coulomb potential. We will again find two fixed points,
a trivial and a non-trivial associated with Weinberg and KSW -like
schemes respectively. The integration constants in the perturbative series
about these fixed points are found to be in a one-to-one correspondence
with the terms in the distorted wave Born \cite{newton} and Coulomb modified
effective range \cite{bethe,j&b,kongr} expansions.

In the remainder of chapter 3 we will consider
two more examples. Firstly, the repulsive inverse square
potential, which is important to the extension of the pionless
KSW EFT to three bodies and study of the power-counting in higher
partial waves. This example is also interesting because it produces novel
power-counting schemes that are very different to the schemes
that are seen in the Coulomb DWRG and the ordinary RG.

Secondly, we shall consider a general example of
``well-behaved'' potentials, which includes among many others
the Yukawa potential that may be used to include one pion
exchange (OPE) in a nucleon-nucleon EFT. The solution of this DWRG has much in
common with the solution of the Coulomb DWRG. This general analysis
allows us to construct a general method for solving DWRG equations
that will prove extremely useful in later applications. We
conclude the chapter with the discussion of how to introduce
explicit pions into an EFT for two nucleons.

After establishing the DWRG we shall look towards using it
in the KSW EFT for three bodies. As outlined above, the
KSW EFT is appropriate for systems with a low energy bound state
or resonance that constitutes a new low energy scale in the
problem. This low energy scale is characterised by an unusually
large scattering length $1/a\sim Q\ll\Lambda_0$. Since this
is a low energy scale its effects must be summed to all orders
\cite{bhvk,hm3brg,brgh} and results in our need to use the DWRG. 

Efimov \cite{efimov} studied systems of three bodies with pairwise
interactions characterised by a large scattering length, $a$, and
zero effective range. He showed that these systems are
similar to a two dimensional problem with an inverse square potential.
This similarity occurs when the characteristic distance between
the three particles $R\ll a$ and the problem essentially becomes
scale free. In the limit of $a\rightarrow\infty$ the system is
described exactly by an inverse square potential.
The strength of the inverse square potential is determined
purely by the statistics in the system. 
In the case of three s-wave Bosons and three s-wave nucleons
in the ${}^3S_1$ channel the potential is attractive and singular.
The singular nature of these potentials has a couple of interesting
effects. Firstly, the Thomas effect \cite{thomas}: the system will be no
ground state, something first noted by Thomas as long ago as 1935.
Secondly, in the limit of infinite scattering length, there will
be an accumulation of geometrically spaced bound states at zero energy,
known as Efimov states \cite{efimov}.

As the first step towards understanding the KSW EFT for three
bodies, in chapter 4 we will look at the DWRG for
the attractive inverse square potential. The singular nature of this
potential means that without special measures the Hamiltonian is
not self-adjoint. To construct a complete set of DWs we must construct
a self-adjoint extension to the Hamiltonian \cite{case,meetz,perelpop}
equivalent to defining some boundary condition near the origin.
The connection to the three-body KSW EFT means that the singular nature of this
potential has once more come under scrutiny \cite{lcbc,camblong,splc},
in particular with respect to understanding the link between the
required self-adjoint extension and the three-body force.

The solution of the DWRG equation for the singular inverse square potential
reveals that the self-adjoint extension is in fact equivalent to
a marginal\footnote{The perturbations around fixed points in the RG
are described as either stable, scaling with positive powers of $\Lambda$,
unstable, scaling with negative powers of $\Lambda$, or marginal
which do not scale with any power of $\Lambda$.} perturbation in
the DWRG. The general method for solving the DWRG outlined at the end of
chapter 3, augmented with some special considerations for the bound states,
will reveal that the RG flow is controlled by limit-cycle solutions
that evolve logarithmically in $\Lambda$.

In chapter 5 we will look at the extension of the DWRG to
three body forces (3BDWRG). The 3BDWRG equation is more complex than the
standard DWRG because of the multi-channelled nature of
the system. The 3BDWRG equation has to describe the coupling of the
three-body force to each of the channels. Fortunately the
lessons learnt in chapters 3 and 4 allow us to find solutions.

We will demonstrate the solution of the 3BDWRG in an example
of ``well-behaved'' pairwise forces and briefly show how the
fixed points correspond to Weinberg and an equivalent KSW counting
for three body forces. More interestingly we will look to the
derivation of the power-counting for the three-body forces in the KSW
EFT.

The nature of the three-body force in this system
is still open to debate. The similarity to the singular inverse square
potential is clear and the need for a self-adjoint extension is
almost universally supported \cite{lcbc,splc}. It was hoped by some that
effective range corrections of the two-body force would resolve the
singular behaviour, however, results in this direction have failed to
match experimental data \cite{gabbiani}.

Bedaque {\it et al} \cite{bhvk,brgh}
provide arguments in which they explicitly use a three-body force to
define a self-adjoint extension and then continue to construct a
power-counting for it. Phillips and Afnan \cite{phill3b} have shown how
the inclusion of one piece of three-body data is enough to define all
the physical variables without explicitly using a three-body force.
These related approaches give results that are consistent with each
other. Although the latter do not specify the actual nature of the
three-body force it is clear that the inclusion of a piece of three-body
data defines a self-adjoint extension of the Hamiltonian and is therefore
equivalent to the leading-order (LO)
three-body force given by Bedaque {\it et al}.

A third view is given by
Gegalia and Blankleider\cite{geg3b} who argue that a unitarity
constraint is sufficient to define physical quantities with no need
for any three-body data at all. However, they have failed to produce any
results that can be used to check the validity of their approach.

Since three-body force terms are not forbidden by the
observed symmetries they must be included in the EFT Lagrangian. The
3BDWRG solution for the three-body force shows that the self-adjoint
extension defining term is equivalent to a LO marginal three-body
force and, consistent with the results of Bedaque {\it et al}
\cite{bhvk,brgh}, acts as limit-cycle. Our results also
support the power-counting suggested by Bedaque {\it et al}.

Having established the power-counting for the three-body force in the
pionless KSW EFT, in chapter 6 we derive expressions for the DWs in this
system at LO and next-to-leading order (NLO). The equations for the
DWs based upon the two-body interactions alone show the singular behaviour
anticipated by Efimov. The insertion of the LO three-body force is
most easily achieved in this equations by demanding some boundary condition
on the DWs close to the origin. Initially we will derive expressions for
a three Boson system but in order to produce physically interesting results
we will extend the equations to deal with a three nucleon system, in particular
a system of two neutrons and a proton which contains the two- and three-
body bound states, the deuteron and the triton.
The LO three-body force can be fixed by matching to the neutron-deuteron
scattering length. The EFT can then be used to construct the triton
wavefunction and neutron-deuteron scattering states below threshold.

\chapter{The Renormalisation Group for Short Range Forces}
\label{Ch1}
\input{ch1/chapter1.tex}

\chapter{The Distorted Wave Renormalisation Group}
\label{Ch2}
\input{ch2/chapter2.tex}

\chapter{The DWRG for Singular Inverse Square Potential}
\label{Ch3}
\input{ch3/chapter3.tex}

\chapter{The DWRG for Three-Body Forces}
\label{Ch4}
\input{ch4/chapter4.tex}

\chapter{The KSW Effective Field Theory of Three Bodies}
\label{Ch5}
\input{ch5/chapter5.tex}

\chapter{Conclusions}
\label{Conc}
\input{misc/conc.tex}

\appendix
\chapter{Analyticity of $\hat J(\hat p,\hat\kappa)$}
\label{app:janalyticity}
\input{app/Appendix0.tex}

\chapter{DWRG Analysis for the Attractive Coulomb Potential}
\label{app:coulatt}
\input{app/AppCoulAtt.tex}

\chapter{The Faddeev Equations for Three Identical Bosons}
\label{app:faddeev}
\input{app/Appfaadeev.tex}

\chapter{Momentum Dependent Perturbations in the DWRG for Three-Bodies}
\label{app:Mompert}
\input{app/Appmompert.tex}

\chapter{Fourier Transforms}
\label{app:Fourier}
\input{app/Appfourier.tex}

\chapter{Zero Energy Wavefunction for the Three Body EFT}
\label{app:zero}
\input{app/Appnicecont.tex}

\label{references}

\input{misc/ref.tex}
\end{document}

%% file: misc/declaration.tex
No portion of the work referred to in this thesis has been submitted in support of an 
application for another degree or qualification at this or any other university or other
institution of learning.

%% file: ch1/chapter1.tex
\section{Introduction}
In this chapter we will introduce the renormalisation group method for
constructing EFTs \cite{bmr}. As already observed,
the two key ingredients of an EFT are a power-counting scheme and a
separation of scales. The first of these allows organisation of the
infinite number of terms that naturally occur in an effective field theory,
the second ensures that these terms may be truncated to achieve
the desired level of precision.

The separation occurs between the scale of the physics of interest, $Q$, 
and that of the underlying high-energy physics, $\Lambda_0$.
The existence of a separation of scales is
usually assumed with a particular physical system in mind. For example,
in nucleon-nucleon scattering at energies well below the pion mass, there
is a natural separation between the momenta of the incoming particles and
the lowest energy component of the strong interaction,
one pion exchange.
Utilisation of such a separation leads to 
what has become known, in nuclear physics, as a pionless EFT, in which the
only fields are those of the asymptotically free particles with no
exchange particles \cite{ksw,uvk,geg,modeft,crs}.

We will consider a system of two non-relativistic
identical particle scattering via an effective Lagrangian.
For weakly interacting systems the terms in the expansion
can be organised according to naive dimensional analysis, a term
proportional to $(Q/\Lambda_0)^d$ being counted as of order $d$.
This is known as Weinberg power-counting and is familiar from chiral
perturbation theory (ChPT) \cite{wein1}.  However, for strongly interacting systems
there can be new low-energy scales which are generated by non-perturbative
dynamics e.g. the very large s-wave scattering length in
nucleon-nucleon scattering. In such cases we need to resum certain terms
to all orders and arrive at a new power-counting scheme, KSW power-counting
\cite{ksw,uvk,geg}.
The origin and use of these schemes becomes quite transparent using the
renormalisation group method. The contents of this chapter largely follows the
work of Birse {\it et al} \cite{bmr,krthesis}

\section{The RG Equation}
In a Hamiltonian formulation, the effective Lagrangian
is written as a potential consisting of contact interactions, which
in coordinate space takes the form of $\delta$-functions and their
derivatives and in momentum space takes the form,
\begin{equation}\label{eq:Vbcs}
V({\bf k',k},p)=C_{00}+C_{200}k'^2+C_{020}k'^2+C_{110}{\bf k.k'}+
C_{002}p^2+\ldots,
\end{equation} 
where $p=\sqrt{ME}$ is the on-shell momentum corresponding to the
total energy $E$ in the centre of mass frame and $M$ is twice the
reduced mass. We shall consider s-wave scattering only at this
stage, then since $V$ must be independent of ${\bf k.k'}$,
$C_{110}=0$. 

Scattering variables are found by solving a Lippmann-Schwinger (LS)
equation. In particular we shall work with the reactance matrix, $K$,
which is related to the phaseshift, $\delta$ by,
\begin{equation}\label{eq:kphase}
\frac{1}{K(p,p,p)}=-\frac{Mp}{4\pi}\cot\delta,
\end{equation}
and solves the LS equation,
\begin{equation}\label{eq:unreglsk}
K(k',k,p)=V(k',k,p)+
\frac{M}{2\pi^2}\fint_0^\infty q^2dq
\frac{V(k',q,p)K(q,k,p)}{p^2-q^2},
\end{equation}
where the bar on the integral sign indicates a principal value prescription
for the pole on the real axis. Unitarity of the $S$-matrix is ensured if the
$K$-matrix is hermitian. The on-shell $T$ and $K$ matrices are related by
\begin{equation}\label{eq:TKrel}
\frac{1}{K(p,p,p)}=\frac{1}{T(p,p,p)}+\frac{iMp}{4\pi}.
\end{equation}

For short range potentials, it is well known that the inverse of the
$K$-matrix may be written in terms of the effective range expansion, \cite{ere}
\begin{equation}\label{eq:ere}
-\frac{4\pi}{M}\frac{1}{K(p,p,p)}=p\cot\delta=
-\frac{1}{a}+\frac12r_ep^2+\ldots
\end{equation}
where $a$ is called the scattering length and $r_e$ the effective
range. The effective range, as its name suggests,
is related in an indirect manner to the range of the potential \cite{newton}
(provided it is ``smooth'').

For contact interactions
the integral in Eqn. (\ref{eq:unreglsk}) is divergent. From the EFT point
of view
this is to be expected, the divergence occurs because we have allowed
high momentum modes to probe the EFT beyond its range of validity. The
solution is to remove the high momentum modes and incorporate their
effects into the effective potential, $V$.
Mathematically, we apply a cut-off, $\Lambda$,\footnote{
Any type of regularisation and
renormalisation should produce the same results, here we just take the 
simple case of a sharp cut-off.} to the integral then derive a
differential
equation for $V(k',k,p,\Lambda)$ based upon the constraint that $K$,
the observable, is independent of that cut-off.
Rather than acting as a UV regulator
the cut-off acts as a separation scale and `floats' between the
low-energy scales, $Q$, and the higher energy scales, $\Lambda_0$. This
leads us the concept of a Wilsonian renormalisation group (RG) \cite{wilson}.
The interesting limit is $\Lambda\rightarrow0$, in which all high
energy physics has been integrated out.

The regularised LS equation can be written in operator
form as,
\begin{equation}\label{eq:lskop}
K(p)=V(p,\Lambda)+V(p,\Lambda)G_0^{P}(p,\Lambda)K(p),
\end{equation}
where $G_0^{P}(p,\Lambda)$ indicates the free Green's
function with standing wave boundary conditions
and sharp cut-off $\Lambda$. 
The first step towards an RG equation for $V$ is
differentiating Eqn. (\ref{eq:lskop}) with respect to $\Lambda$ and then
eliminating $K$ to obtain,
\begin{equation}
\frac{\partial V(p,\Lambda)}{\partial\Lambda}=-V(p,\Lambda)
\frac{\partial G_0^{P}(p,\Lambda)}{\partial\Lambda}V(p,\Lambda).
\end{equation}
Taking matrix elements and expanding the Green's function
this equation takes the form,
\begin{equation}
\frac{\partial V(k',k,p,\Lambda)}{\partial\Lambda}=
\frac{M}{2\pi^2}V(k',\Lambda,p,\Lambda)
\frac{\Lambda^2}{\Lambda^2-p^2}
V(\Lambda,k,p,\Lambda).
\end{equation}

To obtain the RG equation for $V$ we must now rescale all low energy
scales with $\Lambda$. We define the rescaled potential $\hat V$ by,
\begin{equation}
\hat V(\hat k',\hat k,\hat p,\Lambda)=\frac{M\Lambda}{2\pi^2}
V(\Lambda\hat k',\Lambda\hat k\Lambda\hat p,\Lambda).
\end{equation}
where $\hat k=k/\Lambda$, $\hat k'=k'/\Lambda$ and $\hat p=p/\Lambda$.
The resulting RG equation is,
\begin{equation}
\Lambda\frac{\partial\hat V}{\partial\Lambda}=
\hat p\frac{\partial\hat V}{\partial\hat p}+
\hat k\frac{\partial\hat V}{\partial\hat k}+
\hat k'\frac{\partial\hat V}{\partial\hat k'}+
\hat V+\hat V(\hat k',1,\hat p,\Lambda)\frac{1}{1-\hat p^2}
\hat V(1,\hat k,\hat p,\Lambda).
\end{equation}
The reason for this rescaling is that it
allows distinct treatments of the low-energy physics and the
parameterisation of high-energy physics. To explain, let us
consider solutions that are independent of $\Lambda$. These solutions,
known as fixed points, are scale free since they are independent of
the only scale in the problem. Therefore they must be related to the
rescaled low energy physics and be independent of the
unscaled high energy physics.

Fixed-point solutions provide systems in which only the low-energy
physics has been included and it is this that
provides the key to constructing power-counting schemes. Solving
the RG equation in the region of a fixed point by perturbing about it
in powers of $\Lambda$ introduces parameters that must scale in
inverse powers of $\Lambda_0$, the scale of the high energy physics.
The perturbations around a fixed point organise the effective
couplings into a power-counting scheme.

Since a fixed point solution is independent of $\Lambda$, all solutions of
the RG equation tend to one as $\Lambda\rightarrow0$. We may classify
fixed points by examining the stability of the perturbations about them
as $\Lambda\rightarrow0$. If all perturbations tend to zero as $\Lambda
\rightarrow0$, i.e they scale with positive powers of $\Lambda$, the
fixed point is stable. If there are perturbations that scale
with negative powers of $\Lambda$ then the fixed point is unstable.
A perturbation that does not scale with $\Lambda$ is known as marginal,
and as we shall see in later chapters is associated with logarithmic
behaviour in the cut-off.

In this problem, we shall find two fixed points that are of particular
interest. The first of these, which we shall refer to as the trivial
fixed point, is simply the obvious solution $\hat V=0$. This solution
corresponds to the scale free system in which no scattering
occurs. We shall see that the power-counting associated with the fixed
point is the Weinberg scheme. A second fixed point, which we shall
refer to as the non-trivial fixed point, gives a scale free system
in which there is a bound state at exactly zero energy. The power-counting
associated with this fixed point is the KSW scheme.

Eqn. (\ref{eq:Vbcs}) and the assumption of s-wave scattering
implies that $V$ should be an analytic function of $p^2, k^2,$ and
$k'^2$. These constraints constitute a boundary condition upon the
RG equation, physically they follow because the
energy is below all thresholds for production of other particles and that
the effective potential should describe short-ranged interactions.

\section{The Trivial Fixed Point}
The Trivial fixed point solution is the $\Lambda$-independent solution
$\hat V=0$. To examine the perturbations around it we write the
potential as
\begin{equation}
\hat V(\hat k,\hat k',\hat p,\Lambda)=C\Lambda^\nu\phi(\hat k',\hat k,\hat p),
\end{equation}
insert it into the RG equation, linearise, and obtain the eigenvalue
equation
\begin{equation}
\hat p\frac{\partial\phi}{\partial\hat p}+
\hat k\frac{\partial\phi}{\partial\hat k}+
\hat k'\frac{\partial\phi}{\partial\hat k'}+\phi=
\nu\phi.
\end{equation}
The solutions of this equation are easily found to be
\begin{equation}
\phi(\hat k',\hat k,\hat p)=\hat k'^{2l}\hat k^{2m}\hat p^{2n},
\end{equation}
with RG eigenvalues $\nu=2(l+m+n)+1$. The analyticity boundary conditions
imply that $l,m$ and $n$ must be non-negative integers, so
that $\nu$ takes the values $1,3,5,\ldots$. The solution to the
RG equation in the region of the trivial fixed point is
\begin{equation}
\hat V(\hat k',\hat k,\hat p,\Lambda)=
\sum_{l,m,n=0}^\infty\hat C_{lmn}\left(\frac{\Lambda}{\Lambda_0}\right)^
\nu\hat k'^{2l}\hat k^{2m}\hat p^{2n},
\end{equation}
where the coefficients $\hat C_{lmn}$ have been made
dimensionless by extracting the high-energy scale $\Lambda_0^{-\nu}$ and for
a Hermitian potential we must take $\hat C_{lmn}=\hat C_{mln}$. Since all the
RG eigenvalues are positive this fixed
point is stable as $\Lambda\rightarrow0$. 

The RG eigenvalues, $\nu$, provide a systematic way to classify the terms
in the potential. In unscaled units we can see how this provides
us with a power-counting scheme and straight away identify the 
couplings in the naively dimensioned potential suggested in chapter 1:
\begin{equation}
V(k',k,p,\Lambda)=\frac{2\pi^2}{M\Lambda_0}\sum_{l,m,n=0}^\infty
\hat C_{lmn}\frac{k'^{2l}k^{2m}p^{2n}}{\Lambda_0^{2(l+m+n)}}.
\end{equation}
The power-counting scheme is the Weinberg scheme if we assign
an order $d=\nu-1$ to each term in the potential. That this
power-counting scheme is useful for weakly interacting systems
now comes as no surprise considering its association with the
trivial, or zero-scattering, fixed point.

The corresponding $K$-matrix is simply given by the first Born approximation,
as higher-order terms from the LS equation are cancelled by higher-order
terms in the potential from the full, nonlinear RG equation \cite{bmr}. It is
important to notice that terms of the same order in the energy, $p^2$, or
momenta, $k^2$ or $k'^2$, occur at the same order in the power-counting.
It is therefore possible to swap between energy or momentum dependence in the
potential without affecting physical (on-shell) observables.

\section{The Non-Trivial Fixed Point}
\subsection{Momentum Independent Solutions}
Let us consider a fixed point solution, $\hat V_0(\hat p)$,
that depends on the energy, $\hat p$, but not upon the momentum. It
satisfies the RG equation
\begin{equation}
\hat p\frac{d\hat V_0}{d\hat p}+\hat V_0+\frac{\hat V_0^2}{1-\hat p^2}=0.
\end{equation}
A convenient way to solve this equation, as well as other
momentum-independent RG equations, is to divide through by $\hat V_0^2$
and obtain a linear equation for $1/\hat V_0$,
\begin{equation}
\hat p\frac{d}{d\hat p}\left(\frac{1}{\hat V_0}\right)
-\frac{1}{\hat V_0}-\frac{1}{1-\hat p^2}=0.
\end{equation}
This equation is satisfied by the basic loop integral,
\begin{equation}\label{eq:defV0}
\hat J(\hat p)=\fint_0^1d\hat q\frac{\hat q^2}{\hat p^2-\hat q^2}=
-\left[1-\frac{\hat p}{2}\ln\frac{1+\hat p}{1-\hat p}\right].
\end{equation}
Since $\hat J(\hat p)$ is analytic in $\hat p^2$ as $\hat p^2\rightarrow0$
it is a valid solution to the RG equation. Hence we take  $\hat V_0=
1/\hat J$. We shall refer to this as the non-trivial fixed point.
When $\hat V_0$ is inserted into the LS equation we obtain
\begin{equation}
K(p,p,p)^{-1}=0,
\end{equation}
which corresponds to a system with a bound state at zero energy. Since
the energy of the bound state is exactly zero, the system has no scale
associated with it, which is why it is described by a fixed point of
the RG.

To find the power-counting associated with this fixed point we
look for perturbations. This task is considerably easy if we consider
momentum independent solutions. We consider perturbations
of the form
\begin{equation}
\frac{1}{\hat V(\hat p,\Lambda)}=\frac{1}{\hat V_0(\hat p)}
+C\Lambda^\nu\phi(\hat p),
\end{equation}
giving the eigenvalue equation,
\begin{equation}\label{eq:erfppert}
\hat p\frac{\partial\phi}{\partial\hat p}-\phi=
\nu\phi.
\end{equation}
Notice that since the momentum independent RG equation for $1/\hat V$ is
linear, no approximation has been made to obtain eqn.(\ref{eq:erfppert}).
The potential obtained using these perturbations will be an exact solution
to the RG equation. The eigenvalue equation is easy to solve:
\begin{equation}
\phi(\hat p)=\hat p^{2n},
\end{equation}
with RG eigenvalues $\nu=2n-1$.
The analyticity boundary conditions demand that $n$ is a non-negative
integer, so that the eigenvalues are
$\nu=-1,1,3,\ldots$. The first eigenvalue is negative, which means
that the fixed point is unstable as $\Lambda\rightarrow0$. The full,
momentum independent solution is given by
\begin{equation}
\frac{1}{\hat V(\hat p,\Lambda)}=\frac{1}{\hat V_0(\hat p)}+
\sum_{n=0}^{\infty}\hat C_{2n-1}
\left(\frac{\Lambda}{\Lambda_0}\right)^{2n-1}\hat p^{2n}.
\end{equation}
The power-counting about this fixed point is the KSW scheme.
As in the Weinberg case we assign each term an order $d=\nu-1$,
so that the KSW scheme is characterised by the values $d=-2,0,2,\ldots$.

When this potential is inserted into the LS equation,
we obtain an expression for the K-matrix, which upon using 
eqn.(\ref{eq:kphase}) gives the phaseshift,
\begin{equation}
p\cot\delta=-\frac{2\Lambda_0}{\pi}\sum_{n=0}^{\infty}\hat C_{2n-1}
\left(\frac{p}{\Lambda_0}\right)^{2n}.
\end{equation}
We can see that the
terms in the expansion around the non-trivial fixed point are in one-to-one
correspondence with the terms in the effective range expansion (\ref{eq:ere}). In
particular, we find
\begin{equation}\label{eq:scattlength}
\hat C_{-1}=\frac{\pi}{2\Lambda_0a},\qquad\qquad
\hat C_1=-\frac{\pi\Lambda_0r_e}{4}.
\end{equation}
If the theory is natural then we expect the coefficients $\hat C_{2n-1}\sim1$.
In this case the scattering length and
the effective range are of order $a\sim r_e\sim1/\Lambda_0$.
\begin{figure}
\begin{center}
\includegraphics[height=13cm,width=8.5cm,angle=-90]{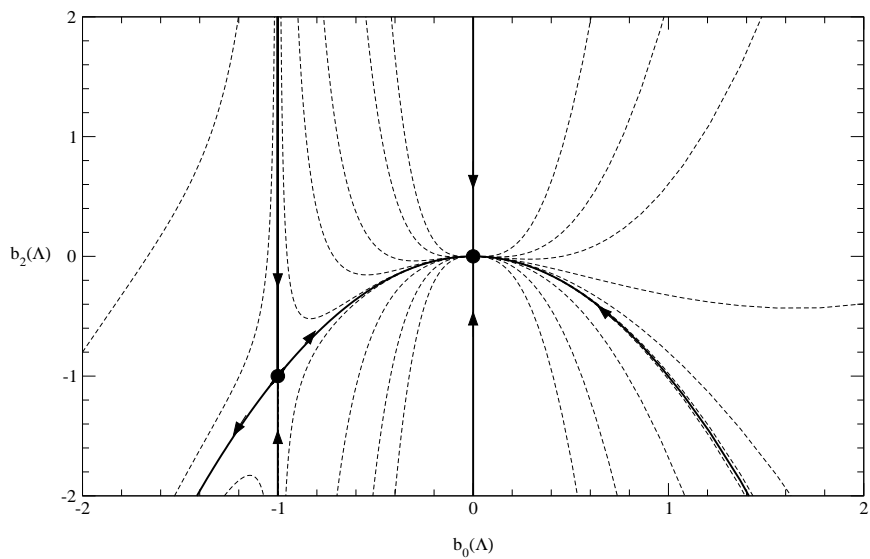}
\caption{The RG flow of the general momentum independent solutions,
$\hat V(\hat p,\Lambda)=\sum_{n=0}^{\infty}b_{2n}(\Lambda)\hat p^{2n}$.
The fixed-points are shown as dots. The eigenvector flows are shown in bold
with the arrows indicating flow as $\Lambda\rightarrow0$. More general flows
are shown as dashed.}\label{fig:SRRGflow}
\end{center}
\end{figure}

To understand which of the two power-counting schemes is appropriate in a given situation,
it is fruitful to examine the RG flow of purely energy dependent RG solutions
illustrated in Fig.\ref{fig:SRRGflow}. This figure shows the RG flow as
$\Lambda\rightarrow0$ in the plane $(b_0(\Lambda),b_2(\Lambda))$ where
\begin{equation}
\hat V(\hat p,\Lambda)=b_0(\Lambda)+b_2(\Lambda)\hat p^{2n}+\ldots.
\end{equation}
The trivial fixed point lies at $(0,0)$. All flows near to it flow into it as
$\Lambda\rightarrow0$, illustrating its already noted stability. The non-trivial
fixed point lies at $(-1,-1)$, only flows lying on the critical line $b_0=-1$
flow into it. All other flows in the region of this critical line flow initially
towards the non-trivial
fixed point before the unstable perturbation becomes apparent and takes the flow
into the domain of the trivial fixed point (possibly `via' infinity).

The values of $\Lambda$
for which the flow is controlled by the non-trivial fixed point is dependent on
the value of the coefficient $\hat C_{-1}$.  In particular if
\begin{equation}\label{eq:condition}
\Lambda>\frac{\pi}{2a}=\Lambda_0\hat C_{-1}
\end{equation} 
then the unstable perturbation is suppressed and the RG flow is controlled
by the expansion around the non-trivial fixed point. As
$\Lambda$ becomes less than $\pi/(2a)$ then the unstable perturbation becomes
important and the flow moves away from the non-trivial fixed point and
into the trivial fixed point.

In a natural theory we have $1/a\sim\Lambda_0$ so that as $\Lambda$
floats between the scale of the high energy physics, $\Lambda_0$, and $0$,
the condition (\ref{eq:condition}) is not met.
This means that the RG flow is in the domain of the trivial fixed
point and that the appropriate power-counting is the one obtained by
perturbing about that point, the Weinberg scheme. 

If there is some fine tuning in the theory so that
$\hat C_{-1}$ is small, then the unstable perturbation is suppressed.
This fine tuning, according to eqn.~(\ref{eq:scattlength}), will
reveal itself in a surprisingly large scattering length, $a\gg1/\Lambda_0$.
As $\Lambda$ varies from $\Lambda_0$ to $1/a$ the condition
(\ref{eq:condition}) is true and the power-counting associated with the
non-trivial fixed point provides a suitable expansion.

For example, in s-wave nucleon-nucleon scattering the
scattering lengths are large\footnote{Naturally the scattering length for NN scattering
should be of the order of the pion mass, $a\sim1/m_\pi$, however this is not what is
empirically observed.} suggesting a finely tuned effective potential with a small
coefficient $\hat C_{-1}$, in this case the RG flow is dominated by the non-trivial
fixed point and the power-counting associated with it, the KSW scheme, 
is the appropriate one to use in organising the terms \cite{ksw,uvk,crs,geg,modeft}.

\subsection{Momentum Dependent Solutions.}
In the analysis above the Weinberg and KSW counting schemes occur quite
naturally. Study of the RG flow leads to conclusions about the usefulness
of each of the schemes in different systems. However, the analysis is not
complete as we have not considered the momentum dependent
solution about the non-trivial fixed point.

Around the trivial fixed point, it was found that the momentum dependent
perturbations provided no new physics since they occurred at the same order
in the power counting as the energy dependent perturbations.
This meant that momentum and energy perturbations could not be
distinguished within on-shell observables.

Around the non-trivial fixed point things are considerably more complicated,
as we shall see the momentum and energy dependent perturbations occur at different
orders in the power counting. This result means that it is not a simple matter to interchange
momentum and energy dependence of physical observables near the
non-trivial fixed point.

To find the momentum dependent perturbations we write the solution
in the vicinity of the fixed point solution as,
\begin{equation}
\hat V(\hat k,\hat k',\hat p,\Lambda)=\hat V_0(\hat p)+C\Lambda^\nu
\phi(\hat k,\hat k',\hat p).
\end{equation}
After linearising, the eigenvalue equation for $\phi$ is
\begin{equation}
\hat p\frac{\partial\phi}{\partial\hat p}+
\hat k\frac{\partial\phi}{\partial\hat k}+
\hat k'\frac{\partial\phi}{\partial\hat k'}+
\frac{\hat V_0(\hat p)}{1-\hat p^2}
\left[\phi(\hat k,1,\hat p)+\phi(1,\hat k',\hat p)\right]
=(\nu-1)\phi.
\end{equation}
Useful solutions to this equation\cite{krthesis} are
\begin{equation}
\varphi_n(\hat k,\hat p)=\left[\hat k^{2n}-\hat p^{2n}+
\sum_{l=0}^{n-1}\frac{\hat p^{2l}}{2(n-l)+1}\hat V_0(\hat p)\right]
\hat V_0(\hat p),
\end{equation}
with eigenvalues, $\nu=2n=2,4,6,\ldots$. 
These solutions lead to two forms of perturbation that are symmetric
in $k$ and $k'$ and so 
may be used within an hermitian momentum dependent potential:
\begin{equation}
\phi(\hat k,\hat k',\hat p)=\hat p^{2m}
\left\{\varphi_n(\hat k,\hat p)+\varphi_n(\hat k',\hat p)\right\},
\end{equation}
with eigenvalues $\nu=2m+2n=2,4,6,\ldots$ and
\begin{equation}
\phi(\hat k,\hat k',\hat p)=\hat p^{2m}\left\{
\varphi_n(\hat k,\hat p)\varphi_{n'}(\hat k',\hat p)\right\},
\end{equation}
with eigenvalues $\nu=2m+2n+2n'+1=5,7,9,\ldots$. 
These perturbations plus the the purely energy dependent perturbations
form a complete set that can be used to expand any perturbation about the
non-trivial fixed point that is well-behaved as $\hat k^2,\hat k'^2,\hat p^2
\rightarrow0$. The momentum dependent perturbations vanish on-shell and
the terms in the on-shell potential are still
in one-to-one correspondence with the terms in the effective range
expansion\cite{krthesis}. The most general
solution in the vicinity of the non-trivial fixed point may be written as
\begin{eqnarray}
&&\displaystyle{
\hat V(\hat k,\hat k',\hat p,\Lambda)=\hat V_0(\hat p)+
\sum_{n=0}^{\infty}\hat C_{2n-1}\left(\frac{\Lambda}{\Lambda_0}\right)^{2n-1}
\hat p^{2n}\hat V_0(\hat p)^2+}\nonumber\\
&&\qquad\displaystyle{
\sum_{n=0,m=1}^{\infty}\hat D_{2(n+m)}\left(\frac{\Lambda}{\Lambda_0}\right)
^{2(n+m)}\hat p^{2n}\left\{\varphi_m(\hat k,\hat p)+\varphi_m(\hat k',\hat p)
\right\}+}\nonumber\\
&&\qquad\qquad\displaystyle{
\sum_{n=0;m,m'=1}^{\infty}\hat E_{2(n+m+m')+1}
\left(\frac{\Lambda}{\Lambda_0}\right)
^{2(n+m+m')+1}\hat p^{2n}
\varphi_m(\hat k,\hat p)\varphi_{m'}(\hat k',\hat p).}
\end{eqnarray}
When the on-shell $K$-matrix is calculated using this general form of
expansion about the non-trivial fixed point, the only terms that contribute
are precisely those that appear in the expansion given in the previous
section, namely the coefficients of the purely energy dependent eigenfunctions,
$\hat C_{\nu}$.
The momentum dependent eigenfunctions do not contribute to
on-shell scattering, i.e.
physical observables are independent of $\hat D_\nu$ and
$\hat E_\nu$ at this order \cite{bmr,krthesis}.

The solution given above results from linearisation of the RG equation
close to the non-trivial fixed point. It is not clear that if
a full solution to the non-linear RG equation was constructed that the
one-to-one correspondence betweens terms in the potential and terms in
the effective range expansion persists. In order to clarify the
situation, Birse {\it et al} \cite{bmr} calculated corrections to the potential
around the non-trivial fixed point due to the
neglected non-linear
terms in the RG equation up to order $\hat C_{-1}\hat D_{2}$ and found the
additional term,
\begin{equation}\label{eq:firstcor}
\frac{\Lambda}{\Lambda_0}\hat C_{-1}\hat D_{2}\left(\hat k'^2+\hat k^2+
A\hat p^2+\frac{4\Lambda^2}{3}\hat V_0(\hat p)\right)
\hat V_0(\hat p),
\end{equation}
where $A$ is a constant of integration. In general this second order piece will
contribute to the effective range term along with the $\hat C_1$ term. Since the
constant $A$ is unfixed, we are free to set it as we wish to ensure a nice
correspondence between terms in the potential and terms in the effective
range expansion. Setting $A=-2$ maintains the
direct correspondence between the effective range expansion and the energy
dependent perturbations around the non-trivial fixed point. 
In general it is assumed that at
each order in the coefficients further degrees of freedom 
will become available to ensure this correspondence is continued to all orders.

The degrees of freedom that become available as the full non-linear
solution is explored allows the possibility of generating the effective range
terms from momentum dependent, rather than energy dependent, perturbations.
For example,
it is possible to generate the effective range, $r_e$, solely from the
momentum dependence in the correction
given in eqn. (\ref{eq:firstcor}). By setting $\hat C_1=0$ and $A=0$
the effective range would be given by,
\begin{equation}
r_e=\frac{4\hat D_2}{\Lambda_0^2 a}.
\end{equation}
However, since the momentum dependent perturbations occur
at different RG eigenvalues, $\nu=2,4,\ldots$, than the energy dependent
perturbations they are replacing, the coefficient $D_\nu$ is forced to
take an unnaturally large value to compensate for the additional factor of
$\Lambda_0a$. 

Unlike the trivial fixed point the possibility of using the equations of motion
to move between momentum and energy dependence is not manifest at any
particular order in the power-counting. This trade has to be between
terms resulting from non-linear corrections in the solution to the RG equation in
the vicinity of the fixed point, making it difficult to see and implement.
Importantly, the momentum dependent perturbations offer no new degrees of freedom
to on-shell observables than the easily elucidated energy dependent ones.

\section{Summary}
In this chapter we have introduced the RG for short range forces and used it
to derive the Weinberg and KSW power-counting schemes. The key to solving the RG equation are
fixed point solutions. Perturbing around the fixed point solutions gives us
a simple recipe for constructing power-counting schemes. We have
identified two fixed point solutions, the trivial fixed point that leads to the Weinberg
scheme and a non-trivial fixed point that leads to the KSW scheme.

The terms in the perturbations around the trivial fixed point are in one to one
correspondence with the terms in the DW Born expansion. Those about the non-trivial
fixed point are in one-to-one correspondence with the terms in the effective
range expansion.

By studying the RG flow with the cut-off $\Lambda$ we may make conclusions about the
usefulness of each of the fixed points. The non-trivial fixed point is unstable and
requires the fine tuning of a parameter in the expansion around it. This
fine tuning results in large scattering lengths. Because of the large scattering lengths
observed in nuclear systems the KSW counting is the appropriate scheme.

%% file: ch2/chapter2.tex
\section{Introduction}
In this chapter we shall introduce the distorted wave renormalisation group (DWRG) \cite{tbmb}.
This is an extension of the RG which allows the summation of some physics to all orders.

The ability to construct an
EFT with non-trivial low-energy physics is important. Consider
proton-proton scattering at low energies. The
Coulomb interaction between the protons is a very long-ranged interaction
with a characteristic momentum scale, $\kappa=\frac{\alpha M}{2}=3.42\text{MeV}$.
If we are interested in scattering between protons of typical energy
$p\sim\kappa$, $\kappa$ must be treated as a low-energy scale,
so we must sum all photon exchange diagrams since they occur at LO
in the EFT \cite{kongr}.

Proton-proton scattering is not the only example that is important to
nuclear physics. Also of interest is nucleon-nucleon scattering at energies
comparable to the pion mass \cite{furnsteele,longandshort,cohan,krthesis}.
In the previous chapter, we constructed a
EFT in which all exchange particles are absorbed into the
effective couplings. In nucleon-nucleon scattering the scale at which
this breaks down is the pion mass, $m_\pi$, which acts as a high
energy scale in the EFT. If we wish to examine nucleon-nucleon scattering at energies
comparable to the pion mass, pion fields must be introduced into the EFT Lagrangian.
Dependent upon our identification of low-energy scales, one approach may be to sum all one
pion exchange diagrams \cite{furnsteele}. There are many issues surrounding how
pions are to be including in an EFT for nucleons \cite{vk1,orvk,wein1,kswapp1,furnsteele},
we shall consider these towards the end of this chapter.

We will introduce the DWRG equation by showing how the
short and long range physics can be neatly separated. Then we will apply
the equation to a number of examples. Our first example, in which the
long-range physics will be modelled by the Coulomb potential, will demonstrate
the methods that will be used in later examples. In the DWRG analysis of this
system we will derive a known result for the distorted wave effective range
expansion\cite{bethe,j&b,kongr}.

Our second example, will be the scale-free, repulsive inverse
square potential. Our interest
in this example is two-fold. Firstly, it will allow examination of short
range forces for higher partial waves in the EFT for short range forces
and secondly, the scale-free nature of the potential makes it very similar to
three body systems\cite{efimov}.

As a final example we shall consider
a more general class of long-range forces, namely non-singular potentials
that facilitate the definition of the Jost function\cite{newton}. This general
analysis is of interest as it includes the Yukawa potential and because the methodology
used in this section acts as inspiration for the DWRG analysis in later chapters.

\section{Separating Short- and Long-Range Physics}\label{sec:rglong}
We shall assume that all non-perturbative EFT diagrams can be summed to give a long range
potential. To this end, we consider a system of two-particles of mass $M$
interacting through the potential,
\begin{equation}
V=V_L+\tilde V_S,
\end{equation}
where $\tilde V_S$ is an effective short-range force that consists
of contact interactions only. The terms in $\tilde V_S$ can be related to the
couplings in the effective Lagrangian.

The full $T$-matrix that describes scattering from both the long and
short range potentials is given by a LS equation,
\begin{equation}\label{eq:fullLS}
T(p)=(V_L+\tilde V_S)+(V_L+\tilde V_S)G_0^+(p)T(p).
\end{equation}
We wish to find an RG equation that allows us to determine the power-counting
for the short range force. Suppose that we were to attempt to find this by
applying the cut-off to the free Green's function in eqn.~(\ref{eq:fullLS}).
The resulting differential equation for $\tilde V_S$, found by demanding
that $T$ is independent of $\Lambda$, is
\begin{equation}
\frac{\partial\tilde V_S}{\partial\Lambda}=-
(V_L+\tilde V_S)\frac{\partial G_0^+}{\partial\Lambda}
(V_L+\tilde V_S).
\end{equation}
After taking matrix elements we obtain an equation
that in general contains complicated $\Lambda$-dependence in terms that
are quadratic, linear and independent of $\tilde V_S$. In essence this complicated
$\Lambda$-dependence is because of the cut-off, which applied to the
free Green's function in the LS equation, not only
regularises the divergent loop integrals between contact interactions
but also cuts-off elements of the long-range potential.
Those parts of the long-range potential ``removed'' by the cut-off then
have to be absorbed into the short-range potential resulting in the
equation above, it then becomes difficult to justify any boundary
condition on the potential, $\tilde V_S$.

To circumvent this complication and
obtain an RG equation containing the short-range potential alone,
it is useful to work in terms of the distorted waves (DWs),
$|\psi_{\bf p}\rangle$, of the long-range potential.
The DWs are simply solutions to the
Schr\"odinger equation containing the long-range potential alone:
\begin{equation}
\left(H_0+V_L-\frac{p^2}{M}\right)|\psi_{\bf p}\rangle=0.
\end{equation}
The $T$-matrix, $T_L$, describing scattering by $V_L$, is simply
given by the LS equation:
\begin{equation}
T_L(p)=V_L+V_L G_0^+(p)T_L(p).
\end{equation}
The corresponding full Green's function is
\begin{equation}\label{eq:fullgreen}
G_L^+(p)=G_0^+(p)+G_0^+(p)T_L(p)G_0^+(p).
\end{equation}

To isolate the effects of the short-range potential
and enable a renormalisation group analysis we write the full
$T$-matrix as \cite{newton}
\begin{equation}\label{eq:sumofTs}
T(p)=T_L(p)+(1+T_L(p)G_0^+(p))\tilde T_S(p)(1+ G_0^+(p)T_L(p)).
\end{equation}
The operator $(1+G_0^+T_L)$ is the
M\"oller wave operator that converts a plane wave into a DW of $V_L$,
\begin{equation}
|\psi_{\bf p}\rangle=(1+G_0^+(p)T_L(p))|{\bf p}\rangle,
\end{equation}
so that the matrix elements of the full $T$-matrix are given by,
\begin{equation}\label{eq:tmatrixelements}
\langle{\bf k}|T(p)|{\bf k'}\rangle=
\langle{\bf k}|T_L(p)|{\bf k'}\rangle+
\langle\psi^-_{\bf k}|\tilde T_S(p)|\psi^+_{\bf k'}\rangle.
\end{equation}
An equation for the operator $\tilde T_S(p)$ can be found by
substituting eqn.~(\ref{eq:sumofTs}) into the full LS
equation (\ref{eq:fullLS}). After identifying and cancelling the terms
in the pure long-range LS equation
and cancelling the M\"oller wave operator on the right hand side
we are left with,
\begin{eqnarray}
&&\displaystyle{
[1+T_L(p)G_0^+(p)]\tilde T_S(p)=
\tilde V_S+\tilde V_SG_0^+(p)[1+T_L(p)G_0^+(p)]\tilde T_S(p)}\nonumber\\
&&\displaystyle{\qquad\qquad\qquad\qquad\qquad\qquad\qquad
+V_LG_0^+(p)[1+T_L(p)G_0^+(p)]\tilde T_S(p).}
\end{eqnarray}
The third term on the right-hand side can be seen to cancel with the
second on the left-hand side after identifying the LS
equation for $T_L$. The second term on the right hand side can be
simplified by identifying the form for the full long-range Green's function,
eqn.~(\ref{eq:fullgreen}). What remains is
a distorted wave Lippmann-Schwinger (DWLS) equation for the $\tilde T_S$,
\begin{equation}\label{eq:dwlst}
\tilde T_S(p)=\tilde V_S+\tilde V_S G_L^+(p) \tilde T_S(p).
\end{equation}
This equation is now a far more promising starting point for
the RG analysis since the effects of the long-range potential have
been dealt with separately. The operator $\tilde T_S$ describes
the interaction between the
short-range potential and the DWs of the long-range potential

We shall assume that $V_L$ is a central potential and work in the
partial wave basis. In particular we shall concentrate
on s-wave scattering. In that case the on-shell 
$T$-matrix may be expressed in terms
of the phaseshift,\footnote{Note that the normalisation used in this chapter
is slightly
different from that implicitly assumed in the previous chapter, hence the
relation between the $T$-matrix and the phaseshift differs by 
a factor of $p^2$.}
\begin{equation}\label{eq:tphase}
\langle p|T(p)|p\rangle=-\frac{4\pi p}{M}\frac{e^{2i\delta}-1}{2i}.
\end{equation}
A similar relation holds between $T_L$ and $\delta_L$, the phaseshift
due to the long range potential alone.
We may write the full phaseshift, $\delta$, as
$\delta_L$ plus a correction, $\tilde\delta_S$, due to the
effect of the short-range potential:
\begin{equation}
\delta=\delta_L+\tilde\delta_S.
\end{equation}
Substituting this form into eqn.~(\ref{eq:tphase}) and subsequently into
eqn.~(\ref{eq:sumofTs}) we obtain a simple relationship between the 
on-shell matrix elements of $\tilde T_S$ and the corrective phaseshift
$\tilde\delta_S$,
\begin{equation}\label{eq:dwtps}
\langle \psi^-_p|\tilde T_S(p)|\psi^+_p\rangle
=-\frac{4\pi p}{M}e^{2i\delta_L(p)}\frac{e^{2i\tilde\delta_S(p)}-1}{2i}.
\end{equation}

\section{The DWRG equation}
In order to derive the DWRG equation for the short-range potential it is once
again more convenient to work with a reactance matrix, $\tilde K_S$, which
satisfies the DWLS equation,
\begin{equation}\label{eq:dwlsk}
\tilde K_S(p)=\tilde V_S(p)+\tilde V_S(p)G_L^P(p)\tilde K_S(p),
\end{equation}
where in this equation $G_L^P$ indicates
the Green's function with standing wave
boundary conditions. The relationship between the on-shell matrix
elements of $\tilde T$ and $\tilde K$ (c.f. eqn.~(\ref{eq:TKrel}) is,
\begin{equation}
\frac{1}{\langle\psi_p|\tilde K_S(p)|\psi_p\rangle}=
\frac{e^{2i\delta_L(p)}}{\langle\psi^-_p|\tilde T_S(p)|\psi^+_p\rangle}
+\frac{iM}{4\pi p},
\end{equation}
where similarly $|\psi_p\rangle$ without superscript indicates
standing wave boundary conditions on the DW.
Hence the on-shell $\tilde K_S$-matrix is given by
\begin{equation}\label{eq:Kphase}
\frac{1}{\langle\psi_p|\tilde K_S(p)|\psi_p\rangle}
=-\frac{M}{4\pi p}\cot\tilde\delta_S.
\end{equation}
To regulate eqn.~(\ref{eq:dwlsk})
we expand $G_L^P$ using the completeness relation for the DWs,
and apply a cut-off to the continuum states,
\begin{equation}
G_L^P(p,\Lambda)
=\frac{M}{2\pi^2}\fint_0^\Lambda {\rm d}q\,\frac{|\psi_q\rangle
\langle\psi_q|}{p^2-q^2}+\frac{M}{4\pi}\sum_n\frac
{|\psi_n\rangle\langle\psi_n|}{p^2+p_n^2}.
\end{equation}
Demanding that $\tilde K_S$ is $\Lambda$-independent we can obtain,
as in the previous chapter, a differential equation for $\tilde V_S$ by
differentiating eqn.~(\ref{eq:dwlsk}) with respect to $\Lambda$ and eliminating $\tilde K_S$
to obtain,
\begin{equation}\label{eq:udwrge}
\frac{\partial \tilde V_S}{\partial\Lambda}(p,\Lambda)
=-\tilde V_S(p,\Lambda)\frac{\partial G_L^P}
{\partial\Lambda}(p,\Lambda)\tilde V_S(p,\Lambda).
\end{equation}

In the remainder of this chapter,
to simplify the analysis, we shall assume that the matrix elements of
$\tilde V_S$ depend only on the energy, $p$ and not
on the momentum.
As shown in chapter \ref{Ch1}, the off-shell momentum dependent solutions to the 
RG equation have more complicated forms but are not needed to describe on-shell 
scattering \cite{bmr,krthesis}. 

In the previous chapter we simply used a contact interaction proportional
to a delta function,
however, such a choice cannot be used in 
combination with some of the long-range potentials of interest.
Long-range potentials that
are sufficiently singular that their DWs, $\psi_p(r)$,
either vanish or diverge 
as $r\rightarrow 0$ will result in a poorly defined DWRG equation if a delta
function contact term is used.

To construct a contact interaction we chose a spherically symmetric potential 
with a short but nonzero range. By choosing the range, $R$, of this potential to be
much smaller than $1/\Lambda$, we ensure that any additional energy or 
momentum dependence associated with it is no larger than that of the physics 
which has been integrated out, and hence the power-counting is not altered by it.
The precise value of this scale is arbitrary and so observables
should not depend on it, we may consider the results to be equivalent 
to those that would be obtained in the limit $R\rightarrow 0$. $R$ should be
thought of nothing more than a tool to avoid a null DWRG equation.
A simple and convenient choice for the form of the potential is the 
``$\delta$-shell" potential,
\begin{equation}
\tilde V_S(r)=V_S\delta(r-R).
\end{equation} 
With this choice, the eqn.~(\ref{eq:udwrge}) for $V_S$ becomes
\begin{equation}\label{eq:udwrge2}
\frac{\partial V_S(p,\kappa,\Lambda)}{\partial\Lambda}=-\frac{M}{2\pi^2}\frac{
|\psi_\Lambda(R)|^2}{p^2-\Lambda^2}
V_S^2(p,\kappa,\Lambda).
\end{equation}

The final step in obtaining the DWRG equation is to rescale each
of the low energy scales, expressing them in terms of the cut-off $\Lambda$.
Dimensionless momentum variables are defined by $\hat p=p/\Lambda$ etc.,
along with a rescaled potential, $\hat V_S$.
The exact nature of the rescaling required for
the potential is dependent on the form of $\psi_\Lambda(R)$.

The solution to the DWRG equation should satisfy
the analyticity boundary conditions in $\hat p^2$ that 
follow from the same arguments given in the previous chapter. 
$\hat V_S$ must also be analytic in any other low energy scale, $\kappa$,
associated with the long-range potential. The exact analyticity
condition for each scale $\kappa$ is case specific. 
In most cases we need to demand only that the effective potential is analytic 
in $\kappa$. An example is the inverse Bohr radius which is the low-energy 
scale associated with the Coulomb potential, and which is proportional to
the fine structure constant, $\alpha$. Since 
the short-distance physics which has been incorporated in the effective potential 
should be analytic in $\alpha$ it should be analytic in $\kappa$.
An important exception is the pion mass. This is proportional to the square root of the
strength of  the chiral symmetry breaking (the current quark mass in the underlying theory, 
QCD) and so, as in ChPT, the effective potential should be analytic in $m_\pi^2$.
Under these restrictions, we see that $\hat V_S$ should have an expansion in non-negative, 
integer powers of $\hat p^2$ and $\hat \kappa$ (or $\hat \kappa^2$).

\section{The Coulomb Potential}
Having discussed the general route to the DWRG equation we shall now
provide a concrete and physically interesting example to see
how the whole procedure fits together.
The Coulomb potential, $V_L(r)=\alpha r^{-1}$, is highly interesting physically for obvious
reasons. The wavefunction for the Coulomb potential may be written in terms of the
confluent hypergeometric function, $\Phi(a,b,z)$,
\begin{equation}
\psi_p^{(+)}(r)=pre^{-\frac12\pi\eta}\Gamma(1+i\eta)
\Phi(1+i\eta,2,-2ipr)e^{ipr},
\end{equation}
where $\eta=\kappa/p$, $\kappa=\alpha M/2$\cite{newton}.
The definition of the phaseshift for the Coulomb potential is unusual because
the long $r^{-1}$ tail results in logarithmic asymptotic contributions.
The Coulomb phaseshift is defined by
writing the asymptotic behaviour of the wavefunction as,
\begin{equation}
\psi_p^{(+)}(r)\rightarrow e^{i\delta_c}\sin(pr-\eta\ln{2pr}+\delta_c),
\end{equation}
which gives the phaseshift $\delta_c={\text Arg}\{\Gamma(1+i\eta)\}$.
Incidentally, this expression also defines the normalisation of the
DWs.
A simple calculation shows that for $R\ll \Lambda^{-1}$
we have,
\begin{equation}\label{eq:smallrcoul}
|\psi_\Lambda(R)|^2\rightarrow \Lambda^2R^2e^{-\pi\hat\kappa}
\Gamma\left(1+i\frac\kappa\Lambda\right)\Gamma\left(1-i\frac\kappa\Lambda\right)
=\frac{2\pi\kappa\Lambda R^2}{e^{2\pi\kappa/\Lambda}-1}=R^2\Lambda^2
{\cal C}(\kappa/\Lambda).
\end{equation}
where
${\cal C}(\eta)$ is known as the Sommerfeld factor.
Hence, in the Coulomb case eqn.~(\ref{eq:udwrge2}) becomes
\begin{equation}
\frac{\partial V_S(p,\kappa,\Lambda)}{\partial\Lambda}=-\frac{MR^2}{2\pi^2}
{\cal C}(\kappa/\Lambda)\frac{\Lambda^2}{p^2-\Lambda^2}
V_S^2(p,\kappa,\Lambda).
\end{equation}
To obtain the DWRG equation we must rescale the variables $p=\hat p\Lambda$
and $\kappa=\hat\kappa\Lambda$. $V_S$ must be rescaled to factor out the scales
$M$ and $R$. Hence we define,
\begin{equation}
\hat V_S(\hat p,\hat\kappa,\Lambda)=\frac{M\Lambda R^2}{2\pi^2}
V_S(\Lambda\hat p,\Lambda\hat\kappa,\Lambda),
\end{equation}
resulting in the DWRG equation for the Coulomb potential:
\begin{equation}\label{eq:dwrgcoul}
\Lambda\frac{\partial\hat V_S}{\partial\Lambda}=
\hat p\frac{\partial\hat V_S}{\partial\hat p}+
\hat\kappa\frac{\partial\hat V_S}{\partial\hat\kappa}+
\hat V_S+\frac{{\cal C}(\hat\kappa)}{1-\hat p^2}\hat V_S^2.
\end{equation}
The boundary conditions that follow from the discussion above are
analyticity about $\hat p^2,\hat\kappa\rightarrow0$.
We shall assume that $\kappa$ is positive, i.e. that the
Coulomb potential is repulsive, the modifications to this discussion for
an attractive Coulomb potential are examined in Appendix \ref{app:coulatt}.

Our study of the Coulomb DWRG equation will mirror the RG analysis of
the previous chapter. In particular we shall concern ourselves
with fixed-point solutions and their corresponding power-counting schemes.
A trivial fixed point, $\hat V_S=0$, is readily identified, but the
existence of a non-trivial fixed point is not obvious. We will see
that no true fixed point other than the trivial one exists. However,
we will find a logarithmically evolving `fixed point' that for the
purpose of constructing a power-counting scheme may be regarded as a true
fixed point. Furthermore, we will find a marginal perturbation in the
RG flow about this fixed point that is associated with its logarithmic evolution.

\subsection{The Trivial Fixed Point}
An obvious fixed point solution of the DWRG equation for the Coulomb
potential is the trivial fixed point solution, $\hat V_S=0$. This solution
corresponds to a vanishing short-range force, i.e. a system in which only
the Coulomb force is active. The power-counting associated with the point
can be obtained by solving the eigenvalue equation for a small perturbation
about it. Putting
\begin{equation}
\hat V_S(\hat p,\hat\kappa,\Lambda)=C \Lambda^\sigma\phi(\hat p,\hat\kappa),
\end{equation}
we obtain an eigenvalue equation that is very similar to that in the previous chapter,
\begin{equation}
\hat p\frac{\partial\phi}{\partial\hat p}+\hat\kappa
\frac{\partial\phi}{\partial\hat\kappa}=
(\nu-1)\phi.
\end{equation}
The solutions that satisfy the boundary conditions are 
$\phi=\hat p^{2n}\kappa^m$, where $n$ and $m$ are positive integers.
The RG eigenvalues are $\nu=2n+m+1=1,2,3,\ldots$. All the RG eigenvalues,
in common with the pure short range case of the previous chapter,
are positive and the fixed-point is stable. The potential in the vicinity
of the fixed-point is given by,
\begin{equation}
\hat V_S=\sum_{n,m=1}^\infty \hat C_{2n,m}\left(\frac{\Lambda}{\Lambda_0}
\right)^{2n+m+1}\hat p^{2n}\hat\kappa^{m}.
\end{equation}
If we assign $d=\nu-1$ then the power-counting scheme is
the Weinberg scheme with additional $\kappa$-dependent terms. Upon
substitution into the Lippmann Schwinger Equation, all
higher order terms vanish and what remains is a distorted wave
Born Expansion,
\begin{equation}\label{eq:dwbe}
\frac{\tan\tilde\delta_S}{p}=-\frac{\pi}{2\Lambda_0}{\cal C}(\eta)
\sum_{n,m=1}^\infty\hat C_{2n,m}\left(\frac{p^{2n}\kappa^m}{\Lambda_0^{2n+m}}
\right),
\end{equation}
which, it should be noted, is independent of the delta-shell range, $R$.
This power-counting scheme, in parallel to the trivial fixed point for purely
short-range potentials, is appropriate for systems with 
weakly interacting short-range potentials that provide only minor
corrections to the Coulomb potential.

\subsection{The Non-Trivial Fixed Point}
The starting point for finding another fixed point solution is to divide
the DWRG equation through by $\hat V_S^2$ to obtain a linear PDE for
$\hat V_S^{-1}$,
\begin{equation}\label{eq:invdwrg}
\Lambda\frac{\partial \hat V_S^{-1}}{\partial\Lambda}=
\hat p\frac{\partial\hat V_S^{-1}}{\partial\hat p}+
\hat\kappa\frac{\partial\hat V_S^{-1}}{\partial\hat\kappa}-
\hat V_S^{-1}-\frac{{\cal C}(\hat\kappa)}{1-\hat p^2}.
\end{equation}
Substitution shows that this equation has a fixed point solution given by
the basic loop integral,
\begin{equation}
\hat J(\hat p,\hat\kappa)=\fint_0^1 \hat q^2
d\hat q\frac{{\cal C}(\hat\kappa/\hat q)}
{\hat p^2-\hat q^2}.
\end{equation}
However, we may not simply take $\hat V_S=\hat J^{-1}$ as the
solution to the RG equation because it does not satisfy the
necessary analyticity boundary conditions. A source of
non-analyticity in $\hat p^2$ are the poles at $\hat q=\pm\hat p$.
As $\hat p\rightarrow0$ these poles move
to the endpoint of the integral, $\hat q=0$,
causing singular behaviour in $\hat J$.
A second source of singular behaviour is the essential singularity
in ${\cal C}(\eta)$ at $\eta=0$. This singularity means that
the integral $\hat J$ only converges for $\text{Re}\{\hat\kappa\}>0$
resulting in singular behaviour about $\hat\kappa=0$.
In order to isolate these non-analyticities we write $\hat J$ as:
\begin{equation}\label{eq:breakupJ}
\hat J(\hat p,\hat\kappa)=-\int_0^1 d\hat q{\cal C}(\hat\kappa/\hat q)
-\hat p^2\int_1^\infty d\hat q \frac{{\cal C}(\hat\kappa/\hat q)}{
\hat p^2-\hat q^2}+\hat{\cal M}(\hat p,\hat\kappa),
\end{equation}
where
\begin{equation}
\hat{\cal M}(\hat p,\hat\kappa)=\hat p^2\fint_0^\infty d\hat q
\frac{{\cal C}(\hat\kappa/\hat q)}{\hat p^2-\hat q^2}.
\end{equation}
The non-analyticity in $\hat p^2$ caused by the $\hat q=0$ endpoint now appears
in $\hat{\cal M}$, whilst the non-analyticity resulting from the essential
singularity in $\cal C$ is contained in both $\hat{\cal M}$ and the first integral
in eqn.~(\ref{eq:breakupJ}). The second integral in eqn.~(\ref{eq:breakupJ})
avoids the troublesome endpoint, $\hat q=0$, and is analytic in
both $\hat\kappa$ and $\hat p$. The first integral in eqn.~(\ref{eq:breakupJ}) may
be written as (see Appendix \ref{app:janalyticity}),
\begin{equation}\label{eq:Jdetail}
-\int_0^1d\hat q\,{\cal C}(\hat\kappa/\hat q)=
-1-\pi\hat\kappa\ln\hat\kappa+\text{Analytic terms in $\hat\kappa$},
\end{equation}
So that overall we have,
\begin{equation}
\hat J(\hat p,\hat\kappa)=\hat{\cal M}(\hat p,\hat\kappa)
-\pi\hat\kappa\ln\hat\kappa+\text{Terms Analytic in $\hat p^2,\hat\kappa$}.
\end{equation}
To construct a solution to the DWRG fixed point equation that satisfies
the analyticity boundary conditions we must remove the non-analytic terms
$\hat{\cal M}$ and $\hat\kappa\ln\hat\kappa$.
The removal of the non-analytic term $\hat{\cal M}$ is simple.
It is not difficult to show that it satisfies the homogeneous form of
eqn.~(\ref{eq:invdwrg}),
\begin{equation}
\hat p\frac{\partial\hat{\cal M}}{\partial\hat p}+
\hat\kappa\frac{\partial\hat{\cal M}}{\partial\hat\kappa}-
\hat{\cal M}=0,
\end{equation}
so that $\hat J-\hat{\cal M}$ satisfies eqn.~(\ref{eq:invdwrg}).
However, the removal of the logarithmic term in eqn.~(\ref{eq:Jdetail}) is not
as simple and cannot be done within the confines of the fixed point
equation. It is because of this term that we are forced to introduce
logarithmic $\Lambda$ dependence into the fixed point solution.
The term,
\begin{equation}\label{eq:logterm}
\hat{\cal L}(\hat\kappa,\Lambda)=-\pi\hat\kappa\ln\frac{\Lambda\hat\kappa}{\mu},
\end{equation}
where $\mu$ is some arbitrary scale,
satisfies the homogeneous form of the eqn.~(\ref{eq:invdwrg}),
\begin{equation}
\Lambda\frac{\partial\hat{\cal L}}{\partial\Lambda}-
\hat\kappa\frac{\partial\hat{\cal L}}{\partial\hat\kappa}+
\hat{\cal L}=0,
\end{equation}
and so may be used to remove the logarithmic term from $\hat J$.
Bringing all the terms together we take
\begin{equation}
\hat V_S^{(0)}(\hat p,\hat\kappa,\Lambda)=\Bigl(\hat J(\hat p,\hat\kappa)-
\hat{\cal M}(\hat p,\hat\kappa)-\hat{\cal L}(\hat\kappa,\Lambda)\Bigr)^{-1}
\end{equation}
as an analytic solution to the DWRG equation with logarithmic dependence on
$\Lambda$. We will refer to this as the non-trivial fixed point
\footnote{Although not
strictly a fixed point, it shall be referred to as such for ease of
nomenclature.}.

The perturbations around the non-trivial fixed point
can be found in the same way as before. Writing,
\begin{equation}
\frac{1}{\hat V_S(\hat p,\hat\kappa,\Lambda)}=
\frac{1}{\hat V_S^{(0)}(\hat p,\hat\kappa,\Lambda)}+
\Lambda^\nu\phi(\hat p,\hat\kappa),
\end{equation}
we obtain a linear equation for $\phi$ which is readily solved,
\begin{equation}
\phi(\hat p,\hat\kappa)=\hat p^{2n}\hat\kappa^m,
\end{equation}
where $n$ and $m$ are positive integers and the RG eigenvalues are
$\nu=2n+m-1$.  The leading order perturbation around this fixed point has a
negative RG eigenvalue and so, like the non-trivial fixed
point seen in the previous chapter, is unstable. Indeed, the power-counting in the
energy dependent terms around this fixed point is simply the KSW scheme.
In contrast to the power-counting observed there, the existence of a zero RG eigenvalue 
means that there is a marginal perturbation that does not
scale with a power of $\Lambda$.
The full solution in the vicinity of the non-trivial fixed point is,
\begin{equation}
\frac{1}{\hat V_S(\hat p,\hat\kappa,\Lambda)}
=\frac{1}{\hat V_S^{(0)}(\hat p,\hat\kappa,\Lambda)}
+\sum_{n,m=0}^\infty\hat C_{2n,m}\left(\frac{\Lambda}{\Lambda_0}\right)^{2n+m-1}
\hat p^{2n}\hat\kappa^m.
\end{equation}
The marginal perturbation is the key to understanding the 
need for logarithmic dependence on $\Lambda$ in the fixed point solution and
the arbitrary scale $\mu$. Since the marginal perturbation is independent of
$\Lambda$ it
cannot be separated unambiguously from the fixed point solution $\hat V_S^{(0)}$.
The degree of freedom associated with the scale $\mu$ is interchangeable with the
coefficient $\hat C_{01}$. Indeed the coefficient, $\hat C_{01}$ can be chosen to
depend on $\mu$ in such a way that the term,
\begin{equation}
\pi\hat\kappa\ln{\frac{\Lambda}{\mu}}+\hat C_{01}(\mu)\hat\kappa,
\end{equation}
and hence the full solution is independent of $\mu$.

\begin{figure}
\begin{center}
\includegraphics[height=13cm,width=8cm,angle=-90]{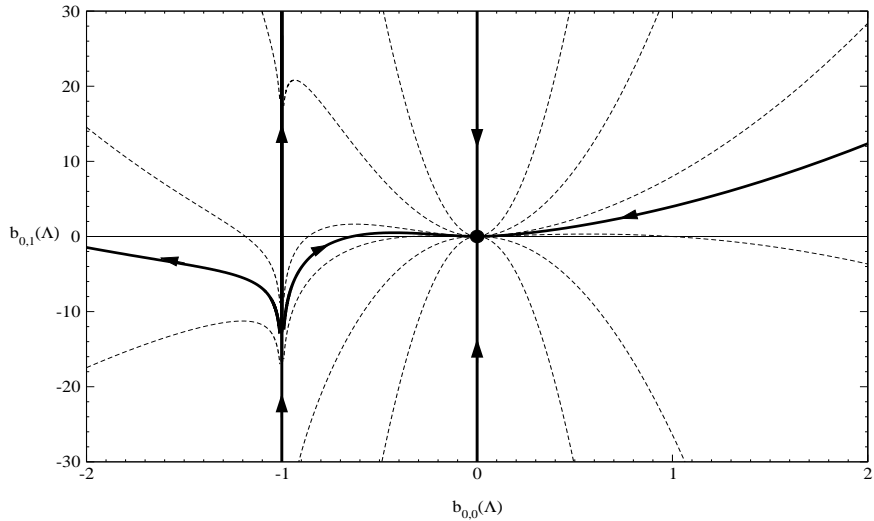}
\caption{The RG flow in the plane $(b_{0,0}(\Lambda),b_{0,1}(\Lambda))$,
where $b_{2n,m}(\Lambda)$ is defined in eqn.~(\ref{eq:Vbexp})}\label{fig:logRGflow}
\end{center}
\end{figure}

This system is very similar to the case of considered in Chapter \ref{Ch1}. The
power-counting in the energy dependent terms around both fixed-points is the
same as observed in the system with just short range forces.  
As a consequence, the RG flow 
and the conclusions drawn from its examination are very similar. Writing,
\begin{equation}\label{eq:Vbexp}
\hat V_S(\hat p,\hat\kappa,\Lambda)=\sum_{n,m=0}^{\infty}
b_{2n,m}(\Lambda)\hat p^{2n}\hat\kappa^{m},
\end{equation}
the flow in the $(b_{0,0}(\Lambda),b_{2,0}(\Lambda))$-plane is as given in Fig.
\ref{fig:SRRGflow}\footnote{The exact position of the non-trivial fixed point may
be different but this does not affect the conclusions.}.
In that diagram the logarithmic
behaviour associated with the non-trivial fixed point is not apparent but is
illustrated in Fig.\ref{fig:logRGflow}, which shows the flow in the plane
$(b_{0,0}(\Lambda),b_{0,1}(\Lambda))$. As before, the fixed point is
shown as a dot, the flows along the RG eigenvectors as bold lines with the
arrows showing the flow as $\Lambda\rightarrow0$ and the dashed lines
showing more general flow lines. The flow associated with the marginal perturbation
carries the potential up the vertical flow line $b_{0,0}=-1$ at a logarithmic rate.
Although the coefficient appears to be tending to infinity along this flow line, it
will eventually become so large that the expansion (\ref{eq:Vbexp}) breaks down.
A general potential, for which $b_{0,0}$ is not exactly $-1$, will flow
into the trivial fixed point. However, provided the coefficient of the
unstable perturbation, $\hat C_{0,0}$ is small, it is still possible to expand
around the logarithmic flow line.

Despite the logarithmic flow we are led
to a similar conclusion arrived at in the previous chapter. The
organisation of terms using the expansion around the non-trivial fixed point
is only suitable if there is some fine-tuning of the parameters, which give an
unnaturally small value for $\hat C_{0,0}$, otherwise the suitable expansion
is that about the trivial fixed point and the distorted wave Born expansion.

\subsection{The Distorted Wave Effective Range Expansion}
In the pure short range case, the terms
in the expansion around the non-trivial fixed point are in one to one
correspondence with the terms in the ERE. The requirement of small
$\hat C_{-1}$ for the suitability of this expansion could be reformulated
as the need for a large scattering length, $1/a\ll\Lambda_0$, as found
in nucleon-nucleon systems. 

The generalisation of the ERE is the DW effective range
expansion (DWERE) \cite{hkok,furnsteele,j&b,bethe}.
We will see that the energy dependent perturbations around the DWRG non-trivial fixed point are
in one-to-one correspondence to the terms in the DWERE.
To show this we insert the solution $\hat V_S$ into the
DWLS equation for $\tilde K_S$ to obtain an expression for the
corrective phaseshift, $\tilde\delta_S$. The DWLS equation can be solved by
expanding the Green's function in terms of a complete set of DWs and
iterating to get a geometric series. This series can then be summed to obtain
\begin{equation}\label{eq:firstdwere}
\frac{\langle\psi_p|\tilde K_S|\psi_p\rangle}{|\psi_p(R)|^2}=
V_S(p,\kappa,\Lambda)\left(1-V_S(p,\kappa,\Lambda)\frac{M}{2\pi^2}
\fint_0^\Lambda dq\frac{|\psi_q(R)|^2}{p^2-q^2}
\right)^{-1}.
\end{equation}
Notice that the integral in the above expression is equal to
$\Lambda R^2\hat J(\hat p,\hat\kappa)$. Inverting the equation and substituting the
expression for the $K$-matrix element in terms of the phaseshift,
eqn.~(\ref{eq:Kphase}), and the expression for
$|\psi_p(R)|^2$ for $R\ll p^{-1}$, eqn.~(\ref{eq:smallrcoul}), gives,
\begin{equation}
{\cal C(\eta)}p\cot{\tilde\delta_S}=\frac{2\Lambda}{\pi}\left(
\hat J(\hat p,\hat\kappa)-
\frac{1}{\hat V_S(\hat p,\hat\kappa,\Lambda)}\right).
\end{equation}
Substituting in the expression for $\hat V_S$ and writing everything in terms of
the physical variables gives the final expression,
\begin{equation}
{\cal C}(\eta)p\cot\tilde\delta_S-{\cal M}(p,\kappa)=
-2\kappa\ln{\frac{\kappa}{\mu}}
-\frac{2\Lambda_0}{\pi}\sum_{n,m}\hat C_{2n,m}\left(\frac{p^{2n}\kappa^{m}
}{\Lambda_0^{2n+m}}\right),
\end{equation}
where \cite{grad},
\begin{eqnarray}
&&\displaystyle{
{\cal M}(p,\kappa)=\frac{2\Lambda}{\pi}\hat{\cal M}(\hat p,\hat\kappa)
=\fint_0^\infty
\frac{dq}{q}\frac{4\kappa}{e^{2\pi\kappa/q}-1}
\frac{p^2}{p^2-q^2}}\nonumber\\
&&\displaystyle{\qquad\qquad\qquad\qquad\qquad\qquad
=2\kappa\text{Re}\left[\ln(i\eta)-\frac{1}{2i\eta}-\psi(i\eta)\right]}
\end{eqnarray}
and $\psi$ is the logarithmic derivative of the $\Gamma$-function.
This expansion is equivalent to the distorted wave effective range
expansion (DWERE) first derived by Bethe \cite{bethe} expanding on earlier work
by Landau and Smorodinski \cite{landaus} and more recently examined from an
EFT viewpoint by Kong and Ravndal \cite{kongr}. In the expansion all non-analytic
behaviour has been isolated in the functions ${\cal C}$ and $\cal M$
allowing an expansion in $p$ and $\kappa$.
The use of renormalisation group
methods in deriving this equation is a new result and provides an
interesting insight into the use of not only this expansion
but also the distorted wave Born approximation.
Beyond that it also provides the full power-counting for this system; more
than was shown in the work of Kong and Ravndal \cite{kongr}.
We may write the distorted wave effective range expansion as,
\begin{equation}
{\cal C}(\eta)p\cot\tilde\delta_S-{\cal M}(p,\kappa)=
-\frac{1}{\tilde a_C}+\frac{1}{2}\tilde r_C p^2+\ldots,
\end{equation}
allowing definition of a Coulomb-modified scattering length and
effective-range, 
\begin{eqnarray}\label{eq:cmodscatt}
\frac{1}{\tilde a_C}&=&\frac{2\Lambda_0\hat C_{0,0}}{\pi}+
\alpha M\left(\ln\frac{\alpha M}{2\mu}+C_{0,1}(\mu)\right)
+\frac{2\Lambda_0}{\pi}
\sum_{m=2}^\infty\hat C_{0,m}\left(\frac{\alpha M}{2\Lambda_0}\right)^m,\\
\tilde r_C&=&
-\frac{4}{\Lambda_0\pi}
\sum_{m=0}^{\infty}\hat C_{2,m}\left(\frac{\alpha M}{2\Lambda_0
}\right)^m,
\end{eqnarray}
These expansions in $\alpha M$ correspond to nucleon loops in photon
exchange diagrams in the EFT.

\subsection{Proton-Proton Scattering.}
The DWERE was first derived specifically to model low-energy
proton-proton scattering and proved very successful in doing so
\cite{bethe,j&b,kongr}.
The RG analysis tells us that the DWERE is only likely to provide a systematic expansion if
the parameter $\hat C_{0,0}$ is small allowing organisation of the terms
around the non-trivial fixed point.  In the case of neutron-neutron
scattering the same criterion was met because of the large scattering length,
$a_{nn}=-18.4$fm. In proton-proton scattering the issue is clouded since
the Coulomb-modified scattering length depends logarithmically upon the
the fine structure constant $\alpha$. If we assume an isospin symmetry for
the strong force, then as $\alpha\rightarrow0$, the DWERE for proton-proton
scattering should become the ERE for neutron-neutron (or proton-neutron)
scattering length, suggesting that eqn.~(\ref{eq:cmodscatt}) can be written as,
\begin{equation}\label{eq:ppnnscats}
\frac{1}{\tilde a_{pp}}=\frac{1}{a_{nn}}+
\alpha M\left(\ln\frac{\alpha M}{2\mu}+\hat C_{0,1}(\mu)\right)
+\ldots.
\end{equation}
Hence, the criteria for the use of the DWERE to systematically describe
proton-proton scattering is met because
$\hat C_{0,0}\sim\pi/(2m_\pi a_{nn})$ is small.

This rather naive analysis
must be taken with caution, the isospin symmetry of the strong interaction is only
approximate. Furthermore, the large separation of scales in the nuclear system
between $\Lambda_0$ and $1/a$ tends to amplify the isospin asymmetry of
the strong interactions due to Coulomb interactions at short scales. For example,
the naive argument above suggests $a_{np}=a_{nn}$. However we find that
$a_{nn}=-18.4$fm and $a_{np}=-23.7$fm in the spin-singlet channel, where
the 25\% difference in the scattering length is a result of a much smaller
difference in the unscaled effective potential.
That said, we can only cite studies that have successfully used the Coulomb
modified ERE in modelling proton-proton scattering \cite{kongr,bethe,j&b}
and the corresponding EFT to model proton-proton fusion \cite{kongr2}

\section{Repulsive Inverse-Square Potential}

The next example in this chapter is the repulsive inverse-square potential,
\begin{equation}\label{eq:invsqpot}
V_L(r)=\frac{\beta}{Mr^2}.
\end{equation}
This potential is of interest because, firstly,
the centrifugal barrier is of this form and
the analysis will show how the power-counting is constructed for
different angular momenta, and secondly, because of the potential's relevance
to the three body problem \cite{efimov}.

The Schr\"odinger equation is easily solved in the case where $\beta>-1/4$.
The DWs satisfying the boundary condition of vanishing amplitude at the
origin are
\begin{equation}
\psi_p(r)=\sqrt{\frac{\pi pr}{2}}J_\nu(pr),
\end{equation}
where $J_\nu(z)$ is Bessel's function and $\nu=\sqrt{\beta+1/4}$.
If $\beta<-1/4$ then $\nu$ becomes imaginary and the boundary condition
at the origin becomes an issue for discussion, this case is considered in the next chapter.
The long-range phaseshift is given by $\delta_L=\pi(1/4-\nu/2)$. The
DWRG equation for the short-range potential is determined
by the value at the wavefunction close to the origin,
\begin{equation}
|\psi_\Lambda(R)|^2=\frac{\pi}{2\Gamma(1+\nu)^{2}}
\left(\frac{\Lambda R}{2}\right)^{2\nu+1},
\end{equation}
giving from eqn.~(\ref{eq:udwrge2}) the differential equation for $V_S$,
\begin{equation}
\frac{\partial V_S}{\partial\Lambda}=\frac{M}{4\pi\Gamma(1+\nu)^{2}}
\left(\frac{\Lambda R}{2}\right)^{2\nu+1}
\frac{V_S^2}{\Lambda^2-p^2}.
\end{equation}
The rescaling of $V_S$ is markedly different to the previous
examples because of the odd dimension in the $\Lambda$ dependence
on the right-hand side. The equation is rescaled via the relationships,
\begin{equation}
\Lambda\hat p=p, \qquad \hat V_S(\hat p,\Lambda)=\frac{M}{4\pi\Gamma(1+\nu)^{2}}
\left(\frac{R}{2}\right)^{2\nu+1}\Lambda^{2\nu}V_S(\Lambda\hat p,\Lambda),
\end{equation}
resulting in the DWRG equation for $\hat V_S$,
\begin{equation}
\Lambda\frac{\partial\hat V_S}{\partial\Lambda}=
\hat p\frac{\partial\hat V_S}{\partial\hat p}+2\nu\hat V_S+
\frac{\hat V_S^2}{1-\hat p^2}.
\end{equation}
The RG analysis of this equation is straightforward but interesting \cite{tbmb,tbmb2}.
There are two fixed points, the trivial fixed point, $\hat V_S=0$, and a non-trivial fixed point.
If $\nu$ is an integer the latter of these has logarithmic $\Lambda$-dependence.
The perturbations around these fixed points scale differently to the cases seen so
far. 
In the two previous cases the trivial fixed point was stable, with the
LO perturbation scaling with $\Lambda$, while the non-trivial
fixed point was unstable, with the LO perturbation scaling
with $\Lambda^{-1}$. In this example, we shall see that the
trivial and non-trivial fixed points are still stable and unstable
respectively (for $\nu\neq0$), but that the
nature of the stable and unstable perturbations are
different, resulting in rather different power-counting schemes.

Perturbations around the trivial fixed point can be determined, as before,
by writing $\hat V_S(\Lambda,\hat p)=C\Lambda^{\sigma}\phi(\hat p)$,
substituting this into the RG equation and linearising. The resulting equation for
$\phi$,
\begin{equation}
\frac{d\phi}{d\hat p}=(\sigma-2\nu)\phi,
\end{equation}
has solutions $\phi(\hat p)=\hat p^{\sigma-2\nu}$, which upon application of
the analyticity boundary condition yields the potential,
\begin{equation}
\hat V_S(\hat p)=\sum_{n=0}^{\infty}
\hat C_{2n}\left(\frac{\Lambda}{\Lambda_0}\right)^{2n+2\nu}\hat p^{2n}.
\end{equation}
As promised, for $\nu\neq0$, the fixed point is stable. However, the LO
perturbation, instead of scaling with $\Lambda$, now scales with
$\Lambda^{2\nu}$. The case of $\nu=0$ is interesting, in this case, the LO
perturbation is marginal, which will be associated with some logarithmic
dependence on $\Lambda$. The term proportional to $p^{2n}$ is of
order $d=2n+2\nu-1$ in the corresponding power-counting.

The non-trivial fixed point is determined by solving the DWRG fixed point equation
\begin{equation}
\hat p\frac{\partial}{\partial\hat p}\left(\frac{1}{\hat V_S}\right)
-\frac{2\nu}{\hat V_S}+\frac{1}{1-\hat p^2}=0,
\end{equation}
which is easily solved by the integral,
\begin{equation}
\hat J(\hat p)=\fint_0^1d\hat q\frac{\hat q^{2\nu+1}}{\hat p^2-\hat q^2}
=\frac{1}{2}\sideset{}{'}\sum_{n=0}^\infty \frac{\hat p^{2n}}{n-\nu}+
\frac{\pi}{2}\hat p^{2\nu}\hat{\cal M}(\hat p,\nu),
\end{equation}
where,
\begin{equation}
\hat{\cal M}(\hat p,\nu)=
\begin{cases}
\cot{\pi\nu}, & \nu\not\in{\mathbb{N}}\\
2\ln{\hat p}, & \nu\in\mathbb N.
\end{cases}
\end{equation}
The prime on the sum here indicates that the term with $n=\nu$ must be omitted when
$\nu$ is an integer. The non-analytic term in the solution $\hat J(\hat p)$ can
be subtracted off in the case of $\nu\notin\mathbb N$ as it satisfies the
homogeneous fixed point DWRG equation. In the case of $\nu\in\mathbb N$
the logarithmic dependence upon $\hat p$ must be removed in the manner
outlined in the Coulomb example
to give an analytic solution which may be expressed as,
\begin{equation}
\frac{1}{\hat V_0(\hat p)}=\hat J(\hat p)-\frac{\pi}{2}
\hat p^{2\nu}{\cal M}\left(
\frac{\hat p\Lambda}{\mu},\nu\right).
\end{equation}
The perturbations around this fixed point are easily found,
\begin{equation}
\frac{1}{\hat V_S(\hat p)}=\frac{1}{\hat V_0(\hat p)}+
\sum_{n=0}^\infty \hat C_{2n}\left(\frac{\Lambda}{\Lambda_0}\right)
^{2n-2\nu}\hat p^{2n}.
\end{equation}
This fixed point is unstable with the number of unstable eigenvectors
being determined by $\nu$. If $\nu$ lies between the integers $N-1$ and $N$ then
the first $N$ perturbations are unstable. If $\nu=N$ then there is also a marginal
eigenvector, $\hat p^{2N}$, which is associated with the logarithmic behaviour
in the usual manner.

The power-counting around the non-trivial fixed point is $d=2n-2\nu-1$ for a
term proportional to $p^{2n}$.  Both of the power-counting schemes are quite
different from the Weinberg or KSW schemes seen so far. Since the inverse-square
potential is scale-free, its strength does not provide an expansion parameter in the
low-energy EFT, instead it appears in the energy power-counting itself.

\begin{figure}
\begin{center}
\includegraphics[height=13cm,width=8cm,angle=-90]{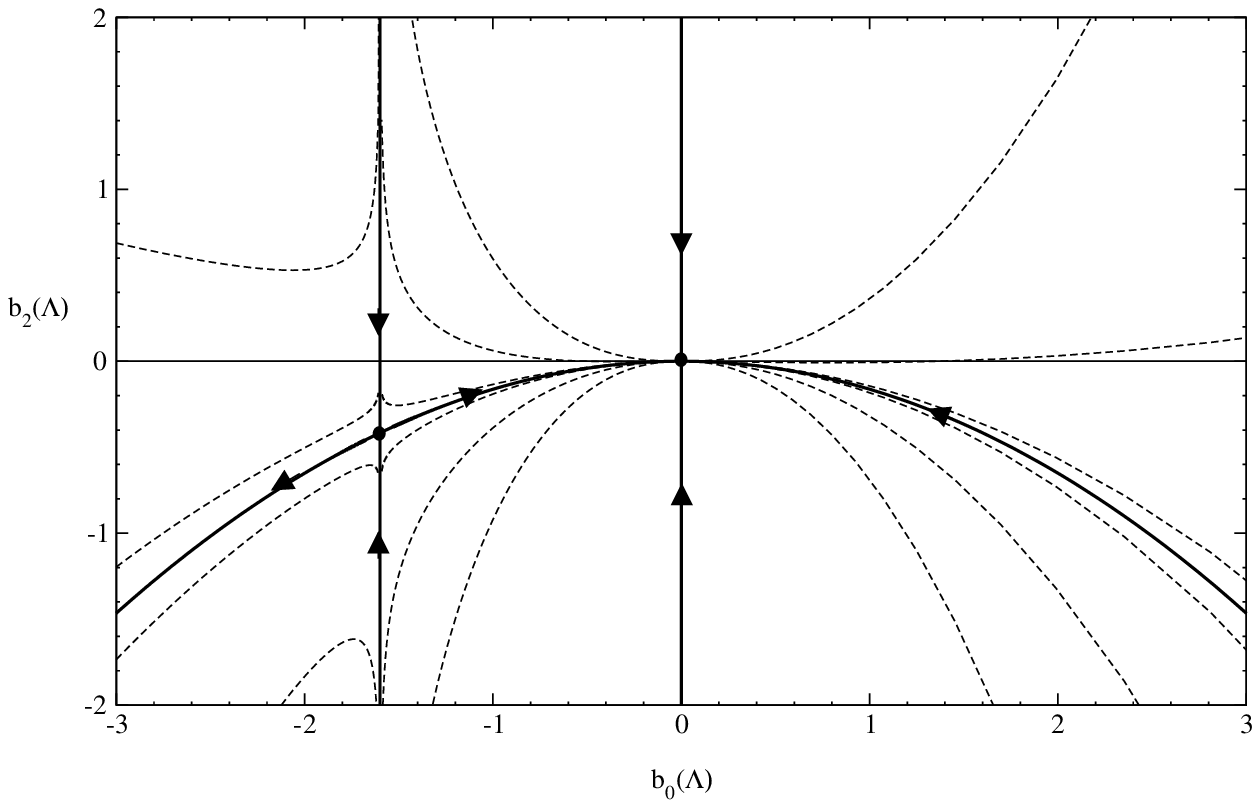}
\caption{The RG flow for $\nu=0.8$ in the plane $(b_{0}(\Lambda),b_{2}(\Lambda))$,
where $\hat V_S(\hat p,\Lambda)=\sum_{n=0}^\infty b_{2n}(\Lambda)\hat p^{2n}$.
The trivial fixed point is stable, the non-trivial fixed point is unstable
with a stable NLO perturbation.}
\label{fig:invRGflow1}
\end{center}
\end{figure}

\begin{figure}
\begin{center}
\includegraphics[height=13cm,width=8cm,angle=-90]{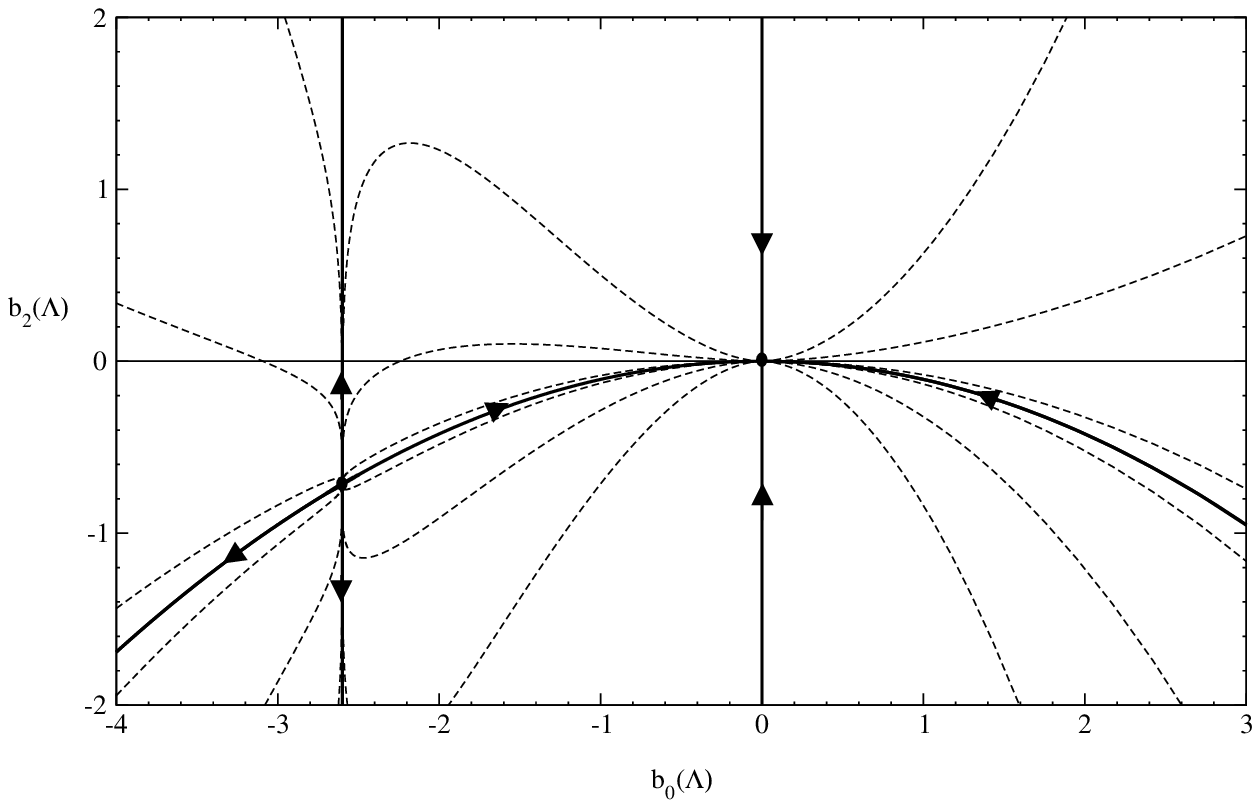}
\caption{The RG flow for $\nu=1.2$ in the plane $(b_{0}(\Lambda),b_{2}(\Lambda))$,
where $\hat V_S(\hat p,\Lambda)=\sum_{n=0}^\infty b_{2n}(\Lambda)\hat p^{2n}$.
The trivial fixed point is stable, the non-trivial fixed point is unstable
with an unstable NLO perturbation.}
\label{fig:invRGflow2}
\end{center}
\end{figure}

Because of the difference in stability in the fixed points from the cases examined
thus far, the RG flow is quite different. The RG flow is illustrated in the familiar way in
Figs.~(\ref{fig:invRGflow1},\ref{fig:invRGflow2}). Fig.~(\ref{fig:invRGflow1}) shows the
flow in the $(b_0(\Lambda),b_2(\Lambda))$ plane with $\nu=0.8$. The trivial fixed point
occurs at $(0,0)$ with the non-trivial fixed point at $(-1/(2\nu),1/(2-2\nu))$.
The solutions flow towards the non-trivial fixed point close to the critical line
as in the short range case (Fig.~\ref{fig:SRRGflow}). As the unstable perturbation
becomes important the RG flow peels away from the critical line very quickly and
into the trivial fixed point. As the strength of the inverse square potential increases
the rate at which the flow peels away from the critical line increases until at $\nu=1$
the flow along that critical line becomes marginal resulting in a flow like that
illustrated in Fig.~\ref{fig:logRGflow}. For $\nu>1$ the
flow on the critical line becomes unstable. Fig.~\ref{fig:invRGflow2} shows the
RG flow for $\nu=1.2$. In this figure any flow in the region of the non-trivial
fixed point is pushed away by the unstable perturbations and into the trivial fixed point.

As the strength of the inverse square potential
increases, the instability of the non-trivial fixed point `increases', i.e.
the number of unstable perturbations and the order $\nu$ of the unstable
perturbation increases. At the same time
the stability of the trivial fixed point also `increases'.
In order to keep the RG flow in the vicinity of the non-trivial fixed point,
all coefficients associated with unstable perturbations, $\hat C_0$ up to $\hat C_{2N-1}$
must be finely tuned. For large values of $\nu$ this fine tuning becomes more and
more contrived.

The expansion around the non-trivial fixed point can be expressed in terms
of the DWERE,
\begin{equation}
p^{2\nu}\left(\cot\tilde\delta_S-{\cal M}(p/\mu,\nu)\right)=
-\frac{2\Lambda_0^{2\nu}}{\pi}\sum_{n=0}^\infty\hat C_{2n}
\left(\frac{p}{\Lambda_0}\right)^{2n}.
\end{equation}
However, in order for this expansion to provide a systematic organisation of
the terms we require fine tuning of all effective couplings associated with
unstable perturbations.

In the case of scattering of a particle with angular momentum $l$ by
a short-range potential, we have $\nu=l+1/2$ and there is no
non-analytic energy dependence in $\cot(\delta+l\pi/2)$. We can
identify two expansions, one based on the trivial fixed point
\begin{equation}
p^{-2l-1}\tan\left(\delta+\frac{l\pi}{2}\right)=
-\frac{\pi}{2\Lambda_0^{2l+1}}\sum_{n=0}^{\infty}\hat C_{2n}
\left(\frac{p}{\Lambda_0}\right)^{2n},
\end{equation}
the other on the non-trivial fixed point,
\begin{equation}
p^{2l+1}\cot\left(\delta+\frac{l\pi}{2}\right)=
-\frac{2\Lambda_0^{2l+1}}{\pi}\sum_{n=0}^{\infty}\hat C_{2n}
\left(\frac{p}{\Lambda_0}\right)^{2n}.
\end{equation}
If the parameters in the latter expansion are natural then it is equivalent
to the expansion around the trivial fixed point as can be seen by
inverting the left and right hand sides.
However, if $\hat C_{2n},\,(n<N)$ are finely tuned to a small amount
then the expansion obtained by inverting the
equation will have increasingly large coefficients and will not
be a systematic.
The fine tuning required for this expansion will show itself
as shallow bound states or resonances. In nucleon-nucleon
scattering there are no shallow bound states or resonances in the higher
partial waves and the suitable power-counting is that associated with the
trivial fixed point.

\section{``Well-behaved'' potentials}
Having looked at two examples in detail, we shall now turn our attention to
a more general class of potentials, those for which the
Jost function exists. That is all potentials for which,
\begin{equation}\label{eq:potconstraints}
\int_0^\infty dr r |V_L(r)|<\infty,\qquad
\int_0^\infty dr r^2 |V_L(r)|<\infty.
\end{equation}
We shall assume that $V_L$ depends on the low energy scales $\kappa_i$.

\subsection{The Jost function}
We present here a quick overview of Jost's solutions to the Schr\"odinger equation
and the Jost function, for a more thorough analysis see Newton \cite{newton}.
We are interested
in two types of solutions of the Schr\"odinger equation, the
regular solution $\varphi(p,\kappa_i,r)$ and the Jost solutions $f_{\pm}(p,\kappa_i,r)$.
These satisfy the boundary conditions:
\begin{equation}\label{eq:regbcs}
\varphi(p,\kappa_i,0)=0,\qquad
\varphi'(p,\kappa_i,0)=1,
\end{equation}
where the prime indicates differentiation with respect to $r$, and
\begin{equation}
\label{eq:irregbcs}f_{\pm}(p,\kappa_i,r)\rightarrow e^{\pm ipr}\,\,\,
\text{as}\,\,\,r\rightarrow\infty.
\end{equation}
These solutions are interesting as their simple boundary conditions
allow us to consider their properties as an analytic function of the
complex variable $p$. By writing these solutions as power series
and assuming the constraints (\ref{eq:potconstraints}) upon the potential
$V_L$, we may show that $\varphi(p,\kappa_i,r)$ is an entire function of $p^2$ and
that $f_+(p,\kappa_i,r)$ $(f_-(p,\kappa_i,r))$ is an analytic function of $p$ in the upper
(lower) half of the complex plane \cite{newton}.

By analytically continuing $f_+(p,\kappa_i,r)$ through the upper half of the
complex $p$-plane we arrive at $f_+(-p,\kappa_i,r)$, which satisfies the same
boundary condition as $f_-(p,\kappa_i,r)$. Hence, for $p>0$ we have,
\begin{equation}
f_-(p,\kappa_i,r)=f_+(pe^{+i\pi},\kappa_i,r).
\end{equation}
Since $f_+$ and $f_-$ are linked by analytic continuation we shall
write $f_+(p,\kappa_i,r)=f(p,\kappa_i,r)$ and $f_-(p,\kappa_i,r)=f(-p,\kappa_i,r)$. From the reality of $V_L$ and the
boundary conditions, it follows that
$\varphi(p,\kappa_i,r)$ is real for real $p$ and that
\begin{equation}\label{eq:conjugate}
[f(p^*,\kappa_i,r)]^*=f(-p,\kappa_i,r).
\end{equation}

We introduce the Jost function, ${\cal F}(p,\kappa_i)$,
by writing the regular solution as a superposition of the Jost solutions,
\begin{equation}\label{eq:regintermsofirreg}
\varphi(p,\kappa_i,r)=\frac{1}{2ip}\Bigl({\cal F}(-p,\kappa_i)f(p,\kappa_i,r)-{\cal F}(p,\kappa_i)
f(-p,\kappa_i,r)\Bigr).
\end{equation}
From the boundary conditions on $\varphi(p,\kappa_i,r)$ and flux conservation it
follows that
\begin{equation}\label{eq:jostasymptote}
{\cal F}(p,\kappa_i)=f(p,\kappa_i,0).
\end{equation}
The analytic properties of ${\cal F}$ as a complex function of $p$ follow from those of $f$.
i.e ${\cal F}$ is analytic in the upper half of the complex plane.
The Jost function also satisfies the conjugate relation (\ref{eq:conjugate}). 
The Jost function is extremely useful as it provides
all the information we require to obtain both the long range phaseshift,
$\delta_L(p)$, and the DWRG equation.

Since the physical wavefunctions must vanish at the origin they must be proportional
to the regular solution, $\varphi(p,\kappa_i,r)$. 
The DWs, $\psi_p(r)$, are found by demanding the normalisation:
\begin{equation}\label{eq:norm}
\psi_p(r)\rightarrow\sin(pr+\delta(p))\,\,\,\text{as}\,\,\,r\rightarrow\infty.
\end{equation}
Comparing the asymptotic form for $\psi_p(r)$ to that for $\varphi(p,\kappa_i,r)$ obtained from
eqns.(\ref{eq:irregbcs},\ref{eq:regintermsofirreg}) we obtain,
\begin{equation}\label{eq:defDW}
\psi_p(r)=p\frac{\varphi(p,\kappa_i,r)}{\sqrt{{\cal F}(p,\kappa_i){\cal F}(-p,\kappa_i)}}.
\end{equation}
From the asymptotic form, eqn.~(\ref{eq:norm}) we can also obtain an expression
for the $S$-matrix:
\begin{equation}\label{eq:jostsmatrix}
e^{2i\delta_L(p)}=\frac{{\cal F}(-p,\kappa_i)}{{\cal F}(p,\kappa_i)}.
\end{equation}

In the case of the bound states we have the
boundary condition of vanishing amplitude as $r\rightarrow\infty$.
For $p$ positive imaginary and for large $r$ we have, from eqn.~(\ref{eq:regintermsofirreg}),
\begin{equation}
\varphi(p,\kappa_i,r)\rightarrow\frac{1}{2ip}\Bigl({\cal F}(-p,\kappa_i)e^{-|p|r}-
{\cal F}(p,\kappa_i)e^{|p|r}\Bigr),
\end{equation}
which will vanish for large $r$ if ${\cal F}(p,\kappa_i)=0$. This shows that the bound states
are given by the zeros of the Jost function on the positive imaginary axis.
The normalisation of the bound state solutions is quite tricky and its proof
(see Newton \cite{newton}) offers no insight so we just state the result:
\begin{equation}\label{eq:boundnorm}
\psi_n(r)=\frac{2ip_n}{\sqrt{{\cal F}(-ip_n,\kappa_i)
\dot{\cal F}(ip_n,\kappa_i)}}\varphi(ip_n,\kappa_i,r),
\end{equation}
where the dot signifies differentiation with respect to $p$.

\subsection{The DWRG equation}
In the DWRG equation we require the magnitude of the DWs near to the origin,
this follows easily from the boundary condition on $\varphi(p,\kappa_i,r)$.
For small $r$, $\varphi(p,\kappa_i,r)\rightarrow r$ so that
for $R\ll\Lambda^{-1}$ we have,
\begin{equation}\label{eq:shortrangeDW}
|\psi_\Lambda(R)|^2=\frac{\Lambda^2 R^2}{{\cal F}(\Lambda,\kappa_i)
{\cal F}(-\Lambda,\kappa_i)}.
\end{equation}
The differential equation (\ref{eq:udwrge2}) becomes
\begin{equation}
\frac{\partial V_S(p,\Lambda)}{\partial\Lambda}=
-\frac{MR^2}{2\pi^2{\cal F}(\Lambda,\kappa_i)
{\cal F}(-\Lambda,\kappa_i)}
\frac{\Lambda^2}{p^2-\Lambda^2}V_S^2(p,\Lambda).
\end{equation}
To rescale this equation we note that the Jost function ${\cal F}$
is dimensionless (this follows from the dimensionless boundary condition on the Jost
solution $f$), so that
the rescaling of the potential is exactly the same as that for the Coulomb potential,
\begin{equation}
\hat V_S(\hat p,\hat\kappa_i,\Lambda)=\frac{M\Lambda R^2}{2\pi^2}
V_S(\Lambda\hat p,\Lambda\hat\kappa_i,\Lambda),
\end{equation}
so that the rescaled DWRG equation is,
\begin{equation}
\Lambda\frac{\partial\hat V_S}{\partial\Lambda}=
\hat p\frac{\partial\hat V_S}{\partial\hat p}+
\sum_i\hat\kappa_i\frac{\partial\hat V_S}{\partial\hat\kappa_i}+
\hat V_S+\frac{{\cal C}(\hat\kappa_i)}{1-\hat p^2}\hat V_S^2,
\end{equation}
where
\begin{equation}
{\cal C}(\hat\kappa_i)=\frac{1}{{\cal F}(\Lambda,\Lambda\hat\kappa_i)
{\cal F}(-\Lambda,\Lambda\hat\kappa_i),}
\end{equation}
which is independent of $\Lambda$ since it is dimensionless.

Once again, we identify the trivial fixed point, $\hat V_S=0$. 
Given the similarity of the DWRG equation to those considered in the
pure-short range case and also in the Coulomb case, the analysis
of this fixed point is identical to those examples. It is stable
with the LO perturbation scaling with $\Lambda$. The power-counting
is the Weinberg scheme augmented with additional terms in $\hat\kappa_i$
that are easily resolved. The correction to the phaseshift is
given by the DW Born expansion, eqn.~(\ref{eq:dwbe}).

\begin{figure}
\begin{center}
\includegraphics[height=8cm,width=8cm,angle=0]{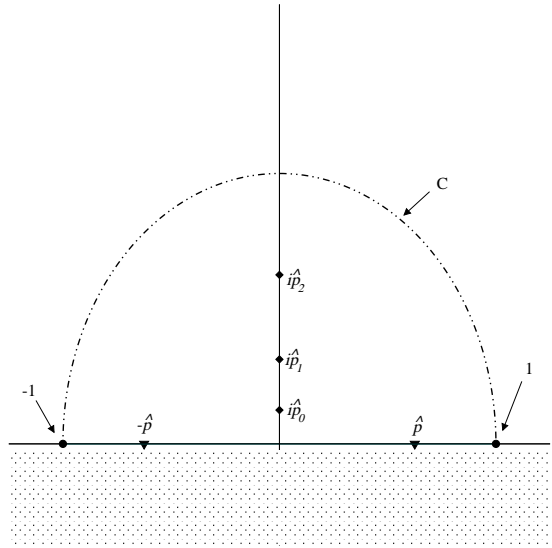}
\caption{The contour for the non-trivial fixed point solution of the DWRG
equation. It must run from $-1$ to $1$ going outside all bound state poles.}
\label{fig:jostpoles}
\end{center}
\end{figure}

Dividing the DWRG equation through by $\hat V_S^2$ and writing it as a linear
PDE in $\hat V_S^{-1}$ we obtain,
\begin{equation}
\Lambda\frac{\partial}{\partial\Lambda}\left(\frac{1}{\hat V_S}\right)=
\hat p\frac{\partial}{\partial\hat p}\left(\frac{1}{\hat V_S}\right)+
\sum_i\hat\kappa_i\frac{\partial}{\partial\hat\kappa_i}\left(\frac{1}{\hat V_S}\right)-
\left(\frac{1}{\hat V_S}\right)-\frac{{\cal C}(\hat\kappa_i)}{1-\hat p^2}.
\end{equation}
In parallel to the solutions for the Coulomb and repulsive inverse square
potentials our starting point for a non-trivial fixed-point solution to this equation is
the basic loop integral,
\begin{equation}\label{eq:bloop}
\hat J(\hat p,\hat\kappa_i)=
\fint_0^1d\hat q\frac{\hat q^2}{\hat p^2-\hat q^2}{\cal C}(\hat\kappa_i/\hat q).
\end{equation}
To isolate the non-analytic behaviour in this integral we need to
understand the analytic properties of ${\cal C}$. Since
\begin{equation}\label{eq:defineC}
{\cal C}(\hat\kappa/\hat q)=\frac{1}{{\cal F}(\hat q,\hat\kappa_i){\cal F}(-\hat q,\hat\kappa_i)},
\end{equation}
and ${\cal F}(p,\kappa_i)$ and ${\cal F}(-p,\kappa_i)$ are only analytic in the upper and lower
half of the complex $p$-plane respectively,
we cannot analytically continue ${\cal C}$ as function of $\hat q$ into the
complex plane at all. To circumvent this problem we note that since,
\begin{equation}\label{eq:niceident}
\varphi'(p,\kappa_i,0)=\lim_{r\rightarrow0}
\frac{1}{2ip}\Bigl({\cal F}(-p,\kappa_i)f'(p,\kappa_i,r)-
{\cal F}(p,\kappa_i)f'(-p,\kappa_i,r)\Bigr)=1,
\end{equation}
we can write
\begin{equation}\label{eq:analC0}
{\cal C}(\hat\kappa_i/\hat q)=\lim_{r\rightarrow0}
\frac{1}{2i\hat q}\left(\frac{f'(\hat q,\hat\kappa_i,r)}
{{\cal F}(\hat q,\hat\kappa_i)}-\frac{f'(-\hat q,\hat\kappa_i,r)}
{{\cal F}(-\hat q,\hat\kappa_i)}\right).
\end{equation}
This form for ${\cal C}$ is more promising as it is now written as the difference
of two functions, one analytic in the upper half of the complex $\hat q$-plane
and the other in the lower half. If the long-range potential does not have
an $r^{-1}$ singularity as $r\rightarrow0$ then we may define the
limit\footnote{See later for dealing with the $r^{-1}$ singularity},
\begin{equation}\label{eq:defM}
\hat{\cal M}(\hat\kappa_i,\hat q)=
\frac{f'(\hat q,\hat\kappa_i,0)}{{\cal F}(\hat q,\hat\kappa_i)},
\end{equation}
so that,
\begin{equation}\label{eq:analC}
{\cal C}(\hat\kappa_i/\hat q)=
\frac{1}{2i\hat q}\left(\hat{\cal M}(\hat q,\hat\kappa_i)-
\hat{\cal M}(-\hat q,\hat\kappa_i)\right),
\end{equation}
and the basic loop integral can be written as,
\begin{equation}\label{eq:jostfixedpoint}
\hat J(\hat p,\hat\kappa_i)=
\frac{1}{2i}\int_{-1}^1 d\hat q\frac{\hat q}{\hat p^2-\hat q^2}
\hat{\cal M}(\hat q,\hat\kappa_i).
\end{equation}
The integrand in this integral is meromorphic in the upper half of the
complex $\hat q$ plane. It has poles at the zeroes of the Jost function,
which correspond to bound states, and propagator poles at $\hat q=\pm\hat p$.
We may ensure analytic properties
of the integral by moving the contour of integration into the complex plane
and not allowing the singularities to `pinch' the contour. We define the
contour of integration, $C$, to run from $-1$ to $1$ and follow a path
in the upper half of the complex $\hat q$-plane that avoids the point
$\hat q=0$ and ensures analyticity in $\hat p$. This means that it must
also go outside all the bound state poles (see Fig.~\ref{fig:jostpoles}).
to avoid getting `pinched' between two of them as
$\hat\kappa_i\rightarrow0$. Hence we write the non-trivial fixed point
solution as,
\begin{equation}\label{eq:contourV}
\frac{1}{\hat V_S^{(0)}(\hat p,\hat\kappa_i)}=\frac{1}{2i}\int_C
d\hat q\frac{\hat q}{\hat p^2-\hat q^2}\hat{\cal M}(\hat q,\hat\kappa_i).
\end{equation}
It is an analytic function of all scales $\hat\kappa_i$ provided the potential
does not violate the constraints (\ref{eq:potconstraints}) or develop an
$r^{-1}$-singularity as each goes to zero. If these conditions are
violated as some scales go to zero then there will be a need to subtract
logarithms using the method outlined in the Coulomb example. These logarithmic
counterterms will be associated with marginal perturbations in the RG.

The construction of the perturbations around $\hat V_S^{(0)}$ 
follows in much the same way as in the Coulomb example,
\begin{equation}
\frac{1}{\hat V_S(\hat p,\hat\kappa,\Lambda)}=\frac{1}{\hat V_S^{(0)}
(\hat p,\hat\kappa,\Lambda)}
+\sum_{n,m_i=0}^\infty\hat C_{2n,m_1,m_2,\ldots}
\left(\frac{\Lambda}{\Lambda_0}\right)^{2n+m_1+m_2+\ldots-1}
{\hat p}^{2n}{\hat\kappa_1}^{m_1}{\hat\kappa_2}^{m_2}\ldots.
\end{equation}
The LO perturbation is unstable as $\Lambda\rightarrow 0$, the NLO
perturbations are marginal and correspond to the logarithmic counter terms\footnote{Not
all low-energy scales will have this perturbation
as they may satisfy an analyticity in $\hat\kappa_j^2$ boundary condition, rather than
simply in $\hat\kappa_j$.}.
All other perturbations are stable as $\Lambda\rightarrow0$. The power-counting
scheme in the $\hat\kappa_i$-independent terms is precisely the KSW scheme.
The RG flow is the same as described in the Coulomb example and the arguments
about the usefulness of each power-counting scheme are the same.

The form for the basic loop integral \ref{eq:contourV} is extremely useful and
will form the basis for non-trivial solutions in later chapters. The important
property it possesses is that it is always analytic in the energy $\hat p$.

\subsection{The DWERE and Interpretation of the Fixed Points} 
\begin{figure}
\begin{center}
\includegraphics[height=8cm,width=8cm,angle=0]{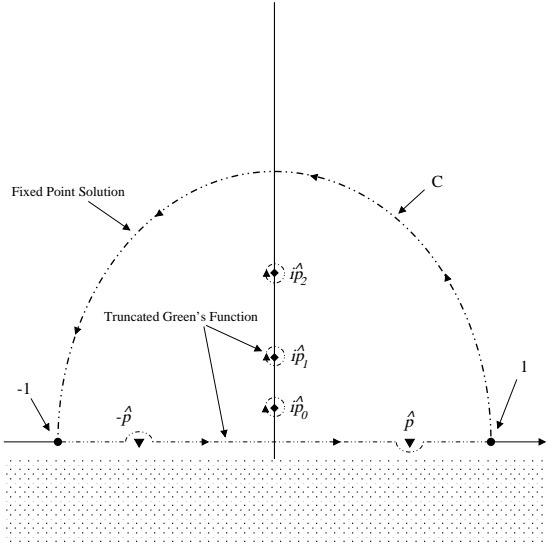}
\caption{Evaluation of the removed non-analytic term in the general DWERE}
\label{fig:jostdwere}
\end{center}
\end{figure}

In all our examples, the physical interpretation of the trivial fixed point has,
for want of a better word, been trivial. In the pure short range case the
non-trivial fixed point corresponded to a system with a bound state at exactly threshold.
Since we began our analysis of the DWRG we have used the non-trivial fixed point as a
tool for constructing power-counting schemes and for understanding the RG flow but have
failed to give an interpretation of it. To remedy this situation let us derive a DWERE by using the
the non-trivial fixed point in this example.
We substitute the potential into the equation for the $\tilde T$-matrix,
\begin{equation}\label{eq:firstjostdwere}
\frac{\langle\psi_p|\tilde T_S|\psi_p\rangle}{|\psi_p(R)|^2}=
V_S(p,\kappa,\Lambda)\left(1-V_S(p,\kappa,\Lambda)\left[\frac{M}{2\pi^2}
\int_0^\Lambda dq\frac{|\psi_q(R)|^2}{p^2-q^2+i\epsilon}
+\frac{M}{4\pi}\sum_n\frac{|\psi_n(R)|^2}{p^2+p_n^2}\right]
\right)^{-1}.
\end{equation}
The term in square brackets on the right hand side may be written as
\begin{equation}
[\ldots]=\frac{M\Lambda R^2}{4i\pi^2}\left[\int_{-1}^1d\hat q\frac{\hat q}{\hat p^2-\hat q^2}
\frac{f'(\hat q,0)}{{\cal F}(\hat q,\hat\kappa_i)}-2\pi i
\,{\cal R}\left\{\frac{\hat q}{\hat p^2-\hat q^2+i\epsilon}
\frac{f'(\hat q,0)}{{\cal F}(\hat q,\hat\kappa_i)},p\rightarrow ip_n\right\}\right],
\end{equation}
where ${\cal R}\{f(z),z\rightarrow z_0\}$ indicates the residue of $f(z)$ at $z=z_0$.
This result follows from definitions (\ref{eq:shortrangeDW},\ref{eq:defineC},\ref{eq:analC})
and from the result for the normalisation of the bound states, (\ref{eq:boundnorm}).
Inverting eqn.~(\ref{eq:firstjostdwere}) and using the expression for the
$\tilde T$-matrix in terms of the phaseshift correction we obtain after some algebra,
\begin{equation}
\frac{p}{|{\cal F}(p,\kappa_i)|^{2}}(\cot\tilde\delta_S-i)+
{\cal M}(p,\kappa_i)=-
\frac{2}{\pi\Lambda_0}\sum_{n,m_i=0}^\infty\hat C_{2n,m_1,m_2,\ldots}
{\Lambda_0}^{-2n-m_1-m_2-\ldots}
{p}^{2n}{\kappa_1}^{m_1}{\kappa_2}^{m_2}\ldots,
\end{equation}
where ${\cal M}(\Lambda\hat p,\Lambda\hat\kappa_i)=\Lambda\hat{\cal M}(\hat p,\hat\kappa_i)$,
we have absorbed any logarithmic counterterms into the marginal couplings
and we have used the result:
\begin{eqnarray}
&&\displaystyle{
{\cal M}(p,\kappa_i)=\frac{2\Lambda}{\pi}
\Biggl[\frac{1}{2i}\int_{-1}^1d\hat q\frac{\hat q}{\hat p^2-\hat q^2+i\epsilon}
\frac{f'(\hat q,0)}{{\cal F}(\hat q,\hat\kappa)}-
\frac{1}{\hat V_S^{(0)}(\hat p,\hat\kappa)}}\nonumber\\
&&\displaystyle{\qquad\qquad\qquad\qquad\qquad\qquad
-\pi
\,{\cal R}\left\{\frac{\hat q}{\hat p^2-\hat q^2}
\frac{f'(\hat q,0)}{{\cal F}(\hat q,\hat\kappa_i)},p\rightarrow ip_n\right\}\Biggr],}
\end{eqnarray}
which follows from evaluating what is now a closed
contour integral using Cauchy's theorem, cancelling the bound state residues
and evaluating the remaining residue at $\hat q=+\hat p$ (See Fig.~\ref{fig:jostdwere}).
The final result is expressed as,
\begin{equation}\label{eq:jostdwere}
|{\cal F}(p,\kappa_i)|^{-2}p(\cot\tilde\delta_S-i)+\frac{f'(p,\kappa_i,0)}{{\cal F}(p,\kappa_i)}=
-\frac{1}{\tilde a}+\frac{1}{2}\tilde r_e p^2+\ldots,
\end{equation}
where the distorted wave scattering length and effective range may
contain logarithmic dependence on some of the scales $\kappa_i$.
Relationship (\ref{eq:analC}) shows that the imaginary parts on the
LHS cancel to ensure a real phaseshift, $\tilde\delta_S$.

The DWERE (\ref{eq:jostdwere}) has been derived before by van Haeringen and Kok \cite{hkok},
following the work of many others (see their references),
in a somewhat more mathematically inspired manner. The advantage of this derivation is
that, because of the EFT philosophy behind the derivation,
we gain a physical rather than mathematical interpretation of how and when it may be used,
i.e. when the series is likely to converge well.

To gain insight into the non-trivial fixed point let us examine what the DWERE
looks like in the limit of all perturbations around the non-trivial fixed
point going to zero. The DWERE when the solution is taken to be the
non-trivial fixed point with no perturbations
(i.e. $\tilde a\rightarrow\infty,\,\tilde r_e\rightarrow0$ etc)
can be written as:
\begin{equation}
\cot\tilde\delta_S=-\frac{1}{p}{\cal F}(p,\kappa_i)f'(-p,\kappa_i,0)+i.
\end{equation}
This may be re-written as an equation for $e^{2i\tilde\delta_S}$:
\begin{equation}
e^{2i\tilde\delta_S}=\frac{{\cal F}(-p,\kappa_i)f'(p,\kappa_i,0)-2ip}{
{\cal F}(-p,\kappa_i)f'(p,\kappa_i,0)}=\frac{{\cal F}(p,\kappa_i)f'(-p,\kappa_i,0)}
{{\cal F}(-p,\kappa_i)f'(p,\kappa_i,0)},
\end{equation}
where the second identity follows from eqn.~(\ref{eq:niceident}).
When this is combined with the form for $e^{2i\delta_L}$, eqn.~(\ref{eq:jostsmatrix}),
we obtain an equation for the full $S$-matrix taking account of both
long and short range physics,
\begin{equation}
e^{2i\delta}=e^{2i\delta_L}e^{2i\tilde\delta_S}
=\frac{f'(-p,\kappa_i,0)}{f'(p,\kappa_i,0)}.
\end{equation}
This equation leads to two interpretations of the non-trivial fixed point,
one mathematical and the other physical.

Mathematically, to understand what this equation says let us look at the resulting
wavefunctions, $\tilde\psi_p(r)$.
Given the form of the $S$-matrix we know that at large $r$ the wavefunctions must
go like
\begin{equation}
\tilde\psi_p(r)\rightarrow \frac{1}{2ip}
\bigl(f'(-p,\kappa_i,0)e^{ipr}-f'(p,\kappa_i,0)e^{-ipr}\bigr)
\end{equation}
and hence are given by (recalling that the short range force acts at zero
radius, so that for $r>0$ the Schr\"odinger equation is unchanged from its
long range counterpart),
\begin{equation}
\tilde\psi_p(r)=\frac{1}{2ip}
\bigl(f'(-p,\kappa_i,0)f(p,\kappa_i,r)-f'(p,\kappa_i,0)f(-p,\kappa_i,r)\bigl).
\end{equation}
As $r\rightarrow0$ these solutions satisfy the Neumann boundary condition,
$\tilde\psi'(p,\kappa_i,r)=0$, rather than the usual Dirichlet boundary condition.
This offers us a new interpretation of the trivial and non-trivial fixed
points in terms of boundary conditions on the DWs. The trivial fixed point
corresponds to the normal boundary condition of vanishing wavefunctions at
$r=0$, while the non-trivial fixed point corresponds to the boundary
condition of vanishing derivative at $r=0$.

The two
solutions to the Schr\"odinger equation with these two different
boundary conditions are linearly independent. Hence, the non-trivial fixed point
changes each DW into a solution linearly independent to it and, in effect,
changes it `by the maximum amount possible'.
This interpretation of the non-trivial fixed point is also appropriate (as
of course it should be) for the example in chapter $2$, in which there
was no long range potential, and had been implicitly noted by van Kolck \cite{uvk}.

Physically, we can understand the non-trivial fixed point in terms of a bound
state at zero energy. As $p\rightarrow0$,
$f'(p,\kappa_i,0)\rightarrow ip$,\footnote{This is only true for potentials without
an $r^{-1}$ singularity and follows from a power series solution in $r$ of the
Schr\"odinger equation.} so that
$e^{2i\delta}\rightarrow -1$ and $\delta(p=0)=(n+1/2)\pi$, which is the condition
for a zero energy bound state. This interpretation is also in parallel with the
pure short-range case.

\subsection{Identification of Scales and the Yukawa Potential}
One particular moot point where the Jost function DWRG analysis is important is
in the inclusion of pions in an EFT for nucleons. We will now look
at the issues surrounding the Yukawa potential and pions in an
EFT for nucleons.

However, before we can discuss the physics of this problem, we must quickly
tidy up one final mathematical issue.
Our definition of $\hat{\cal M}$, eqn.~(\ref{eq:defM}), relies upon the existence
of the limit of the derivative of the Jost function as $r\rightarrow0$. This
limit only exists for potentials that {\it do not} have a $r^{-1}$ divergence as
$r\rightarrow0$. For potentials that do diverge, such as the Yukawa potential, we
must modify the definition of $\hat{\cal M}$. To understand the problem let us
consider the expansion, in powers of $r$, of the Jost solution for the Yukawa potential
for one pion exchange between nucleons of mass $M$.
This potential is defined by,
\begin{equation}\label{eq:yukawa}
MV_L(r)=2\kappa_\pi\frac{e^{-m_\pi r}}{r},
\end{equation}
where,
\begin{equation}
\kappa_\pi=\frac{g_A^2 m_\pi^2 M}{32\pi f_\pi^2},
\end{equation}
the inverse `pionic Bohr radius'\footnote{$f_\pi=93$MeV is the pion decay constant,
$g_A=1.26$ is the axial coupling of the nucleon. The treatment of this scale $\kappa_\pi$
as a low energy scale in an EFT for nucleons with explicit pions is still
open to debate, see below.} and $m_\pi$, the pion mass
are the low energy scales associated with the potential. The Jost solution for the
Yukawa potential is given by
\begin{equation}
f^{\pi}(p,\kappa_\pi,m_\pi,r)={\cal F}^{\pi}(p,\kappa_\pi,m_\pi)\left(1+
{\cal M}_{\text{reg}}^{\pi}(p,\kappa_\pi,m_\pi)r
\label{eq:yukexpandf}
+2\kappa_\pi r\ln m_\pi r+{\cal O}(r^2)\right).
\end{equation}
In this case the term $\sim r\ln m_\pi r$ means that the derivative of the
Jost solution is not defined at $r=0$. However, inserting eqn.~(\ref{eq:yukexpandf})
into eqn.~(\ref{eq:niceident}) shows that the identity (\ref{eq:analC}) holds
with $\hat{\cal M}^{\pi}_{\text{reg}}$ replacing $\hat{\cal M}$.
The rest of the arguments for the
non-trivial fixed point solution hold with the substitution
$\hat{\cal M}$ replaced by $\hat{\cal M}^{\pi}_{\text{reg}}$ which we may write as
\begin{equation}
\hat{\cal M}_{\text{reg}}^{\pi}(\hat q,\hat\kappa_\pi,\hat m_\pi)=\lim_{\hat r\rightarrow0}
\left[\frac{{f^{\pi}}'(\hat q,\hat\kappa_\pi,\hat m_\pi,\hat r)}
{{\cal F}^{\pi}(\hat q,\hat\kappa_\pi,\hat m_\pi)}
-2\hat\kappa_\pi\ln\hat m_\pi\hat r\right].
\end{equation}
Without going into the details this results in a non-trivial fixed-point, $\hat V_S^{(0)}$,
defined by eqn.~(\ref{eq:contourV}) with logarithmic dependence on $\hat m_\pi$,
which can be removed in the usual way. See refs.~\cite{ksw1,ksw,furnsteele} for more details. 

There has been some debate in the literature as to how to handle OPE in an
EFT. One scheme (WvK) proposed by Weinberg \cite{wein1} and further developed by
van Kolck \cite{orvk,vk1}, iterates it to all orders. Another,
proposed by KSW \cite{kswapp1}, treats the force perturbatively.

So far in this section it has been understood that the long range potential
has only low energy scales associated with it. In this case the analysis
is simple and results in a choice of two expansions. A Weinberg scheme
based upon the trivial fixed point and with terms in one-to-one correspondence
with the terms in the DW Born expansion and a KSW scheme based upon the non-trivial
fixed point and with terms in one-to-one correspondence with the terms in
the DWERE. We shall see that it is the identification of the low-energy scales
in the Yukawa potential that has led to such confusion.

Since the DWRG analysis promotes all low-energy physics to a fixed-point,
if it was applied to the Yukawa potential with the
two identified low energy scales $\kappa_\pi$ and $m_\pi$
it would be equivalent to the WvK scheme,
summing the effects of one pion exchange to all orders and leaving
a choice of power-counting schemes. For the strongly interacting nucleon system
we would expect the non-trivial power-counting scheme and the DWERE to
provide a parameterisation that converges up to the mass of the $\rho$-meson,
$m_\rho=770$MeV \cite{furnsteele}. This choice of low energy scales has resulted
in some moderate success \cite{orvk,vk1}.

Unfortunately there is an issue of debate. In the
power-counting around the non-trivial fixed point, there are terms like
\begin{eqnarray}
\frac{m_\pi^{2n}}{\Lambda_0^{2n-1}},\\
\frac{\kappa_\pi^n}{\Lambda_0^{n-1}}.
\end{eqnarray}
The first of these terms is proportional to $m_\pi^{2n}$ and occurs
at order $d=2n-2$ in the power-counting. The second of these terms
occurs at order $d=n-2$ in the power-counting, yet because
$\kappa_\pi\propto m_\pi^2$ it is also proportional to $m_\pi^{2n}$.
Hence, terms of the same order in $m_\pi$ occur at different orders
in this power-counting and the direct link to ChPT has been lost
\cite{bbsvk,kswapp1}.

If we wish the nuclear EFT including OPE to be consistent with ChPT,
then one solution is the KSW scheme \cite{kswapp1}. In this case
we treat the scale,
\begin{equation}
\Lambda_{NN}=\frac{16\pi f_\pi^2}{M g_A^2}=300{\rm MeV},
\end{equation}
as a high energy scale so that the Yukawa potential is given by
\begin{equation}
MV_L(r)=\frac{m_\pi^2}{\Lambda_{NN}}\frac{e^{-m_\pi r}}{r}
\end{equation}
and occurs at the order $(Q/\Lambda_0)^{1}$ in the EFT and so does not require
summing to all orders. From the DWRG point of view, the effects of the
Yukawa potential vanish as the cut-off $\Lambda\rightarrow0$ and so do
not get promoted into the fixed point. Thus the fixed points are the simple
ones of the pure short range case. In this case the rescaled DW Green's function $G_L$
can be expanded in powers of $\Lambda/\Lambda_{NN}$ and treated as perturbations
around the non-trivial fixed point.

Unfortunately, despite consistency with ChPT,
the resulting expansion turns out to be slowly convergent
\cite{furnsteele,krthesis,cohan,fms} because of the small separation of scales
between the pion mass, $m_\pi=140$MeV and $\Lambda_{NN}$.

Although the DWRG analysis gives the possible power-counting schemes
that can be obtained from the non-trivial or trivial fixed point they can only
ever be as good as the information put in. One must consider very carefully
the scales in any particular problem.

\section{Summary}
In this chapter we have introduced the DWRG equation and solved it
for several examples. The analysis for the Coulomb and 
general ``well-behaved'' potentials yielded known results that have
both recently been used in EFTs. Despite these results being well-known,
the method of derivation is novel and in itself is interesting.
The conclusions regarding the usefulness of the two possible
power-counting schemes are very much the same as arrived out for the
pure short-range potential.

The method of solution of the DWRG fixed point equation in the final general example
will prove extremely useful in later chapters and provides a robust
way of solving these types of equation.

The other example considered in this chapter was the repulsive
inverse square potential. As far as we are aware the results
derived here are new. The power-counting schemes that are derived are
novel and depend upon the strength of the potential. The conclusions
about the use of either power-counting scheme are also different.
If the inverse square potential is `strong',
the use of the non-trivial fixed point scheme may require the tuning of
several parameters to unnaturally small values, something that may seem
a little contrived. It is likely that in such a system the trivial
fixed point should provide the correct power-counting scheme.

%% file: ch3/chapter3.tex
\section{Introduction}
The inverse square potential, eqn.~(\ref{eq:invsqpot})
is an interesting example. When $\beta>-1/4$
the potential has a well-defined set of DWs. However, if $\beta<-1/4$
the potential becomes singular with no well-defined set of DWs. The reason
for this is that any two linearly independent solutions of the Schr\"odinger
equation cannot be resolved by a boundary condition at the origin.
Indeed no solution of the Schr\"odinger equation has a well-defined value at
the origin.
Furthermore, it may appear at first glance that there is a continuum of
possible bound states. The solution,
\begin{equation}
\psi\propto \sqrt{r}K_{i\nu}(\sqrt{-ME} r),
\end{equation}
where $\nu=\sqrt{-1/4-\beta}$ and
$K_\mu(z)$ is the modified Bessel function of the third kind,
satisfies vanishing boundary conditions at infinity and the Schr\"odinger equation,
yet the lack of a boundary condition at the origin means these bound states
are not resolved into a discrete spectrum.

Mathematically, the problem with the potential is that the Hamiltonian
is no longer self-adjoint \cite{case,meetz}. The solution is to make it so by introducing an
extra boundary condition to uniquely define a solution \cite{meetz,perelpop}.
This extra boundary condition can be introduced in several ways,
either by requiring vanishing wavefunctions at some radius, $R_0$,
or by defining some bound state, $p_0$ \cite{perelpop}. However the boundary condition
is defined it necessarily introduces a new scale into the problem.
Mathematically, this is equivalent to forming a self-adjoint extension
of the Hamiltonian \cite{meetz,case}.

Once the self-adjoint extension is formed we have a complete set of DWs
including a discrete spectrum of bound states,
however there are still problems to be addressed.
The resulting discrete spectrum consists of an
infinite number of geometrically spaced bound states.
These bound states accumulate at zero energy and have no ground state.
The lack of a ground state shows that even after the self-adjoint
extension is formed the full resolution of the short-range physics has not
been achieved. Without a ground state the system would `implode'
radiating an infinite amount of energy.

The resolution of the attractive inverse square singularity
in an EFT is critical to the understanding of the three-body KSW EFT
\cite{efimov,bhvk,lcbc,brgh,efimovphil,irrglc}. 
Several papers have attempted to resolve the problem
by replacing the singularity with an `effective' potential well
at short range and then
matching the logarithmic derivative of the wavefunctions in the
two regions \cite{camblong,lcbc,splc}. In each case the range, $R$, of the effective potential has 
been assumed to be far less than all inverse momenta.
The matching criteria acts, in effect, as the boundary
condition required to form the self-adjoint extension and does not,
as correctly pointed out by Bawin and Coon \cite{lcbc} but not by Camblong
and Ord\'{o}\~{n}ez \cite{camblong} give a ground state to the system. Further to this,
using this method, the running of the `strength' of the effective well is
multi-valued \cite{splc}.

The DWRG method provides the explanation to these problems. We find
that the choice of a self-adjoint extension is equivalent to the
LO EFT, which in turn corresponds to a marginal perturbation
in the RG.
Because the LO EFT is marginal it cannot distinguish
between high and low energy states, which is why the
self-adjoint extension cannot provide a ground state.

We shall also see that because of the lack of a ground state in the DWs, the
DW Green's function must be truncated in both the continuum and
the bound states to obtain an unique solution to the DWRG.
The multi-valuedness of the effective potential can then be attributed
to different methods of truncating the bound states.
 
\section{Defining the Distorted Waves}\label{sec:defdw}
The distorted waves of the inverse square
potential satisfy the Schr\"odinger equation,
\begin{equation}
\frac{{\rm d}^2}{{\rm d}r^2}\psi_p(r)
-\frac{\beta}{r^2}\psi_p(r)+p^2\psi_p(r)=0,
\end{equation}
of which the general solution can be written as a superposition of
Bessel functions of
order $\pm\sqrt{1/4+\beta}$. When $\beta<-1/4$ the orders of the
Bessel functions are imaginary and the general solution is,
\begin{equation}
\psi_p(r)=\sqrt{r}(A_1 J_{i\nu}(pr)+A_2 J_{-i\nu}(pr)),
\end{equation}
where $\nu=\sqrt{-\beta-1/4}$. It is impossible to find a unique
solution by defining a boundary condition at the origin.
To resolve the ambiguity
and form a self-adjoint extension we begin by considering the Jost solution
$f(p;r)$ that satisfy the Schr\"odinger equation and has the unambiguous
asymptotic boundary
condition, $f(p;r)\rightarrow e^{ipr}$ as $r\rightarrow\infty$. $f$ is
easily found to be
\begin{equation}
f(p;r)=\sqrt{\frac{\pi pr}{2}}e^{-\frac{\pi}{4}(2\nu-i)}
H^{(1)}_{i\nu}(pr),
\end{equation}
where $H^{(1)}_{\mu}(z)$ is the Hankel function of the first kind. We
would now like to define a Jost function, $\cal F$ in terms of the value of the
Jost
solution, $f$, at the origin but this limit does not
exist. Explicitly, in the limit of $r\ll 1/p$ we have,
\begin{equation}
\sqrt{\frac{2}{\pi pr}}f(p;r)\sim
\sin\left(\nu\log\frac{pr}{2}-\theta-i\frac{\pi\nu}{2}\right),
\end{equation}
where $\theta=\arg\{\Gamma(1+i\nu)\}$. The solution is to
define the Jost function in terms
of the value of the Jost solution at the point $r=R_0$ rather than at $r=0$,
resulting in,
\begin{equation}\label{eq:JostF}
{\mathcal F}(p)=
\frac{2e^{-\frac{i\pi}{4}}}{\sqrt{\pi\nu\sinh(\pi\nu)}}
\sin\left(\eta(p)+i\frac{\pi\nu}{2}\right),
\end{equation}
where
\begin{equation}\label{eq:defineeta}
\eta(p)=-\nu\ln{\frac{pR_0}{2}}+\theta=-\nu\ln{\frac{p}{p_0}}.
\end{equation}
Eqn.~(\ref{eq:defineeta}) defines the scale $p_0$, which as we shall see is
simply related to the binding energies of the bound states.
The definition of the Jost function introduced here is equivalent to
applying a boundary condition of vanishing wavefunctions at the small distance
$r=R_0$. With these definitions, the free DWs are given precisely as in eqn.
(\ref{eq:defDW}), from which follows
the usual relation between the $S$-matrix and the
Jost function,
\begin{equation}\label{eq:smatrix}
e^{2i\delta_L(p)}=\frac{{\mathcal F}(-p)}{{\mathcal F}(p)}
=i\frac{\sin(\eta(p)-i\pi\nu/2)}{\sin(\eta(p)+i\pi\nu/2)}.
\end{equation}
With the Jost function we may begin to investigate the bound states. The
eigenvalues, $-p_n^2$, of the Hamiltonian are given by the poles of the
$S$-matrix or equivalently the zeros of the Jost function. 
Using the definition of  ${\mathcal F}(p)$ and $\eta(p)$ we find that these
zeroes are defined by,
\begin{equation}\label{eq:invbdstates}
p_n=p_0 \exp{\left(\frac{n\pi}{\nu}\right)} \qquad n\in{\mathbb Z}.
\end{equation}
Thus, these bound states, $\psi_n(r)$,
form an infinite tower of states with geometrically spaced
energies. It is clear that a ground state does not exist and that the states get
infinitesimally close to zero energy.
Since the distorted waves now form a
complete set,  
the familiar spectral decomposition of the Green's function follows.
The Green's function with standing wave boundary conditions is given by,
\begin{equation}
G_L(p;r,r')=
\frac{M}{2\pi^2}\fint_0^\infty{\rm d}q
\frac{\psi_q(r)\psi_q(r')}{p^2-q^2}
+\frac{M}{4\pi}\sum_{n=-\infty}^{\infty}\frac{\psi_{n}(r)\psi_{n}(r')}
{p^2+p_n^2}.
\end{equation}
Explicitly the DWs, given by eqn. (\ref{eq:defDW}), are
\begin{equation}\label{eq:dwcont}
\psi_p(r)=
\sqrt{\frac{\pi pr}{2}}\frac{1}{2i|\sin(\eta(p)+i\pi\nu/2)|}
\Bigl(e^{i\eta(p)}J_{i\nu}(pr)-e^{-i\eta(p)}J_{-i\nu}(pr)\Bigr),
\end{equation}
and the bound states are,
\begin{equation}\label{eq:dwbound}
\psi_{n}(r)=
\sqrt{\frac{2r\sinh{(\pi\nu)}}{\pi\nu}}p_n K_{i\nu}(p_nr).
\end{equation}

The definition of the Jost function in terms of the Jost solution at $r=R_0$
provides the additional boundary condition, which in turn forms a
self-adjoint extension of the Hamiltonian. However, in obtaining the
Jost function in this way we have implicitly assumed $R_0\ll1/k$ for all
$k$, i.e. $R_0$ is some infinitesimal distance, which makes it difficult
to understand the exact nature of the boundary condition. Fortunately,
this method lends itself to different interpretations. Firstly, and
perhaps most simply, $R_0$ fixes the phase of the trig-log behaviour
close to the origin and hence defines a regular and irregular solution.
Secondly, $p_0$ acts as a choice of a particular bound state, which forces
all other bound states to fall into the exponential tower defined by 
eqn.~(\ref{eq:invbdstates}). Importantly this means that
the relationship between $p_0$ and choice of self-extension is not one-to-one.
Any choice given by $p'_0=p_0e^{n\pi/\nu}$ is equivalent to the choice $p_0$.

\section{The DWRG equation}\label{sec:invsqdwrg}
We now move on to the construction and solution of the DWRG equation
for the short range force. Now that we have arrived at an equivalent of the
Jost function for the attractive inverse square potential it may be hoped
that the general analysis used in section 3.6 will yield
a method of solving the DWRG equation. However, we are immediately
faced with a problem. In the previous analysis the contour of integration
in eqn.~(\ref{eq:jostfixedpoint})
for the DWRG basic loop integral solution had to go outside all the bound state poles
to avoid being `pinched' as the scales associated with the long range potential
went to zero, in this case it is quite impossible for the contour to
go outside all of the bound state poles because there is no ground state.

With a little thought, our inability to define a contour that goes outside the
bound state poles is not a problem. The position of the poles is controlled by the
self-adjoint extension defining scale, $p_0$. As long as we are not concerned with
analyticity of the short-range force as $p_0\rightarrow0$ then there is no worry
with being pinched as $p_0\rightarrow0$.

With this problem resolved a new one
becomes apparent, namely non-unique solutions to the DWRG equation.
To explain, if we are to define our basic loop integral solution in a method like
that outlined in
section 3.6 our contour of integration must cross the imaginary axis between two
particular poles, with no apparent correct choice of which two poles.
Without resolving this issue, any resulting physical observables will invariably
depend upon the choice of contour.

It is clear that this problem is a result of the lack of a ground state which in turn
is a result of the singular behaviour of the long-range potential close to the origin.
The philosophy behind the DWRG is to `replace' all the interactions between high
energy distorted waves with an effective short-range interaction, it therefore
makes sense that the deeply bound states, which are a result of the inverse
square singularity, should be cut-off.
Thus, we define the truncated Green's function to be, 
\begin{equation}\label{eq:truncGreen}
G_L(p,\Lambda;r,r')=
\frac{M}{2\pi^2}\fint_0^\Lambda{\rm d}q
\frac{\psi_q(r)\psi_q(r')}{p^2-q^2}
+\frac{M}{4\pi}\sum_{|p_n|<\Lambda}\frac{\psi_{n}(r)\psi_n(r')}
{p^2+p_n^2},
\end{equation}
in which all distorted waves with energies outside the range $(-\Lambda^2/M,\Lambda^2/M)$
are removed. As we shall see the truncation of bound states and subsequent renormalisation
specifies a non-trivial fixed point solution and leads to well-defined results for physical
observables. The method of truncating the bound states chosen here is far from unique,
other truncation methods will specify different solutions, however
because of the renormalisation procedure, physical observables will be independent of
the truncation method and of the specific loop integral.

The differential equation, (\ref{eq:udwrge2}), for $V_S$ is now given by,
\begin{equation}\label{eq:VSdiff}
\frac{\partial V_S}{\partial\Lambda}=-\frac{M}{2\pi^2}V_S^2\left[
\frac{|\psi_\Lambda(R)|^2}{p^2-\Lambda^2}
+\frac{\pi}{2}\sum_{n=-\infty}^{\infty}
\frac{|\psi_n(R)|^2}{p^2+p_n^2}\delta(\Lambda-p_n)\right],
\end{equation}
where the cutting off of bound states has resulted in a series of
discontinuities expressed in terms of a sum of delta functions. 
Since $R$ is taken as very small, the distorted waves go to,
\begin{equation}\label{eq:shortrangewf}
|\psi_\Lambda(R)|^2\rightarrow
\frac{\Lambda B(R)\sinh{\pi\nu}}{\cosh{\pi\nu}-\cos{2\eta(\Lambda)}},\qquad
|\psi_n(R)|^2\rightarrow \frac{2p_n^2}{\nu}B(R),
\end{equation}
where
\begin{equation}\label{eq:definebeta}
B(R)=\frac{R}{\nu}
\sin^2\left(\nu\ln\left(\frac{p_0R}{2}\right)-\theta\right).
\end{equation}

The final step to obtain the DWRG equation is to rescale. In this problem,
the only low energy scale is the on-shell momentum, $p$, which is rescaled to $\hat
p=p/\Lambda$. $V_S$ is rescaled by the relation,
\begin{equation}\label{eq:potrescale}
\hat V_S(\hat p,\Lambda)=
\frac{MB(R)}{2\pi^2}V_S(\hat p\Lambda,\Lambda),
\end{equation}
finally resulting in the DWRG equation,
\begin{eqnarray}\label{eq:dwrg2}
&&\displaystyle{
\Lambda\frac{\partial}{\partial\Lambda}\left(\frac{1}{\hat V_S}\right)=
\hat p\frac{\partial}{\partial\hat p}\left(\frac{1}{\hat V_S}\right)-
\frac{\sinh{\pi\nu}}{(\cosh{\pi\nu}-\cos{2\nu\ln(\Lambda/p_0)})(1-\hat p^2)}}
\nonumber\\
&&\displaystyle{\qquad\qquad\qquad
+\frac{\pi}{\nu}\sum_{n=-\infty}^{\infty}
\frac{1}{\hat p^2+1}\Lambda\delta(\Lambda-p_n),}
\end{eqnarray}
where we have divided through by $\hat V_S^2$
to obtain a linear equation in $1/\hat V_S$. Unlike all the RG equations seen so
far this one has $\Lambda$ dependence in the inhomogeneous term on the RHS. This
dependence is a result of the scale $p_0$, which may {\it not} be treated as
a low energy scale.

\section{Solving the DWRG equation}\label{sec:solveinvsqrg}
The DWRG equation, as always, yields a trivial fixed point, $\hat V_S=0$.
The perturbations, $\hat V_S=C\Lambda^\nu\phi_\nu(\hat p)$,
satisfy the equation
\begin{equation}\label{eq:triveigen}
\phi'_\nu(\hat p)=\nu\phi_\nu(\hat p).
\end{equation}
We find $\phi_\nu(\hat p)=\hat p^\nu$. The corresponding
RG eigenvalues are $\nu=0,2,4,\ldots$.
Hence, the RG solution in the region of the trivial fixed point is given by,
\begin{equation}
\hat V_S(\Lambda,\hat p)=\sum_{n=0}^\infty \hat C_{2n}
\left(\frac{\Lambda}{\Lambda_0}\right)^{2n}\hat p^{2n}.
\end{equation}
The LO term is marginal, i.e. it does not scale with
any power of $\Lambda$ and is expected to be associated with
some logarithmic behaviour in $\Lambda$. However, before we can give a full
explanation of this we must obtain the full solution to the RG equation.

It is clear, because of the $\Lambda$-dependence on the right-hand side of
equation (\ref{eq:dwrg2}), that no other fixed point solution can be found.
However, since that dependence is only logarithmic we may hope to find a
slowly evolving solution. (Such a solution will be no worse than that encountered
when considering the Coulomb potential, in which logarithmic dependence on $\Lambda$
was introduced to remove logarithmic dependence upon $\hat\kappa$.)
A logarithmically evolving fixed point solution
may still be used to construct a power-counting by perturbing about
it in powers of $\Lambda$.

\begin{figure}
\begin{center}
\includegraphics[height=8cm,width=8cm,angle=0]{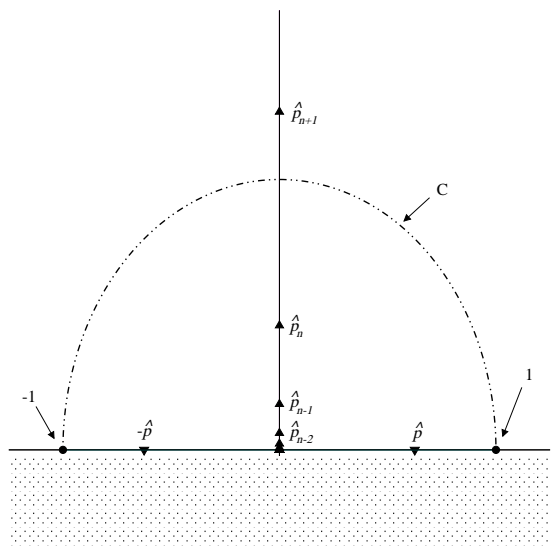}
\caption{The contour in the complex $\hat q$-plane
for the non-trivial fixed point solution of the DWRG
equation. It runs from $-1$ to $1$ and crosses the imaginary axis
at $\hat q=i$. The bound state poles occur at $\hat q=i\hat p_n=ip_0e^{n\pi/\nu}/\Lambda$.
}
\label{fig:invsqpoles1}
\end{center}
\end{figure}

Since (c.f eqn.~(\ref{eq:analC})),
\begin{equation}
\frac{\sinh{\pi\nu}}{\cosh{\pi\nu}-\cos{2\nu\ln{\Lambda/p_0}}}=
\frac{1}{2i}\left[\cot\left(\nu\ln\frac{\Lambda}{p_0}-\frac{i\pi\nu}{2}\right)
-\cot\left(\nu\ln\frac{\Lambda}{p_0}+\frac{i\pi\nu}{2}\right)\right],
\end{equation}
we can immediately write down a solution to the continuous part of the
DWRG equation in parallel to eqn.~(\ref{eq:jostfixedpoint}) considered in section 3.6:
\begin{equation}
\frac{1}{\hat V_S^{(0)}(\hat p,\Lambda)}=\frac{1}{2i}\int_C
d\hat q\frac{\hat q}{\hat p^2-\hat q^2}
\cot\left(\nu\ln\frac{\hat q\Lambda}{p_0}-\frac{i\pi\nu}{2}\right),
\end{equation}
which is clearly analytic as $\hat p\rightarrow0$. The contour $C$ is shown
in Fig.~\ref{fig:invsqpoles1}, it follows a path from $\hat q=-1$ to $\hat q=1$.
The position of bound state poles poles shown in Fig.~\ref{fig:invsqpoles1} are
simply given by,
\begin{equation}\label{eq:rescbound}
i\hat p_n=i\frac{p_n}{\Lambda}=i\frac{p_0}{\Lambda}e^{n\pi/\nu}.
\end{equation}
The position of these poles varies with $\Lambda$. This is in contrast to the
examples considered so far where the rescaling of the scales in the
long-range potential meant that the bound state poles in the complex $\hat q$-plane
were independent of $\Lambda$ (which of course is what one would hope
if looking for a fixed point solution).
The poles in this example move up 
the imaginary $\hat q$ axis as $\Lambda\rightarrow0$ because the scale
that initially controlled their positions, $p_0$, is treated as a high energy
scale and {\it not} absorbed into the fixed-point solution.

Fortunately this migration of the bound state poles with $\Lambda$ is to our
benefit as it enables us to create the necessary discontinuities in the DWRG solution
by allowing the bound state poles to cross the contour of integration.
Dividing eqn.~(\ref{eq:dwrg2}) through by $\Lambda$ and integrating it
with respect to $\Lambda$ from $p_n-\epsilon$ to $p_n+\epsilon$ we
get the discontinuity at $\Lambda=p_n$,
\begin{equation}
\left[\frac{1}{\hat V_S}\right]_{\Lambda=p_n-\epsilon}^{\Lambda=p_n+\epsilon}
=\frac{\pi}{\nu}\frac{1}{\hat p^2+1}.
\end{equation}
If we chose the contour $C$ so that it crosses the imaginary
axis at the point $\hat q=i$ then from eqn.~(\ref{eq:rescbound}) it
follows that the bound state pole at $\hat q=i\hat p_n$ crosses the
contour of integration when $\Lambda=p_n$ and produces
the correct discontinuity, namely
\begin{equation}
\left[\frac{1}{\hat V_S}\right]_{\Lambda=p_n-\epsilon}^{\Lambda=p_n+\epsilon}
=-2\pi i{\cal R}\left\{\frac{1}{2i}\frac{\hat q}{\hat p^2-\hat q^2}
\cot\left(\nu\ln{\hat q}-\frac{i\pi\nu}{2}\right),\hat q\rightarrow
i\right\}=\frac{\pi}{\nu}\frac{1}{\hat p^2+1}.
\end{equation}
Hence, $\hat V_S^{(0)}$ is a ``fixed point'' solution to the DWRG.
From its definition it is readily
observed that $\hat V_S^{(0)}$ is invariant under the transformation
$\Lambda\rightarrow\Lambda e^{n\pi/\nu}$ so that $\hat V_S^{(0)}$ has
log-periodic behaviour in $\Lambda$. Rather than use the rather
stretched notation of a fixed point we shall, more appropriately
refer to it as a limit cycle solution.

The explicit demonstration of limit-cycle behaviour in the RG for this
problem confirms what has been shown at LO in several articles
\cite{splc,bhvk,brgh}. The first example of a RG limit cycle solution 
was given by Wilson and Glazek \cite{wilsonlc} in a toy model. Since then,
because of its importance in the three-body EFT, many authors have taken
an interest in such solutions \cite{wilsonlc2,splc,lcbc,leclairlc,irrglc}.

Perturbations around this limit cycle are obtained in the usual way. 
In fact the eigenvalue equation for the perturbation is exactly the
same as that for the trivial fixed point, (\ref{eq:triveigen}) with the same result.
Because the DWRG equation (\ref{eq:dwrg2}) for $\hat V_S^{-1}$
is linear, these perturbations give an exact solution to the RG equation,
\begin{equation}\label{eq:solution}
\frac{1}{\hat V_S(\hat p,\Lambda)}=\frac{1}{\hat V_S^{(0)}(\hat p,\Lambda)}+
\sum_{n=0}^{\infty}\hat C_{2n}
\left(\frac{\Lambda}{\Lambda_0}\right)^{2n}\hat p^{2n}.
\end{equation}
The LO term is marginal and is associated with the logarithmic flow
in the limit cycle solution. Since there is a marginal 
(or $\Lambda$-independent) perturbation, it cannot
be unambiguously removed from any limit cycle solution. This
means that there is, in fact, a family of limit cycle solutions parameterised
by the value of the marginal perturbation $\hat C_0$.

Previous examples of logarithmic behaviour in the scale $\Lambda$ resulted
from a need to remove logarithmic behaviour in some low-energy scale and
necessitated the introduction of an arbitrary scale, $\mu$. In this example however,
no new scale needs to be introduced since the scale $p_0$ fills its purpose.

Where an arbitrary scale was introduced previously it could be traded off
against the marginal perturbation leaving one remaining degree of freedom.
It is not clear that $p_0$ and $\hat C_0$ represent
the same degree of freedom, resolution of this issue must wait until we consider the
effects of the long and short range potentials in tandem as $p_0$ also
effects the long-range physics.

\begin{figure}
\begin{center}
\includegraphics[height=13cm,width=8cm,angle=-90]{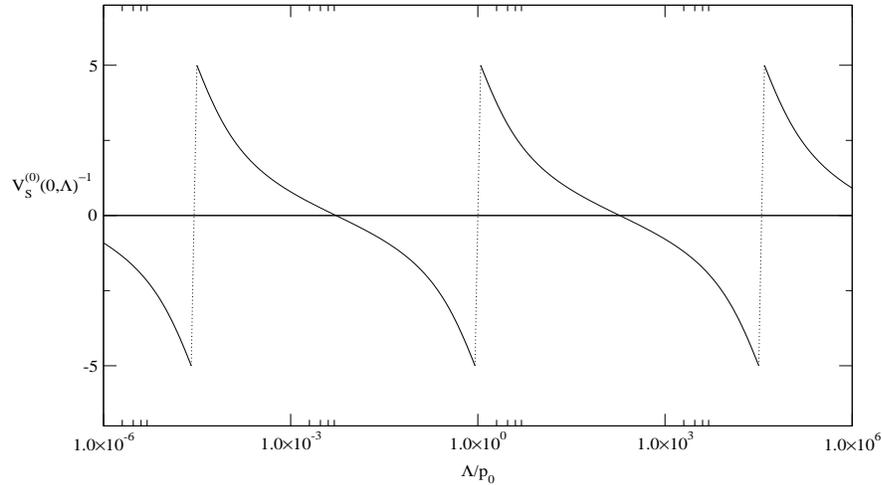}
\caption{The running of $1/\hat V_S(0,\Lambda)$ with $\Lambda$ for
$\nu=0.3$. The dotted lines show discontinuities.}
\label{fig:b0running}
\end{center}
\end{figure}

\begin{figure}
\begin{center}
\includegraphics[height=13cm,width=8cm,angle=-90]{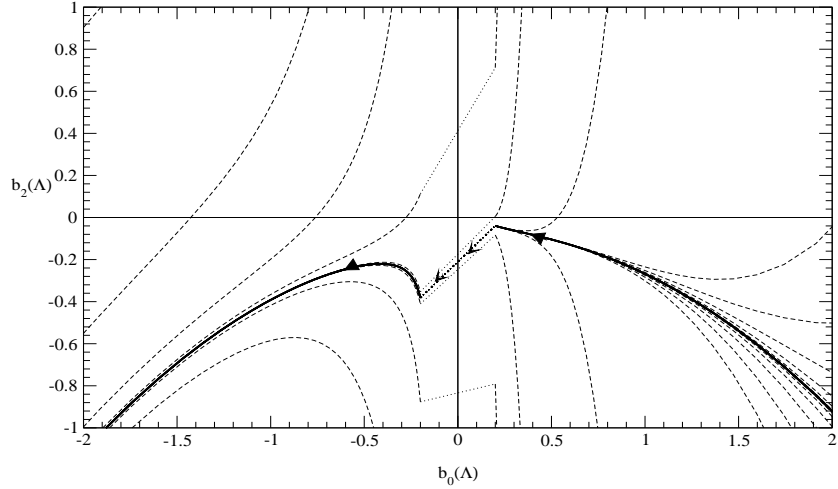}
\caption{The flow of $\hat V_S(\hat p,\Lambda)=\sum b_{2n}(\Lambda)\hat 
p^{2n}$ with $\Lambda$ of $V_S^{(0)}$ (bold line) and stable perturbations 
about that (dashed lines). The dotted lines show discontinuities.}
\label{fig:b0b2flow2}
\end{center}
\end{figure}

The RG flow is illustrated in Figs. \ref{fig:b0running} and \ref{fig:b0b2flow2}.
These diagrams show the flow of solutions $\hat V_S$ with $\Lambda$.
Fig. \ref{fig:b0running} shows the
running of $\hat V_S^{(0)}(\hat p\rightarrow0,\Lambda)^{-1}$.
The bold line shows the solution $1/\hat V_S^{(0)}$,
the dotted lines show the discontinuities.
The solution follows a logarithmically evolving cycle with
discontinuities as each bound state is cut-off.
Any member of the family of limit cycle solutions
can be obtained by adding a marginal term, $\hat C_0$. In Fig.~\ref{fig:b0running}
this would simply appear as a constant shift up or down without change in
form or period. From this diagram alone it is clear that $\hat C_0$ and $p_0$
cannot represent the same degree of freedom in the limit-cycle solution because
variation of $p_0$ will produce a horizontal rather than vertical shift 
in the form of $1/\hat V_S^{(0)}$.

The limit-cycle behaviour is illustrated in fig. \ref{fig:b0b2flow2},
which shows the RG flow in the familiar
slice through the space $(b_0(\Lambda),b_2(\Lambda),\ldots)$.
Again the solution $\hat V_S^{(0)}$ is shown  in bold.
The dashed lines show general solutions generated by perturbing
with $\hat C_2$. The solution $\hat V_S^{(0)}$ acts as a limit-cycle solution that
loops at a  logarithmic rate in $\Lambda$ with a discontinuity as each bound state is cut-off .
All more general solutions move in a cycle (with discontinuities) that tends to
$\hat V_S^{(0)}$ like $\Lambda^2$ as $\Lambda\rightarrow0$.

Having obtained a general solution to the DWRG we may now understand the
nature of the trivial fixed point.
As $\hat C_0\rightarrow\infty$ in eqn.~(\ref{eq:solution}),
the general solution $\hat V_S\rightarrow0$,
i.e the limit cycle solution converges on the trivial fixed point.
This means that the trivial fixed point
is a special case of the more general family of limit-cycle solutions, in which the
whole limit-cycle has been compacted into a point.
If the RG flow is perturbed away from the trivial
fixed point with a marginal term it moves into a nearby limit cycle.

The nature of the renormalisation group flow means that no particular tuning of the
parameter $\hat C_0$ is required. Its exact value merely specifies which limit cycle
the RG flow will converge upon. As we shall see each of these different limit cycles
corresponds to a different `shift' in the self-adjoint phase $\eta$ and that there is
a one-to-one relationship between all possible limit-cycles and all possible
self-adjoint extensions.

\section{Full $S$-matrix and bound states}

Using very much the same procedure that yields the distorted wave effective range
expansion in section 3.6 we may write an expression for the
correction to the phaseshift $\tilde\delta_S$ as,
\begin{equation}\label{eq:dwere}
\sinh{\pi\nu}\cot{\tilde\delta_S}=
\sin{2\eta(p)}-
\frac{2}{\pi}\bigl(\cosh{\pi\nu}-\cos2\eta(p)\bigr)
\sum_{n=0}^\infty \hat C_{2n}\left(\frac{p}{\Lambda_0}\right)^{2n}.
\end{equation}

Equation (\ref{eq:dwere}) still leaves the relation between the marginal perturbation
and the scale $p_0$ clouded. Since $p_0$ also appears in the equation
for the long-range component of the phaseshift, $\delta_L$, the relationship
between $p_0$ and $\hat C_0$ may become more transparent in consideration of
the full $S$-matrix. Study of the full $S$-matrix may also highlight the 
significance of the RG solution $\hat V_S^{(0)}$ and make examination of the
bound states possible. The full $S$-matrix is obtained from equations 
(\ref{eq:smatrix}) and (\ref{eq:dwere}),
\begin{equation}\label{eq:fullsmatrix}
e^{2i\delta}=i\frac{\bar Z^*(p)}{\bar Z(p)},\qquad
\bar Z(p)=\cos\left(\eta(p)+\frac{i\pi\nu}{2}\right)
-\frac{2}{\pi}\sin\left(\eta(p)+\frac{i\pi\nu}{2}\right)
\sum_{n=0}^\infty \hat C_{2n}\left(\frac{p}{\Lambda_0}\right)^{2n}.
\end{equation}
In the limit of all perturbations, $\hat C_{2n}$, going to zero (or equivalently
the RG solution reducing to $\hat V_S^{(0)}$) we have, 
\begin{equation}
\bar Z(p)\rightarrow\cos\left(\eta(p)+\frac{i\pi\nu}{2}\right).
\end{equation}
Comparing this to equation (\ref{eq:smatrix}) for the pure long-range
force, we see that there has been a maximal change in the arbitrary phase $\eta$.
The effect of the solution $\hat V_S^{(0)}$ is to move the choice of self-adjoint
extension, made in constructing the distorted waves, by a factor $\pi/2$.
The significance of the $\hat V_S^{(0)}$ is now resolved, it is the
solution to the RG equation that is`furthest away', in the sense of maximum
possible change of physical observables, from the trivial
solution  $\hat V_S=0$, which of course would leave the full $S$-matrix
unchanged from (\ref{eq:smatrix}). This interpretation is very much in
parallel to that arrived at in the previous chapter for the non-trivial
fixed point for well-behaved potentials. In general we may
say that the non-trivial fixed point solution changes the boundary condition
that defines the regular solution to its linearly independent boundary condition.

Physically one may interpret the limit cycle $\hat V_S^{(0)}$ in terms of its
action on the bound states, which are shifted to the geometric mean of the
original set of states.

The existence of the marginal eigenvalue, $\hat C_0$, and the scale
$p_0$ leads to an embarrassment of parameters that may be resolved if it is apparent that
they do in fact represent the same degree of freedom in the effective theory. This
freedom is not manifest in equation (\ref{eq:fullsmatrix}) but with the correct
rewriting of the parameters, $\hat C_{2n}$,
\begin{equation}\label{eq:changeparams}
\hat C_0=-\frac{\pi}{2}\cot\sigma,\qquad
\frac{2}{\pi}\sum_{n=1}^{\infty}\hat C_{2n}\left(
\frac{p}{\Lambda_0}\right)^{2n} 
=\frac{\csc\sigma\sum_{n=1}^{\infty}\hat C'_{2n}
\left(\frac{p}{\Lambda_0}\right)^{2n}}
{\sin\sigma+\cos\sigma\sum_{n=1}^{\infty}\hat C'_{2n}
\left(\frac{p}{\Lambda_0}\right)^{2n}},
\end{equation}
in terms of the new parameters, $\sigma$ and $\hat C'_{2n}\,,n\geq2$ we are left with a
far more transparent parameterisation,
\begin{equation}\label{eq:bestEFT}
e^{2i\delta}=i\frac{Z^*(p)}{Z(p)},\qquad
Z(p)=\sin\left(\eta(p)+\sigma+\frac{i\pi\nu}{2}\right)
+\cos\left(\eta(p)+\sigma+\frac{i\pi\nu}{2}\right)
\sum_{n=1}^\infty \hat C'_{2n}\left(\frac{p}{\Lambda_0}\right)^{2n}.
\end{equation}
Recalling that $\eta(p)=-\nu\ln(p/p_0)$, eqn.~(\ref{eq:bestEFT})
for the $S$-matrix clearly shows the overlapping roles of $\hat C_0$ and $p_0$.
The scale $p_0$ acted originally to define a self-adjoint extension, through
the marginal term in the short range force we may change it to any other choice of
self-adjoint extension.

The space of all limit-cycle
solutions is mapped out as $\hat C_0$ varies from $-\infty$ to $\infty$ or,
from eqn.~(\ref{eq:changeparams}), as $\sigma$ varies from $0$ to $\pi$. In turn
the variation of $\sigma$ from $0$ to $\pi$ with $p_0$ fixed maps out all possible
self-adjoint extensions. Hence, the relationship between possible limit-cycle
solutions of the DWRG equation and self-adjoint extensions is one-to-one. This is in
stark contrast to the relationship between choices of $p_0$ and self-adjoint
extensions where many equivalent $p_0$'s existed.
In practice, we may ignore
the phase $\sigma$ and allow the self-adjoint extension to be defined by 
our original choice, $p_0$.

The EFT bound states are given by the zeros of $Z(p)$.
The position of the shallower bound states are virtually
unmodified from the original theory and
are controlled by the scale $p_0$ with minimal dependence upon $\hat C_{2n}$.
There is of course still an accumulation of 
bound states at zero energy, though this is expected as these states exist because
of the inverse square tail of the potential, which is unaffected by the short-range
forces. The deepest states within the validity of the EFT, $p<\Lambda_0$,
are strongly affected by the
higher order perturbations. Beyond the range of validity, $p>\Lambda_0$,
we can say nothing about
the bound states except that some other physics must come into play to ensure that a
ground state exists.

\section{Truncation of Bound states and Uniqueness of Solutions.}

Questions still remain about the the truncation of the bound states.
It is obvious that
some sort of regularisation of these states is necessary, yet the method of
truncation of the bound states chosen above, though convenient and intuitive,
is rather arbitrary.
However, with a little
thought it is quite straightforward to show that no matter how the bound
state truncation
is performed, a solution to the RG equation similar to that found above can be
found and
that more importantly the physical results will be identical.

For example, suppose we choose to truncate the bound states with a second cut-off
$\mu$. Then the renormalisation prescription states that physical variables
should be independent of $\mu$ as well as $\Lambda$. The DWRG equation will be the same as
(\ref{eq:dwrg2}) without the $\delta$-function terms. 
It can be solved in a similar way to the discontinuous DWRG equation,
\begin{equation}
\frac{1}{\hat V_S^{(0)}(\hat p,\Lambda,\mu)}
=\frac{1}{2i}\int_{C(\Lambda,\mu)}d\hat q
\frac{\hat q}{\hat p^2-\hat q^2}
\cot\left(\nu\ln\frac{\Lambda\hat q}{p_0}-\frac{i\pi\nu}{2}\right),
\end{equation}
where the contour $C(\Lambda,\mu)$ still runs from $1$ to $-1$
but now follows a $\Lambda$-dependent path that ensures that as $\Lambda$ varies
no bound state poles cross it. The exact path of the contour
(i.e. which two bound state poles it passes between) is then uniquely
specified by the renormalisation condition on $\mu$. The solution to this
DWRG equation is of course quite arbitrary without the second renormalisation condition.

\begin{figure}
\begin{center}
\includegraphics[width=8cm,height=13cm,angle=-90]{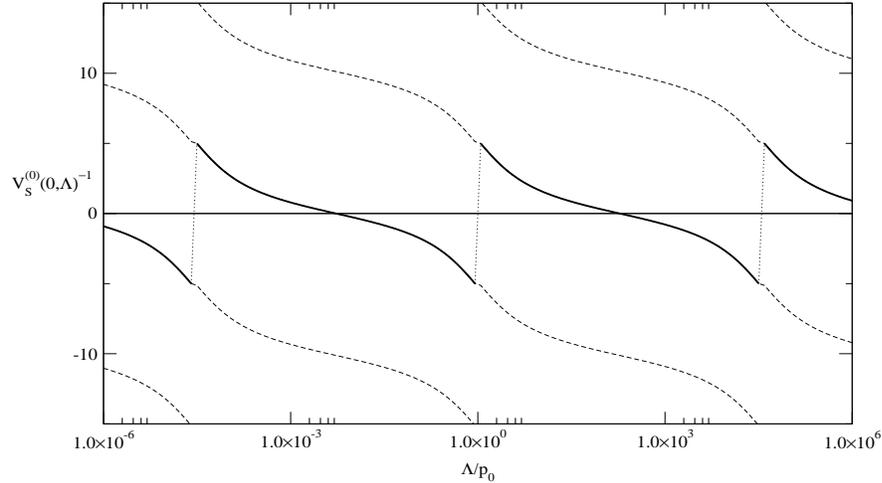}
\caption{The discontinuous solution $1/\hat V_S^{(0)}$ and the family
of continuous solutions that result from an alternative method of truncation of the
bound states ($\nu=0.3$).}
\label{fig:contanddiscont}
\end{center}
\end{figure}

The family of solutions that result are illustrated in Fig.\ref{fig:contanddiscont},
which shows the previous discontinuous solution, $1/\hat V_S^{(0)}$, in bold
with the discontinuities dotted and the various continuous
solutions as dashed lines.
Which solution is correct depends on $\mu$.
Notice, that the  correspondence between $\mu$ and the continuous 
RG solutions is not bijective; 
any two values of $\mu$ between $|p_n|$ and $|p_{n+1}|$ will
map to the same continuous solution.
More generally any prescription for truncating bound states will
give a unique RG solution, jumping to the next branch of the 
solutions as a bound state is cut-off.

\section{Summary}
In this section we have looked at the attractive inverse square potential
and the use of the DWRG equation in resolving its singular behaviour.
The first step was to introduce a self-adjoint extension to the Hamiltonian
by defining a Jost function. This allows definition of the DWs but does not resolve
the singular behaviour. In particular, even after definition of a self-adjoint
extension there is no ground state.

In order to produce a unique result for the
EFT $S$-matrix it was necessary to truncate the
deeply bound states that are a result of the singular behaviour of the long
range potential. The basic loop integral solutions to the DWRG take the form of limit
cycles towards which all more general solutions tend as $\Lambda\rightarrow0$.

The limit cycle solutions are associated with a marginal perturbation in the
DWRG. The degree of freedom embodied in this perturbation is the same
as that in the choice of self-adjoint extension.
Despite the complications in obtaining it, the
EFT $S$-matrix, (\ref{eq:bestEFT}), takes a satisfyingly simple form.

It is difficult to link the limit cycle behaviour seen in our approach to
the those seen by Bawin and Coon \cite{lcbc} and by Beane {\it et al} \cite{splc}.
Their methods rely upon constructing a self-adjoint extension and short-range
force in one step rather than the two step process that our method requires.
So whereas their LO force defines the self-adjoint extension ours serves to
change the original choice. What is particularly pleasing is the common approach
of this chapter and section 3.6. The limit-cycle solutions do not require any
particular special treatment apart from the truncation of bound states to specify
the `branch' of the multi-valued solution.

The truncation of bound states in this problem is forced upon us in seeking
unique physical results. It resolves many of the issues raised about
multi-valued solutions to the RG in refs.~\cite{lcbc,splc}. The need to truncate bound
states in this problem occurs because their position is controlled by a scale, $p_0$,
that cannot be rescaled and absorbed into the limit-cycle.

%% file: ch4/chapter4.tex
In this chapter we shall consider the DWRG equation for three
body forces (3BDWRG). In particular we shall derive the 3BDWRG equation for a general
system for which the two-body forces are known but the three-body force is to be
constructed using the 3BDWRG. The equation is applied to
a system in which the pairwise forces are ``well-behaved'' in the sense of
section 3.6 and then, more interestingly, used to examine the power-counting
for the three body force in the KSW EFT for short range forces.

\section{The Faddeev Equations and the Three-Body Force.}
For a system with three particles the LS equation is
inadequate. Although the three-body $T$-matrix satisfies the
LS equation, it is not possible to use this equation to find it
because its kernel is not compact. This is because a single LS equation
is not enough to specify all the boundary conditions for a three-body DW. The
solution is to use the Faddeev equations (see Appendix \ref{app:faddeev}).

Consider the case of three identical bosons interacting via a
pairwise interaction $V_{2B}$. The wavefunction is written as
\begin{equation}\label{eq:def3bwf}
|\Psi^+\rangle=|\psi_0\rangle+(1+P)|\psi^+\rangle,
\end{equation}
where $|\psi_0\rangle$ is the `in'-state and solves the free
Schr\"odinger equation and $P$ is a permutation operator,
which permutes particle indices and has the matrix elements
\footnote{We will work in the 
coordinate system $(\boldr_{ij},\boldrho_k)$, given by
\begin{equation}
\boldr_{ij}={\bf r}_i-{\bf r}_j,\qquad\boldrho_k=
{\bf r}_k-\frac12({\bf r}_i+{\bf r}_j),
\end{equation}
where $\{i,j,k\}$ is an even permutation of $\{1,2,3\}$ and
${\bf r}_1,{\bf r}_2,{\bf r}_3$ are the absolute
coordinates of the three particles.}
\begin{eqnarray}
&&\displaystyle{
\langle \boldr_{12}, {\boldrho}_{3}|P|\boldr_{12}',\boldrho_3'\rangle=
\delta^3(\boldr_{12}-\boldr_{23}')\delta^3(\boldrho_3-\boldrho_1')
+\delta^3(\boldr_{12}-\boldr_{31}')\delta^3(\boldrho_3-\boldrho_2')}
\nonumber\\&&\displaystyle{\qquad\qquad
=\delta^3\left(\boldr_{12}+\frac12\boldr_{12}'+\boldrho_3'\right)
\delta^3\left(\boldrho_{3}-\frac34\boldr_{12}'+\frac12\boldrho_3'\right)}
\nonumber\\
&&\displaystyle{\qquad\qquad\qquad
+\delta^3\left(\boldr_{12}+\frac12\boldr_{12}'-\boldrho_3'\right)
\delta^3\left(\boldrho_{3}+\frac34\boldr_{12}'+\frac12\boldrho_3'\right)}.
\label{eq:permmatrix}
\end{eqnarray}
The Faddeev equation for the wavefunction is written as,
\begin{equation}
|\psi^+\rangle=G_0^+(p)t^+(p)|\psi_0\rangle+G_0^+(p)t^+(p)P|\psi^+
\rangle,\label{eq:faddeev2}
\end{equation}
where $t^+$ is the $T$-matrix given by the equation,
$t^+(p)=V_{2B}+V_{2B}G_{0}^+(p)t^+(p)$. 
The corresponding equations for the full three-body Green's function and
$T$-matrix now follow from similar decompositions. If we write the
full $T$-matrix, $\cal T$, in terms of components,
\begin{equation}
{\cal T^+}(p)=(1+P)T^+(p),
\end{equation}
then the Faddeev equation for the $T$-matrix component is,
\begin{equation}
T^+(p)=t^+(p)+t^+(p)G_0^+(p)PT^+(p).
\end{equation}

It is possible to incorporate three-body forces directly into the
Faddeev equations, however this is not a suitable method for obtaining the
DWRG for the three-body force as we
wish to separate it from the two-body interactions in order to treat it
distinctly. Fortunately it is quite possible to use the method
outlined in chapter 3 to separate the $T$-matrix. Writing
\begin{equation}
{\cal T^+}(p)={\cal T}_{2B}^+(p)+(1+{\cal T}^+_{2B}(p)G_0^+(p))
\tilde{\cal T}^+_{3B}(p)(1+G_0^+(p){\cal T}^+_{2B}(p)),
\end{equation}
where ${\cal T}_{2B}^+(p)$ is the $T$-matrix in the presence of pair-wise
forces only and must be obtained using Faddeev-like equations,
we obtain the equation for $\tilde{\cal T}^+_{3B}(p)$,
\begin{equation}\label{eq:3bdwls}
\tilde{\cal T}^+_{3B}(p)=V_{3B}+V_{3B}{\cal G}_{2B}^+(p)\tilde{\cal T}^+_{3B}(p).
\end{equation}
Notice that this equation is connected since all three particles must interact
by the very nature of the three-body force.

Eqn.~(\ref{eq:3bdwls}) leads, after
introducing standing wave boundary conditions\footnote{Standing wave boundary
conditions on the full three-body Green's function are quite tricky to
visualise. We use the expression to refer to a Green's function which
uses a principal value prescription on the propagator and avoids any
complications in the DWRG arising from using a complex valued
Green's function.}, to the distorted wave
renormalisation group equation for the three-body force,
\begin{equation}\label{eq:3bdwrg1}
\frac{\partial V_{3B}}{\partial\Lambda}=-
V_{3B}\frac{\partial{\cal G}^P_{2B}}{\partial\Lambda}V_{3B}.
\end{equation}
Solving this equation now offers a recipe for constructing the
power-counting for the three-body force.

\section{Full Green's Function and Projection
Operator.} 
We shall consider the simple case of zero total angular momentum.
We may identify three different types of wave function characterised by
their incoming boundary condition. The wavefunctions with three free
incident particles, $|\Psi_{p,k}^+\rangle$,
\footnote{The wavefunctions are labelled by their total centre of mass
energy, $p^2/M$, ($M$ is the mass of one particle)
and the relative momentum of one pair, $k$.
Of course it doesn't matter which pair since all three particles are identical,
and the wavefunction is symmetric with respect to interchange of any of them.}
are given by eqns.~(\ref{eq:def3bwf},\ref{eq:faddeev2}). 
If there is a two-body bound state with binding momentum $\gamma_n$,
there are wavefunctions, $|\Psi^+_{p,\gamma_n}\rangle$
with the incoming boundary condition of the bound state and one free
particle, the equations that define these are found by taking the residues
of the equations for $|\Psi_{p,k}^+\rangle$ at $k=i\gamma_n$,
\begin{eqnarray}
&&\displaystyle{\qquad
|\Psi_{p,\gamma_n}^+\rangle=(1+P)|\psi^+_{p,\gamma_n}\rangle,}\\
&&\displaystyle{
|\psi^+_{p,\gamma_n}\rangle=|\chi_{p,\gamma_n}\rangle+
G_0^+t^+P|\psi^+_{p,\gamma_N}\rangle,}\label{eq:faddeevbd1}
\end{eqnarray}
where the incoming wavefunction
$|\chi_{p,\gamma_n}\rangle$ satisfies the equation,
$(H_0+V_{2B}+p^2/M)|\chi_{p,\gamma_n}\rangle=0$ and has a bound pair. Finally there are
the three-particle bound states, $|\Psi_n\rangle$; these satisfy the
homogeneous version of eqn.~(\ref{eq:faddeevbd1}) with the boundary condition
of vanishing amplitude at infinity. We normalise these states with the conditions,
\begin{eqnarray}
\langle\Psi_n|\Psi_m\rangle&=&\delta_{nm},\\
\langle\Psi_{p,\gamma_n}|\Psi_{p',\gamma_m}\rangle&=&\frac{\pi}{2}\delta(p-p')
\delta_{nm},\\
\langle\Psi_{p,k}|\Psi_{p',k'}\rangle&=&\frac{\pi^2}{4}\delta(p-p')\delta(k-k').
\label{eq:3bdwnorm}
\end{eqnarray}

We may now define the full Green's function, ${\cal G}^{\pm}_{2B}(p)$,
using the three-body completeness relation,
\begin{eqnarray}
&&\displaystyle{
{\cal G}^\pm_{2B}(p)=\frac{M}{4\pi}\Biggl(\sum_{n}
\frac{|\Psi_n\rangle\langle\Psi_n|}{p^2+p^2_n}
+\frac{2}{\pi}\sum_{\gamma_n}\int_{-\gamma_n^2}^\infty \frac{d(q^2)}{2q}
\frac{|\Psi^+_{q,\gamma_n}\rangle\langle\Psi^+_{q,\gamma_n}|}
{p^2-q^2\pm i\epsilon}}\nonumber\\
&&\displaystyle{\qquad\qquad\qquad\qquad\qquad\qquad\qquad
+\frac{4}{\pi^2}\int_0^\infty dq\int_0^q dk
\frac{|\Psi^+_{q,k}\rangle\langle\Psi^+_{q,k}|}
{p^2-q^2\pm i\epsilon}}\Biggr)\\
&&\displaystyle{\qquad\,\,\,\,\,\,
=\frac{M}{4\pi}\sum_{n}\frac{|\Psi_n\rangle\langle\Psi_n|}{p^2+p_n^2}
+\frac{M}{2\pi^2}\int_{-\gamma_0^2}^\infty \frac{d(q^2)}{2q} \frac{{\cal P}(q)}
{p^2-q^2\pm i\epsilon}},
\label{eq:greenproj}
\end{eqnarray}
where ${\cal P}(q)$ is a projection operator that projects out all
states with energy $q^2/M$ and is given by,
\begin{eqnarray}\label{eq:projop1}
&&\displaystyle{\qquad\,\,\,\,\,\,
{\cal P}(q)=\sum_{\gamma_n>|q|}|\Psi^+_{q,\gamma_n}
\rangle\langle\Psi^+_{q,\gamma_n}|
\qquad\text{if}\,\,-\gamma_0^2<q^2<0,}\\
&&\displaystyle{\label{eq:projop2}
{\cal P}(q)=\sum_{\gamma_n}|\Psi^+_{q,\gamma_n}\rangle
\langle\Psi^+_{q,\gamma_n}|
+\frac{2}{\pi}\int_0^q dk|\Psi^+_{q,k}\rangle\langle\Psi^+_{q,k}|
\qquad\text{if}\,\,q^2>0.}
\end{eqnarray}
The Green's function with standing wave boundary conditions is obtained
by applying a principal value prescription to the pole at $p=q$.
The cut-off to obtain the DWRG may be applied to eqn.~(\ref{eq:greenproj})
by changing the upper limit to $\Lambda^2$,
\begin{equation}\label{eq:3bswgreen}
{\cal G}^{P}_{2B}(p;\Lambda)=\frac{M}{4\pi}\sum_{n}
\frac{|\Psi_n\rangle\langle\Psi_n|}{p^2+p^2_n}
+\frac{M}{2\pi^2}\fint_{-\gamma^2}^{\Lambda^2}
\frac{d(q^2)}{2q} \frac{{\cal P}(q)}{p^2-q^2}.
\end{equation}
We have assumed that the bound states do not need truncating
(which as we shall see is not the case in the KSW EFT).
The resulting equation for $V_{3B}$ from eqn.~(\ref{eq:3bdwrg1}) is
\begin{equation}\label{eq:3bdwrg2}
\frac{\partial V_{3B}}{\partial\Lambda}=
\frac{M}{2\pi^2}V_{3B}\frac{{\cal P}(\Lambda)}{\Lambda^2-p^2}V_{3B}.
\end{equation}
In the 3BDWRG equation the
projection operator replaces the distorted waves.

The three-body effective interaction, which acts at some small
hyperradius $\bar R$ and some arbitrary hyperangle $\bar\alpha$
\footnote{The s-wave wavefunctions
depend only on $x_{ij}=|\boldr_{ij}|$ and $y_k=|\boldrho_k|$.
The hyperpolar coordinates are defined by,
\begin{equation}
R=\sqrt{x_{ij}^2+\frac43y_k^2},\qquad
\alpha_k=\text{Arctan}\left(\frac{2y_k}{\sqrt{3}x_{ij}}\right).
\end{equation}
$R$ is independent of the labelling of particles, while $\alpha$ is not.
Since these are s-wave DWs we have factored out the term
$x_{ij}y_k$ so that they are normalised as,
\begin{equation}
\int_0^\infty RdR\int_0^{\pi/2}d\alpha
\Psi^*_{q,k}(R,\alpha)\Psi_{q',k'}(R,\alpha)=\frac{\pi^2}{4}
\delta(q-q')\delta(k-k'),
\end{equation}
consistent with eqn.(\ref{eq:3bdwnorm}).}, is given by
\begin{equation}\label{eq:3Bpot}
\langle\Psi_{p,k}|V_{3B}(p,\kappa_i,\Lambda)|\Psi_{p,k',}\rangle
=\bar R\,V_{3B}(p,\kappa_i;k,k';\Lambda)\,
\Psi_{p,k}^*(\bar R,\bar\alpha)\Psi_{p,k'}(\bar R,\bar\alpha),
\end{equation}
where the scales $\kappa_i$ are low-energy scales associated with the
two-body long-range potential.
We shall assume, as in the previous two chapters,
that the off-shell matrix elements of the potential are not required
to describe on-shell scattering. However, due to the extra degrees of
freedom in the three body system, the potential must in general
depend upon the unconstrained momentum $k$ and $k'$. To obtain the
matrix elements of the potential between states with incoming and/or
outgoing two body bound states we simply set $k$ and/or
$k'$ to the appropriate $i\gamma_n$. The boundary condition for $V_{3B}$
is the analyticity condition, that is it has an expansion in even powers
of $p$,$k$ and $k'$ and an expansion in positive
(possibly even) powers of $\kappa_i$.

Substituting the form for the potential into eqns.~(\ref{eq:projop2},
\ref{eq:3bdwrg2}) we obtain,
\begin{eqnarray}
&&\displaystyle{
\frac{\partial}{\partial\Lambda}V_{3B}(p,\kappa_i;k,k';\Lambda)=}\nonumber\\
&&\displaystyle{\qquad
\frac{M\bar R}{2\pi^2}\frac{1}{\Lambda^2-p^2}\Biggl[
\sum_{\gamma_n}V_{3B}(p,\kappa_i;k,i\gamma_n;\Lambda)
V_{3B}(p,\kappa_i;i\gamma_n,k';\Lambda)
|\Psi_{\Lambda,\gamma_n}(\bar R,\bar\alpha)|^2}\nonumber\\
&&\displaystyle{\qquad\qquad
+\frac{2}{\pi}\int_0^{\Lambda}dk''
V_{3B}(p,\kappa_i;k,k'';\Lambda)V_{3B}(p,\kappa_i;k'',k';\Lambda)
|\Psi_{\Lambda,k''}(\bar R,\bar\alpha)|^2\Biggr].}\label{eq:3brgdiff}
\end{eqnarray}
The last step to obtaining the 3BDWRG equation is
to rescale all quantities in terms of $\Lambda$, a process that of course
depends upon the explicit form of the distorted waves.

\section{Properties of the Projection Operator.}

The 3BDWRG equation, which comes from rescaling
eqn.~(\ref{eq:3brgdiff}) is obviously going to be very complicated.
This complication arises because the equation describes the coupling
of the three-body force to each of the different distorted waves.
Let us briefly summarise our strategy for solving it. The study
of the trivial fixed point that invariably results in these equations
remains simple since in linearising the equation about the point
we are able to ignore the complicated coupling to all the different
channels. However, the basic loop integral solution which may lead to
a non-trivial fixed point or limit cycle solution and
to a general solution to the DWRG equation will 
depend on these details.

Our strategy in finding a basic loop integral solution
will be to generalise the approach used in section 3.6 and subsequently 
in chapter 4. If we look for solutions to the 3BDWRG equation that are
independent of momentum, then these will satisfy a considerably
simpler equation because the three-body force decouples from
each of the DWs allowing us to divide through by $\hat V_{3B}^2$
to obtain a linear PDE for $\hat V_{3B}^{-1}$.
In unscaled coordinates, eqn.~(\ref{eq:3brgdiff}) becomes,
\begin{equation}
\frac{\partial V_{3B}(p,\kappa_i;\Lambda)}{\partial\Lambda}=
\frac{M\bar R}{2\pi^2}V_{3B}^2(p,\kappa_i;\Lambda)
\frac{\langle\bar R,\bar\alpha|{\cal P}(\Lambda)|\bar R,\bar\alpha\rangle}
{\Lambda^2-p^2}.
\end{equation}

In order to use the methods of section 3.6 to solve the fixed point equation
that results from this simplification we need to know about the analytic
properties of the projection operator, just as we knew the analytic properties
of the Jost functions.

Consider the expression for the full Green's function as a function of
energy, $E$, where $E$ is a general complex number,
\begin{equation}\label{eq:greenenergy}
{\cal G}_{2B}(E)=\frac{1}{4\pi}\sum_n\frac{|\Psi_n\rangle\langle\Psi_n|}
{E+E_n}+\frac{1}{2\pi^2}\int_{-\gamma_0^2/M}^\infty
dE'\frac{\tilde{\cal P}(E')}{E-E'},
\end{equation}
where,
\begin{equation}\label{eq:energyproj}
\tilde{\cal P}(E)=\sqrt{\frac{M}{4E}}{\cal P}(\sqrt{ME}).
\end{equation}
Eqn.~(\ref{eq:greenenergy}) defines the function ${\cal G}_{2B}(E)$ as
an analytic function of $E$ except when $E\in[-\gamma_0^2/M,\infty)$ and
when $E=-E_n$. It is a simple matter to see from its definition
that ${\cal G}_{2B}(E)$ also has simple poles at $E=-E_n$ with residues,
\begin{equation}\label{eq:greenboundres}
{\cal R}\{{\cal G}_{2B}(E),-E_n\}=\frac{
|\Psi_n\rangle\langle\Psi_n|}{4\pi}.
\end{equation}
We can also see directly from its definition that
${\cal G}_{2B}(E)$ has a branch cut running from $-\gamma_0^2/M$ to
infinity along the real axis. Using the result,
\begin{equation}
\int_{-\infty}^\infty dx\frac{f(x)}{x_0-x+i\epsilon}=\fint_{-\infty}^\infty
dx \frac{f(x)}{x_0-x}-i\pi f(x_0)
\end{equation}
and its complex conjugate we may use eqn.~(\ref{eq:greenenergy}) to show
that,
\begin{equation}\label{eq:projgreen}
\tilde{\cal P}(E)=i\pi\Bigl({\cal G}_{2B}(E+i\epsilon)-
{\cal G}_{2B}(E-i\epsilon)\Bigr).
\end{equation}
Hence, for $E\in[-\gamma_0^2/M,\infty)$, the projection operator
$\tilde{\cal P}(E)$ is defined by the discontinuity across this branch
cut. From eqn.~(\ref{eq:projop2}) that defines the projection operator
for real $E$ in terms of the distorted waves it follows that the
discontinuity across the real $E$ axis must in fact be the result of several `overlapping'
branch cuts that run from each two-body threshold $E=-\gamma_n^2/M$
to $E=\infty$ and from the three-body threshold $E=0$ to $E=\infty$. 

\begin{figure}
\begin{center}
\includegraphics[height=8cm,width=8cm,angle=0]{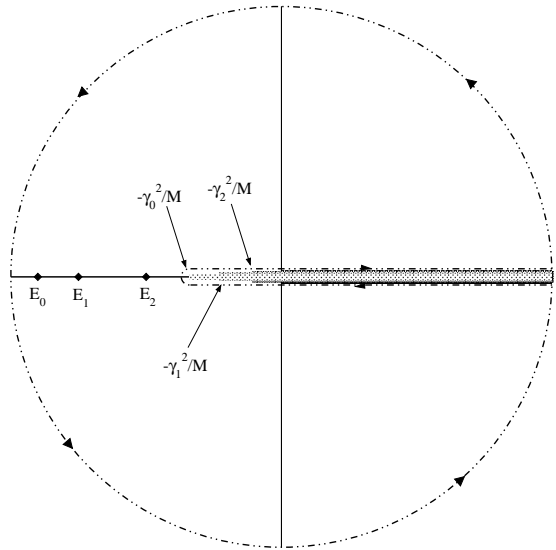}
\caption{The singularity structure of an example of the full Green's function
for complex $E$ with three three-body bound states and three two-body
bound states. Also shown is the contour $C$ used in the dispersion relation.}
\label{fig:G3Bcont}
\end{center}
\end{figure}

Substituting eqns.~(\ref{eq:greenboundres},\ref{eq:projgreen}) into
eqn.~(\ref{eq:greenenergy}) we can see that the spectral decomposition of
the Green's function is equivalent to
a dispersion relation for ${\cal G}_{2B}(E)$,
\begin{equation}
{\cal G}_{2B}(E)+\sum_n\frac{{\cal R}\{{\cal G}_{2B}(E),-E_n\}}{E+E_n}
=\frac{1}{2\pi i}\int_C dE'\frac{
{\cal G}_{2B}(E')}{E-E'},
\end{equation}
where the contour $C$ is illustrated in Fig.~\ref{fig:G3Bcont}. This relation
is consistent with the conclusion that the Green's function is an analytic
function of $E$ with no other singularities than those observed\footnote{There
is the possibility of other symmetric singularities whose contributions
cancel. If these exist they do not effect the validity of the results, which
only rely on the dispersion relation.}.

When we consider the Green's function as an analytic
function of $p=\sqrt{ME}$ the singularity structure found as a function
of energy is mapped into the upper-half of the complex $p$-plane.
The singularity structure in the lower-half of the complex
$p$-plane is difficult to determine, fortunately we do not need to know
any analytic properties in that region.

Now we know the analytic properties of ${\cal G}_{2B}(p)$,
expression (\ref{eq:projgreen}) can be used to obtain an
analytic continuation of the projection operator, ${\cal P}(p)$.
Using eqns.~(\ref{eq:energyproj},\ref{eq:projgreen}) we can
write
\begin{equation}\label{eq:analproj}
{\cal P}(p)=\frac{2\pi ip}{M}\Bigl({\cal G}_{2B}(p)-{\cal G}_{2B}(-p)\Bigr).
\end{equation}
Due to the form of the equation we cannot readily
continue the projection operator into the complex $p$-plane
without assumptions on the analytic properties of ${\cal G}_{2B}(p)$
in the lower half of the complex plane.

Eqn.~(\ref{eq:analproj}) is the generalisation (in unscaled coordinates) of
eqn.~(\ref{eq:analC}) and provides us with the means to solve the 3BDWRG quite generally
\footnote{The extension of the
analysis so far in this chapter to general $N$ body forces is straightforward,
Because of eqn.~(\ref{eq:analproj}), which is entirely independent of the number of
bodies interacting, the method used in finding a non-trivial solution to the
DWRG equation also generalises.}.

\section{The Solution of the 3BDWRG for ``Well-behaved'' Pairwise Forces.}

If the pairwise forces are ``well-behaved'' in the sense of section 3.6 then
the distorted waves must vanish as $R^2$ as $R\rightarrow0$. We can immediately
write down a form for the full Green's function and the distorted waves
at the point at which the three-body force acts $\bar R,\,\bar\alpha$,
\begin{eqnarray}
&&\displaystyle{\qquad
|\Psi_{\Lambda,k}(\bar R,\bar\alpha)|^2=
\Lambda^4\bar R^4|\varphi(\bar\alpha)|^2{\cal D}_3(k/\Lambda,\kappa_i/\Lambda),}\\
&&\displaystyle{\qquad
|\Psi_{\Lambda,\gamma_n}(\bar R,\bar\alpha)|^2=
\Lambda^5\bar R^4|\varphi(\bar\alpha)|^2{\cal D}_2^{n}(\kappa_i/\Lambda),}\\
&&\displaystyle{
\langle\bar R,\bar\alpha|{\cal G}_{2B}(\Lambda)|\bar R,\bar\alpha\rangle
=M\Lambda^4\bar R^4|\varphi(\bar\alpha)|^2{\cal D}_G(\kappa_i/\Lambda),}
\end{eqnarray}
where $\varphi(\bar\alpha)$ is an unimportant functions that describes the
hyperangular behaviour of the DWs close to the origin. Inserting these forms
into eqn.~(\ref{eq:3brgdiff}), the correct rescaling for the three-body force
is,
\begin{equation}
\hat V_{3B}(\hat p,\hat\kappa_i;\hat k,\hat k';\Lambda)=
\frac{M\bar R^5\Lambda^4}{2\pi^2}V_{3B}(\Lambda\hat p,\Lambda\hat\kappa_i;
\Lambda\hat k,\Lambda,\hat k',\Lambda).
\end{equation}
Resulting in the 3BDWRG equation,
\begin{eqnarray}
&&\displaystyle{
\Lambda\frac{\partial \hat V_{3B}}{\partial\Lambda}
=\hat p\frac{\partial\hat V_{3B}}{\partial\hat p}
+\sum_i\hat\kappa_i\frac{\partial\hat V_{3B}}{\partial\hat\kappa_i}
+\hat k\frac{\partial\hat V_{3B}}{\partial\hat k}
+\hat k'\frac{\partial\hat V_{3B}}{\partial\hat k'}+4\hat V_{3B}}\nonumber\\
&&\displaystyle{\,\,\,\,\,\,\,\,
+\frac{1}{1-\hat p^2}\Biggl[\sum_{\gamma_n}
\hat V_{3B}(\hat p,\hat\kappa_i;\hat k,i\hat\gamma_n;\Lambda)
\hat V_{3B}(\hat p,\hat\kappa_i;i\hat \gamma_n,\hat k';\Lambda)
{\cal D}_2^n(\hat\kappa_i)}\nonumber\\
&&\displaystyle{\qquad\,\,\,\,
+\frac{2}{\pi}\int_0^{1}d\hat k''
\hat V_{3B}(\hat p,\hat\kappa_i;\hat k,\hat k'';\Lambda)
\hat V_{3B}(\hat p,\hat\kappa_i;\hat k'',\hat k';\Lambda)
{\cal D}_3(\hat k'',\hat\kappa_i)\Biggr].
\label{eq:3bdwrg}}
\end{eqnarray}

This equation is quite complicated as it has to describe the
coupling of the three-body force to each of the distorted waves.
The trivial fixed point solution, $\hat V_{3B}=0$, corresponds to
a zero three-body force. The perturbations around it,
\begin{equation}
\hat V_{3B}=\sum_{l_i,m,n,n'=0}^\infty\hat C_{l_i,m,n,n'}
\left(\frac{\Lambda}{\Lambda_0}\right)^\nu
\hat\kappa_i^{l_i}\hat p^{2m}\hat k^{2n}\hat k'^{2n'},
\end{equation}
have RG eigenvalues $\nu=2m+2n+2n'+\sum l_i+4=4,(5),6\ldots$. All the eigenvalues
are positive so the fixed point is stable. This fixed point is  simply
Weinberg counting, in which the first three-body force term occurs
at $(Q/\Lambda_0)^4$.

To find a non-trivial fixed point solution, we generalise the arguments of section
3.6. A non-trivial fixed point that is independent of the momentum variables may be
found by solving the equation,
\begin{equation}
\hat p\frac{\partial\hat V_{3B}^{-1}}{\partial\hat p}
+\hat\kappa_i\frac{\partial\hat V_{3B}^{-1}}{\partial\hat\kappa_i}
-4\hat V_{3B}^{-1}
-\frac{2\pi i}{1-\hat p^2}\Bigl({\cal D}_G(\hat\kappa_i)-
{\cal D}_G(-\hat\kappa_i)\Bigr)=0,
\end{equation}
which comes from using the form for the projection operator (\ref{eq:analproj}).
This is solved by the contour integration,
\begin{equation}
\frac{1}{\hat V_{3B}^{(0)}}=2\pi i\int_Cd\hat q\frac{\hat q^5}{\hat p^2-\hat q^2}
{\cal D}_G(\hat\kappa_i/\hat q).
\end{equation}
where $C$ is a contour going from $-1$ to $1$ via the upper-half of the complex
$\hat q$-plane outside all three-body bound state poles and two-body elastic
thresholds on the imaginary axis. This solution is analytic in $\hat p$ but
may have simple logarithmic dependence on some of the scales
that can be removed using counterterms with logarithmic
dependence on $\Lambda$. The momentum-independent
perturbations around this fixed point solution take the form
\begin{equation}
\frac{1}{\hat V_{3B}}=\frac{1}{\hat V_{3B}^{(0)}}+
\sum_{l_i,m=0}^\infty\hat C_{l_i,m,n,n'}
\left(\frac{\Lambda}{\Lambda_0}\right)^\nu
\hat\kappa_i^{l_i}\hat p^{2m},
\end{equation}
with RG eigenvalues $\nu=2m+l_i-4=-4,(-3),-2,\ldots$. This fixed point is unstable
with 2 energy-dependent unstable perturbations and possibly 2 more $\kappa_i$-dependent
unstable perturbations. Like the power-counting schemes in the
two-body higher partial waves
it seems implausible that this fixed point is ever likely to provide a useful
expansion in physical systems because of the fine-tuning required in the unstable
perturbations.

\section{Three Body Force in the KSW EFT for Short Range Forces}

In this section we will return to the EFT considered in the
chapter 2 and its extension to systems of three particles.
When we construct an EFT Lagrangian, we have to include all terms that
do not violate the observed symmetries of the system. In a system in which
all energies are far lower than the inverse range of the interactions
this leads to a Lagrangian containing not only all the four-point
vertex interactions (including derivative couplings) that correspond to
the energy-dependent contact potentials but also to six-point and
higher vertex interactions
needed to model interaction channels that only become available in
the presence of more than two particles. Of course the six
point interactions are only required in a system with at least three
particles, while the eight point terms require four particles and so on.
The two-body couplings are entirely determined by two-body observables
and should be taken as given when considering a three-body system.
The three-body couplings are fixed by three-body
observables.

In reactions between a bound pair and a third particle
the pair-wise interactions result in a long range
force due to the exchange of one particle. The range of this force is
comparable to the scattering length of the two-body interaction.

If the two-body system is weakly interacting, the correct
power-counting in the two-body system is the Weinberg system.
The two-body scattering length
$a_2\sim1/\Lambda_0$, and hence provides a short-range
exchange interaction.
This system does not provide any particular difficulty in the
three-body system since we can use naive dimensional analysis to
determine at which order each diagram occurs. In this expansion
the three-body force occurs at $(Q/\Lambda_0)^4$ as in the previous
example. (There is, of course, the alternative finely-tuned three-body force outlined
in the previous example.)

In a KSW system the fine tuning of the
LO perturbation around the non-trivial fixed point of the RG
results in a large scattering length, $1/a_2\sim Q\ll1/\Lambda_0$, and hence
a long-range exchange interaction. This causes a significant
complication as all the diagrams involving the leading four-point
vertex need to be resummed\cite{quartetbvk,bhvk}. Since naive dimensional analysis is no
longer applicable the power-counting for the three-body forces must
be determined by renormalisation of the diagrams to which they
contribute.

The 3BDWRG allows us to resum the effects
of these diagrams and promote them to a fixed point. In the KSW EFT
we need only resum the effects of the scattering length terms since
it must be treated as a low-energy scale, it is not necessary to
resum the effects of the effective range terms since these are perturbative
$1/r_2\sim\Lambda_0$

The study of this problem has a long history. Skorniakov and Ter-Martirosian
(STM) \cite{stm} were the first to derive
equations for a system of a bound pair and particle with two-body contact interactions
defined solely by a scattering length. There work was followed up by
Danilov\cite{danilov} who realised that in some systems the STM equations did not
have unique solutions and required the input of a single piece of three-body
data, namely the binding energy of the three-body state. The work of both
of these predates that of Faddeev \cite{faddeev}
in resolving the disconnected nature of the
three-body equations. The STM equations rely upon the zero range nature of the
two-body interaction to reduce the problem to a one-dimensional equation. 

An alternative approach to the problem was given by Efimov \cite{efimov}.
Efimov's equations show the singular nature of the system far more transparently
than those of Skorniakov and Ter-Martirosian. We shall look at these equations
below: in a nutshell Efimov's approach allows us to express the system
in terms of a hyperradial inverse square potential. The strength of the potential
depends upon the statistics of the system. In the case of three Bosons and
three nucleons in the ${}^3S_1$ channel the inverse square potential
is attractive and hence singular,
requiring a choice of self-adjoint extension.

Since the conception of EFTs this system and the related singular inverse square
potential have received much attention \cite{bhvk,brgh,lcbc,camblong,splc,phill3b,
geg3b,hm3brg}. In particular, Bedaque {\it et al} \cite{bhvk,brgh} have
used extended versions of the STM equations to create a EFT that has proven
highly effective in describing three-body data. In nuclear systems, the so-called
pionless EFT proves effective in the s-wave $J=3/2$ system \cite{quartetbvk} and in higher
angular momentum systems \cite{highpwnd} without three-body forces. These systems
are repulsive at short distances because of the centrifugal barrier or the Pauli
exclusion principle and are therefore insensitive to short distance physics
and three-body interactions. In Efimov's approach these systems correspond to
repulsive hyperangular inverse square potentials and so, from the DWRG analysis in section
3.5,
the power-countings correspond to a weak system with LO three-body
force scaling with $(Q/\Lambda_0)^{2\nu}$ or a finely-tuned system with a LO
three-body force scaling with $(Q/\Lambda_0)^{-2\nu}$ where $\nu$ is the strength of the
hyperangular potential. Given the apparent success achieved without three-body forces
the first counting is appropriate. These systems are now well understood 
and we shall not concern ourselves with them.

The systems that correspond to singular hyperangular potentials are still being studied.
Bedaque {\it et al} have incorporated three-body forces into the STM equations and
used them to resolve the singular behaviour of these system \cite{bhvk,brgh}.
The effective three-body forces contain free parameters that must be fixed by
fitting to three-body data. The power-counting proposed by Bedaque {\it et al}
has three-body forces occurring at orders $(Q/\Lambda_0)^{2n}$ where $n=0,1,2,\ldots$.
The LO force is marginal. Each of the three-body force terms have cyclic behaviour
in the cut-off, $\Lambda$. Phillips and Afnan \cite{phill3b}
have also looked at the problem, their
solution is to incorporate the 2+1 scattering length into the equations in a manner
that `fixes' the singularity and ensures that they give the correct 2+1 scattering
length. The input of a single piece of three-body data is surely equivalent to the
LO three-body force of Bedaque {\it et al} and both are equivalent to a
particular choice of self-adjoint extension of the Hamiltonian. The difference in their
stances is in attributing it to a three-body force and the need for higher
order three-body forces.

In this section we shall introduce Efimov's equations for three Bosons
and explicitly
demonstrate the connection between the inverse square potential and the KSW EFT
three-body problem. We shall discuss the case of infinite two-body scattering length,
where Efimov's equations may be solved analytically, and show that in this case the
power-counting for the three-body force is the same as proposed by
Bedaque {\it et al}\cite{brgh}.
We shall then use the 3BDWRG to discuss the more general case of systems with
finite two-body scattering length. We will show that the introduction of a finite
scattering length does not change the power-counting for the three-body force and that
the 3BDWRG solution is governed by a limit-cycle solution.

\subsection{Efimov's Equations}
The two-body interaction may be written as a boundary
condition on the wavefunction at zero separation\footnote{Compare
this to the observations about Neumann and Dirichlet boundary
conditions and their connections to the fixed points of
the two-body DWRG.}. The
full three-body wavefunction, $|\Psi\rangle$, is defined by the wave equation
and the three boundary conditions,
\begin{eqnarray}
\label{eq:3bwe}&&\displaystyle{\qquad\qquad
(H_0-E)|\Psi\rangle=0,}\\
&&\displaystyle{
\label{eq:3bbc}\left[\frac{\partial}{\partial x_{ij}}
x_{ij}\langle \boldr_{ij},\boldrho_k|\Psi\rangle
\right]_{x_{ij}=0}
=\left[a_2^{-1}x_{ij}\langle\boldr_{ij},\boldrho_k|\Psi\rangle\right]_{x_{ij}=0}.}
\end{eqnarray}
From here we shall drop the particle number indices by writing the
wavefunction in terms of the permutation operator, $P$,
\begin{equation}
|\Psi\rangle=(1+P)|\psi\rangle.
\end{equation}
Substituting this form into eqn. (\ref{eq:3bbc}) we obtain,
\begin{equation}\label{eq:3bbc2}
\left[\frac{\partial}{\partial x} x\langle\boldr,\boldrho|\psi\rangle+
\langle \boldr,\boldrho|P|\psi\rangle\right]_{x=0}=
\left[a_2^{-1}x\langle\boldr,\boldrho|\psi\rangle\right]_{x=0}.
\end{equation}
We consider the case of zero angular
momentum and write the wavefunction as
$\psi(x,y)=xy\langle\boldr,\boldrho|\psi\rangle$,
so that eqns.(\ref{eq:3bwe},\ref{eq:3bbc2}) become,
\begin{eqnarray}\label{eq:efimov1}
&&\displaystyle{\qquad\qquad
\left(\frac{\partial^2}{\partial x^2}+\frac34\frac{\partial^2}{\partial y^2}+ME
\right)\psi(x,y)=0},\\
&&\displaystyle{\label{eq:efimov2}
\left[\frac{\partial\psi(x,y)}{\partial x}\right]_{x=0}
+\frac{4}{y}\psi\left(y,\frac12y\right)
=a_2^{-1}\psi(0,y).}
\end{eqnarray}
In addition we have the boundary condition which follows from the definition
of $\psi$: $\psi(x,0)=0$.
Eqns.(\ref{eq:efimov1},\ref{eq:efimov2})
are the wave equation with non-trivial boundary condition
and are, in general, extremely difficult to solve. To examine them more closely
we rewrite them
in terms of the hyperpolar coordinates, $x=R\cos\alpha,y=\sqrt{3}/2R\sin\alpha$,
\begin{eqnarray}\label{eq:efimovhyp1}
&&\displaystyle{\,\,\,\,
\left(\frac{\partial^2}{\partial R^2}+\frac{1}{R}\frac{\partial}{\partial R}
+\frac{1}{R^2}
\frac{\partial^2}{\partial \alpha^2}+ME\right)\psi(R,\alpha)}=0,\\
&&\displaystyle{\label{eq:efimovhyp2}
\left[\frac{\partial\psi(R,\alpha)}{\partial\alpha}\right]_{\alpha=\pi/2}
+\frac{8}{\sqrt{3}}\psi\left(R,\frac{\pi}{6}\right)=
Ra_2^{-1}\psi\left(R,\frac\pi2\right)}.
\end{eqnarray}
These provide us with the information
we require to study the DWRG in this problem. 

\subsection{The Infinite Scattering Length Limit.}
Efimov's equations, (\ref{eq:efimovhyp1},\ref{eq:efimovhyp2}) may be
solved analytically in the limit of infinite scattering length.
We shall turn our attention to this case as it is considerably
simpler than the more general case of finite scattering length.
In the limit $a_2\rightarrow\infty$ the boundary condition
becomes separable. Writing,
\begin{equation}
\psi(R,\alpha)=F_s(R)\sin s\alpha,
\end{equation}
we obtain,
\begin{eqnarray}
&&\displaystyle{\label{eq:efimov}
\left(\frac{\partial^2}{\partial R^2}+\frac{1}{R}\frac{\partial}{\partial R}
-\frac{s^2}{R^2}+p^2\right)F_s(R)}=0,\\
&&\displaystyle{\qquad\label{eq:danilov}
-s\cos\frac{s\pi}{2}+\frac{8}{\sqrt{3}}\sin{\frac{s\pi}{6}}=0
},
\end{eqnarray}
where $p=\sqrt{ME}$.
These equations are now a simple matter to solve. The transcendental equation
(\ref{eq:danilov}) must be solved numerically. It has two imaginary solutions
$s_0=\pm i1.006237\ldots$ and an infinite number of real solutions, the smallest
of which is $s_1=4$. The imaginary roots of Eqn.(\ref{eq:danilov})
result in an attractive inverse square potential for the hyperradial equation
(\ref{eq:efimov}) while the real roots give repulsive potentials.

The hyperradial part of the wavefunction is given by,
\begin{equation}
F_{s}(R;p)=
\begin{cases}
\sqrt{\frac{\pi p}{2}}J_{s}(pR)\qquad\text{if}\qquad s\in{\mathbb R}\\
\sqrt{\frac{\pi p}{2}}(2i|\sin(\eta(p)+i\pi\bar s_i/2)|)^{-1}
\Bigl(e^{i\eta(p)}J_{i\bar s}(pR)-e^{-i\eta(p)}J_{-i\bar s}(pR)\Bigr)
\qquad\text{if}\qquad s\not\in{\mathbb R},
\end{cases}
\end{equation}
where $\bar s=\text{Im}{s}$. In the imaginary hyperangular momentum channel,
since this is an attractive channel, we have, as was necessary,
defined a self-adjoint extension of the Hamiltonian by choosing a phase,
$\eta(p)$.
The bound states of the system only occur in this channel and
are given, in parallel to the bound states of the attractive inverse square
potential, by
\begin{equation}
F_{s}^{(n)}(R)=\sqrt{\frac{2\sinh(\pi\bar s)}
{\pi\bar s}}p_n K_{i\bar s}(p_nR).
\end{equation}
These states, like the attractive inverse square potential, form an infinite
tower of bound states. The lack of a ground state in three body system acting under such
forces was first noted by Thomas\cite{thomas}
and is known as the Thomas effect; of course such a state of
affairs is a result of an inexact treatment of short-range physics and will
be resolved by the EFT precisely as it was in the chapter \ref{Ch3}. The
accumulation of bound states at zero energy is due to the tail of the
inverse square potential, and hence a result of this special case of infinite
scattering length. These states are known as Efimov states \cite{efimov} and are not
a result of any mistreatment of physics but are genuine phenomena
\footnote{It is hoped that Efimov states may be observed experimentally
by tuning a diatomic bound state in a Bose Einstein Condensate.\cite{bose}}.

Since the Hamiltonian conserves hyperangular momentum we can expand any
three body scattering amplitude in terms of the
hyperangular eigenfunctions and corresponding hyper-phaseshifts.
This makes the DWRG analysis particularly simple since the equations
become one-dimensional.
The effect of the three-body force may be analysed
in each hyperangular channel, in other words dependence upon the
unconstrained
external momentum terms $k$ and $k'$, favoured earlier in the chapter,
can be dropped in favour of
dependence upon the hyperangular momentum $s$. The differential
equation for the three body force in each channel is then,
\begin{eqnarray}
&&\displaystyle{\qquad\qquad
\frac{\partial V_{3B}^{(s)}(p,\Lambda)}{\partial\Lambda}=\frac{M\bar R}{2\pi^2}
\frac{|\Psi_{\Lambda,s}(\bar R,\bar\alpha)|^2}
{\Lambda^2-p^2}V_{3B}^{(s)}(p,\Lambda)^2\qquad\text{ if }s\in{\mathbb R},}\\
&&\displaystyle{
\frac{\partial V_{3B}^{(s)}(p,\Lambda)}{\partial\Lambda}=\frac{M\bar R}{2\pi^2}
\left(\frac{|\Psi_{\Lambda,s}(\bar R,\bar\alpha)|^2}{\Lambda^2-p^2}
-\frac{\pi}{2}\sum_n\delta(\Lambda-|p_n|)
\frac{|\Psi_s^{(n)}(\bar R,\bar\alpha)|^2}{p^2+p_n^2}\right)
V_{3B}^{(s)}(p,\Lambda)^2}\nonumber\\
&&\displaystyle{\qquad\qquad\qquad\qquad\qquad\qquad\qquad\qquad\qquad
\qquad\qquad\qquad\qquad\text{ if }s\not\in{\mathbb R}.}
\end{eqnarray}

In the repulsive channels (real $s$) the resulting DWRG equation
is precisely that considered in section 3.5
and the analysis there can be followed to obtain the power-counting
for the three-body force in those channels. The effects of the force can
be expressed as a distorted wave Born or effective range expansion for the
correction to the hyper-phaseshift. Following the discussion in section 3.5
the fine-tuning required for a non-trivial fixed point power-counting for
these forces would be rather contrived. For example, the least repulsive
channel has the eigenvalue $s_1=4$ and would require three unnaturally small
parameters to ensure the RG flow remained in the region of the non-trivial
fixed point. Hence, the appropriate power-counting for the three-body
force in these channels is based upon the trivial fixed point and provides
only very small (at least of order $(Q/\Lambda_0)^4$) contributions to the scattering
observables.

In the attractive channels we must turn to the analysis considered
of chapter \ref{Ch3}. The DWRG analysis is precisely as found there
and is very different from that in the repulsive
channels and results in a three-body force that contributes at a surprisingly
low order, $(Q/\Lambda_0)^0$, to the scattering observables.
In this role it is more important than effective range corrections
which occur at an order, $(Q/\Lambda_0)^1$. This power-counting for the
three-body force agrees with Bedaque {\it et al} \cite{brgh}.

The implications, for both EFT and potential modelling of the three-body
problem, of such a prominent three-body force will be examined later.
Before then we shall show how the DWRG analysis for the three body force
needs to be modified after the introduction of a finite two-body
scattering length.

The method outlined in this section is useful because everything may be
dealt with analytically. However, in order to fully grasp the
general problem with finite scattering length, our general analysis must be
based upon the the 3BDWRG. This not only allows us to deal with
the problem of the different channels but also allows us to presents results
for observables in more practical format.

\subsection{Finite Scattering Length}

\subsubsection{The Distorted Waves and the DWRG Equation}

In reaching eqns.(\ref{eq:efimov},\ref{eq:danilov}) we took the two-body
scattering length to infinity. This limit is of course also reached if we
take the hyperradius to zero, meaning that at small hyperradii the general
distorted waves have the same functional form (in $R$)
as their infinite scattering length counterparts. In particular
we must have for any DW,
\begin{equation}\label{eq:asymp}
\Psi(R,\alpha)={\cal D}\varphi(\alpha)\sin(\bar s_0\ln p_*R),
\end{equation}
where we must chose the scale $p_*$ in order to define a self-adjoint
extension. The magnitude of the DWs close to the origin will in general
depend upon the two-body scattering length. As in the attractive inverse
square case, the relationship between self-adjoint extensions and
$p_*$ is not one-to-one. All observables are invariant under the transformation,
\begin{equation}\label{eq:selfadjinv}
p_*\rightarrow p_*e^{\pi n/\bar s_0}.
\end{equation}
where $\bar s_0=1.00623\ldots$ is the magnitude of the imaginary solution
of eqn.~(\ref{eq:danilov}).

The Green's function, since it solves the Schr\"odinger equation,
also observes the behaviour (\ref{eq:asymp}).
We write the Green's function matrix elements evaluated
at small hyperradius as
\begin{equation}
\langle\bar R,\bar\alpha|{\cal G}_{2B}(p)|\bar R,\bar\alpha\rangle
=M{\cal D}_{G}\left(\gamma/p,\ln p/p_*\right)
|\varphi(\bar\alpha)|^2\sin^2\left(\bar s_0\ln{p_*\bar R}\right),
\end{equation}
where $\gamma=1/a_2$ is the binding momentum of the two-body bound state.
Because of the
relationship (\ref{eq:selfadjinv}), ${\cal D}_G(x,y)$ is periodic in $y$
with period $2\pi/\bar s_0$. All the analytic properties of
${\cal G}_{2B}(p)$ as a function of $p$, described earlier,
are conferred on the function ${\cal D}_{G}$. Namely, there is the elastic
two-body scattering cut at $p=i\gamma$ that runs down the imaginary $p$ axis
then along the real axis and there are poles at each of the bound states.

The singular behaviour at small hyperradii that leads to the trig-log
behaviour in the DWs also means that there is no ground state. At very
high momentum the two-body scattering length becomes insignificant and
the DWs take the form they would in the limit $\gamma\rightarrow0$. This means
that the deeply bound state take on the geometrically spaced form of
the attractive inverse square potential's states. At low energies the
value of $\gamma$ is important and disrupts the exponential ladder of
bound states; the Efimov effect no longer occurs.

The lack of a ground state means that, as in chapter 4, we must truncate
the full Green's function in both the continuum and bound states. Notice
there is no need to continue this truncation into the elastic threshold
since this is controlled
by the low-energy scale $\gamma$. The result of truncating the bound states is
slightly more complicated than in the attractive inverse square example because
the position of each bound state is not only controlled by the high-energy scale
$p_*$ but also by the low-energy scale $\gamma$. The effect this has on the
renormalisation group flow is important and requires careful consideration.

For the individual distorted waves we write down the forms for small $\bar R$ as
\begin{eqnarray}
&&\displaystyle{
|\Psi_{p,k}(\bar R,\bar\alpha)|^2=
{\cal D}_3\left(\frac{k}{p},\frac{\gamma}{p},\ln\frac{p}{p_*}\right)
|\varphi(\bar\alpha)|^2
\sin^2\left(\bar s_0\ln p_*\bar R\right),}\\
&&\displaystyle{
|\Psi_{p,\gamma}(\bar R,\bar\alpha)|^2=
p{\cal D}_2\left(\frac{\gamma}{p},\ln\frac{p}{p_*}\right)|\varphi(\bar\alpha)|^2
\sin^2\left(\bar s_0\ln p_*\bar R\right),}\\
&&\displaystyle{
|\Psi_n(\bar R,\bar\alpha)|^2=
p_n(\gamma,p_*)^2{\cal D}_B\left(\gamma,p_n(\gamma,p_*)\right)
|\varphi(\bar\alpha)|^2
\sin^2\left(\bar s_0\ln{p_*\bar R}\right),}
\end{eqnarray}
where we have explicitly shown the dependence of the binding momenta, $p_n$, on
$\gamma$ and $p_*$.
From eqn.~(\ref{eq:greenboundres}) it follows that ${\cal D}_B(\gamma,p_n)$
can simply be found by taking the residue of the Green's function at the
appropriate bound state pole to give,
\begin{equation}\label{eq:bounddisc}
{\cal D}_B(\gamma,p_n)=\frac{8\pi i}{p_n}
{\cal R}\{{\cal D}_G(\gamma/p,\ln p/p_*),p\rightarrow ip_n\}.
\end{equation}
The projection operator matrix element for small $\bar R$ 
and $p^2>0$ follow from its definition (\ref{eq:projop2}),
\begin{equation}
\langle\bar R,\bar\alpha|{\cal P}(p)|\bar R,\bar\alpha\rangle=
p{\cal C}\left(\frac{\gamma}{p},\ln\frac{p}{p_*}\right)
|\varphi(\bar\alpha)|^2
\sin^2\left(\bar s_0\ln{p_*\bar R}\right),
\end{equation}
where
\begin{equation}\label{eq:CintermsofDWs}
{\cal C}\left(\frac{\gamma}{p},\ln\frac{p}{p_*}\right)=
{\cal D}_2\left(\frac{\gamma}{p},\ln\frac{p}{p_*}\right)+\frac{2}{\pi}\int_0^pdk
{\cal D}_3\left(\frac{k}{p},\frac{\gamma}{p},\ln\frac{p}{p_*}\right).
\end{equation}
The function ${\cal C}$ may also be written using eqn.~(\ref{eq:analproj})
as,
\begin{equation}\label{eq:CintermsofDg}
{\cal C}\left(\frac{\gamma}{p},\ln\frac{p}{p_*}\right)=
2\pi i\left[{\cal D}_G\left(\frac{\gamma}{p},\ln \frac{p}{p_*}\right)
-{\cal D}_G\left(\frac{-\gamma}{p},\ln\frac{p}{p_*}+i\pi\right)\right].
\end{equation}
We are now in a position to rescale the differential equation for
$V_{3B}$ to obtain the DWRG equation. The scales $p,\gamma,k$ and
$k'$ rescale in the usual way. The three-body force is rescaled by
the relation,
\begin{equation}
\hat V_{3B}(\hat p,\hat\gamma;\hat k,\hat k';\Lambda)=
\frac{M \bar R}{2\pi^2}|\varphi(\bar\alpha)|^2
\sin^2\left(\bar s_0\ln{p_*\bar R}\right)
V_{3B}(\Lambda\hat p,\Lambda\hat\gamma;\Lambda\hat k,\Lambda\hat k';\Lambda).
\end{equation}
Using eqn.(\ref{eq:3brgdiff}) with the addition of the truncated bound states
the resulting DWRG for the three-body force is,
\begin{eqnarray}
&&\displaystyle{
\Lambda\frac{\partial \hat V_{3B}}{\partial\Lambda}
=\hat p\frac{\partial\hat V_{3B}}{\partial\hat p}
+\hat\gamma\frac{\partial\hat V_{3B}}{\partial\hat\gamma}
+\hat k\frac{\partial\hat V_{3B}}{\partial\hat k}
+\hat k'\frac{\partial\hat V_{3B}}{\partial\hat k'}}\nonumber\\
&&\displaystyle{\,\,\,\,
+\frac{1}{1-\hat p^2}\Biggl[
\hat V_{3B}(\hat p,\hat\gamma;\hat k,i\hat\gamma;\Lambda)
\hat V_{3B}(\hat p,\hat\gamma;i\hat \gamma,\hat k';\Lambda)
{\cal D}_2(\hat\gamma,\ln \Lambda/p_*)}\nonumber\\
&&\displaystyle{\qquad
+\frac{2}{\pi}\int_0^{1}d\hat k''
\hat V_{3B}(\hat p,\hat\gamma;\hat k,\hat k'';\Lambda)
\hat V_{3B}(\hat p,\hat\gamma;\hat k'',\hat k';\Lambda)
{\cal D}_3(\hat k'',\hat\gamma,\ln \Lambda/p_*)\Biggr]}\nonumber\\
&&\displaystyle{
-\frac{\pi}{2}\sum_n
\hat V_{3B}(\hat p,\hat\gamma;\hat k,i/3;\Lambda)
\hat V_{3B}(\hat p,\hat\gamma;i/3,\hat k';\Lambda)
\frac{\delta(\Lambda-p_n(\Lambda\hat\gamma,p_*))\Lambda}{1+\hat p^2}
{\cal D}_B(\hat\gamma/\hat p_n,\ln p_n/p_*).}\nonumber\\
\label{eq:3bdwrg3}
\end{eqnarray}
where we have suppressed the obvious functional dependence of $\hat V_{3B}$
in the terms on the first line to save space.

The perturbations around the trivial fixed point solution,
$\hat V_{3B}=0$, are given by,
\begin{equation}
\hat V_{3B}=\sum_{k,l,m,n}\hat C_{l,m,n,n'}\hat p^{2m}\hat k^{2n}
\hat k'^{2n'}\hat\gamma^l
\left(\frac{\Lambda}{\Lambda_0}\right)^{l+2m+2n+2n'}.
\end{equation}
The leading order perturbation is marginal, this is consistent
with the trivial fixed point solution in the attractive inverse square
(zero $\hat\gamma$)
solution. Like the attractive inverse square DWRG this solution
means very little by itself and in order to determine the nature
of the marginal perturbation and the general flow of the DWRG
we must examine a more general solution. Like the attractive inverse
square DWRG the marginal perturbation occurs because the basic loop integral
solution is an example of a limit-cycle solution.

\subsubsection{The Limit-Cycle DWRG solution.}

\begin{figure}
\begin{center}
\includegraphics[height=8cm,width=8cm,angle=0]{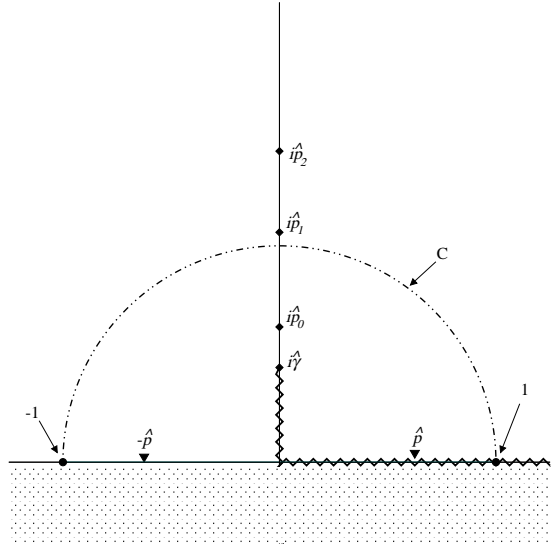}
\caption{The singularity structure of the integrand in the limit cycle
solution. The bound state poles occur at $\hat q=i\hat p_n$, the
two-body elastic scattering threshold branch point occurs
at $\hat q=i\hat\gamma$. The propagator poles occur at $\hat q=\pm\hat p$.
The contour $C$ runs in a semi-circle from $-1$ to $1$ through the
upper half of the complex plane and crosses the imaginary axis at $\hat q=i$.}
\label{fig:poles5}
\end{center}
\end{figure}

We look now for a solution to the 3BDWRG equation that depends on
$\Lambda$ logarithmically and that is independent of the momentum
terms. This satisfies the 3BDWRG equation,
\begin{equation}
\Lambda\frac{\partial}{\partial\Lambda}\frac{1}{\hat V_{3B}}
=\hat p\frac{\partial}{\partial\hat p}\frac{1}{\hat V_{3B}}
+\hat\gamma\frac{\partial}
{\partial\hat\gamma}\frac{1}{\hat V_{3B}}
-\frac{{\cal C}(\hat\gamma,\ln\Lambda/p_*)}{1-\hat p^2}
+\frac{\pi}{2}\sum_n
\frac{\delta(\Lambda-p_n(\Lambda\hat\gamma,p_*))\Lambda}{1+\hat p^2}
{\cal D}_B(\hat\gamma,\ln p_n/p_*).
\end{equation}
The solution to this equation with logarithmic dependence upon $\Lambda$
may be constructed in much the same way as the solution to the attractive
inverse square potential DWRG. We write the solution
as the integral,
\begin{equation}\label{eq:initsol}
\frac{1}{\hat V_{3B}^{(0)}(\hat p,\hat\gamma,\Lambda)}=2\pi i\int_C
d\hat q\frac{\hat q}{\hat p^2-\hat q^2}
{\cal D}_{G}\left(\frac{\hat\gamma}{\hat q},\ln\frac{\Lambda\hat q}{p_*}\right),
\end{equation}
where the contour $C$ is illustrated in Fig.\ref{fig:poles5},
it runs from $-1$ to $1$ through the upper half of the complex plane
and crosses the imaginary axis at $\hat q=i$. Substitution
and use of eqn.~(\ref{eq:CintermsofDg})
shows that it satisfies the continuous part of the 3BDWRG equation.
The integral is clearly analytic about $\hat p\rightarrow0$ since the propagator
poles to do not approach any part of the contour $C$ in this limit. The integral
is also analytic as $\hat\gamma\rightarrow0$ since it matches
continuously onto the attractive inverse square solution in that limit.
The discontinuities caused by the truncation of bound states result from allowing
the bound state poles to cross the contour. That the residues of these poles produce
the correct discontinuities is readily confirmed using eqn.~(\ref{eq:bounddisc}).

The migration of the bound state poles up the imaginary axis is more
complicated in this example than in the attractive inverse square case.
The position of each bound state in the unscaled theory is determined
by the scales $\gamma$ and $p_*$. In the rescaled theory they are determined
by $\hat\gamma$ and the ratio $p_*/\Lambda$. This has the interesting consequence
that as $\Lambda$ varies only that element of the rescaled poles position controlled
by the high energy scale $p_*$ changes. The element controlled by $\gamma$, as
low energy physics, remains fixed.

We know that the transformation $p_*\rightarrow p_*e^{\pi/\bar s_0}$
leaves the bound state spectrum unchanged.
This transformation is equivalent, in the rescaled spectrum, to
\begin{equation}\label{eq:cyclicsym}
\Lambda\rightarrow \Lambda e^{-\pi/\bar s_0},
\end{equation}
which must therefore leave the rescaled spectrum unchanged.

As $\Lambda$ decreases the poles move up the imaginary $\hat q$ axis. To
ensure the rescaled bound states poles are invariant under transformation
(\ref{eq:cyclicsym}), new poles must `condense' out of the elastic
threshold cut when necessary. 

The solution (\ref{eq:initsol}) describes a logarithmically evolving
limit cycle solution. This can be seen by observing that since
${\cal D}_G(x,y)={\cal D}_G(x,y+2\pi/\bar s_0)$,
\begin{equation}\label{eq:cyclic}
\hat V_{3B}^{(0)}(\hat p,\hat\gamma,\Lambda)=
\hat V_{3B}^{(0)}(\hat p,\hat\gamma,\Lambda e^{2\pi n/\bar s_0}).
\end{equation} 
This cyclic behaviour means that there must be an infinite number of geometrically
spaced discontinuities as $\Lambda$ goes to either zero or infinity.
At first it seems strange that we should be able to produce an infinite
number of discontinuities as $\Lambda\rightarrow0$ because for any 
particular $\Lambda$ there are only a finite number of bound state poles inside the
contour $C$ in the complex $\hat q$-plane. However, bearing in mind the discussion
above we can see that as $\Lambda\rightarrow0$ more rescaled bound states appear
and, because of eqn.~(\ref{eq:cyclicsym}), these will cross the contours in
a strictly geometrically spaced pattern.

The limit cycle behaviour is only exact in the rescaled theory. The unscaled
potential will approximate the limit cycle when $\Lambda\gg\gamma$ but as $\Lambda$
becomes less than $\gamma$ it will deviate significantly from it.

The momentum independent perturbations around the limit-cycle solution take the
form observed in chapter 4,
\begin{equation}
\frac{1}{\hat V_{3B}}(\hat p,\hat\gamma,\Lambda)=
\frac{1}{\hat V_{3B}^{(0)}}(\hat p,\hat\gamma,\Lambda)+
\sum_{n,m=0}^\infty\hat C_{2n,m}\left(\frac{\Lambda}{\Lambda_0}\right)^{2n+m}
\hat p^{2n}\gamma^m.
\end{equation}
The marginal perturbation $\hat C_{0,0}$ must, as in the inverse square case,
embody the same degree of freedom as the self-adjoint extension defining scale, $p_*$.
The DWRG flow is very much as described in section 3.3 and as illustrated
in Figs.~\ref{fig:b0running},\ref{fig:b0b2flow2}.

This momentum independent solution may be substituted into the three-body
distorted wave Lippmann-Schwinger equation (\ref{eq:3bdwls}) to
obtain results for the $\tilde T_{3B}$-matrix. The analysis follows
much the same form of that in section 2.6. In this case the truncated Green's
function also includes the integral around the branch cut along the
imaginary axis, which corresponds to elastic bound pair and particle scattering.
The result is
\begin{eqnarray}
&&\displaystyle{
\left[{\cal D}_3\left(\frac kp,\frac\gamma p,\ln\frac pp_*\right)
{\cal D}_3\left(\frac {k'}p,\frac\gamma p,\ln\frac pp_*\right)\right]^{1/2}
\langle \Psi_{p,k}|\tilde{\cal T}_{3B}|\Psi_{p,k'}\rangle^{-1}
-M{\cal D}_G\left(\frac\gamma p,\ln\frac pp_*\right)}\nonumber\\
&&\displaystyle{\qquad\qquad\qquad\qquad\qquad\qquad\qquad\qquad\qquad
=\frac{M}{2\pi^2}\sum_{m,n=0}{\hat C}_{2n,m}\frac{p^{2n}\gamma^m}{\Lambda_0^{2n+m}}}
\label{eq:dwere3b}
\end{eqnarray}
for a general $3\rightarrow3$ scattering. For elastic scattering
of a bound pair and a single particle we may write the $\tilde{\cal T}_{3B}$
matrix element in terms of a correction to the phaseshift in accordance with
eqn.~(\ref{eq:dwtps}). The DWRG limit-cycle potential then yields the
DWERE,
\begin{equation}\label{eq:elastic3b1}
{\cal D}_2\left(\frac{\gamma}{p},\ln\frac pp_*\right)
(\cot\tilde\delta_{3B}-i)+4\pi
{\cal D}_G\left(\frac\gamma p,\ln\frac pp_*\right)
=-\frac{2}{\pi}\sum_{m,n=0}{\hat C}_{2n,m}\frac{p^{2n}\gamma^m}{\Lambda_0^{2n+m}}.
\end{equation}
If we have $p^2<0$ then by eqns.~(\ref{eq:CintermsofDWs},\ref{eq:CintermsofDg})
we have,
\begin{equation}\label{eq:3bunit}
4\pi\,\text{Im}{\cal D}_G\left(\frac{\gamma}{p},\ln\frac pp_*\right)
={\cal D}_2\left(\frac\gamma p,\ln\frac pp_*\right),
\end{equation}
so that the imaginary parts on the LHS of eqn.~(\ref{eq:elastic3b1}) cancel and
the correction to the phaseshift, $\tilde\delta_{3B}$ 
is real. If $p^2>0$ then eqn.(\ref{eq:3bunit})
no longer holds because of the $3\rightarrow3$ states.

The results above are for the momentum independent solutions of the DWRG
only. More general solutions to the DWRG equations can be obtained
by perturbing about the limit cycle solution with momentum dependent
perturbations. These perturbations allow the three-body potential to differentiate
between different types of DW. It is clear that they will be important in
amplitudes like (\ref{eq:dwere3b}). However, in amplitudes such as 
(\ref{eq:elastic3b1}) they will not provide any extra degrees
of freedom in the power-counting since the external momenta $k$ and $k'$
are set to $i\gamma$ in this amplitude. For completeness the momentum
perturbations are found in Appendix \ref{app:Mompert}.

In any particular system that the EFT is describing the two-body
bound state binding momentum, $\gamma$, is not adjustable. It is therefore impossible to
determine the three-body force terms that correspond to these
perturbations around the limit-cycle solution. For practical
purposes these terms can be absorbed into the energy dependent
perturbations to obtain for elastic scattering,
\begin{equation}\label{eq:elastic3b}
{\cal D}_2\left(\frac{\gamma}{p},\ln\frac pp_*\right)
(\cot\tilde\delta_{3B}-i)+4\pi
{\cal D}_G\left(\frac\gamma p,\ln\frac pp_*\right)
=-\frac{2}{\pi}\sum_{m,n=0}
{\hat C}_{2n}\left(\frac{p}{\Lambda_0}\right)^{2n}.
\end{equation}

The power-counting in the energy dependent terms for the three-body
force about the limit-cycle solution is precisely that found by Bedaque
{\it et al}. The leading three-body force term is marginal and occurs at
order $(Q/\Lambda_0)^0$. This leading order force fixes the phase of the
three-body DWs close to the origin, or equivalently forms a self-adjoint
extension. The use of the DWRG method in deriving the power-counting for the
three-body force in the EFT is a new result. It provides support
for the work of Bedaque {\it et al}\cite{bhvk}, which relies on introducing a
dimer\footnote{A field with the quantum numbers of two particles, essentially
used as a bound state.} field into the EFT Lagrangian and then practical use of
the three-body force to renormalise the theory order by order. Despite the
power of their approach Bedaque {\it et al} cannot provide a
simple algebraic statement of the power-counting nor a simple formula like (\ref{eq:elastic3b}).

When looked at from a different angle, our analysis means that the introduction of any
three-body data to fix a self-adjoint extension such as was done by Phillips and
Afnan \cite{phill3b} is equivalent to a LO three-body force. It is not necessary to
explicitly calculate the three-body force. However, if we were to do so after fixing
a self-adjoint extension we would find that the degree of freedom associated with that
three-body force would be equivalent to that associated with the self-adjoint extension
choice.

The RG methods used here are very different from those used by Hammer and Mehen \cite{hm3brg},
which start with the STM equation and supplement the approach of Bedaque {\it et al}
\cite{bhvk,brgh}. The STM equation can only describe three-body bound states
and 2+1 scattering and so does not include all of the three-body physics.
Furthermore, since the equations of Hammer and Mehen start from a truncated STM equation,
part of the role of their three-body force is to restore the truncated part of the STM equation,
a complication the DWRG method was designed to avoid. For similar reasons the form of
our three-body force can not be directly compared to that given by Bedaque {\it et al}.

A limit cycle form of the three-body force in this EFT has been
suggested by many previous papers\cite{lcbc,splc} that have essentially tackled the
infinite scattering length case of the attractive inverse square problem.
Bedaque {\it et al} also found a limit cycle solution but because of
their approach could only find the force order by order \cite{brgh}.

Our result is also applicable to the three nucleon force in the pionless KSW
EFT \cite{brgh,bhvk,irrglc}. Braaten and Hammer have discussed the tuning of the quark masses
in QCD required for an infra-red limit cycle in the three nucleon problem \cite{irrglc}.
They assume that to obtain the scale invariant relationships that define a limit-cycle,
such as (\ref{eq:cyclicsym}), the system must be tuned to the infinite scattering length limit
to avoid the introduction of scales that break the symmetry.
The analysis here, which uses a Wilsonian RG approach, readily produces limit cycles
in the rescaled potential since the low-energy quantities
do not affect the symmetry. The discussion of Braaten and Hammer \cite{irrglc} applies
to our unscaled potential, which only exhibits infra-red limit cycle behaviour in the
infinite scattering length limit.

The complete 3BDWRG analysis in this system provides a concrete
grounding for what has been intuitively understood. Beyond that,
because of the common method of solving the DWRG equations in
in this chapter and the previous two, the connection between all these
results can be seen.

\section{Summary}
In this chapter we have examined the DWRG for three body forces. The approach
used can be generalised quite simply to $N$-body forces. The 3BDWRG
equation is complicated by this multi-channelled problem because the
3BDWRG has to describe the coupling of the three-body force
to each of the different distorted waves.

The application of the 3BDWRG to the case of ``well-behaved'' potentials
was included as an example. The two fixed points found correspond
to Weinberg counting and to the three body equivalent of the KSW
scheme. The instability of the non-trivial fixed point leads us to
conclude that in almost all systems of this kind Weinberg counting and
naive dimensional analysis is appropriate.

The 3BDWRG was also used to study three body forces in the
KSW EFT. The solutions to the equation take the form of limit cycles
and have much in common with the solutions found in chapter 4. The power
counting found is marginal at leading order and matches that found by Bedaque
{\it et al} \cite{brgh}.

Although our analysis has only considered the system of three Bosons, it
is also appropriate for three nucleons in the ${}^3S_1$ channel as we shall see
in the next chapter.

%% file: ch5/chapter5.tex
A pionless KSW EFT for three nucleons has been successfully described by Bedaque
{\it et al} \cite{brgh,quartetbvk,highpwnd}. Their work also covers
the related system of three Bosons.
We have seen in the previous chapter how the 3BDWRG analysis of the
three-body force in this problem supports their conclusions. We shall now look
to complement this work by deriving equations for the three body DWs.

All the predictive literature on the three body KSW EFT is concerned with scattering
observables\cite{brgh,phill3b}.
However, some applications of the EFT require knowledge of the
wavefunctions. For example, if we want to model scattering of an electron from
the triton we need to know its charge density, which can be ascertained from the
wavefunction.

The explicit solution of the equations for the DWs will also serve to illustrate
the `phase-fixing' role of the LO three-body force. We have seen in the
3BDWRG the equivalence of the
choice of self-adjoint extension and the LO marginal term in the short range force.
Because of this equivalence, the introduction of the three-body force at LO
is a trivial matter of choosing a boundary condition for the poorly defined
DWs.

As an introduction we shall consider the example of three Bosons. We shall
then move on to the physically more interesting example of three nucleons.

\section{The EFT for Three Bosons}
The distorted waves for elastic scattering of a bound pair and a particle
are not particularly easy to find using Efimov's
boundary conditions, although Fedorov and Jensen have attempted to develop a method
of doing so \cite{fedorov}. We will derive equations for the DWs using a useful
property of systems with zero range interactions. Namely, that the value of the
DW may be simply determined from the value of the DW when the relative
coordinate of a pair of particles is zero.
Because the potential is zero at any point other than where two particles
are `touching', the value of the DW anywhere else may be found
by integrating the wave equation.
Skorniakov and Ter-Martirosian (STM)\cite{stm} were the first to
take advantage of this property in their derivation of equations for the
2+1 scattering amplitude for zero range forces.

Our equations for the DWs will also use this property. We will derive equations that may be
used to find all DWs below the three-body threshold.
At one point in the derivation we will be able to link our method to the STM equations.
Unlike Skorniakov and Ter-Martirosian, we will start our derivation with the Faddeev equation
(\ref{eq:faddeevbd1}), which we rewrite here for convenience,
\begin{eqnarray}
&&\displaystyle{\qquad\,\,\,\,\,\,
|\Psi ^+\rangle=(1+P)|\psi^+_{{\bf k}_0}\rangle},\\
&&\displaystyle{
|\psi^+_{{\bf k}_0}\rangle=|\chi_{{\bf k}_0}\rangle+
G_0^+(p)t^+(p)P|\psi^+_{{\bf k}_0}\rangle,}\label{eq:faddeevbd2}
\end{eqnarray}
where we have now labelled the wavefunctions by ${\bf k}_0$, the vector of the third
particle in the centre of mass frame which has magnitude,
$|{\bf k}_0|=\sqrt{4(p^2+\gamma^2)/3}$, and $p^2=ME<0$ is the centre of
mass energy, relative to the three-body threshold.
Initially we will derive equations for the LO EFT, hence the two-body T-matrix  
is given simply by the scattering length term in the
effective range expansion\footnote{${\boldr}$ and ${\boldrho}$ are defined,
as before, by
\begin{equation}
\boldr_{ij}={\bf r}_i-{\bf r}_j,\qquad\boldrho_k=
{\bf r}_k-\frac12({\bf r}_i+{\bf r}_j),
\end{equation}
where $\{i,j,k\}$ is an even permutation of $\{1,2,3\}$ and
${\bf r}_1,{\bf r}_2,{\bf r}_3$ are the absolute
coordinates of the three particles. The momentum conjugate to these we
label as ${\bf l}_{ij}$ and ${\bf k}_k$ and are given by
\begin{equation}
{\bf l}_{ij}={\bf q}_i-{\bf q}_j,\qquad{\bf k}_k=
{\bf q}_k-\frac12({\bf q}_i+{\bf q}_j),
\end{equation}
where $\{i,j,k\}$ is an even permutation of $\{1,2,3\}$ and
${\bf q}_1,{\bf q}_2,{\bf q}_3$ are the absolute
momenta of the three particles},
\begin{eqnarray}
&&\displaystyle{
\langle {\bf k,l}|t(p)|{\bf k',l'}\rangle
=(2\pi)^3\delta^3({\bf k-k'})\tau(p,k),}\\
&&\displaystyle{
\qquad\tau(p,k)=-\frac{4\pi}M\left(-\gamma+i\sqrt{p^2-\frac34k^2}\right)^{-1}.
}
\end{eqnarray}
To make the importance of the zero separation value of the DWs
apparent we will work in a mixed co-ordinate system 
$({\bf k},\boldr)$, i.e. in terms of the distance between two particles and 
the momentum of the third particle in the centre of mass frame.
Two matrix elements that will be needed are:
\begin{eqnarray}
&&\displaystyle{
\langle{\bf k,x}|G_0^+(p)t^+(p)|{\bf k',x'}\rangle=
\tau(p,k')L_1(p,{\bf k,k',x})\delta^3({\bf x'}),}\\
&&\displaystyle{\qquad\qquad\qquad
L_1(p,{\bf k,k',x})=
-(2\pi)^3\delta^3({\bf k-k'})\frac{M}{4\pi x}e^{ix\sqrt{p^2-\frac34k^2}},
}\label{eq:defL1}\\
&&\displaystyle{
\langle{\bf k,x}|PG_0^+(p)t^+(p)|{\bf k' x'}\rangle=
\tau(p,k')L_2(p,{\bf k,k',x})\delta^3({\bf x'}),}\\
&&\displaystyle{\qquad\qquad\qquad
L_2(p,{\bf k,k',x})=
-2M\frac{\cos[{\bf x}.({\bf k'}+\frac{1}{2}{\bf k})]}
{k^2+k'^2+{\bf k.k'}-p^2-i\epsilon},}\label{eq:defL2}
\end{eqnarray}
The correctly normalised two-body bound state wavefunction is simply given by
\begin{equation}
\chi({\bf k},{\bf x})=(2\pi)^3\delta^3({\bf k-k_0})
\frac{\sqrt{2\gamma}}{4\pi x}e^{-\gamma x}=-\frac{\sqrt{2\gamma}}{M}
L_1\left(p,{\bf k,k_0,x}\right).
\end{equation}

\subsection{Momentum Space Equations}
As a starting point, we may use the Faddeev equations to write an integral 
equation for $\psi({\bf k,x})$. Inserting the values for the matrix elements
given above we get,
\begin{equation}
\psi({\bf k,x})=
-\frac{\sqrt{2\gamma}}{M}L_1\left(p,{\bf k},{\bf k}_0,{\bf x}\right)
+\int\frac{{\rm d}^3{\bf k'}}{(2\pi)^3}\tau\left(p,k'\right)
L_1\left(p,{\bf k,k',x}\right)\xi({\bf k'}),
\label{eq:intcomp}
\end{equation}
where
\begin{equation}
\xi({\bf k})=\Psi ({\bf k},{\bf x}=0)-\psi({\bf k},{\bf x}=0).
\end{equation}
Already, the importance of the DWs at ${\bf x}=0$ is apparent. 
$\Psi ({\bf k,x})$ is found by operating on (\ref{eq:faddeevbd2}) with $1+P$. This gives,
\begin{equation}
\Psi ({\bf k,x})=
-\frac{\sqrt{2\gamma}}{M}L\left(p,{\bf k,k}_0,{\bf x}\right)
+\int\frac{{\rm d}^3{\bf k'}}{(2\pi)^3}\tau\left(p,k'\right)
L\left(p,{\bf k,k',x}\right)\xi({\bf k'}),
\label{eq:intfull1}
\end{equation}
where $L=L_1+L_2$. These equations may be used to find a one-dimensional equation 
for $\xi$. Subtracting (\ref{eq:intcomp}) from (\ref{eq:intfull1}) and  setting ${\bf x}=0$ we get,
\begin{eqnarray}
\xi({\bf k})=-\frac{\sqrt{2\gamma}}{M}L_2\left(p,{\bf k,k}_0,0\right)
+\int\frac{{\rm d}^3{\bf k'}}{(2\pi)^3}\tau\left(p,k'\right)
L_2\left(p,{\bf k,k'},0\right)\xi({\bf k'}).
\label{eq:intxi}
\end{eqnarray}
Equations (\ref{eq:intxi}) and (\ref{eq:intfull1}) may now be solved to find 
$\Psi$. Our aim is to derive equations for the DWs in the coordinate space
representation but these equations are not in a convenient form for converting into
coordinate space. To find more suitable equations,
let us define the projection of the wavefunction as
\begin{equation}
\Phi({\bf k})=\lim_{x\rightarrow 0} x\Psi ({\bf k,x})
=\frac{\sqrt{2\gamma}}{4\pi}(2\pi)^3
\delta^3({\bf k-k}_0)-\frac{M}{4\pi}\tau\left(p,k\right)\xi({\bf k}),
\end{equation}
where the second identity follows from eqns.~
(\ref{eq:defL1},\ref{eq:defL2},\ref{eq:intfull1}). Substituting this
into equations (\ref{eq:intfull1}) and (\ref{eq:intxi}) we get,
\begin{eqnarray}
&&\displaystyle{\label{eq:STM}
\Psi ({\bf k,x})=
-\frac{4\pi}{M}\int\frac{{\rm d}^3{\bf k'}}{(2\pi)^3}
L(p,{\bf k,k',x})\Phi ({\bf k'}),}\\
&&\displaystyle{
\tau(k)^{-1}\Phi ({\bf k})=
\int\frac{{\rm d}^3{\bf k'}}{(2\pi)^3}
L_2(p,{\bf k,k'},0)\Phi ({\bf k'}).}\label{eq:intfull2}
\end{eqnarray}
These equations were first derived by Skorniakov and Ter-Martirosian
(STM)\cite{stm} in a somewhat different manner. The projection of the wavefunction, $\Phi$, is
the value of $\Psi$ when two of the particles are at zero separation. It can, with
caution, be thought of as a wavefunction that represents the penetration of the third
particle into the bound pair. Mathematically, $\Phi$ is simply a tool that allows
the full DW, $\Psi$, to be found using a one-dimensional equation. 
An equation for the elastic scattering amplitude
can be obtained by assuming an asymptotic form for $\Phi$,
\begin{equation}
\Phi ({\bf k})\rightarrow\,\sqrt{\frac{\gamma}{2\pi}}\left((2\pi)^3\delta^3({\bf k-k_0})+
\frac{4\pi}{M}\frac{a({\bf k_0,k})}{k^2-k_0^2+i\epsilon}\right).
\end{equation}
The resulting equation has been studied by Danilov\cite{danilov} and has more
recently been derived by Bedaque {\it el al}\cite{brgh,bhvk} from the KSW EFT Lagrangian.

\subsection{Coordinate Space Equations}
The STM equations may be Fourier transformed to obtain equations that may be
solved to find the distorted waves in the coordinate representation (see
Appendix \ref{app:Fourier}). We define,
\begin{eqnarray}
&&\displaystyle{
W(\boldrho,\boldrho',\boldr)=
-\frac{4\pi}{M}\int\frac{{\rm d}^3{\bf k'}}{(2\pi)^3}
\int\frac{{\rm d}^3{\bf k'}}{(2\pi)^3}L(p,{\bf k,k',x})
e^{i({\bf k.y-\bf k'.y'})}}\nonumber\\
&&\displaystyle{\qquad\qquad\qquad=
\frac{\sqrt{3}\kappa^2}{4\pi^2}\left\{
\frac{K_2(\kappa{\cal Q}_0)}{{{\cal Q}_0}^2}+
\frac{K_2(\kappa{\cal Q}_+)}{{\cal Q}_+^2}+
\frac{K_2(\kappa{\cal Q}_-)}{{\cal Q}_-^2}\right\},}\\
&&\displaystyle{\label{eq:LOpot}
V_0({\bf y,y'})=-\frac{16\pi}{3M}
\int\frac{{\rm d}^3{\bf k}}{(2\pi)^3}
\int\frac{{\rm d}^3{\bf k'}}{(2\pi)^3}
\left[\gamma+\sqrt{\frac34k^2-p^2}\right]L_2(p,{\bf k,k'},0)
e^{i({\bf k.y-\bf k'.y'})}}\nonumber\\
&&\displaystyle{\qquad\qquad\qquad
=\frac{\kappa^2}{2\sqrt{3}\pi^2}\left\{
3\kappa y'\frac{K_3(\kappa {\cal R})}{{\cal R}^3}+
4\left(\gamma-\frac{1}{y'}\right)\frac{K_2(\kappa {\cal R})}{{\cal R}^2}
\right\},}
\end{eqnarray}
where,
\begin{eqnarray}
&&\displaystyle{
{\cal Q}_0=\sqrt{\frac34x^2+({\bf y-y'})^2},\qquad
{\cal Q}_\pm=\sqrt{\frac34x^2+y^2+y'^2+{\bf y.y'}\pm\frac32{\bf x.y'}},
}\nonumber\\
&&\displaystyle{\qquad\qquad\,\,\,\,\,\,
{\cal R}=\sqrt{y^2+y'^2+\boldrho.\boldrho'},\qquad
\kappa=\sqrt{-\frac43p^2}.}\nonumber
\end{eqnarray}
$K_n(z)$ is the modified Bessel function of the third kind. 
In the case of positive energy, the $i\epsilon$ prescription gives 
$\kappa=-2ip/\sqrt{3}$ so that each of the modified Bessel functions go to
\begin{equation}
K_n(\kappa{\cal R})\rightarrow\frac{\pi}{2}i^{n+1}
H^{(1)}_n\left(\kappa{\cal R}\right).
\end{equation}
Using the transforms we may find equations for $\Psi({\bf x,y})$ and 
$\Phi({\bf y})$.  A simple Fourier transform of eqn.~(\ref{eq:STM})
yields,
\begin{equation}
\Psi ({\bf x,y})=\int{\rm d}^3{\bf y'}W({\bf y,y',x})
\Phi ({\bf y'})\label{eq:wfcoord1}.
\end{equation}
In order to Fourier transform eqn.~(\ref{eq:intfull2}) we first multiple
both sides by
\begin{equation}\nonumber
\frac{16\pi}{3M}\bigl(\gamma+\sqrt{3k^2/4-p^2}\bigr),
\end{equation}
and then transform to obtain,
\begin{equation}
\left(\nabla^2+k_0^2\right)\Phi ({\bf y})
+\int{\rm d}^3{\bf y'}V_0({\bf y,y'})\Phi ({\bf y'})
=0\label{eq:wfcoord}.
\end{equation}
This looks very much like the Schr\"odinger equation with
a non-local potential. This form is very suggestive of the interpretation
of $\Phi $ mentioned above, namely a wavefunction describing the penetration
of the third particle into the bound pair.
However since the potential is not symmetric,
and so is not hermitian, this interpretation should be used with caution.

The s-wave wavefunctions are simply found by integrating over all angles,
\begin{equation}
\Psi(x,y)=xy\int\frac{{\rm d}\Omega_x}{4\pi}
\int\frac{{\rm d}\Omega_y}{4\pi}\Psi ({\bf x,y}),\qquad
\Phi(y)=y\int\frac{{\rm d}\Omega_y}{4\pi}\Phi ({\bf y}).
\end{equation}
By integrating eqns.~(\ref{eq:wfcoord1},\ref{eq:wfcoord}) over all angles we can 
obtain equations for the s-wave wavefunctions:
\begin{eqnarray}
&&\displaystyle{\qquad\,\,\,\,\,\,
\Psi(x,y)=\int_0^\infty{\rm d}y'w(y,y',x)\Phi(y')},\label{eq:swavestm2}\\
&&\displaystyle{
\label{eq:swavestm}
\left(\frac{{\rm d}^2}{{\rm d}y^2}+k_0^2\right)\Phi(y)
+\int_0^\infty {\rm d}y'v_0(y,y')\Phi(y')=0,}
\end{eqnarray}
where,
\begin{eqnarray}
&&\displaystyle{
w(y,y',x)=\frac{\sqrt{3}\kappa x}{2\pi}\left(
\frac{K_1(\kappa Q_0^-)}{Q_0^-}-\frac{K_1(\kappa Q_0^+)}{Q_0^+}\right)}\nonumber\\
&&\displaystyle{\qquad\qquad\qquad\qquad
+\frac{4}{\sqrt{3}\pi y'}\Biggl(K_0(\kappa Q_+^+)+
K_0(\kappa Q_-^-)-K_0(\kappa Q_-^+)-K_0(\kappa Q_+^-)\Biggr),}\\
&&\displaystyle{
v_0(y,y')=\frac{2\sqrt{3}\kappa^2y'}{\pi}
\left(\frac{K_2(\kappa R_-)}{R_-^2}-\frac{K_2(\kappa R_+)}{R_+^2}\right)}\nonumber\\
&&\displaystyle{\qquad\qquad\qquad\qquad
+\frac{8\kappa}{\sqrt{3}\pi}\left(\gamma-\frac{1}{y'}\right)
\left(\frac{K_1(\kappa R_-)}{R_-}-\frac{K_1(\kappa R_+)}{R_+}\right),}
\end{eqnarray}
and
\begin{eqnarray}
&&\displaystyle{\qquad\qquad\qquad\qquad\qquad
R_\pm=\sqrt{y^2+y'^2\pm yy'},}\label{eq:defineR}\\
&&\displaystyle{
Q_0^{\pm}=\sqrt{\frac34x^2+y^2+y'^2\pm2yy'},\qquad
Q_\pm^\pm=\sqrt{\frac34x^2+y^2+y'^2\pm yy'\pm\frac32xy'}.}
\label{eq:defineQ}
\end{eqnarray}
These s-wave results follow from the indefinite integral result\cite{grad},
\begin{equation}
\int dz \frac{K_n(z)}{z^{n-1}}=\frac{K_{n-1}(z)}{z^{n-1}}.
\end{equation}

These equations can be used to find all s-wave DWs to LO in the EFT with total
centre of mass energy less than zero, i.e. all bound states and all
bound pair and particle interactions below the three-body threshold.
If the centre of mass energy becomes greater than zero, $\kappa$
becomes imaginary and the modified Bessel functions in the `potential'
become Hankel functions.
Since $v_0(y,y')$ will then have oscillatory behaviour for large $y$,
solution of the equations becomes extremely difficult.

\subsection{Analytic Solution}
In general eqns.~(\ref{eq:swavestm2},\ref{eq:swavestm}) have to be solved numerically.
Before we look at their general solution we shall look at their analytic solution
in certain limits.
Eqns.~(\ref{eq:swavestm2},\ref{eq:swavestm}) may be solved analytically in the limit 
$\kappa,\gamma\rightarrow0$ as they take on the far simpler forms:
\begin{eqnarray}
\Psi(x,y)&=&\frac{\sqrt{3}x}{2\pi}\int_0^\infty{\rm d}y'\Phi(y')\left\{
\frac{{Q_0^+}^2-{Q_0^-}^2}{ {Q_0^+}^2
 {Q_0^-}^2}-\frac{8}{3xy'}\log\left[\frac{ Q_+^+
 Q_-^-}{ Q_-^+ Q_+^-}\right]\right\},\nonumber\\
\frac{{\rm d}^2\Phi(y)}{{\rm d}y^2}
&+&\frac{16y}{\sqrt{3}\pi}\int_0^{\infty}{\rm d}y'\Phi(y')
\frac{2y'^4+2y^2y'^2-y^4}
{(y^4+y^2y'^2+y'^4)^2}=0.\label{eq:projzero}
\end{eqnarray}
The solutions to these equations correspond to the short range behaviour of the
DWs.
To solve them we assume an ansatz, $\Phi(y)=y^s$, which we justify 
by observing that since no other scales exist in the equation, the 
wavefunction must satisfy a power law. This argument may be put into a more 
rigorous form by considering Mellin transforms. Eqn.~(\ref{eq:projzero})
gives the possible values of $s$ which we find, by evaluating the integral,
must  satisfy the equation,
\begin{equation}\label{eq:danilov2}
\cos\left(\frac{\pi s}{2}\right)-\frac{8}
{\sqrt{3}s}\sin\left(\frac{\pi s}{6}\right)=0.
\end{equation}
This equation, as it should be, is precisely that obtained from Efimov's approach.
It is in fact possible to show that, in accordance with the analytic results of
Efimov's approach, in the limit of $\gamma\rightarrow0$, but non-zero
$\kappa$ eqn.~(\ref{eq:swavestm}) has the solutions $K_{s}(\kappa y)$ where
$s$ is a root of eqn.~(\ref{eq:danilov2}).

As noted before, eqn.~(\ref{eq:danilov2}) has two imaginary
solutions, $s=\pm i\bar s_0$, with $s=4$ the smallest real solution. Therefore,
the solutions of
eqn.~(\ref{eq:swavestm}) will be dominated by the $s_0$ solution at small $y$.

If we take 
$\Phi(y)=y^{s}$, the integral to find $\Psi(x,y)$ is far more involved. 
After some work (see Appendix \ref{app:zero}) 
we find we may write the result in terms of the hyperradius 
and the hyperangle:
\begin{eqnarray}\label{eq:analsol}
\Psi(R,\theta)&&=\left(\frac{\sqrt{3}R}{2}\right)^{s}\varphi(\alpha),\\
\varphi(\alpha)&&=
\begin{cases}
\sin(s\alpha)\cos\left(\frac{\pi}{3}s\right)
\csc\left(\frac{\pi}{6}s\right)&\text{when $\alpha<\frac{\pi}{6}$,}\\
\cos\left(s\left(\frac{\pi}{2}-\alpha\right)\right)
&\text{when $\alpha>\frac{\pi}{6}$.}
\end{cases}
\end{eqnarray}
This result confirms that there is no good boundary condition for the DWs as $R\rightarrow0$.
The angular behaviour is continuous at $\alpha=\pi/6$ but has a glitch. Since
$\alpha=\pi/6$ is where the third particle meets either of the particles in the pair,
this glitch corresponds to their interactions. 

The behaviour of the solutions, $\Phi$, of eqn.~(\ref{eq:swavestm}) for large $y$ 
depend simply on the relative values of $\kappa$ and $\gamma$. If $k_0>0$ and
$\kappa>0$ then the system is above the threshold for elastic scattering of a bound pair and particle
and below the three-body threshold.
The projection, $\Phi(y)\rightarrow\sin(k_0y+\delta)$, where $\delta$ is the
phaseshift. The `potential', $v_0(y,y')$, is real in this limit so the phaseshift
is also real. Using eqn.~(\ref{eq:swavestm2}) we may
find $\Psi(x,y)$ in this limit. We find,
\begin{equation}
\Psi(x,y)\rightarrow\sin(k_0y+\delta)e^{-\gamma x},
\end{equation}
which describes a non-interacting free particle far from a bound pair.

If $k_0^2<0$ then the system is below the two-body threshold
and the only solutions are the bound state solutions $\Phi(y)\rightarrow e^{-|k_0|y}$.

\subsection{Numerical solution}
Solving eqn.~(\ref{eq:swavestm}) with some fixed phase in the trig-log behaviour
close to the origin is equivalent to the three boson
EFT to order $(Q/\Lambda_0)^0$. The LO three-boson force
is taken to be the mechanism by which the phase is fixed
in accordance with the analysis in chapters 4 and 5.

In order to solve eqn.~(\ref{eq:swavestm}) numerically we must
find some way to deal with the singular behaviour close to the
origin and to fix the phase of the solution there. We introduce
some scale $\Omega\ll\Lambda_0$ and define the solution to
be,
\begin{equation}
\Phi(y)=
\begin{cases}
\sin(\bar s_0\ln(\gamma y)+\eta), & y<\Omega\\
\Phi_\Omega(y), & y>\Omega.
\end{cases}
\end{equation}
The three-body force may now be chosen by simply choosing the phase $\eta$
of the solution. Inserting this into eqn.~(\ref{eq:swavestm}) gives
a  homogeneous equation for $\Phi_\Omega$,
\begin{equation}\label{eq:STMSENum}
\frac{{\rm d}^2\Phi_\Omega(y)}{{\rm d}y^2}+k_0^2\Phi_\Omega(y)+
\int_\Omega^\infty{\rm d}y'\Phi_\Omega(y')v_0(y,y')
=-\int^\Omega_0{\rm d}y'\sin(\bar s_0\ln(\gamma y')+\eta)v_0(y,y'),
\end{equation}
with the boundary conditions
\begin{eqnarray}
&&\displaystyle{\,\,\,\,\,\,
\Phi_\Omega(\Omega)=\sin\left(\bar s_0\ln\gamma\Omega+\eta\right),}\nonumber\\
&&\displaystyle{
\left[\frac{d\Phi_\Omega(y)}{dy}\right]_{y=\Omega}=
\frac{\bar s_0}{\Omega}\cos\left(\bar s_0\ln\gamma\Omega+\eta\right).}
\end{eqnarray}
$\Omega$ is simply a
numerical tool and, up to numerical noise, results should not depend upon it.

This equation may be solved using linear algebra methods, in practice the upper limit
in the integral must be made finite, however, because of the exponential suppression of the
kernel $v_0(y,y')$ for large $y$ the effect upon the result is minimal.
The first step in solving numerically is discretisation of the variable $y$. 
Since the solutions of the eqn.~(\ref{eq:swavestm})
depend logarithmically on $y$ at small $y$ and linearly at large $y$, the discretisation
of the variable must reflect this. We define the points $y_n$ by the relation,
\begin{equation}
u_n=\bar s_0\ln\left(\gamma y_n+e^{k_0y_n/{\bar s_0}}-1\right),
\end{equation}
where $u_n$ varies linearly on $n$. At small $y_n$ we have,
\begin{equation}
u_n\approx \bar s_0\ln\gamma y_n,
\end{equation}
so that the $y_n$'s are logarithmically spaced allowing them to describe the trig-log
behaviour of the solutions in that region. At large $y_n$ we have,
\begin{equation}
u_n\approx k_0y_n,
\end{equation}
so that the $y_n$ are linearly spaced allowing them to describe the ordinary trigonometric
or exponential behaviour there. In order to set the two boundary conditions at the points
$y_0$ and $y_1$ we set
$y_2=\Omega$ as the first point at which the value of the solution is unknown.

Using this discretisation of the variable $y$, eqn.~(\ref{eq:STMSENum}) can be written as
an inhomogeneous matrix equation. We define $\Phi_n=\Phi(y_{n+1})$ 
so that the derivative term is found using the difference,
\begin{eqnarray}
\frac{d^2\Phi(y_n)}{dy^2}&=&\frac{1}{2(\Delta_n+\Delta_{n-1})}\left(
\frac{1}{\Delta_n}\Phi(y_{n+1})-\frac{1}{(\Delta_n+\Delta_{n-1})}\Phi(y_{n})
+\frac{1}{2\Delta_{n-1}}\Phi(y_{n-1})\right)\\
&=&{\cal D}_{n,n-2}\Phi_{n-2}+{\cal D}_{n,n-1}\Phi_{n-1}+{\cal D}_{n,n}\Phi_{n},
\end{eqnarray}
where $\Delta_n=y_{n+1}-y_{n}$. The last line defines the derivative matrix ${\cal D}_{i,j}$,
$1\leq i,j \leq N$ in which all other elements are set to zero. Notice that our definition
also defines three other elements, ${\cal D}_{1,0},{\cal D}_{1,-1}$ and ${\cal D}_{2,0}$
that do not appear in the matrix; these terms are required to set the boundary conditions
at $y_0$ and $y_1$.

The integral term can simply be written as
\begin{equation}
\int_{y_2}^{y_N}dy' v_0(y_n,y')\Phi(y')=\sum_{i=1}^{N}{\cal V}_{n,i}\Phi_i,
\end{equation}
where the coefficients, ${\cal V}_{ij}$ form the matrix ${\cal V}$ and can most easily
be found using Simpson's rule, in which case $N$ must be chosen to be odd.

Altogether the eqn.~(\ref{eq:swavestm}) is written in matrix form as,
\begin{eqnarray}
&&\displaystyle{
\sum_{j=1}^N\left({\cal D}_{i,j}+\delta_{i-1,j}k_0^2+{\cal V}_{i,j}\right)\Phi_j=I_i}\nonumber\\
&&\displaystyle{\qquad\qquad-
\delta_{i,1}\Bigl(({\cal D}_{1,0}+k_0^2)\sin(\bar s_0\ln\gamma y_1+\eta)+{\cal D}_{1,-1}
\sin(\bar s_0\ln\gamma y_0+\eta)\Bigr)}\nonumber\\
&&\displaystyle{\qquad\qquad\qquad\qquad\label{eq:nummatrix}
-\delta_{i,2}{\cal D}_{2,0}\sin(\bar s_0\ln\gamma y_1+\eta),}
\end{eqnarray}
where,
\begin{equation}
I_n=-\int_0^{y_2}dy' v_0(y_{n+1},y')\sin(\bar s_0\ln\gamma y'+\eta).
\end{equation}
The second and third lines of eqn.~(\ref{eq:nummatrix}) define the boundary conditions
by ensuring that the solution matches on to the form chosen for $y<y_2$.

For scattering solutions, $k_0^2>0$, we solve the equation numerically for two
values of $\eta$, $\eta=0$ and $\eta=\pi/2$ for instance. All solutions with different
values of $\eta$ can be constructed by superposition.

Bound state solutions can be found for any energy by simply solving the
equation for the two different $\eta$'s then finding the superposition
that vanishes for large $y$. This gives a constraint on the interior phase of the
bound state solution at that energy. Only the bound states with the correct interior
phase - chosen by the LO three-body force are the physical ones.

\begin{figure}
\begin{center}
\includegraphics[height=8cm,width=12.5cm,angle=0]{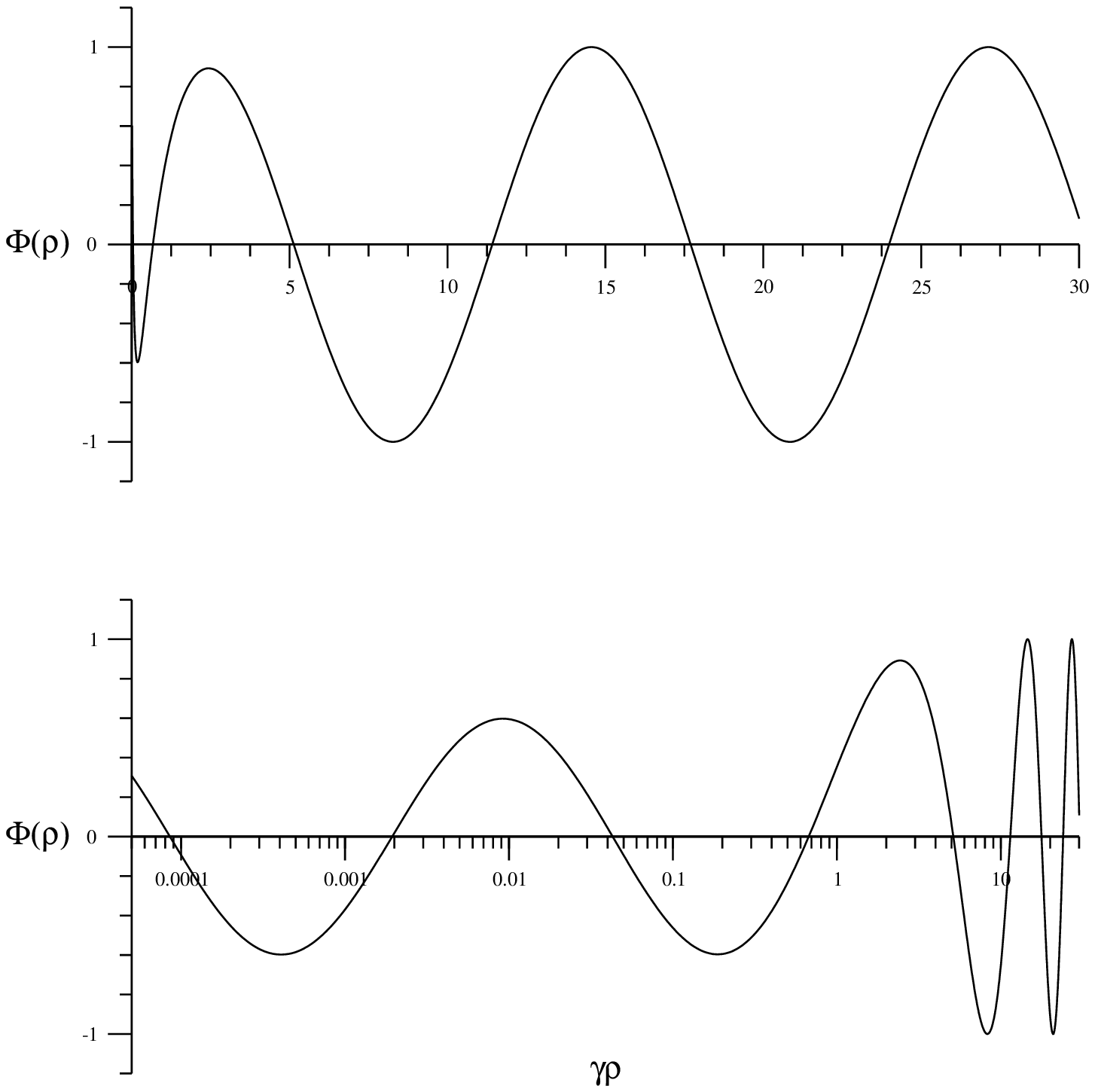}
\caption{
Numerical 2+1 scattering solution to eqn.~(\ref{eq:swavestm})
with relative momenta $k_0=\gamma/2$ and total centre of mass energy
$E=-13\gamma^2/(16M)$}
\label{fig:dwscatt}
\includegraphics[height=8cm,width=12.5cm,angle=0]{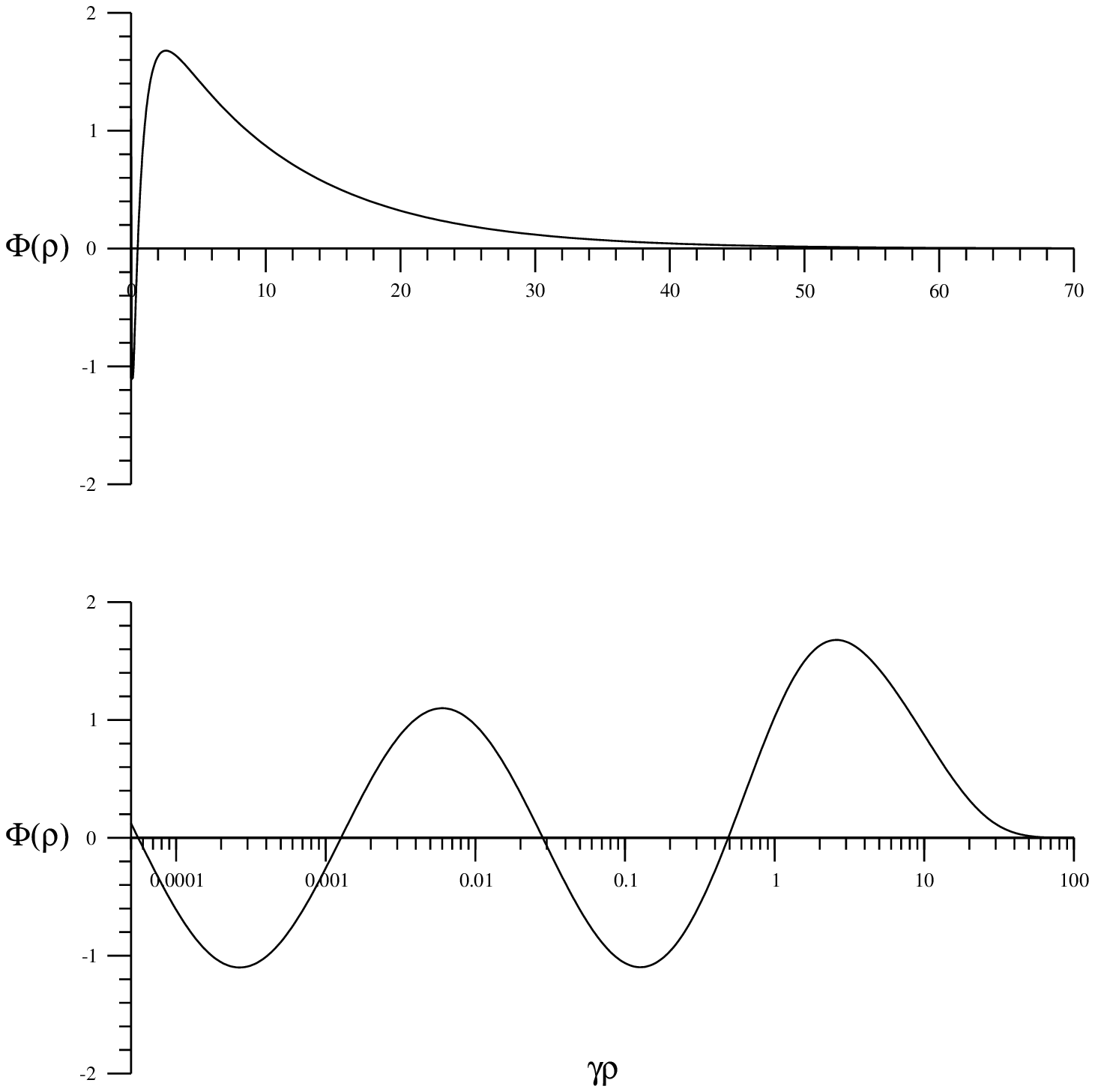}
\caption{Numerical solution to eqn.~(\ref{eq:swavestm}).
Example of a shallow bound state solution with centre of mass energy
$E=-101\gamma^2/(100M)$.}
\label{fig:dwbound}
\end{center}
\end{figure}

\begin{figure}
\begin{center}
\includegraphics*[100,430][480,715]{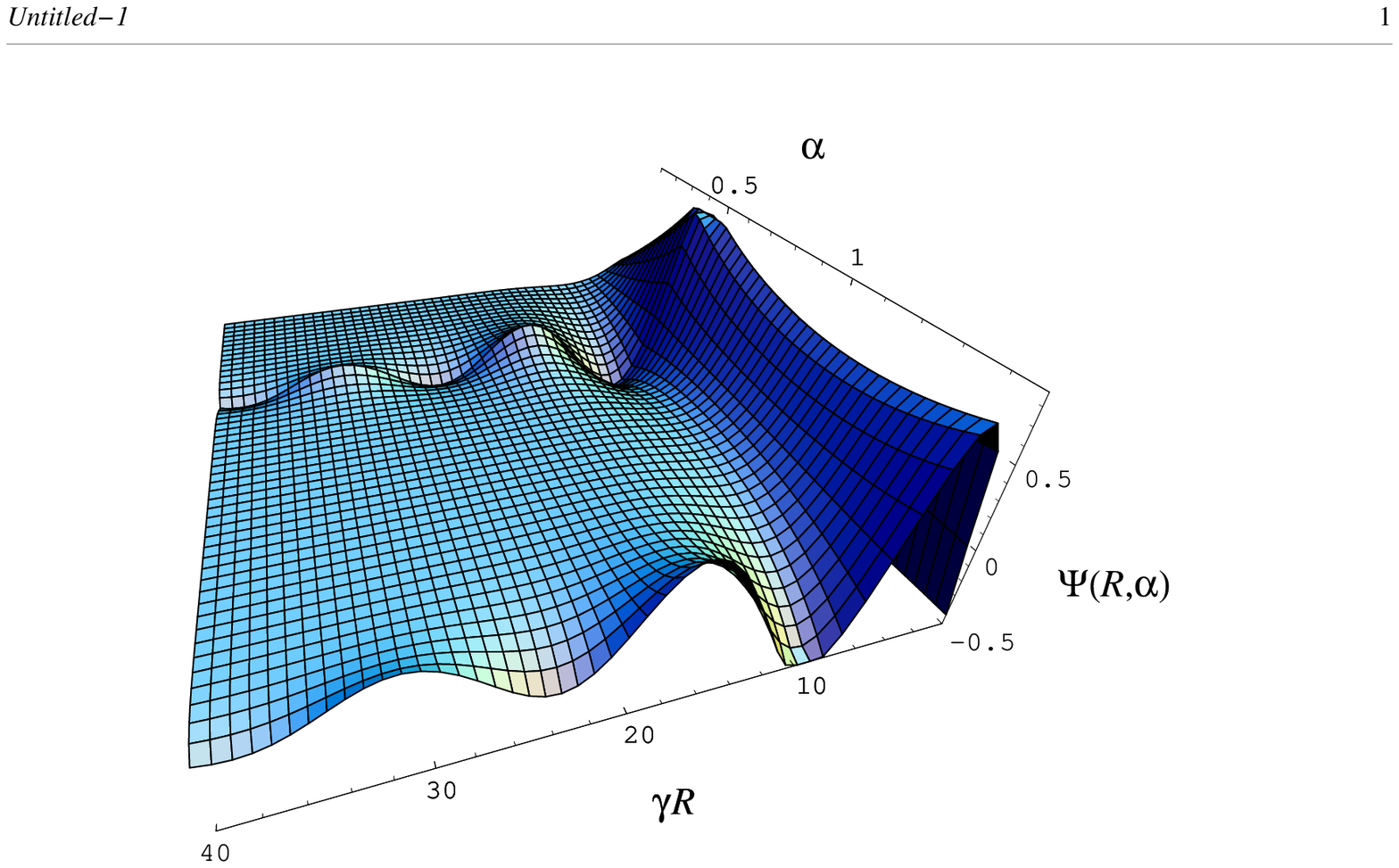}
\caption{Three-body distorted wave, numerical solution to eqns.~(
\ref{eq:swavestm2},\ref{eq:swavestm}) with relative momenta $k_0=\gamma/2$
and total centre mass energy $E=-13\gamma^2/(16M)$.}
\label{fig:3ddws1}
\includegraphics*[95,430][480,710]{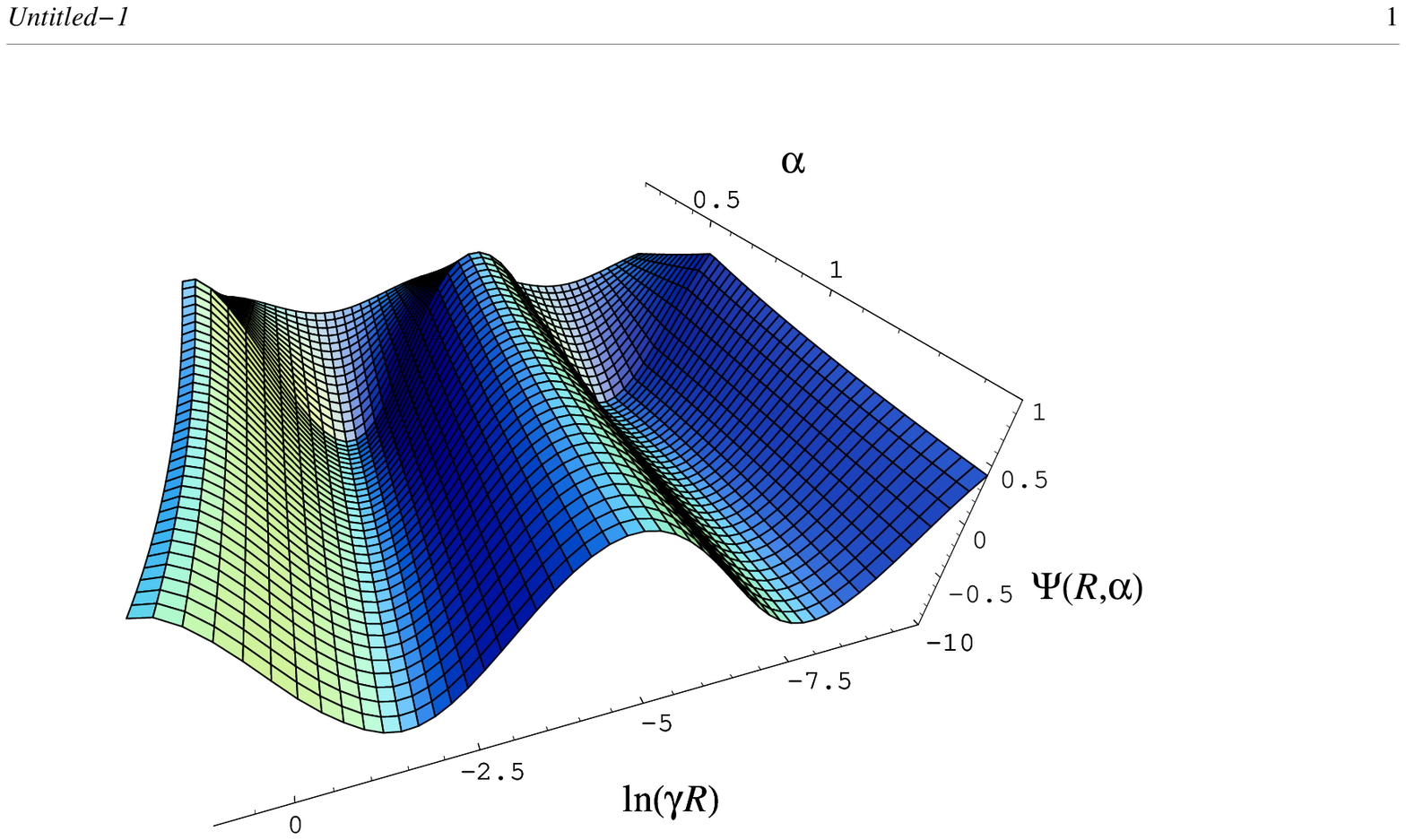}
\caption{The small $R$ detail of fig.~\ref{fig:3ddws1}, matching the
anticipated result, eqn.~(\ref{eq:analsol}).}
\label{fig:3ddws2}
\end{center}
\end{figure}

Two sample solutions of eqn.~(\ref{eq:swavestm}) are shown in Figs.~\ref{fig:dwscatt},
\ref{fig:dwbound}. Fig.~\ref{fig:dwscatt} shows a scattering solution to the equation with
incoming momenta $k_0=\gamma/2$. The figure
shows the solution on both log and linear plots so that the interior and exterior behaviour
of the wavefunction is visible. In the interior the trig-log behaviour is evident, externally
the solution goes to $\sin(k_0y+\delta)$ where $\delta$ is the phaseshift for the
elastic scattering amplitude. The full DW, $\Psi(x,y)$ which results from this
solution and was obtained using eqn.~(\ref{eq:swavestm2}) is shown in
Figs.~\ref{fig:3ddws1},\ref{fig:3ddws2}.

The full DW solution is plotted on a hyperpolar scale $(R,\alpha)$.
The hyperradius $R$ can be considered the mean distance between the particles
and the hyperangle $\alpha$ describes the configuration of the three particles.
Evident in the plots is the symmetry of the wavefunction with respect to the interchange
of particles. In the fig.~\ref{fig:3ddws1} the hyperradial scale is linear. There are two outgoing
states at $\alpha=\pi/2$, corresponding to the `chosen'\footnote{Chosen in the sense of our
definition of $\alpha$.} bound pair, and at $\alpha=\pi/6$
corresponding to the other two possible pairs. The discontinuity across the derivative
at $\alpha=\pi/6$,
anticipated in eqn.~(\ref{eq:analsol}) is also evident. Fig.~\ref{fig:3ddws2} shows the
small $R$ detail of the DWs, which exactly matches the analytic result, (\ref{eq:analsol}).

Fig.~\ref{fig:dwbound} shows a bound state solution to eqn.~(\ref{eq:swavestm}),
again showing detail on both log and linear plots.
The interior phase of this solution was determined by the choice of binding energy.
From an EFT perspective the three-body force determines the
interior phase which in turn determines the binding energy.

\begin{figure}
\begin{center}
\includegraphics[height=15cm,width=15cm,angle=-90]{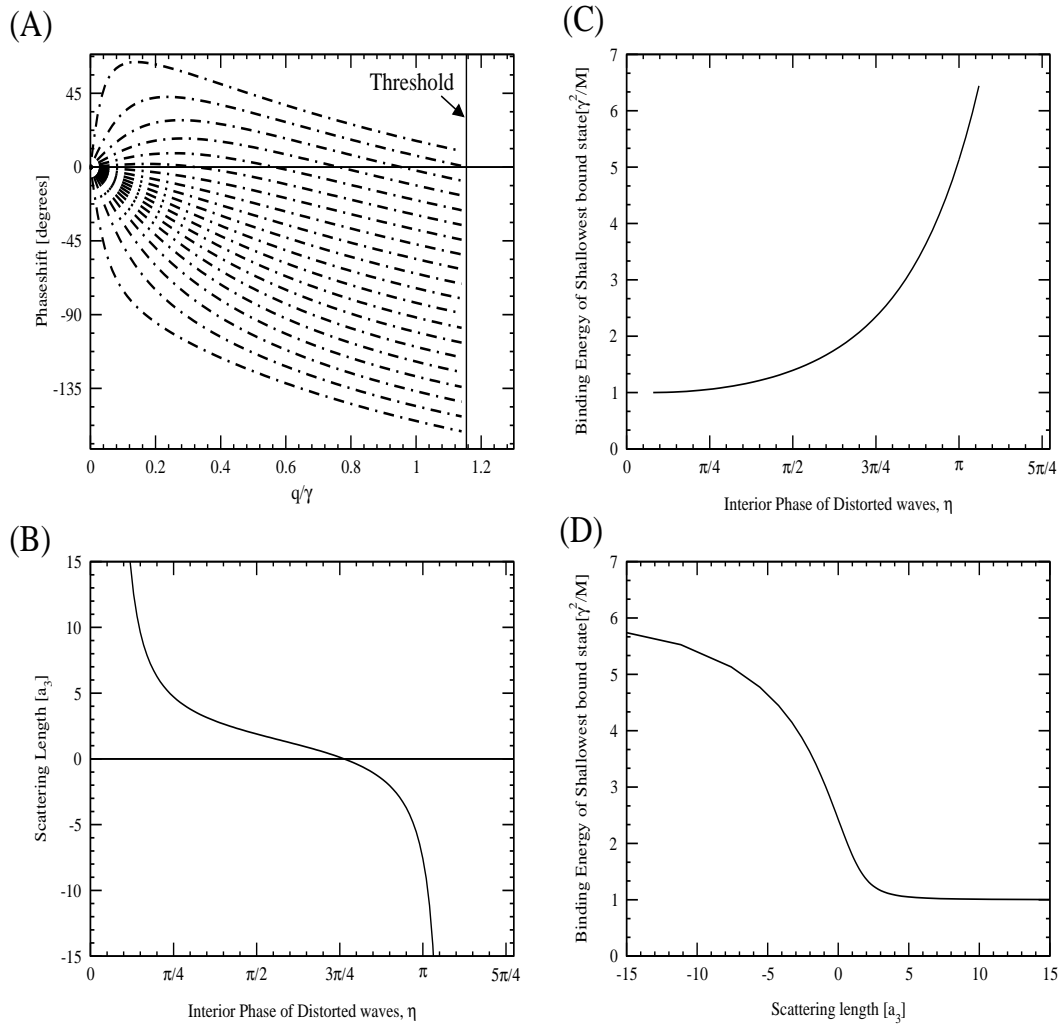}
\caption{Numerical Results of the LO EFT for three-bosons.
(A) All possible phaseshift curves that result from different values
of the leading three body force. Each value for the LO three body force
corresponds to a different interior phase $\eta$ and results in a
different scattering length for 2+1 elastic scattering (B) and the
shallowest bound state (C). Together these imply a single parameter
relationship between the scattering length and binding energy of the
bound states (D).}
\label{fig:physresults}
\end{center}
\end{figure}

Fig.~\ref{fig:physresults} shows physical observables determined from solutions to
the LO EFT equation (\ref{eq:swavestm}). Fig.~\ref{fig:physresults}(A) shows the
possible different phaseshift curves for the elastic amplitude
that result from different choices for the LO
three-body force. Each of these curves corresponds to a different value for the
scattering length (Fig.~\ref{fig:physresults}(B)). The range of possible
scattering lengths that can be obtained from different three-body forces is
$-\infty<a_3<\infty$.
Notice that it is only because the
three-body force occurs at such a low order, $(Q/\Lambda_0)^0$ that it is
able to affect the scattering length of these amplitudes so completely.
Fig.~\ref{fig:physresults}(C) shows the shallowest bound states
that result from our choice of three-body force.  Figs.~\ref{fig:physresults}(B)
and \ref{fig:physresults}(C) together imply that there is a single parameter relationship
between the elastic scattering length and the binding energy of the shallowest bound
state, which is illustrated in  Fig.~\ref{fig:physresults}(D). This curve
cannot be seen experimentally as the physical world corresponds to just one point on it.
However, the equivalent curve in the three nucleon problem, known as the Phillips line
\cite{bhvk,brgh,phillips} has very interesting implications.

\subsection{NLO Equations}

At NLO, $(Q/\Lambda_0)^{1}$, in accordance with eqn.~(\ref{eq:elastic3b1}) there is no extra three-body
force term\footnote{Recall that the general expansion of the three-body force contained a term
at this order but proportional to $\gamma$. This will in fact be seen as a small correction to our LO
`phase-fixing' three body force.}
 but there is an effective range correction in the two-body force.
The equations for the effective range corrections can be found using eqn.~(\ref{eq:intfull1})
and the form for the two-body $T$-matrix with effective range corrections,
\begin{eqnarray}
&&\displaystyle{
\tau(p,k)=-\frac{4\pi}{M}\frac{1}{-\gamma+
\frac12\rho_e\left(\gamma^2+p^2-\frac34k^2\right)+\sqrt{\frac34k^2-p^2}}}\nonumber\\
&&\displaystyle{\qquad\,\,\,\,\,\,\,
=\frac{4\pi}{M}\frac{1}{\gamma^2-\frac34k^2+p^2}\Biggl[
\left(\gamma+\sqrt{\frac34k^2-p^2}\right)}\nonumber\\
&&\displaystyle{\qquad\qquad\qquad\qquad\qquad\qquad\qquad
+\frac12\rho_e\left(\gamma+\sqrt{\frac34k^2-p^2}\right)^2
+\ldots\Biggr].}
\end{eqnarray}
This modified form of the effective range expansion ensures that there is a bound state pole at
$i\gamma$. Substituted into eqn.~(\ref{eq:intfull1}) and keeping just terms of order
$(Q/\Lambda_0)$ (i.e. neglecting terms like $\rho_e^2$) gives,
\begin{eqnarray}
&&\displaystyle{
\left(k_0^2-k^2\right)\Phi ({\bf k})-
\Biggl[\frac{16\pi}{3M}\left(\gamma+\sqrt{\frac34k^2-p^2}\right)}
\nonumber\\
&&\displaystyle{\qquad\qquad
+\frac{8\pi}{3M}\rho_e\left(\gamma+\sqrt{\frac34k^2-p^2)}\right)^2\Biggr]
\int\frac{{\rm d}^3{\bf k'}}{(2\pi)^3}
L_2(p,{\bf k,k'},0)\Phi ({\bf k'})=0.}
\end{eqnarray}
Fourier transforming this equation as before we
obtain,
\begin{equation}
\left(\nabla^2+k_0^2\right)\Phi ({\bf y})
+\int{\rm d}^3{\bf y'}\bigl(V_0({\bf y,y'})
+\rho_eV_1({\bf y,y'})\bigr)\Phi ({\bf y'})
=0,\label{eq:wfcoordNLO}
\end{equation}
where (see Appendix \ref{app:Fourier}),
\begin{eqnarray}
&&\displaystyle{\label{eq:NLOpot}
V_1(y,y')=-\frac{8\pi}{3M}
\int\frac{{\rm d}^3{\bf k}}{(2\pi)^3}
\int\frac{{\rm d}^3{\bf k'}}{(2\pi)^3}
\left[\gamma+\sqrt{\frac34k^2-p^2}\right]^2L_2(p,{\bf k,k'},0)
e^{i({\bf k.y-\bf k'.y'})}}\nonumber\\
&&\displaystyle{\qquad\,\,\,\,\,\,\,\,\,\,
=\frac{\sqrt{3}\kappa^2}{2\pi^2}\Biggl\{
\frac{3\kappa^2y'^2}{8}\frac{K_4(\kappa {\cal R})}{ {\cal R}^4}
+\kappa\left(\gamma y'-\frac32\right)\frac{K_3(\kappa {\cal R})}{ {\cal R}^3}}
\nonumber\\
&&\displaystyle{\qquad\qquad\qquad\qquad\qquad\qquad\qquad\qquad\qquad
+\frac{2\gamma}{3}\left(\gamma-\frac{2}{y'}\right)\frac{K_2(\kappa {\cal R})}
{ {\cal R}^2}\Biggr\}.}
\end{eqnarray}
The s-wave equivalent follows quite simply,
\begin{equation}
\label{eq:NLOswavestm}
\left(\frac{{\rm d}^2}{{\rm d}y^2}+k_0^2\right)\Phi(y)
+\int_0^\infty {\rm d}y'(v_0(y,y')+\rho_e v_1(y,y')\Phi(y')=0,
\end{equation}
where
\begin{eqnarray}
&&\displaystyle{
v_1(y,y')=\frac{2\sqrt{3}\kappa}{\pi}\Biggl\{
\frac{3\kappa^2y'^2}{8}\left(\frac{K_3(\kappa R_-)}{R_-^3}
-\frac{K_3(\kappa R_+)}{R_+^3}\right)
+\kappa\left(\gamma y'-\frac32\right)\left(
\frac{K_2(\kappa R_-)}{R_-^2}-\frac{K_2(\kappa R_+)}{R_+^2}
\right)}
\nonumber\\
&&\displaystyle{\qquad\qquad\qquad\qquad\qquad\qquad
+\frac{2\gamma}{3}\left(\gamma-\frac{2}{y'}\right)
\left(\frac{K_1(\kappa R_-)}{R_-}-\frac{K_1(\kappa R_+)}{R_+}
\right)\Biggr\}.}
\end{eqnarray}
We can solve eqn.~(\ref{eq:NLOswavestm}) up to order $\rho_e$ using first order perturbation
theory. If we define $\Phi(y)=\Phi^{(0)}(y)+\rho_e\Phi^{(1)}(y)$ and $\Phi^{(0)}(y)$
satisfies the LO equation (\ref{eq:swavestm}) then up to order $\rho_e$, $\Phi^{(1)}(y)$
satisfies
\begin{equation}
\left(\frac{{\rm d}^2}{{\rm d}y^2}+k_0^2\right)\Phi^{(1)}(y)
+\int_0^\infty {\rm d}y'v_0(y,y')\Phi^{(1)}(y')=
-\int_0^\infty v_1(y,y')\Phi^{(0)}(y').
\end{equation}
This equation may be solved numerically by adapting the method used for
the LO equation. Since no new parameters are entered at this order the results of
solving the equations to this order give a refinement to the one parameter relationship
between the 2+1 scattering length and the three particle binding energies.

\section{The Pionless EFT for Three Nucleons.}

One area where the two-body EFT for short range forces has proven
very successful is nuclear physics. As noted in the earlier chapters, nuclear physics
provides an excellent and physically important example of the KSW power-counting
scheme in action. In this chapter we will extend the results of the previous
section to look at the pionless KSW EFT for three nucleons.

The spin and isospin degrees of freedom available to the nucleons
as well as Fermi statistics makes the problem slightly more complicated then the analysis for Bosons.
To avoid the complication of electromagnetic interactions we
shall concentrate on the system of two neutrons and a proton\footnote{This system is clearly more
interesting than the other electromagnetically neutral
possibility of three neutrons as it contains a well-known two nucleon bound state,
the deuteron, and three nucleon bound state, the triton.}.

If the total spin of the three nucleons is $3/2$ then the three spins are aligned
and the Pauli exclusion principle acts as a repulsive force between the two neutrons.
This system is insensitive to short range physics and has been successfully described
by an EFT without three-body forces \cite{quartetbvk}. 
However, if the total spin is $1/2$ then, because it is possible for the three nucleons,
now non-identical, to approach each other, the system is
sensitive to short range physics, it is upon this system that we shall concentrate
in this chapter.

\subsection{Wavefunction Symmetry}
We assume that the pairwise nuclear force conserves both isospin and spin so that these 
may be taken as constants, and we put the total spin and isospin and isospin component
to $S=T=T_z=\frac12$.  If we identify one particular pair of nucleons, $1$ and $2$,
then we can identify two linearly independent
states of three particles with spin $1/2$ that may be used as a basis.
Firstly, the singlet state in which the pair of nucleons are in the spin singlet state, $s=0$,
which we denote by $\omega_s^{12}$ and secondly, the triplet state in which the
pair are in a spin triplet state, $s=1$, which we denote by $\omega_t^{12}$. 
The singlet state is antisymmetric and the triplet state is symmetric under
interchange of $1$ and $2$.
A spin basis may also be constructed using an alternative pair, $2$ and $3$ say,
as the starting point for the basis. The transformations between the different possible
bases are given by,
\begin{equation}
\left(\begin{array}{c}\omega^{23}_s\\\omega^{23}_t\end{array}\right)
=\left(\begin{array}{cc}-\frac12 &-\frac{\sqrt{3}}{2}\\-\frac{\sqrt{3}}{2} &
-\frac12\end{array}\right)
\left(\begin{array}{c}\omega^{12}_s\\\omega^{12}_t\end{array}\right)\qquad
\left(\begin{array}{c}\omega^{31}_s\\\omega^{31}_t\end{array}\right)
=\left(\begin{array}{cc}-\frac12 &\frac{\sqrt{3}}{2}\\\frac{\sqrt{3}}{2} &
-\frac12\end{array}\right)
\left(\begin{array}{c}\omega^{12}_s\\\omega^{12}_t\end{array}\right).\label{eq:transf}
\end{equation}
Similarly for isospin we have the singlet and triplet states,
$\vartheta_s^{12}$ and $\vartheta_t^{12}$ which satisfy the same transformations and
symmetries.

Since the pairwise potential is symmetric under interchange of the two interacting
particles the continuous part of the wavefunction must be symmetric under
interchange of the particles in the pair. Thus a general state which is antisymmetric
in $1$ and $2$ can be written in the form,
\begin{equation}
|\varphi^{12}\rangle\rangle=\omega_s^{12}\vartheta_t^{12}|\psi_s^{12}\rangle+
\omega_t^{12}\vartheta_s^{12}|\psi_t^{12}\rangle.
\end{equation}
The full wavefunction, which must be antisymmetric in the interchange of any pair
can now be written as,
\begin{equation}\label{eq:nucdecomp}
|\Psi\rangle\rangle=|\varphi^{12}\rangle\rangle+
|\varphi^{23}\rangle\rangle+|\varphi^{31}\rangle\rangle
=(1+P)|\varphi\rangle\rangle,
\end{equation}
where $P$ is the permutation operator.
To see that $|\Psi\rangle\rangle$ is totally antisymmetric notice that
\begin{equation}
|\varphi^{23}\rangle\rangle\rightarrow|\varphi^{13}\rangle\rangle=
-|\varphi^{31}\rangle\rangle,
\end{equation}
under the interchange of $1$ and $2$. Since we have now introduced the permutation
operator, $P$, we shall drop particle labels. Using the decomposition of $|\Psi\rangle\rangle$
we can write down the Faddeev equation for $|\varphi\rangle\rangle$,
\begin{equation}
(H_0+\hat V-E)|\varphi\rangle\rangle=-\hat VP|\varphi\rangle\rangle,
\label{eq:nucfad}
\end{equation}
where the hat on the potential signifies that it is an operator in spin-isospin space.

We shall assume that the potential $\hat V$ is diagonal in spin and isospin.
(This assumption is approximate but is true to the orders at which we shall be
working \cite{bhvk}.) In the centre of mass frame of the two interacting particles the
Pauli exclusion principle restricts $\hat V$ to just two non-zero matrix elements:
\begin{equation}
(\omega_s\vartheta_t,\hat V\omega_s\vartheta_t)=V_s,\qquad
(\omega_t\vartheta_s,\hat V\omega_t\vartheta_s)=V_t.
\end{equation}
The matrix elements, in spin-isospin space, of $\hat VP$ now follow from the
transformation relationships (\ref{eq:transf}),
\begin{eqnarray}
(\omega_s\vartheta_t,\hat VP\omega_s\vartheta_t)=\frac14V_sP,\nonumber\\
(\omega_s\vartheta_t,\hat VP\omega_t\vartheta_s)=\frac34V_sP,\nonumber\\
(\omega_t\vartheta_s,\hat VP\omega_t\vartheta_s)=\frac14V_tP,\nonumber\\
(\omega_t\vartheta_s,\hat VP\omega_s\vartheta_t)=\frac34V_tP.
\end{eqnarray}

Taking the inner product of eqn.~(\ref{eq:nucfad}) with $\omega_s\vartheta_t$
and $\omega_t\vartheta_s$ yields coupled Faddeev equations for the
$|\psi_s\rangle$ and $|\psi_t\rangle$,
\begin{eqnarray}
(H_0+V_s-E)|\psi_s\rangle=-V_sP\left(\frac14|\psi_s\rangle+\frac34|\psi_t\rangle\right),
\nonumber\\
(H_0+V_t-E)|\psi_t\rangle=-V_tP\left(\frac34|\psi_s\rangle+\frac14|\psi_t\rangle\right),
\end{eqnarray}
which can be written in the form,
\begin{eqnarray}
&&\displaystyle{
|\psi_s\rangle=G_0^+(E)t_s(E)P\left(\frac14|\psi_s\rangle+\frac34|\psi_t\rangle\right),
}\nonumber\\
&&\displaystyle{
|\psi_t\rangle=|\chi_t\rangle+
G_0^+(E)t_t(E)P\left(\frac34|\psi_s\rangle+\frac14|\psi_t\rangle\right),}
\end{eqnarray}
where $|\chi_t\rangle$ is the asymptotic state with a bound pair (the deuteron) and 
free particle. There is no bound pair in the singlet channel so there is
no asymptotic state.

Using eqns.~(\ref{eq:transf},\ref{eq:nucdecomp}) we can now write the full wavefunction
in its final form,
\begin{eqnarray}
&&\displaystyle{
|\Psi\rangle\rangle=\omega_s\vartheta_t|\Psi_s\rangle+
\omega_t\vartheta_s|\Psi_t\rangle+(\omega_s\vartheta_s-\omega_t
\vartheta_t)|\Psi_u\rangle},\nonumber\\
&&\displaystyle{\qquad\qquad
|\Psi_s\rangle=\left(1+\frac P4\right)|\psi_s\rangle+\frac{3P}{4}|\psi_t\rangle,}\nonumber\\
&&\displaystyle{\qquad\qquad
|\Psi_t\rangle=\left(1+\frac P4\right)|\psi_t\rangle+\frac{3P}{4}|\psi_s\rangle,}\nonumber\\
&&\displaystyle{\qquad\qquad
|\Psi_u\rangle=\frac{\sqrt{3}}{4}\tilde P(|\psi_s\rangle+|\psi_t\rangle),}
\end{eqnarray}
where, $\langle\boldr_{12},\boldrho_{3}|\tilde P=\langle\boldr_{23},\boldrho_{1}|-
\langle\boldr_{31},\boldrho_{2}|$. From these equations we may now parallel
the analysis for three Bosons to obtain the equations for the full wavefunction
which are,
\begin{eqnarray}
&&\displaystyle{
\Psi_a({\bf k,x})=-\frac{4\pi}{M}\int\frac{{\rm d}^3{\bf k'}}{(2\pi)^3}
\left[L_{1}(p,{\bf k,k',x})\Phi_a({\bf k'})
+L_{2}(p,{\bf k,k',x})\left\{\frac{1}{4}\Phi_a({\bf k'})
+\frac{3}{4}\Phi_b({\bf k'})\right\}\right],}\nonumber\\
&&\displaystyle{\qquad\qquad\,\,\,\,\,
\Psi_u({\bf k,x})=-\frac{\sqrt{3}\pi}{M}
\int\frac{{\rm d}^3{\bf k'}}{(2\pi)^3}L_{3}(p,{\bf k,k',x})
\Bigl[\Phi_s({\bf k'})+\Phi_t({\bf k'})\Bigr],}\nonumber\\
&&\displaystyle{\qquad\qquad
\tau_a\left(p,k\right)^{-1}\Phi_a({\bf k})=
\int\frac{{\rm d}^3{\bf k'}}{(2\pi)^3}L_{2}(p,{\bf k,k'},0)\left[
\frac{1}{4}\Phi_a({\bf k'})
+\frac{3}{4}\Phi_b({\bf k'})\right],}
\end{eqnarray}
where $a,b\in\{s,t\}$ with $a\neq b$, and $L_3$ is given by,
\begin{equation}
L_3(p,{\bf k,k',x})=
-2iM\frac{\sin[{\boldr}.({\bf k'}+\frac{1}{2}{\bf k})]}
{k^2+k'^2+{\bf k.k'}-p^2-i\epsilon}.
\end{equation}

\subsection{The Coordinate Space Equations.}
These equations can as before be readily converted into coordinate space.
The equations for $\Psi(\boldr,\boldrho)$ do not depend upon the
two-body $T$-matrices and are given by,
\begin{eqnarray}
&&\displaystyle{
\Psi_a(\boldr,\boldrho)=\int d^3y'\left[
W_1(\boldrho,\boldrho',\boldr)\Phi_a(\boldrho')+
W_2(\boldrho,\boldrho',\boldr)\left(\frac14\Phi_a(\boldrho')
+\frac34\Phi_b(\boldrho')\right)\right],}\label{eq:make3d}\\
&&\displaystyle{\qquad\qquad
\Psi_u(\boldr,\boldrho)=\int d^3y'
W_3(\boldrho,\boldrho',\boldr)\left(\frac14\Phi_a(\boldrho')
+\frac34\Phi_b(\boldrho')\right),}
\end{eqnarray}
where
\begin{eqnarray}
&&\displaystyle{\qquad\,\,\,\,
W_1(\boldrho,\boldrho',\boldr)=
\frac{\sqrt{3}\kappa^2}{4\pi^2}\frac{K_2(\kappa{\cal Q}_0)}{{\cal Q}_0^2},}\\
&&\displaystyle{
W_2(\boldrho,\boldrho',\boldr)=\frac{\sqrt{3}\kappa^2}{4\pi^2}
\left\{\frac{K_2(\kappa{\cal Q}_+)}{{\cal Q}_+^2}+\frac{K_2(\kappa{\cal Q}_-)}{{\cal Q}_-^2}
\right),}\\
&&\displaystyle{
W_3(\boldrho,\boldrho',\boldr)=\frac{3\kappa^2}{16\pi^2}
\left\{\frac{K_2(\kappa{\cal Q}_+)}{{\cal Q}_+^2}-\frac{K_2(\kappa{\cal Q}_-)}{{\cal Q}_-^2}
\right).}
\end{eqnarray}
In order to get the equations for $\Phi(y)$ we must define the two-body
$T$-matrices. The $T$-matrix in the triplet channel is given by,
\begin{eqnarray}
&&\displaystyle{
\tau_t(p,k)=-\frac{4\pi}{M}\frac{1}{-\gamma_t+
\frac12\rho_t\left(\gamma_t^2+p^2-\frac34k^2\right)+\sqrt{\frac34k^2-p^2}}}\nonumber\\
&&\displaystyle{\qquad\,\,\,\,\,\,\,
=\frac{4\pi}{M}\frac{1}{\gamma_t^2-\frac34k^2+p^2}\Biggl[
\left(\gamma_t+\sqrt{\frac34k^2-p^2}\right)}\nonumber\\
&&\displaystyle{\qquad\qquad\qquad\qquad\qquad\qquad\qquad
+\frac12\rho_t\left(\gamma_t+\sqrt{\frac34k^2-p^2}\right)^2
+\ldots\Biggr].}
\end{eqnarray}
This has a pole at $p=i\gamma_t$, which corresponds to the deuteron, hence
$\gamma_t$ is given by the deuteron binding momenta
$\gamma_t=45.70$MeV. The effective range is $\rho_t=1.764$fm. In the
singlet channel there is no bound state so the $T$-matrix is given by,
\begin{eqnarray}
&&\displaystyle{
\tau_s(p,k)=-\frac{4\pi}{M}\frac{1}{-\gamma_s+
\frac12r_s\left(p^2-\frac34k^2\right)+\sqrt{\frac34k^2-p^2}}}\nonumber\\
&&\displaystyle{\qquad\,\,\,\,\,\,\,
=\frac{4\pi}{M}\frac{1}{\gamma_s^2-\frac34k^2+p^2}\Biggl[
\left(\gamma_s+\sqrt{\frac34k^2-p^2}\right)}\nonumber\\
&&\displaystyle{\qquad\qquad\qquad
+\frac12r_s\left\{\left(\gamma_s+\sqrt{\frac34k^2-p^2}\right)^2
-\gamma_s^2\frac{\gamma_s+\sqrt{\frac34k^2-p^2}}
{\gamma_s-\sqrt{\frac34k^2-p^2}}\right\}
+\ldots\Biggr].}
\end{eqnarray}
The difference in the expansion makes the NLO term slightly
more complicated. Here we have $\gamma_s^{-1}=a_s=-23.714$fm and
$r_s=2.73$fm. The Fourier transforms now follow in much the same way
as in the Boson case. We get,
\begin{eqnarray}
&&\displaystyle{
\left(\nabla^2+\frac43\gamma_a^2-\kappa^2\right)\Phi_a({\bf y})
+\int{\rm d}^3{\bf y'}\Biggl(V_0^a({\bf y,y'})}\nonumber\\
&&\displaystyle{\qquad\qquad\qquad
+r_a(V_1^a({\bf y,y'})
+\delta_{as}\tilde V_1^s({\bf y,y'})\Biggr)\left(\frac14\Phi_a({\bf y'})+
\frac34\Phi_b({\bf y'}\right)
=0,\label{eq:nuccoordNLO}}
\end{eqnarray}
where $V^a_0(\boldrho,\boldrho')$ is given by eqn.~(\ref{eq:LOpot}) with
the substitution $\gamma\rightarrow\gamma_a$, $V^a_1(\boldrho,\boldrho')$
is given by eqn.~(\ref{eq:NLOpot}) with the substitution $\gamma\rightarrow\gamma_a$
and
\begin{equation}\label{eq:extraterm}
\tilde V^s_1(y,y')=\frac{8\pi\gamma_s^2}{3M}
\int\frac{{\rm d}^3{\bf k}}{(2\pi)^3}
\int\frac{{\rm d}^3{\bf k'}}{(2\pi)^3}
\left[\frac{\gamma_s+\sqrt{\frac34(k^2+\kappa^2)}}{\gamma_s-\sqrt{\frac34(k^2+\kappa^2)}}
\right]L_2(p,{\bf k,k'},0)
e^{i({\bf k.y-\bf k'.y'})}.
\end{equation}
We were unable to evaluate this final integral analytically. It can be reduced to a one
dimensional equation as in appendix \ref{app:Fourier} and then evaluated numerically.
The s-wave equivalents of these follow easily. The term in eqn.~(\ref{eq:extraterm})
is neglected by Bedaque {\it et al} \cite{brgh}, who simply use the modified form
of the ERE in the singlet channel and take $\rho_s=r_s$. The difference is comparable
to NNLO corrections.

\subsection{Analytic Results.}

It is useful to rewrite the equations for $\Phi_s$ and $\Phi_t$ in terms of
$\Phi_\pm=\Phi_s\pm\Phi_t$. The LO s-wave equations
for these are then given by,
\begin{eqnarray}
&&\displaystyle{
\left(\frac{\partial^2}{\partial y^2}+\frac{2}{3}\left(\gamma_s^2+\gamma_t^2\right)
-\kappa^2\right)\Phi_\pm(y)+\frac{2}{3}\left(\gamma_s^2-\gamma_t^2\right)
\Phi_\mp(y)}\nonumber\\
&&\displaystyle{\qquad\qquad\qquad\qquad
+\int_0^\infty dy'\left(v_0^\pm(y,y')\Phi_+(y')-
\frac12v_0^\mp(y,y')\Phi_-(y')\right)=0,}\label{eq:swavenucstm}
\end{eqnarray}
where,
\begin{equation}
v_0^\pm(y,y')=\frac12(v_0^s(y,y')\pm v_0^t(y,y')).
\end{equation}

The equations for $\Phi_\pm$ are useful because at small $y$ they decouple to give
\begin{equation}
\frac{{\rm d}^2\Phi_\pm(y)}{{\rm d}y^2}
+\frac{16y\lambda_\pm}{\sqrt{3}\pi}\int_0^{\infty}{\rm d}y'\Phi_\pm(y')
\frac{2y'^4+2y^2y'^2-y^4}
{(y^4+y^2y'^2+y'^4)^2}=0,
\end{equation}
where $\lambda_+=1$ and $\lambda_-=-1/2$.  Substituting $\Phi_\pm=y^s$ as before
we obtain the equation for $s$,
\begin{equation}\label{eq:danilov3}
\cos\left(\frac{\pi s}{2}\right)-\frac{8\lambda_\pm}{\sqrt{3}s}\sin\left(\frac{\pi s}{6}\right)=0.
\end{equation}
This equation for $s$ when $\lambda=1$ is the same as the equation that describes the Bosonic
wavefunction at short distances and hence has the imaginary solutions $s=\pm\bar s_0$
\cite{efimov,bhvk,brgh}.
The equation for $\lambda=-1/2$ only has
real solutions, the smallest of which is $s=2$.
In solving eqn.~(\ref{eq:swavenucstm}) generally we may identify solutions by their
behaviour at small $y$. There are solutions in which the interior behaviour is determined
by a boundary condition on $\Phi_+(y)$, for example,
\begin{eqnarray}\label{eq:intplusbc1}
&&\displaystyle{\qquad
\Phi_+(y)\rightarrow\sin(\bar s_0\ln\gamma_t y+\eta)+\ldots,}\\
&&\displaystyle{\label{eq:intplusbc2}
\Phi_-(y)\rightarrow{\cal A}(\gamma_s-\gamma_t)\sin(\bar\ln\gamma_t y+\eta+\sigma)+\ldots,}
\end{eqnarray}
where the constants ${\cal A}$ and $\sigma$ can be determined by solving eqn.~(\ref{eq:swavenucstm})
order-by-order in $y$. There are also
solutions in which the interior behaviour is determined by a boundary condition
on $\Phi_-(y)$, e.g.
\begin{eqnarray}\label{eq:intminusbc}
&&\displaystyle{\qquad
\Phi_-(y)\rightarrow y^2+\ldots,}\\
&&\displaystyle{
\Phi_+(y)\rightarrow{\cal B}(\gamma_s-\gamma_t)y^3+\ldots.}
\end{eqnarray}
These solutions must be uniquely combined in a manner determined by the
boundary conditions at $y\rightarrow\infty$. All the physical distorted waves will be dominated
at short distances by the trig-log behaviour of the $\Phi_+$ solution.

The three-body force will prove to be very important in this system. The 3BDWRG analysis for the
three Boson EFT may be extended to this example and its conclusions are very much the
same. Namely
the LO three-body force will be marginal and will correspond to a self-adjoint extension
of the Hamiltonian.
It will in effect fix the phase of the distorted waves close to the origin. The next
three-body force term that can be determined\footnote{Recall that the
terms that scale with $\gamma$ cannot be measured since we cannot change the value of $\gamma$,
they are therefore absorbed into the energy-dependent force couplings.} 
occurs at the order $(Q/\Lambda_0)^2$. The eqns.~(\ref{eq:swavenucstm})
plus a phase fixing condition correspond to an EFT to order $(Q/\Lambda^0)^0$.

At large $y$, below the three-body threshold but above the deuteron binding energy,
the singlet solution should vanish.
This constitutes an additional boundary condition on the equation that determines
the combination of solutions described above.
After applying this boundary condition, the triplet wavefunction has the
long-distance form,
\begin{equation}
\Phi_t(y)\rightarrow \sin(k_0y+\delta),
\end{equation}
where $k_0^2=4/3\gamma_t^2+\kappa^2$ is the momentum of the free neutron in the centre of
mass frame and $\delta$ is the phaseshift for neutron-deuteron scattering

In solving the NLO equations no new three-body force term is required. The $\gamma_t,\gamma_s$
dependent
terms at this order in the three-body force will appear as refinements in the LO `phase-fixing'
three-body force. At this order there are two equations to solve, corresponding to the perturbations,
$r_s$ and $\rho_t$. Writing,
\begin{equation}
\Phi_\pm(y)=\Phi_\pm^{(0)}(y)+(r_s+\rho_t)\Phi_\pm^{(1+)}+(r_s-\rho_t)\Phi_\pm^{(1-)}+\ldots,
\end{equation}
where $\Phi_\pm^{(0)}$ satisfy eqn.~(\ref{eq:swavenucstm}), using first order perturbation
theory we obtain the equations,
\begin{eqnarray}
&&\displaystyle{
\left(\frac{\partial^2}{\partial y^2}+\frac{2}{3}\left(\gamma_s^2+\gamma_t^2\right)
-\kappa^2\right)\Phi_\pm^{(1+)}(y)+\frac{2}{3}\left(\gamma_s^2-\gamma_t^2\right)
\Phi_\mp^{(1+)}(y)}\nonumber\\
&&\displaystyle{\qquad\qquad\qquad
+\int_0^\infty dy'\left(v_0^\pm(y,y')\Phi_+^{(1+)}(y')-
\frac12v_0^\mp(y,y')\Phi_-^{(1+)}(y')\right)=}\nonumber\\
&&\displaystyle{\qquad\qquad\qquad\qquad\qquad\qquad
-\int_0^\infty\left(
v_1^{+}(y,y')\Phi_+^{(0)}(y')-\frac12v_1^{-}(y,y')\Phi_-^{(0)}(y')\right),}
\label{eq:NLOswavenucstm1}\\
&&\displaystyle{
\left(\frac{\partial^2}{\partial y^2}+\frac{2}{3}\left(\gamma_s^2+\gamma_t^2\right)
-\kappa^2\right)\Phi_\pm^{(1-)}(y)+\frac{2}{3}\left(\gamma_s^2-\gamma_t^2\right)
\Phi_\mp^{(1-)}(y)}\nonumber\\
&&\displaystyle{\qquad\qquad\qquad
+\int_0^\infty dy'\left(v_0^\pm(y,y')\Phi_+^{(1-)}(y')-
\frac12v_0^\mp(y,y')\Phi_-^{(1-)}(y')\right)=}\nonumber\\
&&\displaystyle{\qquad\qquad\qquad\qquad\qquad\qquad
-\int_0^\infty\left(
v_1^{-}(y,y')\Phi_+^{(0)}(y')-\frac12v_1^{+}(y,y')\Phi_-^{(0)}(y')\right),}\label{eq:NLOswavenucstm2}
\end{eqnarray}
where,
\begin{equation}
v_1^{\pm}=(v_1^s\pm v_1^t+\tilde v_1^s)/4.
\end{equation}
Taking $\kappa$ and $\gamma$ to zero in these equations
gives equations for the behaviour of the corrections for small $y$. Since these equations become
scale free at small $y$ it means that the two corrections
$\Phi_\pm^{(1\pm)}(y)\sim y^{s-1}$ as $y$ becomes small. Hence, these `corrections'
radically change the
form of the wavefunctions at small $y$ but provide only minor corrections at large $y$.

\subsection{Numerical Results}

We take the mass of each nucleon to the be isospin averaged mass, $M=938.9$MeV.
Eqn.~(\ref{eq:swavenucstm}) may be solved numerically using a method adapted
from that used in the three Boson example.
To set the interior phase defined by the LO three-body force, we apply a cut-off, $\Omega$,
that allows us to `feed' the desired phase into the equation. Below the cut-off, $y<\Omega$,
we use the form given in eqns.~(\ref{eq:intplusbc1},\ref{eq:intplusbc2}), for $y>\Omega$ the
solution is to be determined from the resulting homogeneous equation.
In order to set the exterior boundary condition on the
singlet solution we must also solve eqn.~(\ref{eq:swavenucstm}) to obtain a solution that
has the small $y$ asymptotic form given by eqn.~(\ref{eq:intminusbc}).

Numerically, the solutions are again most easily found using linear algebra.
The discretisation of the $y$ variable is done as before, however, since we are solving
coupled equations, the matrix equation to be solved is now of order $2n$.

At NLO, the numerical solution of eqns.~(\ref{eq:NLOswavenucstm1},\ref{eq:NLOswavenucstm2})
is achieved similarly but is more complicated. Since these equations are still differential
equations they require boundary conditions. The correct boundary conditions are those that
remove all solutions to the homogeneous equations, which are identical to the LO equations.
That is, the solutions of the NLO equations should be the particular integral.
To find these boundary conditions we must solve
eqns.~(\ref{eq:NLOswavenucstm1},\ref{eq:NLOswavenucstm2}) order-by-order in $y$.
The results give,
\begin{eqnarray}
\Phi_+^{(1+)}\rightarrow{\cal A}_+^+ y^{-1}\sin(\bar s_0\ln y\gamma+\eta+\sigma_+^+),\\
\Phi_-^{(1+)}\rightarrow{\cal A}_-^+ y^{-1}\sin(\bar s_0\ln y\gamma+\eta+\sigma_-^+),\\
\Phi_-^{(1-)}\rightarrow{\cal A}_-^-(\gamma_s-\gamma_t)\sin(\bar s_0\ln y\gamma+\eta+\sigma_-^-),\\
\Phi_+^{(1-)}\rightarrow{\cal A}_+^-(\gamma_s-\gamma_t)\sin(\bar s_0\ln y\gamma+\eta+\sigma_+^-),
\end{eqnarray}
where the exact values of the constants ${\cal A}$ and $\sigma$ are not important for this
discussion.
These boundary conditions are now fed into the equation by using the cut-off $\Omega$.

\begin{figure}
\begin{center}
\includegraphics[height=20cm,width=15cm,angle=0]{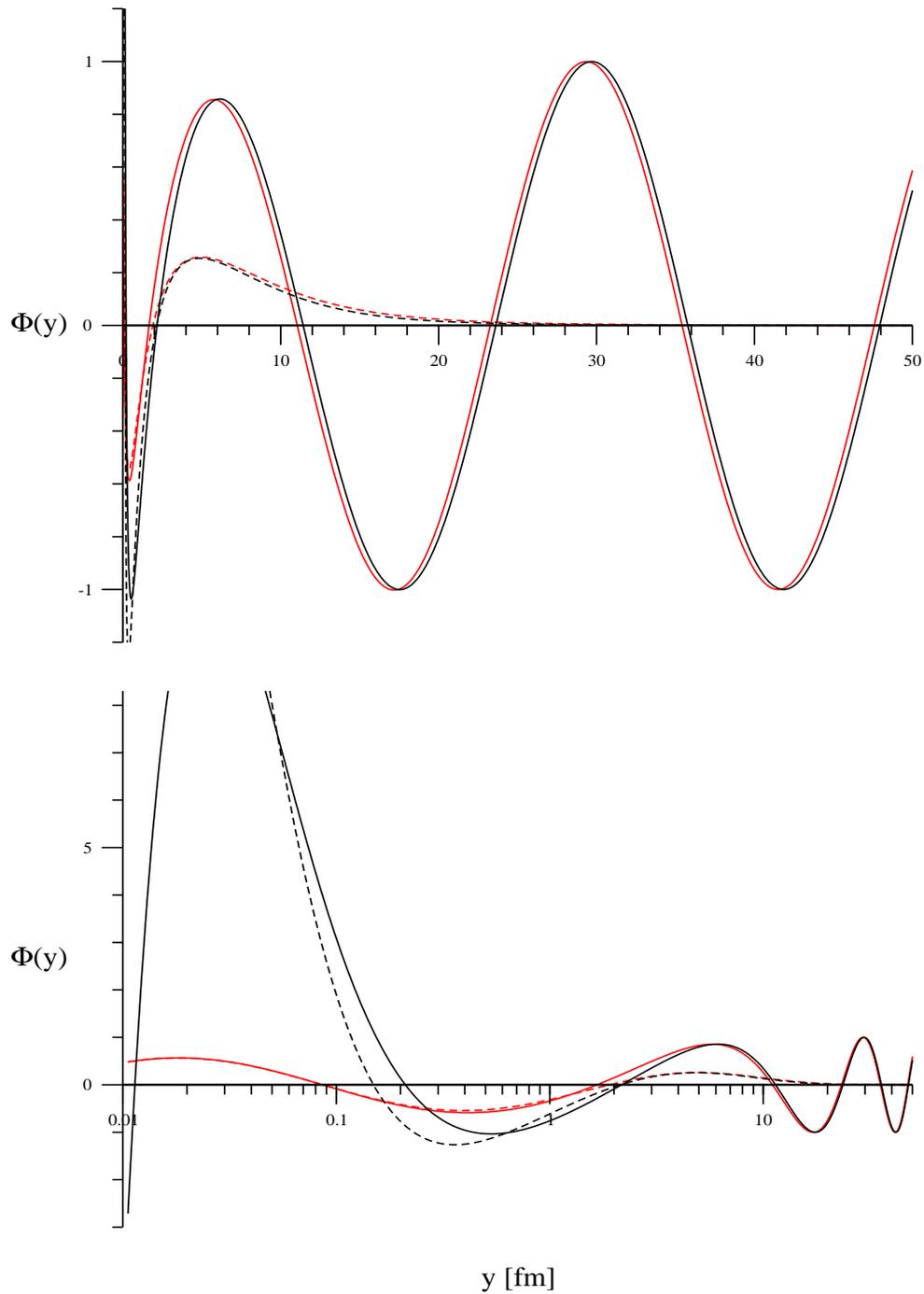}
\caption{Scattering solution to eqn.~(\ref{eq:nuccoordNLO})
with $q_0=51.0$MeV ($\kappa=13.6$MeV) at LO (red) and
NLO (black). The solid lines are the triplet wavefunction $\Phi_t(y)$,
the dashed lines are the singlet wavefunction $\Phi_s(y)$}
\label{fig:nucscatter}
\end{center}
\end{figure}

Fig.~\ref{fig:nucscatter} shows the projection of the wavefunction, $\Phi(y)$,
of an elastic scattering wavefunction at LO and NLO. The solid lines
show the triplet wavefunction and dashed lines show the singlet wavefunction. The LO
result is shown in red and the NLO result in black. Since there is no outgoing bound state
in the singlet state, the singlet wavefunction vanishes as $y\rightarrow\infty$. The
wave in the triplet channel corresponds to an outgoing/incoming deuteron and neutron.
At small $y$ the trig-log behaviour is apparent in the LO solution. The NLO solution
also displays this behaviour with an extra factor of $y^{-1}$. The LO and NLO
solutions become distinct at around $2{\rm fm}\sim r_s,\rho_t$.

To obtain physical variables, the LO three-body force must be fixed. This can be done
by matching to the deuteron neutron scattering length, $a_{nd}=0.65\pm0.04$fm\cite{ndscat}.
With this single input the EFT becomes predictive for all other scattering observables
including the triton binding energy. The three-body force at NLO will differ from the
three-body force at LO by a term of order $\rho_e\gamma$.

\begin{figure}
\begin{center}
\includegraphics[height=12.5cm,width=8cm,angle=-90]{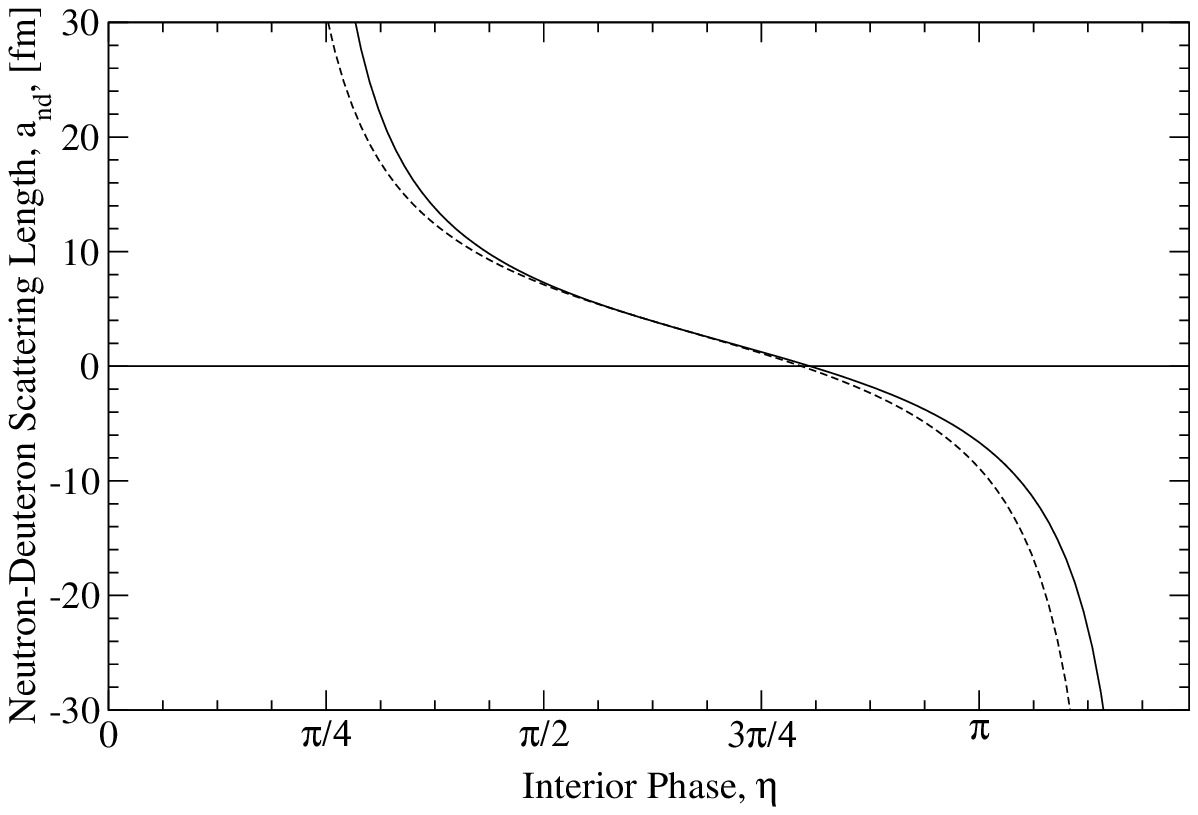}
\caption{Neutron-Deuteron scattering lengths that result from choice
of LO three-body force, corresponding to a different interior phase,
$\eta$, for the trig-log behaviour of the wavefunctions. The solid
and dashed lines show the LO and NLO results respectively
}
\label{fig:Scatts}
\includegraphics[height=12.5cm,width=8cm,angle=-90]{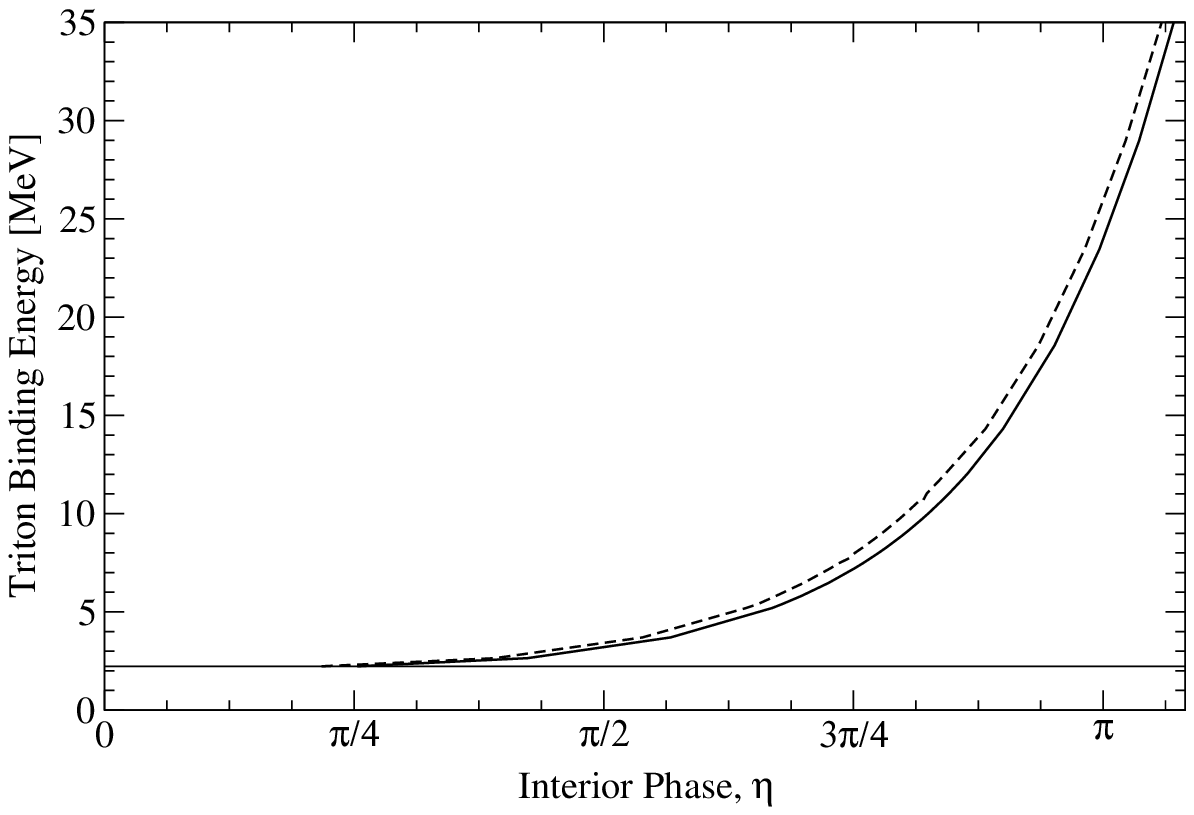}
\caption{Triton binding energies that result from different choices
of the LO three-body force. The solid
and dashed lines show the LO and NLO results respectively
}
\label{fig:Bounds}
\end{center}
\end{figure}

\begin{figure}
\begin{center}
\includegraphics[height=12.5cm,width=8cm,angle=-90]{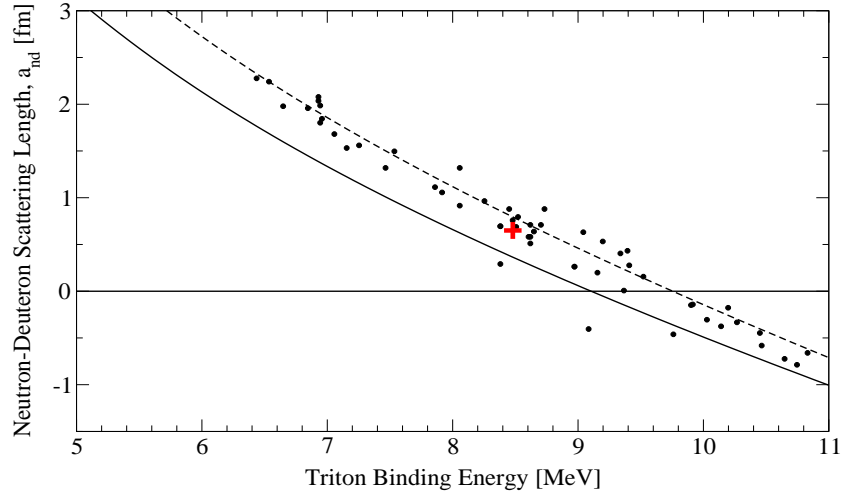}
\caption{The Phillips line as predicted by the pionless EFT at LO (bold line)
and NLO (dashed line). The dots correspond to the predictions for the
triton binding energy and neutron-deuteron scattering length in different
models with the same two-body inputs \cite{efimovphil}. The cross is the experimental result.}
\label{fig:phillips}
\end{center}
\end{figure}

The relationship between the choice of three-body force, the neutron-deuteron scattering
length and the triton binding energy is shown in figs.~\ref{fig:Scatts},\ref{fig:Bounds}
and \ref{fig:phillips}. Fig.~\ref{fig:Scatts} shows the values of the neutron
deuteron scattering length that correspond to different interior phases of the DWs
and hence to different three-body forces. Fig.\ref{fig:Bounds} shows how the binding
energy of the Triton (the shallowest bound state)
depends upon our choice of three body force. Many of these shallowest bound states actually
lie outside the validity of the EFT defined by the pion mass.

Bringing the results of figs. \ref{fig:Scatts} and \ref{fig:Bounds} together gives a
one variable relationship between the neutron-deuteron scattering length
and the triton binding energy, fig.~\ref{fig:phillips}.
Although this curve cannot be seen experimentally
as the physical values correspond to a single point in this space,
shown by a cross, it does
have interesting implications for the study of three nucleon problems with
potential models. The dots in fig.~\ref{fig:phillips} correspond to different
predictions for the triton binding energy and the neutron-deuteron scattering
length obtained from different pair wise nucleon potentials with the
same two body scattering lengths and effective ranges as inputs \cite{efimovphil}. Instead of
clustering randomly around the physical values, these form a curve through
the space.. This relationship was first noted by Phillips \cite{phillips}
in looking at the predictions for three-body variables by nuclear pairwise potential models.
In fig.~\ref{fig:phillips} the NLO result (dashed)
is a clear improvement upon the LO result (solid). The connection between the Phillips line
and the three-body force in the three-body EFT has already been illustrated by Bedaque
{\it et al} \cite{brgh}, our NLO curve differs slightly from theirs because of the
additional term \ref{eq:extraterm}.

\begin{table}
\begin{center}
\begin{tabular}{|c|c|c|c|}\hline
& LO
& NLO & Physical\\ \hline\hline
Interior Phase of DWs, $\eta$, where
& & &\\
$\Phi\rightarrow\sin(\bar s_0\ln{\mu y}+\eta)$ & 2.442 & 2.419 &\\
and $\mu=100$MeV. & & &\\ \hline
& & & \\
Triton Binding Energy & 8.08MeV & 8.68MeV & 8.48MeV\\ 
& & & \\ \hline
\end{tabular}
\caption{The interior phase of the DWs and the Triton binding
energy at LO and NLO found by matching to the neutron-deuteron
scattering length.}
\label{table}
\end{center}
\end{table}

\begin{figure}
\begin{center}
\includegraphics[height=20cm,width=15cm,angle=0]{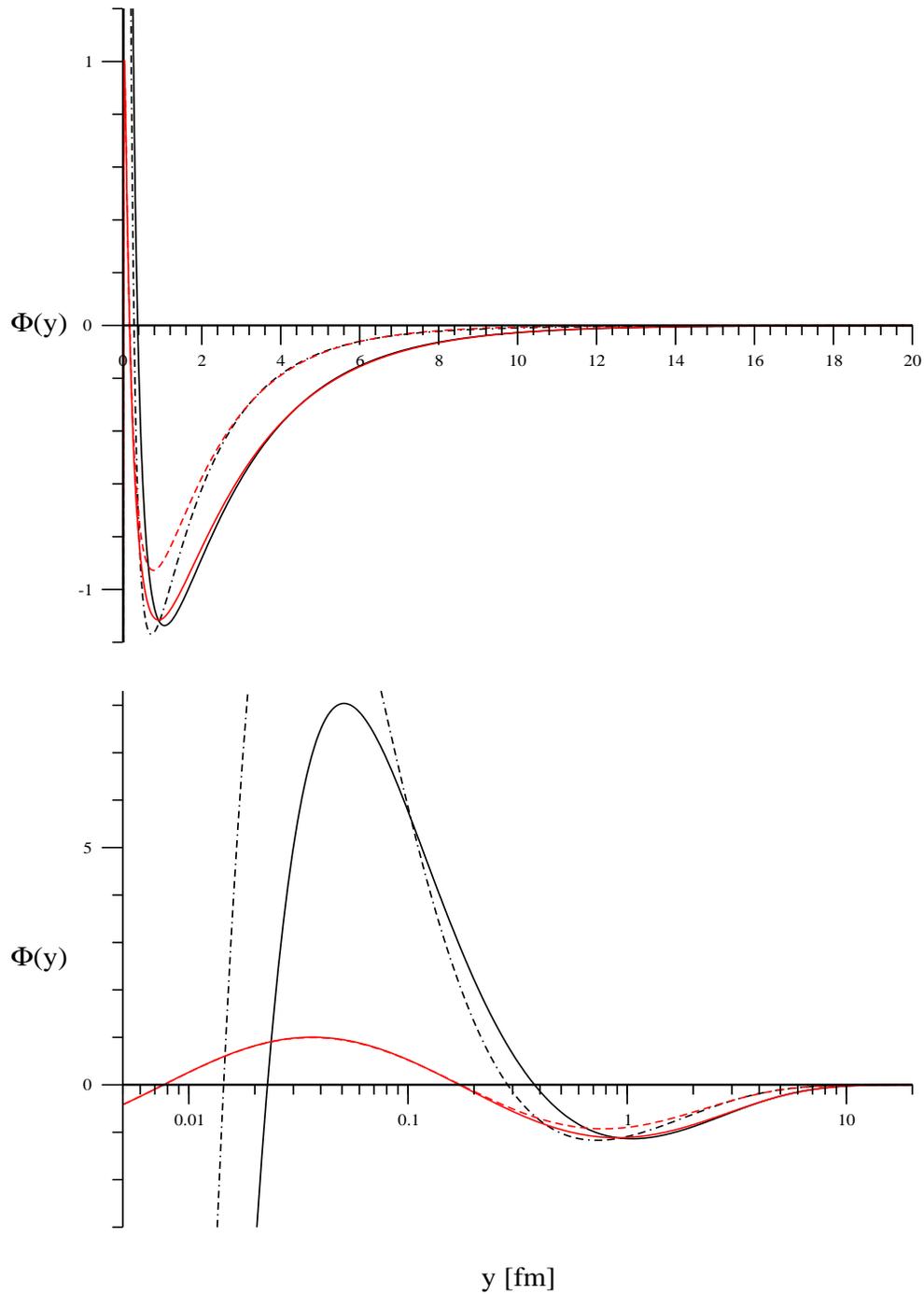}
\caption{The triton wavefunction's projection $\Phi(y)$
as found by matching the three-body
force to the neutron-deuteron scattering length at LO (red) and
NLO (black). The solid lines are the triplet wavefunction $\Phi_t(y)$,
the dashed lines are the singlet wavefunction $\Phi_s(y)$}
\label{fig:tritonproj}
\end{center}
\end{figure}

\begin{figure}
\begin{center}
Singlet state $\Psi_s(R,\alpha)$

$\alpha$
\includegraphics*[90,490][323,715]{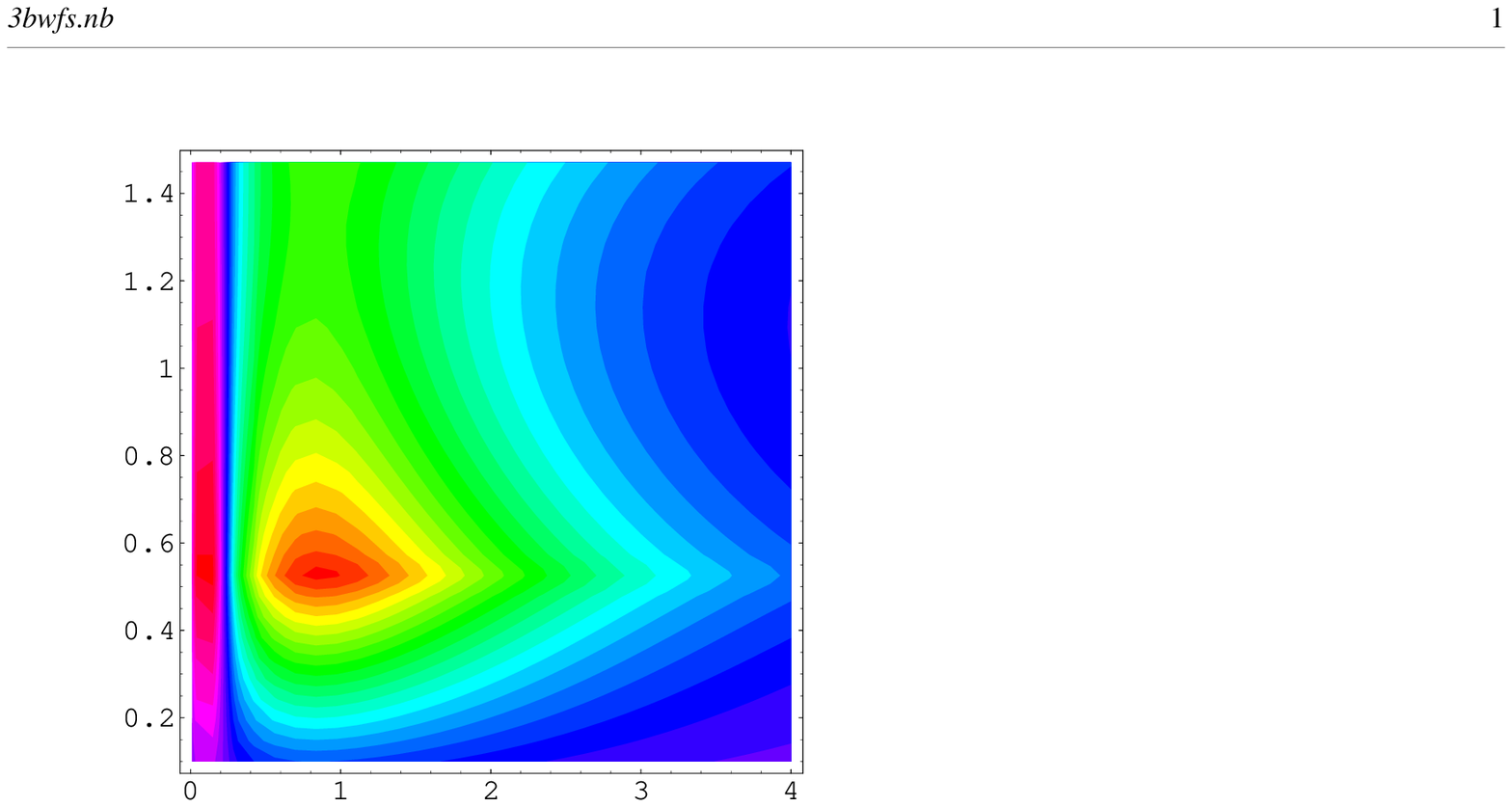}

R [fm]
\vspace{1cm}

Triplet state $\Psi_t(R,\alpha)$

$\alpha$
\includegraphics*[90,490][323,715]{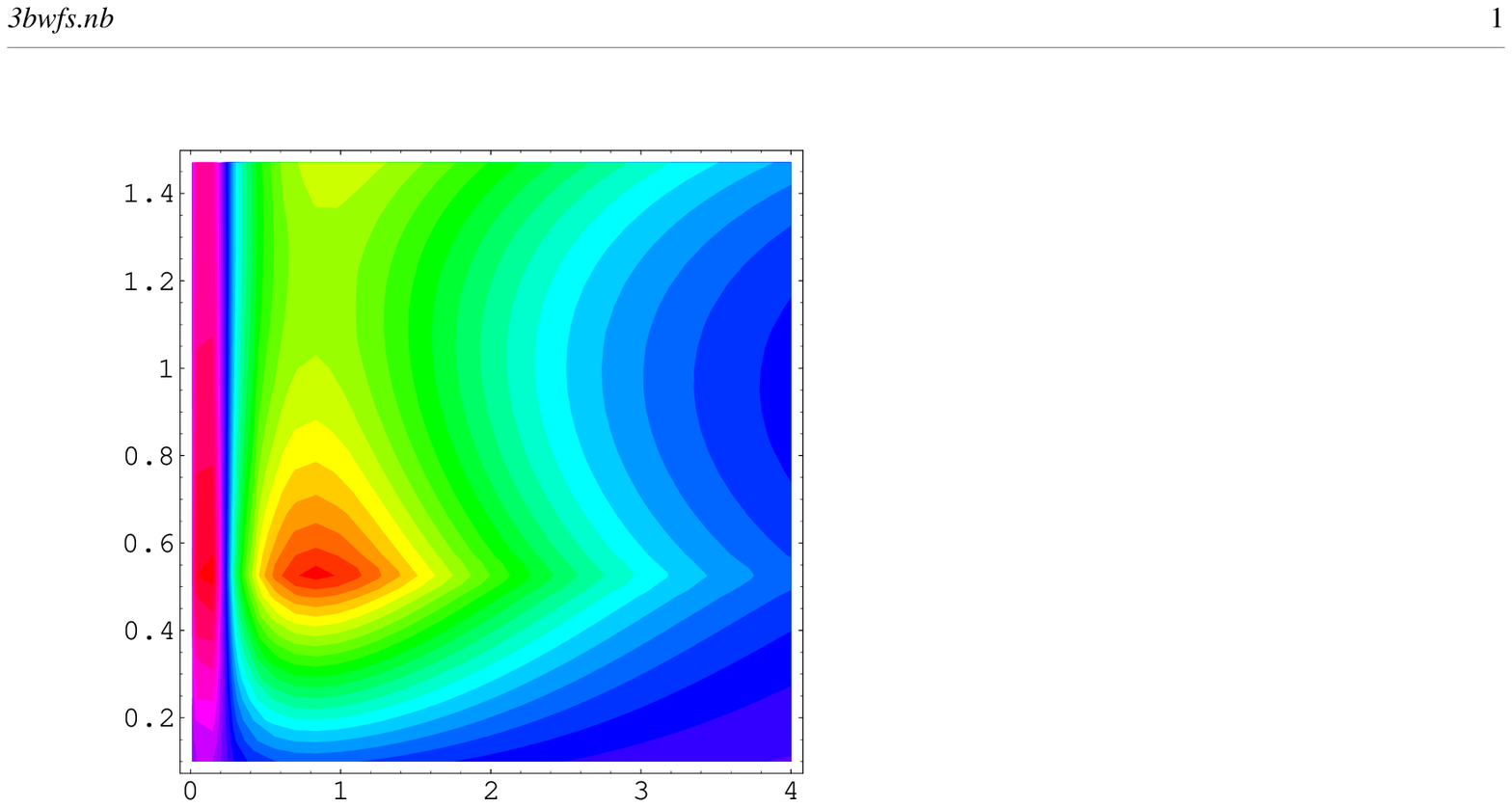}

R [fm]

\caption{Contour plots showing the triton wavefunction at LO. The highest
particle density occurs at $\alpha=\pi/6$, $R\sim 1$fm in both the
triplet and singlet wavefunctions. This corresponds to a proton and
a neutron on top of each other with the third neutron 1fm away.
In the triplet wavefunction the $\alpha=\pi/2$ `tail' is longer and peaks
higher as this also corresponds to a proton neutron pair with a 
distinct neutron.}
\label{fig:3dtriton}
\end{center}
\end{figure}

The interior phase of the DWs and the prediction for the Triton
binding energy are shown in table~\ref{table}.
Fig.~\ref{fig:tritonproj} shows the projection of the triton wavefunction at LO and NLO.
The full LO Triton wavefunction obtained from eqn.~(\ref{eq:make3d}) are
shown in fig.~\ref{fig:3dtriton} on hyperpolar plots. The configuration of
the three particles, parameterised in $\alpha$, depends upon our choice
of pair in defining $\alpha$.
In the triplet wavefunction, the `chosen' pair
have isospin 0 and so correspond to a neutron and proton, in the singlet
wavefunction the `chosen` pair are the two neutrons. 

The highest value
of the wavefunction occurs as a proton and neutron are paired with a second
neutron around 1fm away, which occurs at $\alpha=\pi/6$, $R=1$fm in both the
singlet and triplet wavefunctions and also at $\alpha=\pi/2$, $R=1$fm in the
triplet wavefunction. This configuration also allows the three nucleons to
be spread over the greatest volume, i.e the highest density at large values of
$R$ corresponds to the values $\alpha=\pi/6$ in both singlet and triplet states and also
to $\alpha=0$ in the triplet state. For $R<1$fm the wavefunction
does not distinguish much between the different configurations of the three
particles (different $\alpha$'s) except for $\alpha\rightarrow0$ where the
wavefunction vanishes. This value, $\alpha=0$, corresponds to the situation where
the third nucleon is directly between the other two.

\begin{figure}
\begin{center}
\includegraphics[height=12.5cm,width=8cm,angle=-90]{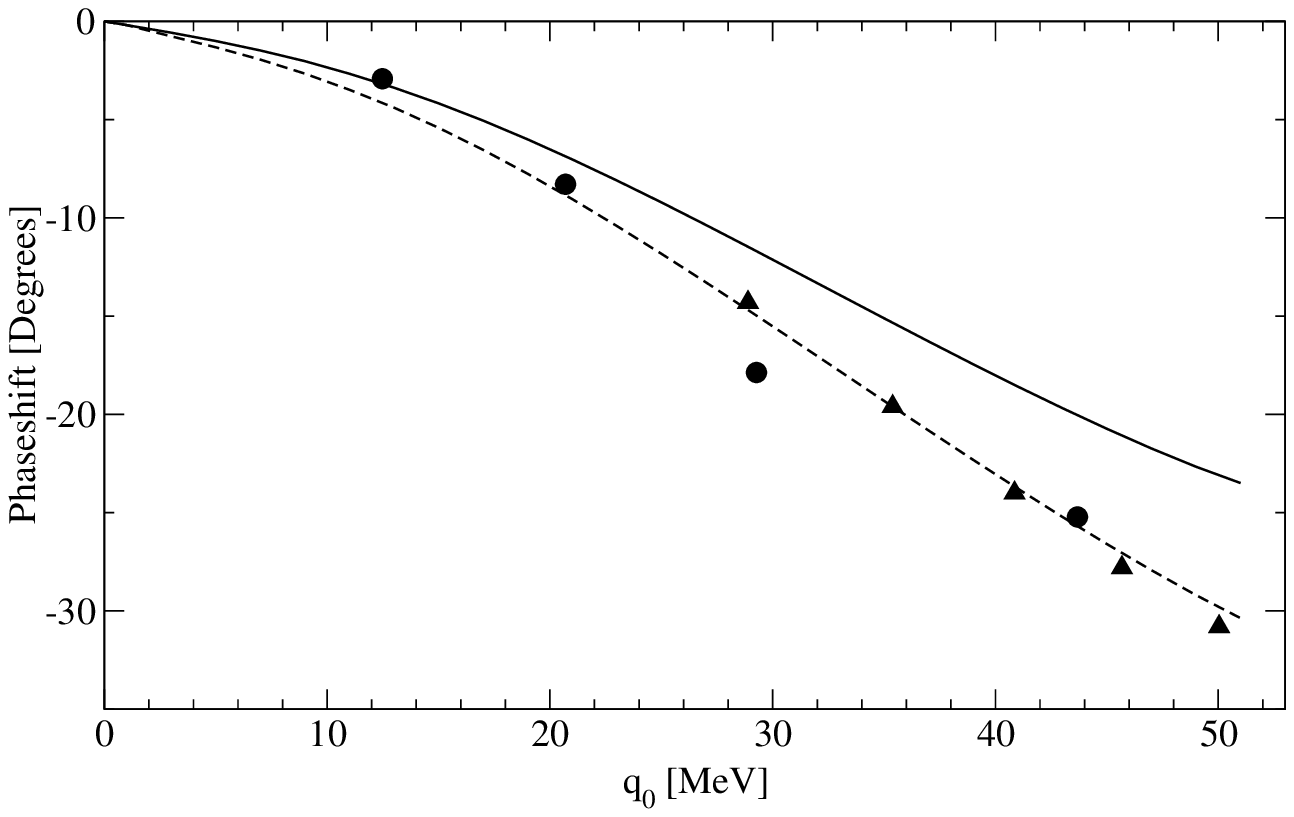}
\caption{Physical phaseshifts for neutron-deuteron scattering
($q_0=\sqrt{4/3\gamma_t^2-\kappa^2}$)
with the LO three-body force fitted to the neutron-deuteron
scattering length. The solid and dashed lines are the LO and
NLO result respectively. The circles are the most recent experimental
results from 1967\cite{oerssea} and the triangles are the results obtained
using the Argonne V18+Urbana IX two and three-body forces \cite{kievsky}. 
}
\label{fig:Physical}
\end{center}
\end{figure}

Fig.~\ref{fig:Physical} shows the predictions for the elastic neutron-deuteron
phaseshift after matching the three-body force to the neutron-deuteron
scattering length at LO (solid line) and NLO (dashed line). The 
circles show the most recent experimental results (circa 1967) \cite{oerssea}
and the triangles show the results obtained from a combination of V18 and Urbana
IX two and three-body forces \cite{kievsky}. Given the age of the experimental
results, in makes sense to compare the data to that obtained from sophisticated
potential models, rather than the experimental data. These models are fitted to
many different variables in two- and three-body systems and can be expected to
produce a reasonable curve \cite{quartetbvk,brgh,phill3b}.

The EFT curve fits the potential model curve well at NLO. The computing effort in
solving the EFT equations for the wavefunctions and the physical variables is minimal
compared with the effort required for the more sophisticated approaches.

To go to higher orders than NLO requires considerably more effort. A new three-body
force is required, isospin symmetry violation must be considered and the computational
effort in calculating the four second-order perturbations is considerable.
However, the agreement achieved with very few parameters and at the relatively low
order of $(Q/\Lambda_0)^1$ is pleasing.

\subsection{Summary}

In this chapter we have derived and solved equations for the
three body DWs for three Bosons and three nucleons at LO and NLO
in the KSW EFT. In each case a `phase-fixing' three-body force is required.

The implications of the LO three-body force in the case of three nucleons
is embodied in the Phillip's line, which shows the one parameter relation
between the neutron-deuteron scattering length and the triton binding energy
as predicted by pairwise nuclear potentials.
The NLO results show convergence upon the potential model results in both
the neutron-deuteron phaseshift and the Phillip's line.

The equations, once the three-body force is determined by fitting to the
neutron-deuteron scattering length, yield the triton wavefunction and all
DWs describing elastic scattering of a neutron and a deuteron below
breakup. The triton wavefunction result can be used in applications,
such as electron triton scattering.

%% file: misc/conc.tex
In this thesis we have looked at the use of the renormalisation
group in the development of effective field theories. Our three
main results are the development of a tool, the DWRG and general
methods of solving the DWRG equation; the use of the 3BDWRG equation
in deriving the limit-cycle solution for the three-body force
in the KSW EFT for short-range forces and the corresponding
power-counting; the development and solution of equations that can
be used to practically evaluate three body DWs in the KSW EFT
for Bosons and the physically and mathematically
interesting system of three nucleons in the ${}^3S_1$ channel.

The DWRG is tool that enables us to resum some diagrams to all orders.
The general method of solving the DWRG equation relies upon the
basic loop integral, perturbations about which give an alternate
power-counting scheme to naive dimensional analysis that corresponds
to a trivial fixed point in the DWRG. What is pleasing about the results
we have obtained, using the DWRG, is that established results
such as the DWERE for well-behaved potentials share a common method with
new results such as the limit-cycle solutions for short range forces with
an attractive inverse square potential.

The solution of the DWRG for the attractive inverse square
problem required a little care. The system is not properly defined without the
input of a scale, $p_0$, that defines a self-adjoint extension. The
solutions of the DWRG are in the form of limit-cycles. The power-counting
associated with the limit-cycle solutions have a LO marginal term, which
corresponds to the degree of freedom offered by the choice of $p_0$.
The solution of the DWRG answers many issues raised in refs. \cite{lcbc,splc}
concerning the connection between the short range effective force and the
need for a self-adjoint extension, the possibility of multi-valued solutions
of the RG, and the lack of a ground state.

Our study of the KSW EFT three-body problem was motivated by the
recent realisation that the three-body force is extremely important in some
systems \cite{bhvk,brgh}. 
Two systems that are particularly interesting are the three s-wave Bosons and
three nucleons in the ${}^3S_1$ channel. These systems have no net repulsion due to the
pair-wise forces and result in singularities similar to the attractive inverse
square potential. Because of this similarity, the 3BDWRG solutions exhibit
limit-cycle behaviour and a three-body force that occurs at order $(Q/\Lambda_0)^0$.
Our results confirm the power-counting proposed by Bedaque {\it et al}
\cite{bhvk,brgh}. The power of our approach is that we are able to give a simple
algebraic statement of that counting.

Our results also illustrate the limit cycle behaviour of the three-body force
that has been suggested based on analysis of the attractive
inverse square potential \cite{lcbc,splc} and the STM equation \cite{bhvk,brgh}.

The generalisation of the 3BDWRG to four or more bodies is straightforward,
our method of solution involving the basic loop integral is easily
generalised. The scale-free systems studied in this thesis and the novel
power-counting schemes that correspond to them are likely to be important
in N-body forces in the KSW EFT due to the scale-free nature of these systems
at hyperradii far smaller than the two-body scattering length.

The final chapter offers some physical results for the KSW EFT. The convergence
of the results using a limit-cycle three-body force and their agreement with
the results of sophisticated potential methods has already been shown,
\cite{brgh,phill3b}. To complement these results and to offer new possibilities of
testing the EFT we have derived equations that allow calculation of the DWs
for the three-body KSW EFT below threshold. Our equations are the coordinate
space equivalent of the STM equations \cite{stm}. 
Taking advantage of what we had shown earlier,
the LO three-body force can be chosen by simply fixing the phase of the DWs
at small hyperradii.

Our results for three nucleons in the ${}^1S_1$ channel agree well with those
of Bedaque {\it et al} (up to a difference in definition of NLO corrections).
Our method allows us to calculate particle density of the triton wavefunction
that may be tested by looking at electron-triton scattering data.

In summary, the DWRG has shown itself to be a versatile tool and has been
used to shed light on several issues currently under debate. This thesis has
produced several results that provide insight into current issues in the
EFT community and open new doors for testing the KSW EFT for short-range forces.

%% file: app/Appendix0.tex
We want to study the analyticity of the integral,
\begin{equation}
I=\int_0^1 d\hat q{\cal C}(\hat\kappa/\hat q)=
2\pi\kappa\int_0^1 \frac{d\hat q}{\hat q(e^{\frac{2\pi\kappa}{\hat q}}-1),}
\end{equation}
which we can write as,
\begin{equation}
I=2\pi\hat\kappa\int_{2\pi\hat\kappa}^\infty \frac{dx}{x(e^x-1)}=
2\pi A\hat\kappa-\int^{2\pi\hat\kappa}\frac{dx}{x(e^x-1)},
\end{equation}
where $A$ is some constant. Since we are investigating the behaviour
of the integral near $\hat\kappa\rightarrow0$ we can expand the
integrand in the final term in a series in $x$ and assume $\kappa$
to be within the radius of convergence to get,
\begin{equation}
I=2\pi A\hat\kappa-\int^{2\pi\hat\kappa} dx\left(
\frac{1}{x^2}-\frac{1}{2x}+\frac{1}{12}+\ldots\right)
=1+\pi\hat\kappa\bigl(2A+\ln\hat\kappa+\ln2\pi\bigr)-\frac{\pi\hat\kappa}{6}
+\ldots.
\end{equation}

%% file: app/AppCoulAtt.tex
In this appendix we shall look quickly at the modifications to the
DWRG analysis for an attractive Coulomb potential,
i.e. when $\kappa$ is negative.

Clearly, the power-counting around each fixed point is unchanged
as these result from the form of the DWRG equation rather than details. 
The trivial fixed point is also unchanged, the expansion around it yields
the distorted wave Born expansion, eqn.~(\ref{eq:dwbe}).

\begin{figure}
\begin{center}
\includegraphics[height=13cm,width=8cm,angle=-90]{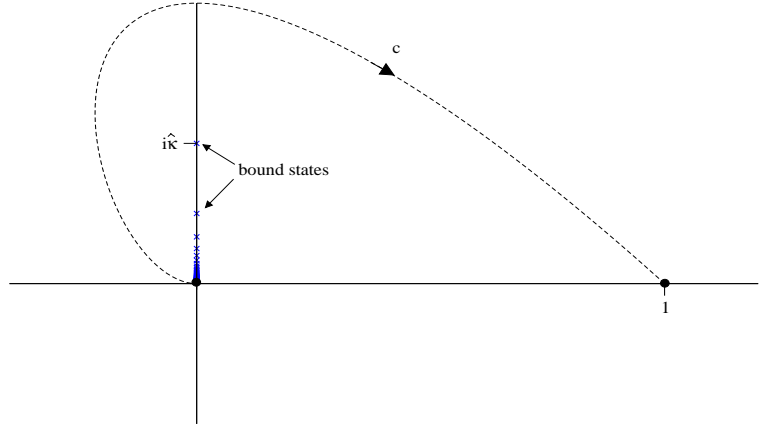}
\caption{The contour of integration in the complex $\hat q$-plane
for the non-trivial fixed point in the attractive Coulomb case.
The bound states appear as poles at $\hat q=i\hat\kappa/n$.}
\label{fig:coulcont}
\end{center}
\end{figure}

The two differences that result from having an attractive Coulomb
potential is the existence of bound states that occur in the
truncated Green's function and the definition of the non-trivial
fixed point. In the repulsive case the starting point for
the non-trivial fixed point solution was the basic loop integral,
\begin{equation}
\hat J(\hat p,\hat\kappa)=\fint_0^1\hat q^2 d\hat q
\frac{{\cal C}(\hat\kappa/\hat q)}{\hat p^2-\hat q^2}.
\end{equation}
However, in the repulsive case this integral no longer
converges because of the essential singularity at $\hat q=0$.
To remedy this, we now define $\hat J$ as
\begin{equation}
\hat J(\hat p,\hat\kappa)=\text{Re}\int_{c}\hat q^2 d\hat q
\frac{{\cal C}(\hat\kappa/\hat q)}{\hat p^2-\hat q^2},
\end{equation}
where $c$ is the contour shown in fig.~\ref{fig:coulcont}.
This integral is convergent because it approaches the essential
singularity from the other direction.

The integral $\hat J$ can now be split up as before to isolate the
singularities:
\begin{eqnarray}
&&\displaystyle{
\hat J(\hat p,\hat\kappa)=-\text{Re}\int_c d\hat q \,{\cal C}(\hat\kappa/\hat q)
-\hat p^2\int_1^\infty d\hat q\frac{{\cal C}(\hat\kappa/\hat q)}
{\hat p^2-\hat q^2}+\hat{\cal M}(\hat p,\hat\kappa),}\\
&&\displaystyle{\qquad\qquad\qquad
\hat{\cal M}(\hat p,\hat\kappa)=\hat p^2\,\text{Re}\int_{c'}
 d\hat q\frac{{\cal C}(\hat\kappa/\hat q)}
{\hat p^2-\hat q^2}},
\end{eqnarray}
where the contour $c'$ extends the contour $c$ along the real axis to
infinity. As in the repulsive case, the second term is analytic, the
term $\hat\cal M$ is not analytic but satisfies the homogeneous DWRG equation
and the first term contains logarithmic dependence, which is still given
by the analysis in the previous appendix. Hence we can write the non-trivial
fixed point solution precisely as before,
\begin{equation}
\hat V_S^{(0)}=\Bigl(\hat J(\hat p\hat\kappa)-\hat{\cal M}(\hat p,\hat\kappa)-
\hat{\cal L}(\hat\kappa,\Lambda)\Bigr)^{-1}.
\end{equation}
where ${\cal L}$ is defined in eqn.~(\ref{eq:logterm}). Now we have found
a non-trivial fixed point solution the comments on the RG flow are the same as
in the repulsive case.

We will find that when writing down the distorted wave effective range expansion
there is the additional bound state terms in the truncated Green's function.
Eqn.~(\ref{eq:firstdwere}) should now read,
\begin{equation}\label{eq:newdwere}
\frac{\langle\psi_p|\tilde K_S|\psi_p\rangle}{|\psi_p(R)|^2}=
V_S(p,\kappa,\Lambda)\left(1-V_S(p,\kappa,\Lambda)\left\{\frac{M}{2\pi^2}
\fint_0^\Lambda dq\frac{|\psi_q(R)|^2}{p^2-q^2}
+\frac{M}{4\pi}\sum_{n=1}^\infty \frac{|\psi_n(R)|^2}{p^2+p_n^2}\right\}\right)^{-1},
\end{equation}
where $\psi_n(r)$ is the bound state with binding energy $E_n=|\kappa|^2/n^2$
and is given by,
\begin{equation}
\psi_n(r)=\sqrt{\frac{\kappa^3}{n^5}}2re^{-\kappa r/n}L_{n-1}^1(2\kappa r/n)
\rightarrow 2r\left(\frac{\kappa}{n}\right)^3 \text{as $r\rightarrow0$},
\end{equation}
where $L^\alpha_n$ is the associated Laguerre Polynomial.
The term in braces in eqn.~(\ref{eq:newdwere}) may be written as,
\begin{equation}
\frac{M\Lambda R^2}{2\pi^2}\left(\int_0^1 d\hat q 
\frac{\hat q^2{\cal C}(\hat\kappa/\hat q)}{\hat p^2-\hat q^2}
-\sum_{n=1}^\infty
{\cal R}\left\{\frac{\hat q^2{\cal C}(\hat\kappa/\hat q)}{\hat p^2-\hat q^2},
\hat q\rightarrow i|\kappa|/n\right\}\right)
\end{equation}
and so by Cauchy's theorem is equal to $M\Lambda^2\hat J(\hat p,\hat\kappa)$.
The analysis follows as before, resulting in the DWERE,
\begin{equation}
{\cal C}(\eta)\cot\tilde\delta_S-{\cal M}(p,\kappa)=
-2\kappa\ln{\frac{\kappa}{\mu}}
-\frac{2\Lambda_0}{\pi}\sum_{n,m}\hat C_{2n,m}\left(\frac{p^{2n}\kappa^{m}
}{\Lambda_0^{2n+m}}\right),
\end{equation}
where,
\begin{eqnarray}
&&\displaystyle{
{\cal M}(p,\kappa)=\frac{2\Lambda}{\pi}\hat{\cal M}(\hat p,\hat\kappa)
=\text{Re}\int_c'
\frac{dq}{q}\frac{4\kappa}{e^{2\pi\kappa/q}-1}
\frac{p^2}{p^2-q^2}}\nonumber\\
&&\displaystyle{\qquad\qquad\qquad\qquad\qquad\qquad
=-2\kappa\text{Re}\left[\ln(-i\eta)+\frac{1}{2i\eta}-\psi(-i\eta)\right].}
\end{eqnarray}

%% file: app/Appfaadeev.tex
For a system with three particles the Lippmann-Schwinger equation is
inadequate. Although the three-body $T$-matrix satisfies the
Lippmann-Schwinger equation, it is not possible to use this equation to find it
because its kernel is not compact. This may be understood in several
ways. Most simply in coordinate space it is a result of a non-vanishing
potential as the mean separation of the three particles is taken to infinity but
two of the particles remain close. This in turn is related to the possibility
of incoming and outgoing boundary conditions involving
bound states of particles and also to the possibility of
so-called disconnected diagrams, in which one of the three-particles
fails to interact.

In short, a wave-function with an incoming
two-body bound state and a single particle satisfies the homogeneous version
of the Lippmann-Schwinger equation resulting in non-unique solutions.
The resolution to this quandary is well-known. We must ensure that disconnected
diagrams are not allowed. One favoured method is to use the Faddeev equations.

Consider the case of three identical bosons interacting via a
pairwise interaction $V_{2B}$.
The wavefunction must be
symmetric with respect to transposition of any two particle indices,
hence the wavefunction, $|\Psi^+\rangle$
with incoming boundary conditions of three free
particles (rather than a bound state of any two) can be written as
\begin{equation}
|\Psi^+\rangle=|\psi_0\rangle+(1+P)|\psi^+\rangle,
\end{equation}
where $|\psi_0\rangle$ is the `in'-state and solves the free
Schr\"odinger equation and $P$ is a permutation operator,
which permutes particle indices and has the matrix elements defined
in eqn.~(\ref{eq:permmatrix}).
The Schr\"odinger equation is written as,
\begin{eqnarray} 
&&\displaystyle{\qquad\qquad
(H_0+(1+P)V_{2B}-E)|\Psi^+\rangle=0,}\\
&&\displaystyle{
(1+P)(H_0+V_{2B}-E)|\psi^+\rangle=-(1+P)V_{2B}(|\psi_0\rangle+P|\psi^+\rangle,}
\label{eq:faddeev1}
\end{eqnarray}
where to obtain the second line, we have used the fact that $H_0$
commutes with the permutation operator, $P$. Cancelling the 
$(1+P)$ on the left hand side and using
$G_{2B}^+=(E-V_{2B}-H_0+i\epsilon)^{-1}$ we obtain,
\begin{eqnarray}
|\psi^+\rangle&=&G_{2B}^+(E)V_{2B}|\psi_0\rangle
+G_{2B}^+(E)V_{2B}^+P|\psi^+\rangle,\\
&=&G_0^+(E)t^+(E)|\psi_0\rangle+G_0^+(E)t^+(E)P|\psi^+
\rangle,
\end{eqnarray}
where $t^+$ is the $T$-matrix given by the equation,
$t^+(E)=V_{2B}+V_{2B}G_{2B}^+(E)V_{2B}$. Eqn.~(\ref{eq:faddeev2}) is the
Faddeev equation for the wavefunction component $|\psi^+\rangle$. It is
connected because of the $P$-operator which ensures that each interaction,
given by the two-body $T$-matrix, is not between the same
two particles as the preceding one.

The corresponding equations for the full three-body Green's function and
$T$-matrix now follow from similar decompositions. If we write the
full $T$-matrix, $\cal T$ in terms of components,
\begin{equation}
{\cal T^+}(E)=(1+P)T^+(E),
\end{equation}
then from the equation defining the full wavefunction in terms of the
M\"oller wave operator $1+G_0^+(E){\cal T^+}(E)$,
\begin{equation}
|\Psi^+\rangle=(1+G_0^+(E){\cal T^+}(E))|\psi_0\rangle
=|\psi_0\rangle+(1+P)G_0^+(E)T^+|\psi_0(E)\rangle,
\end{equation}
and from the definition of $|\Psi^+\rangle$ in terms of its components,
eqn.~(\ref{eq:def3bwf}), we obtain the relation, 
$|\psi^+\rangle=G_0^+(E)T^+(E)|\psi_0\rangle$, which when substituted into the
Faddeev eqn.~(\ref{eq:faddeev2}) yields the Faddeev equation for the $T$-matrix
component,
\begin{equation}
T^+(E)=t^+(E)+t^+(E)G_0^+(E)PT^+(E).
\end{equation}
We may find a similar relation for the full Green's function, ${\cal G^+}(E)=
G_0^+(E)+(1+P)G^+(E)$, by using the equation ${\cal G^+}(E)=
G_0^+(E)+G_0^+(E){\cal T}^+(E)G_0^+(E)$.

%% file: app/Appmompert.tex
Momentum dependent perturbations about the limit cycle solution are found
using the ansatz
\begin{eqnarray}
&&\displaystyle{
{\hat V}_{3B}(\hat p,\hat\gamma,\hat k,\hat k',\Lambda)=
{\hat V}_{3B}^{(0)}(\hat p,\hat\gamma,\Lambda)+C\Lambda^\nu
\phi(\hat p,\hat\gamma,\Lambda),}\\
&&\displaystyle{\,\,\,\,\,\,\,\,\,
\phi(\hat p,\hat\gamma,\Lambda)=
k^{2n}\phi_1(\hat p,\hat\gamma,\Lambda)+\phi_2(\hat p,\hat\gamma,\Lambda)
,}
\end{eqnarray}
where $\phi_1$ and $\phi_2$ only depend upon $\Lambda$ logarithmically.
Substituting this into the DWRG equation (\ref{eq:3bdwrg3}), neglecting for
now the discontinuities, then linearising
gives a partial differential equation for $\phi_1$ and $\phi_2$,
\begin{eqnarray}
&&\displaystyle{
\hat k^{2n}\left[-\Lambda\frac{\partial\phi_1}{\partial\Lambda}
+\hat p\frac{\partial\phi_1}{\partial\hat p}
+\hat\gamma\frac{\partial\phi_1}{\partial\hat\gamma}
+(2n-\nu)\phi_1
+V_{3B}^{(0)}\phi_1\frac{{\cal C}(\hat\gamma,\ln\Lambda/p_*)
}{1-\hat p^2}\right]}\nonumber\\
&&\displaystyle{\qquad\qquad
+\left[-\Lambda\frac{\partial\phi_2}{\partial\Lambda}
+\hat p\frac{\partial\phi_2}{\partial\hat p}
+\hat\gamma\frac{\partial\phi_2}{\partial\hat\gamma}
-\nu\phi_2
+V_{3B}^{(0)}\phi_2\frac{2{\cal C}(\hat\gamma,\ln\Lambda/p_*)}{1-\hat p^2}\right]}
\nonumber\\
&&\displaystyle{\qquad\qquad\qquad\qquad\qquad\qquad\qquad\qquad\qquad
+V_{3B}^{(0)}\phi_1\frac{\tilde{\cal C}(\hat\gamma,\ln\Lambda/p_*)}{1-\hat p^2}=0,}
\end{eqnarray}
where $\tilde C(\hat \gamma,\ln\Lambda/p_*)$ is given by,
\begin{equation}\label{eq:tCintermsofDWs}
\tilde{\cal C}\left(\hat\gamma,\ln\frac{\Lambda}{p_*}\right)=-\hat\gamma^{2n}
{\cal D}_2\left(\hat\gamma,\ln\frac{\Lambda}{p_*}\right)+\int_0^1d\hat k
\,\hat k^{2n}{\cal D}_3\left(\hat k,\hat\gamma,\ln\frac{\Lambda}{p_*}\right).
\end{equation}
From the first line we obtain the RG eigenvalue $\nu=2n$ and the solution
$\phi_1=\hat V_{3B}^{(0)}$. So that a with the further substitution,
\begin{equation}
\phi_2=\left[\phi_3\hat V_{3B}^{(0)}-\hat p^{2n}\right]\hat V_{3B}^{(0)}
\end{equation}
we obtain a PDE for $\phi_3$,
\begin{equation}
-\Lambda\frac{\partial\phi_3}{\partial\Lambda}
+\hat p\frac{\partial\phi_3}{\partial\hat p}
+\hat\gamma\frac{\partial\phi_3}{\partial\hat\gamma}
-2n\phi_3+\frac{\tilde{\cal C}(\hat\gamma,\ln\Lambda/p_*)
-\hat p^{2n}{\cal C}(\hat\gamma,\ln\Lambda/p_*)}{1-\hat p^2}=0.
\end{equation}
This equation should of course have discontinuities on the RHS. The solution
of this PDE, including those discontinuities, is straight forward and can be
reached in much the same way as the limit-cycle solution. The momentum
perturbations occur at orders $2,4,6,\ldots$ at the same orders
of the energy perturbations.

%% file: app/Appfourier.tex
We wish to perform the Fourier transform,
\begin{eqnarray}
&&\displaystyle{
W(\boldrho,\boldrho',\boldr)=
-\frac{4\pi}{M}\int\frac{{\rm d}^3{\bf k'}}{(2\pi)^3}
\int\frac{{\rm d}^3{\bf k'}}{(2\pi)^3}L(p,{\bf k,k',x})
e^{i({\bf k.y-\bf k'.y'})}}\nonumber\\
&&\displaystyle{\qquad\qquad\,\,\,=
\frac{1}{x}\int \frac{d^3{\bf k}}{(2\pi)^3}
e^{i{\bf k}.({\bf y-y'})}e^{ix\sqrt{p^2-\frac34k^2}}}\nonumber\\
&&\displaystyle{\qquad\qquad\qquad
+8\pi\int \frac{d^3{\bf k}}{(2\pi)^3}\int \frac{d^3{\bf k'}}{(2\pi)^3}
e^{i({\bf k.y-k'.y'})}\frac{\cos[{\boldr.({\bf k}'+\frac12{\bf k})}]}{
k^2+k'^2+{\bf k.k'}-p^2-i\epsilon}.}\\
&&\displaystyle{\qquad\qquad\,\,\,=I_1+I_2}
\end{eqnarray}
Doing the integral over all angles in $I_1$ gives,
\begin{equation}
I_1=\frac{1}{2\pi^2 x|{\bf y-y'}|}\int_0^\infty
dk\, k\sin(k|{\bf y-y'}|)e^{-\frac{\sqrt{3}}{2}x\sqrt{\kappa^2+k^2}},
\end{equation}
where we have put $\kappa=-2p/\sqrt{3}$. Using the result
\begin{equation}\label{eq:besselint1}
\int_0^\infty dx\, xe^{-\beta\sqrt{\gamma^2+x^2}}\sin{bx}=
\frac{\beta b\gamma^2}{\beta^2+b^2}K_2(\gamma\sqrt{\beta^2+b^2})
\end{equation}
we get,
\begin{equation}
I_1=\frac{\sqrt{3}\kappa^2}{4\pi^2}\frac{K_2(\kappa{\cal Q}_0)}{{\cal Q}_0^2}
\end{equation}
where ${\cal Q}_0$ is given by eqn.~(\ref{eq:defineQ}). The second integral, $I_2$, is more involved,
making the change of variable ${\bf k''}={\bf k'+k/2}$ we obtain,
\begin{equation}
I_2=4\pi\int \frac{d^3{\bf k}}{(2\pi)^3}\int \frac{d^3{\bf k''}}{(2\pi)^3}
e^{i{\bf k}.(\boldrho+\frac12\boldrho')}
\frac{(e^{-i{\bf k''.(y'+x)}}+e^{-i{\bf k''.(y'-x)}})}
{k''^2+\frac34(k^2+\kappa^2)}.
\end{equation}
Performing the ${\bf k''}$ integral gives$I_2=I_++I_-$ where,
\begin{equation}
I_\pm=
\frac{1}{|{\bf y'\pm x}|}
\int \frac{d^3{\bf k}}{(2\pi)^3}
e^{i{\bf k}.(\boldrho+\frac12\boldrho')}
e^{-\frac{\sqrt{3}}{2}|{\boldrho'\pm\boldr}|\sqrt{k^2+\kappa^2}}.
\end{equation}
Then the angular ${\bf k}$ integral can be done to give,
\begin{eqnarray}
&&\displaystyle{
I_\pm=
\frac{1}{2\pi^2|{\boldrho+\frac12\boldrho'}||{\boldrho'\pm \boldr}|}\int_0^\infty
dk\,k\sin(k|{\boldrho+\frac12\boldrho'}|)
e^{-\frac{\sqrt{3}}{2}|{\boldrho'\pm \boldr}|\sqrt{k^2+\kappa^2}}}\\
&&\displaystyle{\qquad
=\frac{\sqrt{3}\kappa^2}{4\pi^2}\frac{K_2(\kappa{\cal Q}_\pm)}{{\cal Q}_\pm^2},}
\end{eqnarray}
where we have again used result (\ref{eq:besselint1}). The result quoted in the
text follows.

\vspace{2cm}

The second integral to be done is
\begin{eqnarray}
&&\displaystyle{
V_0({\bf y,y'})=-\frac{16\pi}{3M}
\int\frac{{\rm d}^3{\bf k}}{(2\pi)^3}
\int\frac{{\rm d}^3{\bf k'}}{(2\pi)^3}
\left[\gamma+\sqrt{\frac34k^2-p^2}\right]L_2(p,{\bf k,k'},0)
e^{i({\bf k.y-\bf k'.y'})}}\nonumber\\
&&\displaystyle{\qquad\,\,\,\,\,\,\,\,\,
=\frac{32\pi\gamma}{3}\int\frac{{\rm d}^3{\bf k}}{(2\pi)^3}
\int\frac{{\rm d}^3{\bf k'}}{(2\pi)^3}
\frac{e^{i({\bf k.y-\bf k'.y'})}}{k^2+k'^2+{\bf k.k'}+\frac34\kappa^2}}
\nonumber\\
&&\displaystyle{\qquad\qquad\qquad+
\frac{16\pi}{\sqrt{3}}\int\frac{{\rm d}^3{\bf k}}{(2\pi)^3}
\int\frac{{\rm d}^3{\bf k'}}{(2\pi)^3}\sqrt{k^2+\kappa^2}
\frac{e^{i({\bf k.y-\bf k'.y'})}}{k^2+k'^2+{\bf k.k'}+\frac34\kappa^2}}\\
&&\displaystyle{\qquad\,\,\,\,\,\,\,\,\,=I_3+I_4.}
\end{eqnarray}
$I_3$ but for an overall numerical factor
is the same of the $I_2$ with $x=0$. Hence we have,
\begin{equation}
I_3=\frac{2\kappa^2\gamma}{\sqrt{3}\pi^2}\frac{K_2(\kappa{\cal R})}
{{\cal R}^2}.
\end{equation}
$I_4$ can be done in much the same way as the $I_2$ integral
(with $x=0$) all the way up to the final $k$ integral. Hence we get,
\begin{equation}
I_4=\frac{2}{\sqrt{3}\pi^2|{\boldrho+\frac12\boldrho'}|y'}\int_0^\infty
dk\,k\sqrt{k^2+\kappa^2}\sin(k|{\boldrho+\frac12\boldrho'}|)
e^{-\frac{\sqrt{3}}{2}y'\sqrt{k^2+\kappa^2}}.
\end{equation}
Differentiating the result (\ref{eq:besselint1}) on both sides w.r.t.
$\beta$ gives
\begin{equation}\label{eq:besselint2}
\int_0^\infty dx\,x\sqrt{\gamma^2+x^2}e^{-\beta\sqrt{\gamma^2+x^2}}
\sin{bx}=\frac{\beta^2b\gamma^3}{(\beta^2+b^2)^{3/2}}K_3(\gamma\sqrt{\beta^2+b^2})
-\frac{b\gamma^2}{\beta^2+b^2}K_2(\gamma\sqrt{\beta^2+b^2}),
\end{equation}
which can be used to obtain,
\begin{equation}
I_4=\frac{\sqrt{3}\kappa^3y'}{2\pi^2}\frac{K_3(\kappa{\cal R})}{{\cal R}^3}-
\frac{2\kappa^2}{\sqrt{3}y'}\frac{K_2(\kappa{\cal R})}{{\cal R}^2}.
\end{equation}
The quoted result follows.

\vspace{2cm}

The final integral is
\begin{eqnarray}
&&\displaystyle{
V_1({\bf y,y'})=-\frac{8\pi}{3M}
\int\frac{{\rm d}^3{\bf k}}{(2\pi)^3}
\int\frac{{\rm d}^3{\bf k'}}{(2\pi)^3}
\left[\gamma+\sqrt{\frac34k^2-p^2}\right]^2L_2(p,{\bf k,k'},0)
e^{i({\bf k.y-\bf k'.y'})}}\nonumber\\
&&\displaystyle{\qquad\,\,\,\,\,\,\,\,\,
=\frac{16\pi\gamma^2}{3}\int\frac{{\rm d}^3{\bf k}}{(2\pi)^3}
\int\frac{{\rm d}^3{\bf k'}}{(2\pi)^3}
\frac{e^{i({\bf k.y-\bf k'.y'})}}{k^2+k'^2+{\bf k.k'}+\frac34\kappa^2}}
\nonumber\\
&&\displaystyle{\qquad\qquad\qquad+
\frac{16\pi\gamma}{\sqrt{3}}\int\frac{{\rm d}^3{\bf k}}{(2\pi)^3}
\int\frac{{\rm d}^3{\bf k'}}{(2\pi)^3}\sqrt{k^2+\kappa^2}
\frac{e^{i({\bf k.y-\bf k'.y'})}}{k^2+k'^2+{\bf k.k'}+\frac34\kappa^2}}
\nonumber\\
&&\displaystyle{\qquad\qquad\qquad\,\,\,\,\,\,
4\pi\int\frac{{\rm d}^3{\bf k}}{(2\pi)^3}
\int\frac{{\rm d}^3{\bf k'}}{(2\pi)^3}(k^2+\kappa^2)
\frac{e^{i({\bf k.y-\bf k'.y'})}}{k^2+k'^2+{\bf k.k'}+\frac34\kappa^2}}
\nonumber\\
&&\displaystyle{\qquad\,\,\,\,\,\,\,\,\,=\gamma\left(\frac12I_3+I_4\right)+I_5.}
\end{eqnarray}
The results for $I_3$ and $I_4$ have already been established.
$I_5$, like $I_4$, can be done in much the same way as the $I_2$ integral
(with $x=0$) all the way up to the final $k$ integral. Hence we get,
\begin{equation}
I_5=\frac{1}{2\pi^2|{\boldrho+\frac12\boldrho'}|y'}\int_0^\infty
dk\,k(k^2+\kappa^2)\sin(k|{\boldrho+\frac12\boldrho'}|)
e^{-\frac{\sqrt{3}}{2}y'\sqrt{k^2+\kappa^2}}.
\end{equation}
Differentiating the result (\ref{eq:besselint2}) on both sides w.r.t.
$\beta$ gives
\begin{equation}\label{eq:besselint3}
\int_0^\infty dx\,x(\gamma^2+x^2)e^{-\beta\sqrt{\gamma^2+x^2}}
\sin{bx}=\frac{\beta^3b\gamma^4}{(\beta^2+b^2)^{2}}K_4(\gamma\sqrt{\beta^2+b^2})
-\frac{3\beta b\gamma^3}{(\beta^2+b^2)^{3/2}}K_3(\gamma\sqrt{\beta^2+b^2}),
\end{equation}
which can be used to get,
\begin{equation}
I_5=\frac{3\sqrt{3}\kappa^4y'^2}{16\pi^2}\frac{K_4(\kappa{\cal R})}{{\cal R}^4}
-\frac{3\sqrt{3}\kappa^3}{4\pi^2}\frac{K_3(\kappa{\cal R})}{{\cal R}^3}.
\end{equation}
All the results brought together give the result in the text.

%% file: app/Appnicecont.tex
\begin{figure}
\begin{center}
\includegraphics[height=16cm,width=8cm,angle=-90]{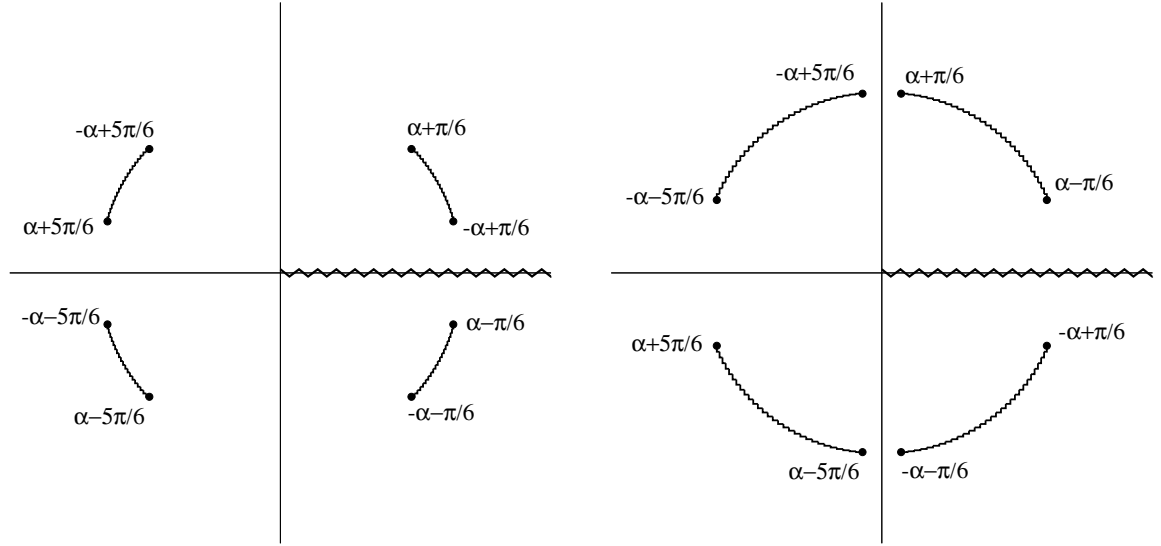}
\caption{The branch structure of the integrand in $I_2$, the labels show the
argument of the branch point.
On the left the case of $0<\alpha<\pi/6$ and on the right $\pi/6<\alpha<\pi/2$.}
\label{fig:nicecont}
\end{center}
\end{figure}

We wish to evaluate the integral
\begin{equation}
I=\frac{\sqrt{3}}{2\pi}\int_0^\infty{\rm d} y' y'^s\left\{
\frac{x({\mathcal R}_+^2-{\mathcal R}_-^2)}{{\mathcal R}_+^2
{\mathcal R}_-^2}-\frac{8}{3 y'}\log\left[\frac{{\mathcal Q}_+^+
{\mathcal Q}_-^-}{{\mathcal Q}_-^+{\mathcal Q}_+^-}\right]\right\},
\end{equation}
where,
\begin{equation}
{\mathcal R}_{\pm}=\sqrt{\frac34x^2+ y^2+ y'^2\pm2 y y'},\qquad
{\mathcal Q}_\pm^\pm=\sqrt{\frac34x^2+ y^2+ y'^2\pm  y y'\pm
\frac32x y'}.
\end{equation}
In order to use residue calculus to evaluate the integral we write
it as
\begin{eqnarray}
&&\displaystyle{\qquad\qquad\qquad
I=I_1+I_2,}\\
&&\displaystyle{\,\,\,\,
I_1=\frac{\sqrt{3}}{2\pi(1-e^{i\pi s})}\int_{-\infty}^\infty{\rm d} y' y'^s
\frac{x({\mathcal R}_+^2-{\mathcal R}_-^2)}{{\mathcal R}_+^2
{\mathcal R}_-^2},}\\
&&\displaystyle{I_2=-
\frac{4}{\sqrt{3}\pi(1-e^{i\pi s})}\int_{-\infty}^\infty{\rm d} y' y'^{s-1}
\log\left[\frac{{\mathcal Q}_+^+
{\mathcal Q}_-^-}{{\mathcal Q}_-^+{\mathcal Q}_+^-}\right].}
\end{eqnarray}
The poles of the integrand of $I_1$ are given by the zeroes of ${\cal R}_\pm$
which may be written in terms of the hyperpolar coordinates as,
\begin{equation}
y'=\pm i\frac{\sqrt{3}}{2}Re^{\pm i\alpha}.
\end{equation}
Evaluating the integral using the two residues
in the upper half-plane (recalling that $0<\alpha<\pi/2$) we get,
\begin{equation}
I_1=\left(\frac{\sqrt{3}R}{2}\right)^s \csc\left(\frac{\pi s}{2}\right)
\sin(\alpha s).
\end{equation}
$I_2$ can be evaluated by closing the contour in the upper half plane
then collapsing it onto the branch cuts there. The branch points are given by
the zeroes of ${\cal Q}_\pm^\pm$ which are simply,
\begin{equation}
 y'=\frac{\sqrt{3}}{2}Re^{\pm i(\alpha\pm\pi/6)}
\qquad\text{and}\qquad y'=\frac{\sqrt{3}}{2}Re^{\pm i(\alpha\pm5\pi/6)}.
\end{equation}
How the branch points up are joined up depends on the value of $\alpha$.
We do not want any of the branch cuts to cross the real axis. The branch
structures are shown in Fig.~\ref{fig:nicecont}.

Simply evaluating the integrals around the branch cuts gives,
\begin{equation}
I_2=
\begin{cases}
\left(\frac{\sqrt{3}R}{2}\right)^s
\frac{16}{s\sqrt{3}}\frac{\cos\left(\frac{\pi s}{6}\right)\sin(\alpha s)}
{1+2\cos\left(\frac{\pi s}{3}\right)}, &\text{when $\alpha<\frac{\pi}{6}$,}\\
\left(\frac{\sqrt{3}R}{2}\right)^s
\frac{8}{s\sqrt{3}}\frac{\sin\left(\frac{\pi s}{2}-\alpha s\right)}
{1+2\cos\left(\frac{\pi s}{3}\right)} &\text{when $\alpha>\frac{\pi}{6}$.}
\end{cases}
\end{equation}
Finally using the equation for $s$ (\ref{eq:danilov2}), and combining the
two results we obtain the quoted result.

%% file: thesis.bbl
\begin{thebibliography}{99}
\bibitem{eck}G. Ecker, Prog. Part. Nucl. Phys. \textbf{35}, 1 (1995) 
[hep-ph/9501357].
\bibitem{bkm95}V. Bernard, N. Kaiser and U.-G. Meissner, 
Int. J. Mod. Phys. \textbf{E4}, 193 (1995) [hep-ph/9501384].
\bibitem{uvkrev}U. van Kolck, Prog. Part. Nucl. Phys. \textbf{43}, 337 (1999) 
[nucl-th/9902015].
\bibitem{eft1}R. Seki, U. van Kolck and M. J. Savage (editors), {\it Nuclear 
Physics with Effective Field Theory} (World Scientific, Singapore, 1998).
\bibitem{eft2}P. F. Bedaque, M. J. Savage, R. Seki and U. van Kolck (editors),
{\it Nuclear Physics with Effective Field Theory II} (World Scientific, 
Singapore, 1999).
\bibitem{border}S. R. Beane, P. F. Bedaque, W. C. Haxton, D. R. Phillips and 
M. J. Savage, nucl-th/0008064.
\bibitem{weinchpt}S. Weinberg, Physica \textbf{A96}, 327 (1979).
\bibitem{wein1}S. Weinberg, Phys. Lett. \textbf{B251}, 288 (1990);
Nucl. Phys. \textbf{B363}, 3 (1991).
\bibitem{ksw} D. B. Kaplan, M. J. Savage, and M. B. Wise, Phys. Lett. \textbf{B424}, 
390 (1998) [nucl-th/9801034]; Nucl. Phys. \textbf{B534}, 329 (1998)
[nucl-th/9802075].
\bibitem{uvk}U. van Kolck, Nucl. Phys. \textbf{A645}, 273 (1999) 
[nucl-th/9808007].
\bibitem{geg}J. Gegelia, nucl-th/9802038; Phys. Lett. \textbf{B429}, 227 (1998);
J. Phys. G: Nucl. Part. Phys. \textbf{25}, 1681 (1999) [nucl-th/9805008].
\bibitem{egm}E. Epelbaum, W. Gl\"ockle and U.-G. Meissner, Nucl. Phys. \textbf{A637},
107 (1998); Nucl. Phys. \textbf{A671}, 295 (2000) [nucl-th/9910064].
\bibitem{bcp}S. R. Beane, T. D. Cohen and D. R. Phillips, Nucl. Phys. \textbf{A632},
445 (1998) [nucl-th/9709062].
\bibitem{bmr}M. C. Birse, J. A. McGovern and K. G. Richardson, 
Phys. Lett. \textbf{B464}, 169 (1999) [hep-ph/9807302].
\bibitem{wilson}K. G. Wilson and J. G. Kogut, Phys. Rep. \textbf{12}, 75 (1974); 
J. Polchinski, Nucl. Phys. \textbf{B231}, 269 (1984); R. D. Ball and R. S. Thorne,
Ann. Phys. \textbf{236}, 117 (1994); T. R. Morris, Prog. Theor. Phys. Suppl.
\textbf{131}, 395 (1998).
\bibitem{ere}J. Schwinger, Phys. Rev. \textbf{72}, 742A (1949).
\bibitem{tbmb} T. Barford and M. C. Birse, Phys. Rev. 
\textbf{C60}:064006 (2003) [hep-ph/0206146].
\bibitem{newton}R. G. Newton, {\it Scattering Theory of Waves and Particles.}
2nd Edition. (Springer, New York, 1982).
\bibitem{bethe}H. A. Bethe, Phys. Rev. \textbf{76}, 38 (1949).
\bibitem{j&b} J. D. Jackson and J. M. Blatt, Rev.  Mod. Phys. \textbf{22}, 77 
(1950); J. M. Blatt and J. D. Jackson, Phys. Rev. \textbf{76}, 18 (1949).
\bibitem{kongr}X. Kong and F. Ravndal, Phys. Lett. \textbf{B450}, 320 (1999)
[nucl-th/9811076]; Nucl. Phys. \textbf{A665}, 137 (2000) [hep-ph/9903523].
\bibitem{bhvk}P. F. Bedaque, H.-W. Hammer and U. van Kolck, 
Phys. Rev. Lett. \textbf{82}, 463 (1999) [nucl-th/9809025];
Nucl. Phys. \textbf{A646}, 444 (1999) [nucl-th/9811046];
Nucl. Phys. \textbf{A676}, 357 (2000) [nucl-th/9906032].
\bibitem{hm3brg} H.-W. Hammer and T. Mehen, Nucl. Phys. \textbf{A690},
535 (2001) [nucl-th/0011024]; Phys. Lett. \textbf{B516},
353 (2001) [nucl-th/0105072].
\bibitem{brgh} P. F. Bedaque, G. Rupak, H. W. Griesshammer and H.-W. Hammer,
Nucl. Phys. \textbf{A714}, 589 (2003) [nucl-th/0207034]
\bibitem{efimov}V. N. Efimov, Sov. J. Nucl. Phys. \textbf{12}, 589 (1971).
\bibitem{thomas} L. H. Thomas, Phys. Rev. \textbf{47}, 903 (1935).
\bibitem{case} K. M. Case, Phys. Rev. \textbf{80}, 797 (1950).
\bibitem{meetz} K. Meetz, Il Nuovo Cimento \textbf{34}, 690 (1964).
\bibitem{perelpop}A. M. Perelomov and V. S. Popov, Thoer. Math. Phys. \textbf{4}, 664 
(1970).
\bibitem{lcbc} M. Bawin and S. A. Coon, [quant-ph/0302199]
\bibitem{camblong} H. E. Camblong and C. R. Ord\'{o}\~{n}ez, [hep-th/0305035].
\bibitem{splc} S. R. Beane, P. F. Bedaque, L. Childress, A. Kryjevski,
J. McGuire and U. van Kolck, Phys. Rev. \textbf{A64}:042103, (2001)
[quant-ph/0010073].
\bibitem{phill3b} I. R. Afnan and D. R. Phillips [nucl-th/0312021]
\bibitem{geg3b}B. Blankleider and J. Gegelia,[nucl-th/0009007].
\bibitem{modeft} D. R. Phillips, G. Rupak and M. J. Savage, Phys. Lett.
\textbf{B473}, 209 (2000) [nucl-th/9908054]
\bibitem{crs}J.-W. Chen, G. Rupak and M. J. Savage, Nucl. Phys. \textbf{A653}, 
386 (1999) [nucl-th/9902056].
\bibitem{krthesis}K. G. Richardson, Ph.~D. thesis, University of Manchester 
(1999) [hep-ph/0008118].
\bibitem{cohan}T. D. Cohen and J. M. Hansen, Phys. Rev. \textbf{C59}, 13 (1999)
[nucl-th/9808038]; Phys. Rev. \textbf{C59}, 3047 (1999) [nucl-th/9901065].
\bibitem{longandshort} D. B. Kaplan and J. V. Steele, Phys. Rev.
\textbf{C60}:064002 (1999) [nucl-th/9905027]
\bibitem{vk1}U. van Kolck, Nucl. Phys. \textbf{A631}, 56c (1998) [hep-ph/9707228].
\bibitem{grad} L. S. Gradshteyn and I. M. Ryzhik, {\it Table of Integrals,
Series, and Products.} 6th Edition. 
(Academic Press, San Diego, San Fransisco, New York, Boston,
London, Sydney and Tokyo, 2000) 
\bibitem{kongr2} X. Kong and F. Ravndal, Nucl. Phys. \textbf{A665}, 137 (2000);
Phys. Rev. \textbf{C64}:044002 (2001).
\bibitem{hkok}H. van Haeringen and L. P. Kok, Phys. Rev. \textbf{A26}, 1218 
(1982).
\bibitem{ksw1}D. B. Kaplan, M. J. Savage and M. B. Wise, 
Nucl. Phys. \textbf{B478}, 629 (1996) [nucl-th/9605002].
\bibitem{furnsteele}J. V. Steele and R. J. Furnstahl, Nucl. Phys. \textbf{A645}, 439
(1999) [nucl-th/9808022].
\bibitem{fms}S. Fleming, T. Mehen and I. W. Stewart, Nucl. Phys. \textbf{A677},
313 (2000) [nucl-th/9911001].
\bibitem{landaus} L. D. Landau and J. Smorodinski,
J. Phys. Acad. Sci. USSR, \textbf{8}, 219 (1944).
\bibitem{tbmb2} T. Barford and M. C. Birse, AIP Conf. Proc. \textbf{603},
229 (2001) [nucl-th/0108024].
\bibitem{orvk}C. Ord\'o\~nez, L. Ray and U. van Kolck, Phys. Rev. \textbf{C53}, 
2086 (1996) [hep-ph/9511380].
\bibitem{kswapp1}D. B. Kaplan, M. J. Savage, and M. B. Wise, Phys. Rev. \textbf{C59}, 
617 (1999) [nucl-th/9804032].
\bibitem{bbsvk}S. R. Beane, P. F. Bedaque, M. J. Savage and U. van Kolck,
Nucl. Phys. \textbf{A700}, 377 (2002) [nucl-th/0104030].
\bibitem{efimovphil} V. N. Efimov and E. G. Tkachenko, Few-Body Systm. \textbf{4},
71 (1988).
\bibitem{wilsonlc} St. D. Glazek and K. G. Wilson, Phys. Rev. \textbf{D47},
4657 (1993)
\bibitem{wilsonlc2} St. D. Glazek and K. G. Wilson, Phys. Rev. Lett.
\textbf{89}:230401 (2002) [hep-ph/0203088].
\bibitem{leclairlc} A. LeClair, J. M. Roman and G. Sierra, hep-ph/0312141
\bibitem{stm} G. V. Skorniakov and K. A. Ter-Martirosian,
Sov. Phys. JETP \textbf{4}, 648 (1957).
\bibitem{danilov} G. S. Danilov and V. I. Lebedev, Sov. Phys. JETP \textbf{44},
1509 (1963).
\bibitem{faddeev}L. D. Faddeev, {\it Mathematical Aspects of the Three-body
Problem in Quantum Scattering Theory.} (Steklov Math. Inst., Leningrad Publ. 69,
transl. D. Davey and Co., New York, 1965)
\bibitem{highpwnd} F. Gabbiani, P. F. Bedaque and H. W. Griesshammer,
Nucl. Phys. \textbf{A675}, 601 (2000) [nucl-th/9911034]
\bibitem{fedorov} D. V. Fedorov and A. S. Jensen Phys. Rev. Lett.
\textbf{71}, 4103 (1993); Phys. Rev. \textbf{A63}:063608 (2001);
Nucl. Phys. \textbf{A697}, 783 (2002).
\bibitem{phillips} A. C. Phillips, Nucl. Phys. \textbf{A107}, 209 (1968).
\bibitem{kievsky} A. Kievsky, S. Rosati, W. Tornow and M. Viviani,
Nuc. Phys. \textbf{A607}, 402 (1996).
\bibitem{ndscat} W. Dilg, L. Koester and W. Nistler, Phys. Lett.
\textbf{B36}, 208 (1971).
\bibitem{oerssea} W. T. H. van Oers and J. D. Seagrave,
Phys. Lett. \textbf{B24}, 562 (1967).
\bibitem{irrglc} E. Braaten and H.-W. Hammer, Phys. Rev. Lett. \textbf{91}:102002
(2003) [nucl-th/0303038].
\bibitem{quartetbvk}P. F. Bedaque and U. van Kolck, Phys. Lett. \textbf{B428}, 221
(1998) [nucl-th/9710073]; Phys. Rev. \textbf{C58}, 641 (1998) [nucl-th/9802057]. 
\bibitem{gabbiani} F. Gabbiani, [nucl-th/0104088].
\bibitem{bose} E. Braaten, H.-W. Hammer and M. Kusunoki, Phys. Rev. Lett \textbf{90}:
120402 (2003) [cond-mat/0206232]; E. Braaten, H.-W. Hammer and T. Mehen,
Phys. Rev. Lett. \textbf{88}:040401 (2002) [cond-mat/0108380];
P. F. Bedaque, E. Braaten and H.-W. Hammer,
Phys. Rev. Lett. \textbf{85}, 908 (2000), [cond-mat/0002365]







\end{thebibliography}
